\documentclass[12pt,fleqn,a4paper,groupedaddress,nofootinbib]{thesis}

\usepackage[english]{babel}
\usepackage{graphicx}

\usepackage{here,epsf,titlepag}
\usepackage{latexsym}		
\usepackage{amsfonts}
\usepackage{amssymb}
\usepackage{epsfig}
\usepackage{amsmath}
\usepackage{fancyhdr}
\usepackage{textcomp}
\usepackage{setspace}
\usepackage[numbers]{natbib}

\usepackage{slashbox}
\usepackage{multirow}

\def\a{\alpha}
\def\b{\beta}

\def\d{\delta}

\def\i{\iota}

\def\O{\Phi}

\def\p{\partial}

\def\t{\tau}

\def\w{\omega}

\def\={\nonumber &=}

\def\&{{}&}

\def\({\left(}
\def\){\right)}
\def\[{\left[}
\def\]{\right]}
\def\<{\left\langle}
\def\>{\right\rangle}

\def\uk{{\bf \hat{k}}}

\def\ux{{\bf \hat{x}}}
\def\bk{{\bf k}}
\def\bn{{\bf n}}

\def\bx{{\bf x}}
\def\bK{{\bf K}}
\def\by{{\bf y}}

\def\curl{\mathcal}

\def\eq{\begin{align}}
\def\qe{\end{align}}
\def\eqa{\begin{eqnarray}}
\def\qea{\end{eqnarray}}

\def\and{\quad \mbox{and} \quad}

\def\fnl{f_\textrm{NL}}

\def\Fnl{ F_\textrm {NL}}

\def\bfnl{\kern2pt\overline{\kern-2ptf}_\textrm{NL}}

\def\lmax{l_\textrm{max}}

\def\Blll{B_{l_1l_2l_3}}

\def\lb{\mathbf{l}}

\def\kall{k_1,k_2,k_3}

\def\aone{a_{l_1 m_1}}
\def\atwo{a_{l_2 m_2}}
\def\athree{a_{l_3 m_3}}
\def\afour{a_{l_4 m_4}}

\def\Qn{\curl{Q}_n}
\def\Qm{\curl{Q}_m}

\def\Rn{\curl{R}_n}
\def\Rm{\curl{R}_m}

\def\barQ{\kern2pt\overline{\kern-2pt\curl{Q}}}

\def\barR{\kern2pt\overline{\kern-2pt\curl{R}}}

\def\nmax{n_\textrm{max}}
\def\konkmax{\frac{k}{k_\textrm{max}}}
\def\konkmaxb{\left(\frac{k}{k_\textrm{max}}\right)}

\def\aR{\alpha^{\scriptscriptstyle{\cal R}}}
\def\aRn{\aR_n}

\def\aQ{\alpha^{\scriptscriptstyle{\cal Q}}}
\def\aQn{\aQ_n}

\def\baQ{\bar{\alpha}^{\scriptscriptstyle{\cal Q}}}

\def\bR{\beta^{\scriptscriptstyle{\cal R}}}
\def\bRn{\bR_n}

\def\bQ{\beta^{\scriptscriptstyle{\cal Q}}}
\def\bQn{\bQ_n}

\def\bbQ{\bar{\beta}^{\scriptscriptstyle{\cal Q}}}

\def\setsize{\csname @setfontsize\endcsname \setsize}

\def\setsize{\csname @setfontsize\endcsname \setsize}

\input epsf
\begin{document}
\setcounter{secnumdepth}{4}
\pagenumbering{roman}
\pagestyle{headings}

\begin{titlepage}
\setcounter{page}{1}
\begin{center}
\vspace*{1.2cm}
{\Huge \bf Measuring CMB non-Gaussianity }\\
\vspace{0.3cm} {\Huge \bf as a probe of}\\
\vspace{0.3cm}
{\Huge \bf Inflation and Cosmic Strings}\\
\vspace{2cm}
{\Large by}\\
\vspace{0.5cm}
{\Large\bf Donough Regan}\\
\vspace{0.5cm}
{\Large Selwyn College}\\
\vspace{2cm}
\begin{figure}[H]
\centering
\includegraphics[width=50mm]{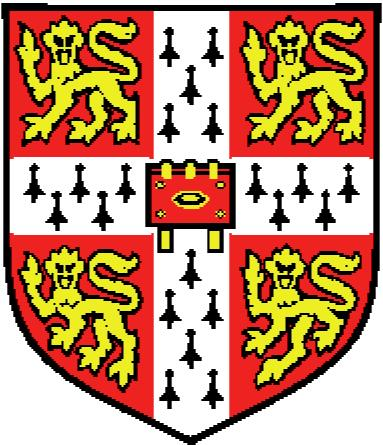}
\end{figure}
\vspace{1.8cm}
{\Large Submitted in partial fulfilment of the requirements}\\
\vspace{0.2cm}
{\Large for the degree of Doctor of Philosophy in the }\\
\vspace{0.2cm}
{\Large\bf University of Cambridge}\\
\vspace{0.2cm}
{\Large\bf 2011}\\
\end{center}
\end{titlepage}

\newpage
\addcontentsline{toc}{chapter}{Declaration}
\chapter*{Declaration}

\vspace*{3cm}
I declare that this thesis is entirely my own work, carried out
at the University of Cambridge, and has not been submitted for a degree
to this or any other University and that the contents are
original unless otherwise stated.
\par
Chapter $4$ contains work done with Paul Shellard and published in Physical Review D (PRD) \cite{Regan:2009hv}. Chapters  $5$ was completed in collaboration with Paul Shellard and James Fergusson and is published in PRD \cite{RSF10}. Chapters $6$ and $7$ are based primarily on work  also done with Paul Shellard and James Fergusson and large parts of these chapters are available as arXiv e-prints \cite{FRS10,FRS10Trisp}.

\vspace*{4cm}
\hspace*{9cm} Signed,

\vspace*{3.5cm}\hspace*{9cm}\rule[0pt]{6cm}{0.01cm}\\
\hspace*{9cm}Firstname Lastname

\vspace*{2cm}
\hspace{9cm}4$^{\rm th}$ of January 2011

\newpage
\addcontentsline{toc}{chapter}{Abstract}
\chapter*{Abstract}
The leading candidate for the very early universe is described by a period of rapid expansion known as inflation. While the standard paradigm invokes a single slow-rolling field, many different models may be constructed which fit the current observational evidence. In this work we outline theoretical and observational studies of non-Gaussian fluctuations produced by models of inflation and by cosmic strings - topological defects that may be generated in the very early universe during a phase transition. In particular, we consider the imprint of cosmic strings on the cosmic microwave background (CMB) and describe a formalism for the measurement of general four-point correlation functions, or trispectra, using the CMB. In addition we describe the application of our methodology to non-Gaussian signals imprinted in the large scale structure of the universe. Such deviations from Gaussianity are generally expressed in terms of the so-called bispectrum and trispectrum.
\newline
\newline
In our study of cosmic strings we have extended and developed a formalism by Hindmarsh to investigate higher order correlators of the density perturbations due to cosmic strings as functions of two-point functions of the string
network. Previously such a formalism had only been applied to small angular scales. Our extension allows the investigation of large angular scales. Despite uncertainties due to recombination this
approach appears quite accurate. We've considered the power spectrum, bispectrum and trispectrum here. The shape and normalisation of the power spectrum agrees well with simulations, while measurements of the skewness and kurtosis agree well with our predictions for the bispectrum and trispectrum respectively. The predicted CMB bispectrum and trispectrum due to cosmic strings are compared to data from the WMAP satellite giving bounds on the tension of cosmic strings.
\newline
\newline
We have extended methods to efficiently measure the bispectrum implied by these alternative models to the measurement of the trispectrum. The methods we have developed allow for vast improvements in computational time and, therefore, allow the potential to constrain many more models of the early universe. In particular we've shown how a separable approximation to the trispectrum can
be used to greatly improve computational speed. We relate the CMB trispectrum to the primordial version
but the techniques developed here are also applicable to the CMB trispectrum alone (and so are applicable to cosmic
strings for instance). We've shown how to calculate an efficient estimator, how to measure the correlation between different trispectra, and how to construct maps for general
trispectra (and bispectra). 
\newline
\newline
The next part of the thesis describes the numerical implementation of the methodology for analysing the trispectrum to a special class of models for which the computational speed is further improved, beyond the general prescription. Equilateral models and the cubic term for the local model fall into this class. Constraints are derived by comparison to data from the WMAP satellite. While trispectrum constraints for the cubic term of the local model exist in the literature, the methodology employed here involves no simplifying assumptions, while the constraints derived for the equilateral model are entirely novel.
\newline
\newline
Finally we investigate the application of methods similar to those used in our trispectrum analysis to probes of the large scale structure of the universe. The separable approach has been adapted to describe an efficient algorithm for the implementation of
initial conditions for N-body codes and the reconstruction of the bispectrum and trispectrum. We have also
shown that this approach allows for quick calculation of, for example, the bispectrum contribution to the galaxy power
spectrum. In addition we have investigated the calculation of general estimators for the bispectrum and trispectrum, the correlation between two spectra, as well as an approach for non-Gaussian parameter estimation which remains valid in the non-linear regime. This methodology is expected to allow for a vast improvement in the range of models investigated using such datasets.

\newpage
\addcontentsline{toc}{chapter}{Acknowledgements}
\chapter*{Acknowledgements}

\vspace*{4.5em}

Firstly, I would like to thank my supervisor, Paul Shellard, for his endless support, guidance and encouragement. I could not have wished for a better mentor. I would also like to offer my gratitude to James Fergusson for all the help he has given me with my research, and who has made our collaboration such an enjoyable experience.
\par
 For their advice on my research and for the numerous discussions I have greatly appreciated, I am indebted to Michele Liguori, Xingang Chen and Anastasios Avgioustidis. I am also grateful for expert computational help from Andrey Kaliazin.
\par
My office mates past and present - Mark, Jacques, Hiro and Andrew - deserve gracious acknowledgement for tolerating the lack of tidiness I have brought to our office, and more especially for their friendship. I am very grateful to the other friends I've been fortunate to meet in the last few years - while I would like to name each of them here such a list would only be marked by an unintentional omission! Thanks also to Amanda and Julie for all their administrative help.
\par
I am indebted to my parents John and Claire, my brother and sisters for their backing over the years.
\par
Finally I would like to thank Mair\'ead whose patience and support seemingly knows no bounds. \it{M}$\rm{\acute{\i}}$\it{le bu}$\rm{\acute{\i}}$\it{ochas duit as do ghr\'a is as do thaca}$\rm{\acute{\i}}$\it{ocht}. \it{Is tusa mo chro}$\rm{\acute{\i}}$.\rm
\newpage

\addcontentsline{toc}{chapter}{Contents}
\tableofcontents
\newpage

\addcontentsline{toc}{chapter}{List of Figures}
\listoffigures
\newpage

\addcontentsline{toc}{chapter}{List of Tables}
\listoftables


\addcontentsline{toc}{chapter}{Acronyms and Abbreviations}
\chapter*{Acronyms and Abbreviations}
\markright{Acronyms and Abbreviations}

\begin{tabbing}			
CMB	\hspace*{2cm}\=	Cosmic Microwave Background	\\
WMAP	\hspace*{1.5cm}\=Wilkinson Microwave Anisotropy Probe\\
BAO\hspace*{2.2cm}\= Baryon Acoustic Oscillations\\
SN\hspace*{2.6cm}\=Supernova\\
SDSS\hspace*{2.1cm}\=Sloan Digital Sky Survey\\
COBE\hspace*{1.9cm}\=Cosmic Background Explorer\\
GUT\hspace*{2.2cm}\=Grand Unified Theory\\
UETC\hspace*{1.9cm}\=Unequal Time Correlator\\
VOS\hspace*{2.2cm}\=Velocity-dependent One Scale model\\
\end{tabbing}

\newpage


\pagenumbering{arabic}
\setcounter{page}{1}
\chapter{Introduction}
\markright{Introduction}
\label{chapter:intro}
In recent years the study of the very early universe has entered an era of precision cosmology. The cosmic microwave background (CMB) radiation has been used to differentiate the many conflicting theories for the evolution of the universe. In particular, the CMB power spectrum has provided substantial evidence supporting the standard paradigm of a hot big bang followed by a period of inflation. However, there remains a plethora of inflationary theories compatible with many of the observations considered  to date. Measurements of the CMB power spectrum are not sensitive to non-Gaussianities. Therefore, a study of non-Gaussianity offers the possibility of distinguishing such theories.
A comprehensive study of non-Gaussianity involves measurements of the three-point correlator of density perturbations, known as the bispectrum, and measurements of the four-point correlator of density perturbations, known as the trispectrum. A comprehensive study of the bispectrum was carried out in \cite{FLS09,FLS10} with considerable success. The study of the trispectrum represents a large part of this thesis. 
\par
Cosmic strings are line-like discontinuities which may be formed during a phase transition in the very early universe \cite{KibbleMech}. They were once regarded as a viable model for seeding large scale structure. However, observations of the CMB power spectrum have ruled out this possibility. Nevertheless, it remains possible that cosmic strings may seed up to $10\%$ of the density perturbations \cite{Bevis}. Cosmic strings are also of interest since they may be formed at the end of brane inflation \cite{TyeSarangi}. However, such strings have somewhat different network properties and are generally referred to as cosmic superstrings. The possibility of using non-Gaussian CMB signatures to investigate cosmic strings is particularly interesting since such a study may offer a testable prediction of string theory. In this thesis the imprint of cosmic strings on the CMB is studied with particular focus on their non-Gaussian signal.
\par
While the CMB offers the best current hope for testing different inflationary models, the possibility of using large-scale structure to provide complementary constraints has received much attention in recent times. The study of large-scale structure is somewhat more complicated due to issues in dealing with non-linear evolution. However, with improving galaxy surveys over a growing fraction of the sky, it is expected that measurements from such three-dimensional data sets will provide the best and most comprehensive information about non-Gaussianity in the near future. The final part of this thesis focusses on the extension of methods developed to investigate the CMB to the study of large-scale structure.
\par
In Chapter $2$ we review the standard model of cosmology and discuss the motivation for an inflationary era in the very early universe. We discuss the canonical inflation model and describe the resultant spectrum of density perturbations. We summarise briefly the concordance model of inflation and its observational evidence. Next, we define the non-Gaussian primordial measures known as the bispectrum and trispectrum. Having described the primordial inflationary perturbations, we describe how inhomogeneities in the CMB may be generated from primordial perturbations. This process is described by the Boltzmann equation. Next, we define the CMB power spectrum, bispectrum and trispectrum and establish the relation to their respective primordial counterparts. We also outline briefly how large-scale structure may be used to investigate the signatures of inflation.
\par
Chapter $3$ details briefly the topic of topological defects, with particular focus on cosmic strings. The process of cosmic string formation is elucidated along with an effective action description of the dominant string dynamics. Cosmic strings are objects of extremely high density and, as such, are most likely to be discovered via their gravitational interaction. The gravitational field near a long straight string is shown to be described by a conical metric. Next, we describe the dominant properties of a network of cosmic strings. A simple, but effective, description of such a network is described well by the so-called `one-scale model' which is also summarised. Some of the different observational consequences are described, along with a depiction of the temperature discontinuity induced by cosmic strings known as the Gott-Kaiser-Stebbins effect. A short review of the imprint cosmic strings would leave on the CMB is then sketched. We finish the chapter with a discussion about cosmic superstrings and the possibility of distinguishing cosmic strings from superstrings is elucidated.
\par
Our investigation of the CMB (poly)spectra induced by cosmic strings is presented in Chapter $4$. This study makes use of a flat sky approximation and an assumption that the temperature discontinuity induced by cosmic strings is described entirely by the Gott-Kaiser-Stebbins effect. Specifically, the contribution due to decoupling is neglected. The temperature (poly)spectra are then derived as functions of two-point correlators of the string network parameters. Summing up contributions from strings between last scattering and today allows an extension of the flat-sky approximation to an almost full-sky description. It is verified that, in spite of the approximations made, the power spectrum correlates well with full numerical analyses provided in the literature. Analytic formulae for the bispectrum and trispectrum are also provided along with estimates of the level of non-Gaussianity which may be expected from cosmic strings. It appears from this study that an investigation of the CMB trispectrum could provide competitive constraints on cosmic strings to the power spectrum.
\par
The next part of the thesis, given in Chapter $5$, details a formalism for the analysis of a general CMB trispectrum. This work extends a methodology for the study of the CMB bispectrum \cite{FLS09}. A general estimator which accounts for the effects of inhomogeneous noise and masking is derived. This estimator provides a general measure of the CMB trispectrum and is used to define an integrated measure of non-Gaussianity. The study exploits the use of an eigenmode decomposition, either at primordial times or today. Such a decomposition allows us to write the primordial or CMB trispectrum in a separable form. Separability is vital for an efficient analysis of the trispectrum and the use of a separable expansion allows models, previously thought to be intractable, to be numerically analysed. The creation of CMB maps which include an arbitrary bispectrum and trispectrum is also discussed in this chapter, as is the possibility of recovering the primordial or CMB trispectrum from observational data.
\par
In Chapter $6$ we outline the application of the formalism, described in Chapter $5$, to the class of trispectra which are independent of their diagonal term. Use of the separable approach ensures that analysis of such trispectra may be achieved in $\mathcal{O}(l_{\mathrm{max}}^4)$ operations as opposed to $\mathcal{O}(l_{\mathrm{max}}^7)$ operations for general trispectra without use of such an expansion\footnote{$l_{\mathrm{max}}$ denotes the maximum angular multipole attainable by the experiment in question. In the case of WMAP data $l_{\mathrm{max}}=\mathcal{O}(700)$, while for Planck data a maximum multipole $l_{\mathrm{max}}=\mathcal{O}(2000)$ is expected to be achieved.}. We summarise the methodology introduced in the previous chapter for the particular case of a diagonal-free trispectrum. We proceed to investigate the cubic term of the local model, the equilateral model and the constant model, deriving constraints by comparison to WMAP $5$ year data. We also derive a constraint on the cosmic string tension and establish earlier indications that the trispectrum may be expected to provide, perhaps, the best probe for cosmic strings in the near future.
\par
As already alluded to, large-scale structure is expected to provide competitive and ultimately superior constraints on non-Gaussianity to the CMB in the near future. In Chapter $7$ we describe the application of the eigenmode expansion formalism to studies of large scale structure. The approach allows for an efficient estimation and reconstruction of the bispectrum and trispectrum from observational (or simulated) large-scale structure data. Of particular interest in studies of non-Gaussianity using large scale structure is the generation of arbitrary non-Gaussian initial conditions for use in N-body codes. It is shown that the separable method allows for an efficient production of initial conditions for arbitrary bispectra and trispectra. The key issues involved in parameter estimation are also described.
\par
In Chapter $8$ we summarise the results of the thesis and present our concluding remarks.

\newpage

\newpage
\thispagestyle{empty}
\mbox{}
\newpage
\chapter{Background I : Primordial Cosmology and the CMB}
\markright{Background}
\label{chapter:bgd}
\textbf{Summary}\\
\textit{In this section we present a review of the basics of cosmology. We begin with a description of the spacetime used to describe the expanding universe. Motivated by conflicts between observations and this picture of the very early universe, we introduce the notion of inflation. The standard paradigm for inflation, known as slow-roll inflation, is described. The perturbations produced during this inflationary era are then analysed and the resulting power spectrum evaluated. Next, we discuss current cosmological observations and the concordance model of cosmology, $\Lambda$CDM. In order to distinguish alternative models for inflation we introduce higher order correlators such as the primordial bispectrum and trispectrum. Of course, we do not observe these primordial quantities directly. One source of observation is provided by the cosmic microwave background. We describe the evolution of primordial perturbations forward to the CMB temperature fluctuations. The relationship between the primordial spectra and the CMB versions is then elucidated. Finally, we describe how the distribution of large scale structure may also be used to investigate primordial cosmological models. For a broader review of the concepts discussed in this section the interested reader is referred to refs.~\cite{Mukhanov,liddle,Challinor,dodelson,peebles,PDG}.}
\vspace{15pt}

\section{Hot Big Bang and FLRW Cosmology}
Modern cosmology has developed from the basic tenet known as the `Copernican principle', which states that the Earth is not the centre of the universe and that we do not live in a special location within the universe. The generalisation of this - the cosmological principle - posits that there is no preferred direction and no preferred location in the universe. This may be expressed mathematically by imposing that the universe is homogeneous and isotropic at every point. While it is clear from observations of the night-sky that the universe is neither homogeneous nor isotropic on small scales due to the non-linear evolution which resulted in structures such as galaxies and planets, there is however strong evidence that the principle is obeyed on large scales. One source of such evidence is provided by the cosmic microwave background, smooth to at least one part in $10^5$.
\par
By imposing homogeneity and isotropy on a general $4$-dimensional metric, we obtain the line element of the FLRW metric with coordinates $(t,r,\theta,\phi)$:
\begin{align}\label{FLRW}
ds^2=-dt^2+a^2(t)\left(\frac{dr^2}{1-k r^2} +r^2(d\theta^2+\sin^2\theta d\phi)	\right),
\end{align}
where $a(t)$ is known as the scale factor, $t$ the proper time and $k= -1,0$ or $1$ depending on whether the universe is open, flat or closed. The rate of change of the logarithm of the scale factor is known as the Hubble parameter, i.e. 
\begin{align}
H=\dot{a}/a\,,
\end{align}
where overdot is used to denote differentiation with respect to time, i.e. $\dot{a}\equiv da/dt$.
Using the Einstein equations 
\begin{align}
R_{\mu \nu}-(R/2+\Lambda)g_{\mu \nu}=8\pi GT_{\mu \nu},
\end{align}
(where $R_{\mu \nu}$ denotes the Ricci tensor, $R$ the Ricci scalar, $\Lambda$ the cosmological constant, $g_{\mu \nu}$ the metric, $G$ Newton's constant, and where we set the speed of light $c$ to unity) two evolution equations for the universe may be obtained. In particular, assuming the universe is filled with a perfect fluid with four-velocity $u^{\mu}$ satisfying $u^{\mu}u_{\mu}=-1$ and the stress-energy tensor is given by
\begin{align}\label{stress}
T_{\mu \nu}=(\rho+P)u_{\mu}u_{\nu}+P g_{\mu \nu},
\end{align}
 where the total energy density of the constituents of the universe is $\rho=\sum_i \rho_i$ and $P=\sum_i P_i$ the total pressure, we find
\begin{align}\label{hubblesq}
H^2&=\frac{8\pi G}{3}\rho - \frac{k}{a^2}+\frac{\Lambda}{3},\\\label{fried2}
\dot{H}+H^2&=\frac{\ddot{a}}{a}=-\frac{4\pi G}{3}(\rho+3 P)+\frac{\Lambda}{3}.
\end{align}
From \eqref{hubblesq} it is clear that, for an expanding universe with positive curvature $k$ and $\Lambda=0$, the scale factor has a turning point at $\frac{8\pi G}{3}\rho = \frac{k}{a^2}$. There also exists a critical density at which the curvature of the universe is zero, i.e. $\rho_{c}=3H^2/(8\pi G)$. In particular, if $\rho>\rho_c$ then $k>0$ and the universe has a closed geometry, i.e. the expansion of the universe will halt at some point and turn to re-collapse. If $\rho<\rho_c$ the universe will expand forever and in the far future matter and radiation will be diluted away. Writing $\Omega_m=\rho/\rho_c$ we may rewrite the Friedmann equation, \eqref{hubblesq}, as
\begin{align}\label{FriedOmega}
1=\Omega_m+\Omega_{\Lambda}+\Omega_k
\end{align}
where $\Omega_{\Lambda}=\dfrac{\Lambda}{3 H^2}$ and $\Omega_k=-\dfrac{k}{a^2 H^2}$. 
\par
The discovery by Hubble in 1929 that galaxies are moving away from us at a speed proportional to their distance implies, via the relation $\mathbf{v}=\dot{a}\mathbf{x}=H\mathbf{r}$, that the universe is expanding\footnote{We note that the physical separation between stationary objects is given by $\mathbf{r}=a(t)\mathbf{x}$, with $\mathbf{x}$ a constant comoving Lagrangian coordinate.}. The consequence of this observation is that there must be a beginning where all the constituents of the universe were compressed into small volume and that the universe began with a `hot big bang'. An observational relic of this hot big bang is the nucleosynthesis of light elements predicted by Gamow, Alpher and Bethe which has been verified to high precision (see review \cite{PDG}). In addition a relic cosmic radiation remaining from the period when the universe cooled to a degree such that radiation and matter decoupled was predicted by Alpher and Herman to pervade the universe. This was later observed by Penzias and Wilson in their observation of the cosmic microwave background, \cite{Penzias}, offering further credence to the theory.

\section{Problems with the hot big bang model}
Despite the success of the hot big bang model some open questions remain. Amongst these include the following:

\subsubsection*{Flatness Problem}
In conformal time $d\tau=dt/a$ we define $\mathcal{H}=a'/a=a H$ we rewrite equations \eqref{hubblesq}, \eqref{fried2} in the form
\begin{align}
\mathcal{H}^2(\Omega-1)=&k,\\
\mathcal{H}'=&-\frac{1}{2}\mathcal{H}^2 \Omega,
\end{align}
where we assume a universe with pressureless matter (i.e. dust) and no cosmological constant. The equations may be combined to give
\begin{align}
\Omega'=\mathcal{H}\Omega (\Omega-1).
\end{align}
From observations of WMAP data we have $|\Omega_0-1|\sim 0.02$. Since we also observe $\mathcal{H}>0$, any deviations from $\Omega =1$ will grow in time. Thus the universe is required to be flat ($k=0$) to extremely high precision in the very early universe. In particular, we have that during the radiation dominated era in the very early universe that $\Omega -1 \propto a^2 \propto t$, and during the dust-dominated era $\Omega -1 \propto a \propto t^{2/3}$, implying that at nucleosynthesis we require $|\Omega -1|\sim 10^{-16}$ and at the electroweak scale $|\Omega -1|\sim 10^{-27}$. This fine tuning is known as the $flatness$ problem.
 
\subsubsection*{Horizon Problem}
While the observation of the CMB was seen as a success of the hot big bang model, the isotropy of the signal cannot be explained within the standard FLRW picture. The particle or comoving horizon defines the maximum separation between two points that could have been in causal contact in the past. This comoving horizon is given by
\begin{align}
\eta=\int_{a_i}^{a} \frac{da}{a}\frac{1}{a H(a)} 
\end{align}
where $a_i=a(t_i)$ corresponds to the scale factor at the initial time $t_i$. Particles separated by more than the comoving horizon cannot have been in causal contact. Measurements of the CMB show that the temperature is uniform to one part in $10^{5}$, while calculation of the comoving horizon at the time of decoupling implies that it extends an angle on the sky of only about $2^o$. Hence the CMB sky is made up of many causally disconnected regions, each with the same temperature to a high degree of accuracy. This is known as the $horizon$ problem.

\subsubsection*{Magnetic Monopole Problem}
Spontaneous symmetry breaking predicts an abundance of magnetic monopoles and other exotic objects, such as cosmic strings, domain walls and textures, would be created from topological defects. In addition supergravity and superstring theories predict the existence of moduli fields.  Monopoles, should they exist, would dominate the energy density of the universe. The hot big bang picture is unable to explain why such relics are not observed.

\subsubsection*{Inflation as a solution}
Consider a universe consisting of a purely positive cosmological constant. From equation \eqref{fried2}
we have
\begin{align}
\ddot{a}=\frac{\Lambda}{3}a \implies a=e^{t\sqrt{\Lambda/3}},
\end{align} 
implying the universe will expand exponentially. From equation \eqref{hubblesq} this implies 
\begin{align}
\Omega -1 \propto e^{-2t\sqrt{\Lambda/3}}.
\end{align}
Hence, any deviation from flatness is suppressed by this period of exponential expansion. In fact the value $\Omega=1$ becomes an attractor solution during such an inflationary phase. In order to solve the horizon and flatness problems the inflationary era must be sufficiently long. In particular, an inflationary period of $50-70$ efolds is required. The cosmological constant cannot, of course, be the mechanism for inflation since such a period of expansion would never end. Instead we require a field, known as the {\textit{inflaton}}, that may play the role of the cosmological constant in the very early universe, which decays once the universe has undergone sufficient expansion. The decay of the inflaton reheats the universe, which at the end of inflation is extremely dilute of particles due to the exponential expansion.
\par
As we shall establish, a consequence of this mechanism for inflation is the generation of small perturbations. During inflation the inflaton undergoes small fluctuations as its position moves slightly along the potential. This slight movement along the potential results in regions with slightly more or less efolds of inflation, thus becoming less or more dense respectively. Hence, inflation converts quantum fluctuations in the inflaton into classical density perturbations. The amplitude of these perturbations is given by $\delta_H=H^2/(2\pi \dot{\phi})$.

\section{Canonical Inflation Model}
\begin{figure}[htp]
\centering 
\includegraphics[width=102mm]{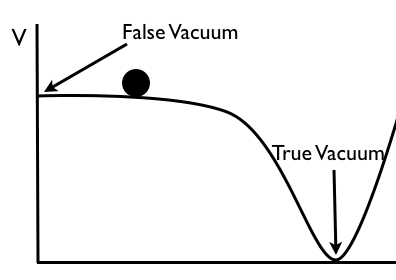}
\caption{Slow-roll potential. The inflaton rolls from a false vacuum state to the true vacuum. The end of inflation is marked by a period of reheating during which the potential energy is converted into standard model particles.}
\label{fig:slowroll}
\end{figure}

We consider a flat universe that is isotropic and homogeneous filled with a perfect fluid with four-velocity $u^\mu$, so that the metric is flat FLRW, i.e. $ds^2$ is given by equation \eqref{FLRW} with $k=0$ and the stress energy tensor is given by equation \eqref{stress}.
In order to seed an inflationary period we require $\ddot{a}>0$. From equation~\eqref{fried2} we observe that (considering $\Lambda=0$) this condition is satisfied only if $\rho+3P<0$. The simplest model of inflation is that of a single scalar field, $\phi$. The action for this scalar field is given by
\begin{align}
S=\int d^4 x \sqrt{-g}\mathcal{L},
\end{align}
where $g=\rm{det}(g_{\mu\nu})$ and 
\begin{align}
\mathcal{L}=\frac{1}{2}\partial_{\mu}\phi\partial^{\mu}\phi -V(\phi).
\end{align}
By varying the action with respect to $\phi$ we obtain the equation of motion
\begin{align}\label{eom1}
\Box \phi +V_{,\phi}=0,
\end{align}
where $\Box \phi=\partial_{\mu}(\sqrt{-g}g^{\mu\nu}\partial_{\nu}\phi)$ and $V_{,\phi}\equiv dV/d\phi$. In the FLRW metric, where we may assume that $\phi=\phi(t)$, i.e. the scalar field does not vary spatially, this equation reads
\begin{align}\label{eom}
\ddot{\phi}+3 H\dot{\phi}+V_{,\phi}=0.
\end{align}
Using Noether's principle we may also obtain the energy momentum tensor,
\begin{align}
T_{\mu \nu}=\frac{2}{\sqrt{-g}}\frac{\delta S}{\delta g^{\mu \nu}}=\partial_{\mu}\phi\partial_{\nu}\phi-g_{\mu \nu}\mathcal{L}.
\end{align}
By comparison with \eqref{stress} we find that
\begin{align}
\rho=&\frac{\dot{\phi}^2}{2}+V\nonumber\\
P=&\frac{\dot{\phi}^2}{2}-V.
\end{align}
Therefore the Friedmann equations \eqref{hubblesq},\eqref{fried2} read
\begin{align}
H^2=&\frac{8\pi G}{3}\left( \frac{\dot{\phi}^2}{2}+V \right),\nonumber\\  \label{adotdot}
\frac{\ddot{a}}{a}=&-\frac{4\pi G}{3}(\dot{\phi}^2-V).
\end{align}
It is clear from equation \eqref{adotdot} that inflation may be thought of as a period when the
potential energy dominates over the (canonical) kinetic energy, i.e. we require $V>\dot{\phi}^2$. In order that inflation solves the flatness and horizon problems we require that it lasts a sufficiently long period of time. In order that $V>\dot{\phi}^2$ for at least $50-70$ efolds we, therefore, require that $\ddot{\phi}$ is small compared to the change in the potential. From the equation of motion \eqref{eom} we observe that in order for inflation to persist for  enough efolds we require $\ddot{\phi}\ll V_{,\phi}$. Hence during the period of inflation the Hubble damping term dominates and we have the following $slow$-$roll$ approximation,
\begin{align}
\dot{\phi}\approx& -\frac{V_{,\phi}}{3H},\\
H^2\approx& \frac{8\pi G}{3}V.
\end{align}
Under the slow-roll approximations $V\approx \rm{constant}$ and, therefore, $H^2\approx \rm{constant}$.   Hence, $a\propto e^{H t}$ implying that we have an exponential expansion. Inflation ends when the slow-roll approximations break down. Figure~\ref{fig:slowroll} shows a typical slow roll potential with the inflaton rolling from a false vacuum state to the true vacuum where inflation ends and reheating occurs. This process of reheating converts the potential energy of the inflaton into standard model particles.

\section{Perturbations}
Clearly the assumption of an exactly homogeneous scalar field does not fit our observations of the universe today. In order to seed structure formation, an initial fluctuation must be present in the early universe. In this section we present a discussion of first order perturbation theory which will be used to calculate the observable quantities important for inflation. We shall assume purely scalar perturbations of the metric and the scalar field $\phi\rightarrow \tilde{\phi}=\phi_0+\delta \phi_1$.
\par
The perturbed flat FLRW metric may be written as follows
\begin{align}
 ds^2=a^2\left[-(1+2\Psi)d\tau^2    +2B_{,i}dx^i d\tau-\left((1-2\Phi)\delta_{ij}-2 E_{,ij}dx^i dx^j\right)     \right].
\end{align}
Under a gauge transformation $x^\mu\rightarrow x^{\mu'}=x^{\mu}+\xi^{\mu}=x^{\mu}+(\xi^0,\xi_{,i})$ the perturbed part of the metric $g_{\mu \nu}+\delta g_{\mu \nu}$ becomes transformed to give
\begin{align}
\tilde{g}_{\mu \nu}(x^{\mu'})&={g}_{\mu \nu}(x^{\mu})+\delta g_{\mu \nu}-g_{\mu \rho } \xi^{\rho}_{,\nu}-g_{\rho \nu } \xi^{\rho}_{,\mu},\nonumber\\
\implies \delta\tilde{g}_{\mu \nu}(x^{\mu'})&=\delta g_{\mu \nu}-g_{\mu \rho } \xi^{\rho}_{,\nu}-g_{\rho \nu }\xi^{\rho}_{,\mu}.
\end{align}
This gauge transformation allows us to relate different gauges to linear order. In terms of the perturbed FLRW metric we have
\begin{align}\label{metricPert}
\tilde{\Psi}&=\Psi+\mathcal{H}\xi_0 +\xi'\nonumber\\
\tilde{B}&=B-\xi_0+\xi',\nonumber\\
\tilde{\Phi}&=\Phi-\mathcal{H}\xi_0,\nonumber\\
\tilde{E}&=E+\xi.
\end{align}
where $'$ denotes differentiation with respect to the conformal time $\tau$ and $\mathcal{H}=a'/a$.
\par
Using the gauge transformations we can set any two of the four functions $\Psi, \Phi, B, E$ to zero. In this section we will consider the flat gauge where spatial hypersurfaces are unperturbed, i.e. $\Phi=0=E$. Under a gauge transformation the scalar field is also transformed to give
\begin{align}\label{inflatPert}
\delta \tilde{\phi}_1=\delta \phi_1 +  \phi_0'\xi^0.
\end{align}

 Using equations \eqref{metricPert} and \eqref{inflatPert} we form the following gauge invariant quantity
\begin{align}
v=\delta \phi_1+\phi_0' \frac{\Phi}{\mathcal{H}},
\end{align}
known as the Sasaki-Mukhanov variable. In the flat (zero curvature) gauge this variable is given by $\delta \phi_1$. It is often more useful to write this variable in the form
\begin{align}\label{curv}
\mathcal{R}=\frac{\mathcal{H}}{\phi_0'}v.
\end{align}
This quantity is known as the curvature variable \cite{Bardeen}. This is a useful parameter since it tracks the curvature and energy density perturbations throughout the history of the universe. It applies equally well during inflation and today. 
\par
Using equation \eqref{eom1}, the equation of motion for $\phi_0$ and the perturbed quantity $\delta \phi_1$ becomes
\begin{align}
&\phi_0''+2\mathcal{H}\phi_0'+a^2 V_{,\phi}=0,\\
&\delta \phi_1''+2\mathcal{H}\delta \phi_1'+k^2 \delta \phi_1+a^2\left(\frac{8\pi G}{\mathcal{H}}\right)^2(\phi_0')^2V(\phi_0)\delta \phi_1+ a^2\left[V_{,\phi\phi}+\frac{16\pi G}{\mathcal{H}}  \phi_0' V_{,\phi}  \right]\delta \phi_1=0.
\end{align}
We may neglect the term proportional to $V_{,\phi \phi}$ since the slow-roll conditions requires it to be small.
Setting $u=a\delta \phi_1(\equiv a v)$ and using the slow-roll approximations we obtain the perturbation equation
\begin{align}
u'' +\left(k^2-\frac{a''}{a}\right) u=0.
\end{align}
We solve this equation at early times, on quantum scales, where the field satisfies Minkowski space quantum theory in the subhorizon limit $k/(aH)\rightarrow \infty$. In particular, we may write the perturbation as the quantum superposition of creation and annihilation operators,
\begin{align}
u(x)=\int \frac{d^3 k}{(2\pi)^3 2\omega_k}\left(a_{\mathbf{k}}u_k e^{i\bk.\bx}+a_{\mathbf{k}}^{\dagger}u^*_k e^{-i\bk.\bx}\right),
\end{align}
where $\omega$ the frequency corresponding to mode $k$. We define $|0\rangle$ to be the zero particle quantum state with
\begin{align}
a_{\mathbf{k}}^{\dagger}|0\rangle=|1\rangle,\qquad a_{\mathbf{k}}^{}|0\rangle=0.
\end{align}
These operators satisfy the commutation relations
\begin{align}
[a_{\mathbf{k}}^{},a_{\mathbf{k}'}^{\dagger}]&=(2\pi)^3\delta(\mathbf{k}-\mathbf{k}'),\nonumber\\
[a_{\mathbf{k}}^{\dagger},a_{\mathbf{k}'}^{\dagger}]&=0=[a_{\mathbf{k}}^{},a_{\mathbf{k}'}^{}].
\end{align}
The modes $u_k$ satisfy the equation of motion
\begin{align}
u_k'' +\left(k^2-\frac{a''}{a}\right) u_k=0.
\end{align}
In the subhorizon limit $k/(aH)\rightarrow \infty$ this equation is approximately $u_k''+k^2 u_k=0$, which has the plane wave solution
\begin{align}
u_k=\frac{1}{\sqrt{2 k}}e^{-ik\tau}.
\end{align}
This choice of initial condition is due to the imposition that the field asymptotically satisfies the Bunch-Davies (Minkowski) vacuum. More generally if we consider the equation during the inflationary era during which we regard $H\equiv \mathcal{H}/a\approx \rm{constant}$ we find
\begin{align}
u_k'' +\left(k^2-\frac{1}{\tau^2}\right) u_k&=0,\\
\implies u_k &= c_1\frac{e^{-i k\tau}}{\sqrt{2 k}}\left(1-\frac{i}{k\tau}\right)+c_2\frac{e^{i k\tau}}{\sqrt{2 k}}\left(1+\frac{i}{k\tau}\right).
\end{align}
The matching condition with the solution for subhorizon modes implies $c_1=1, c_2=0$, i.e.
\begin{align}
u_k &= \frac{e^{-i k\tau}}{\sqrt{2 k}}\left(1-\frac{i}{k\tau}\right).
\end{align}
Due to inflation the mode will grow from subhorizon scales to superhorizon scales. At horizon crossing, the mode ceases to oscillate and its amplitude freezes. At horizon crossing we have $k=aH$, where the mode wavelength equals the Hubble radius. At late times the fluctuations on scales well outside the horizon, $k\tau\rightarrow 0^-$, have magnitude given by
\begin{align}\label{ampl}
\lim_{k\tau\rightarrow 0^-}|v_{k}|=\lim_{k\tau\rightarrow 0^-}\left|\frac{u_k}{a}\right|=\frac{H}{\sqrt{2}k^{3/2}},
\end{align}
where we use the expression $a=-1/(H\tau)$ to simplify the final expression.
In terms of the curvature variable \eqref{curv} we have
\begin{align}\label{amplcurv}
|{\mathcal{R}}_{k}|=\frac{H^2}{\dot{\phi_0}\sqrt{2}k^{3/2}}.
\end{align}
The curvature remains constant until after inflation ends when the Hubble radius grows relative to the wavelength of the mode. When the modes re-enter the horizon they begin to evolve again.

\section{Power Spectrum}
The power spectrum, $P_{\delta \phi_1}(k),$  is defined as the two-point correlator of the variable, $\delta \phi_1$ ($=v$ in the zero curvature gauge),
\begin{align}
\langle \delta \phi_1(\mathbf{k}_1 )\delta \phi_1(\mathbf{k}_2 )\rangle=(2\pi)^3 \delta(\mathbf{k}_1+\mathbf{k}_2)P_{\delta \phi}(k_1),
\end{align}
where $\langle \dots\rangle$ denotes the ensemble average.
The density power spectrum today may be interpreted as the Fourier transform of the galaxy two-point correlation function $\xi(\mathbf{x})=\langle \delta(\mathbf{x}')\delta(\mathbf{x}+\mathbf{x}')\rangle$. However, we note that it is the galaxy number density that is actually observed. The relation of $\delta n$ to $\delta$ is parametrised by a bias factor $b$. We will refer to this issue later in our discussion of large scale structure.
\par
In the context of inflation, the perturbation is created from a quantum fluctuation upon horizon crossing and hence from equation \eqref{ampl} we find
\begin{align}
P_{\delta \phi}(k)=\langle 0| |\delta \phi_k |^2 |0\rangle=\frac{H^2}{2 k^3}.
\end{align}
The dimensionless power spectrum is then given by
\begin{align}
{\Delta}_{\Phi} =\frac{k^3}{2\pi^2}P_{\delta \phi}(k)=\left(\frac{H}{2\pi}\right)^2.
\end{align}
\par
In a similar fashion to the above we define the power spectrum of the curvature variable ${\mathcal{R}}$. From equation $\eqref{amplcurv}$ we have
\begin{align}
P_{{\mathcal{R}}}(k)=\langle 0| |{\mathcal{R}}_k |^2 |0\rangle=\left(\frac{\mathcal{H}}{{\phi_0'}}\right)^2 P_{\delta \phi}(k)=\left(\frac{{H}}{{\dot{\phi}_0}}\right)^2 P_{\delta \phi}(k).
\end{align}
Hence the dimensionless curvature power spectrum is given by
\begin{align}
\Delta_{{\mathcal{R}}} =\frac{k^3}{2\pi^2}P_{{\mathcal{R}}}(k)=\left(\frac{H^2}{2\pi \dot{\phi}_0}\right)^2.
\end{align}
While this quantity is (almost) scale-free in the simple model of single-field slow-roll inflation, in other models of inflation there may be a non-trivial scaling. Therefore, we define the spectral index
\begin{align}
n_s-1=\frac{d \ln\Delta_{{\mathcal{R}}}}{d \ln k},
\end{align}
with $n_s=1$ indicating a scale invariant spectrum. The possible running of the spectral index is measured by $\alpha_s=d n_s/d \ln k$.
We note here that in addition to scalar perturbations, tensor perturbations may also be produced\footnote{Vector fluctuations within the inflationary picture are generally shown to redshift away.}. The tensor perturbation $h_{i j}$ has two polarisations $h_s$, where $s=+,\times$. The analysis follows similar lines to that carried out in the previous section. The overall tensor perturbation is given by
\begin{align}
\Delta_{T}(k)=\Delta_{h_+}(k)+\Delta_{h_\times }(k)=\frac{2 H^2}{M_{pl}^2 \pi^2}.
\end{align}
We may also define a tensor spectral index $n_T=d \ln\Delta_{T}/d \ln k$ and a spectral running 
$d n_T/d \ln k$. Of particular observational interest is the tensor-to-scalar ratio
\begin{align}
r=\frac{\Delta_{T}(k)}{\Delta_{{\mathcal{R}}}(k)}.
\end{align}
For slow-roll inflation this quantity is expected to be of order $r\lesssim \mathcal{O}(0.15)$.
\par
Since, in this simplest model of inflation the inflaton fluctuates approximately as a free field, such that its different frequencies oscillate independently, the resulting density perturbations follow a Gaussian distribution. Therefore, in single-field slow-roll inflation, higher order contributions are expected to be negligible. Non-Gaussianity appears when the dynamics of inflaton fluctuations is affected by sizable non-linearities. In particular, violation of any of the assumed conditions of the standard picture presented here, namely single field, slow roll, canonical kinetic energy and Bunch-Davies vacuum, is expected to produce sizable non-Gaussianity. In addition most alternatives to the model of inflation are expected to produce large deviations from Gaussianity.

\section{Concordance Model and Current Observations}
In recent years we have entered the era of precision cosmology. Measurements of the cosmic microwave background have improved greatly in resolution since the launch of the COBE experiment in 1989, to the WMAP satellite in 2001, and most recently the Planck mission in 2009. Additional data sets have been provided by balloon-borne microwave telescopes such as BOOMERANG. Due to the degeneracy involved in the measurement of some of the cosmological parameters within the CMB, complementary data sets are required. These are provided by measurements of Baryon Acoustic Oscillations (BAO) in galaxy distributions \cite{Bassett} and by surveys of supernovae provided by the Hubble telescope \cite{Clocch}.
\par
In addition to the inflationary parameters, cold dark matter and dark energy are required to fit observations. In order to explain the discrepancy between the rotation speeds of matter in the disks of spiral galaxies and the prediction for the rotational velocity given the visible mass (baryonic matter) dark matter is required. This dark matter must travel at non-relativistic speeds to satisfy the requirements for galaxy formation and hence is known as cold dark matter. Cold dark matter is parametrised by the density parameter, $\Omega_c$. In particular, we write equation \eqref{FriedOmega} with $\Omega_m=\Omega_b+\Omega_c$, where $\Omega_b$ is the baryon density\footnote{Within the $\Lambda$CDM model we assume a flat universe. This assumption is relaxed for general analysis.}. Dark energy is necessary to explain current observations of an accelerating rate of expansion of the universe. Within the standard $\Lambda$CDM model this is expected to be sourced by the cosmological constant. Dark energy may be described by the equation of state $P=\omega \rho$ with $\omega=-1$ for the cosmological constant. With the additional assumption of a flat universe, such that $\Omega_k=0$ we may infer $\Omega_{\Lambda}$ from equation \eqref{FriedOmega} and measurements of $\Omega_b$ and $\Omega_c$. The Hubble constant is often expressed with the variable $h$ where $H_0=100 h (km/s)/Mpc$. With this variable we express the matter parameters $\omega_b=\Omega_b h^2, \omega_c=\Omega_c h^2$. The concordance model of cosmology, known as the $\Lambda$CDM model, assumes an adiabatic, flat, Gaussian power law cosmology and is parametrised effectively by the six parameters $(\omega_b, \omega_c, \tau, n_s, \Delta_{{\mathcal{R}}}, h)$, where $\tau$ describes the reionisation optical depth. In Table~\ref{table1} the current constraints on these parameters are given.

\begin{table}
\vskip-5pt
\centering
\begin{tabular}{|c||c||c|}
 \hline

    Parameter  &          WMAP $5$-year    &               WMAP+BAO+SN  \\

    \hline
    \hline

$\omega_b$ &$(2.273\pm 0.062)\times 10^{-2}$ &$(2.267^{+ 0.058}_{0.059})\times 10^{-2}$ \\
& &\\
$\omega_c$ & $0.1099\pm 0.0062$&$0.1131 \pm 0.0034$\\
& &\\
$\tau$ & $0.087\pm 0.017$&  $0.084\pm 0.016$\\
& &\\
$n_s$ &$ 0.963^{+ 0.014}_{-0.015}$&$ 0.960 \pm 0.013 $\\
& &\\
$\Delta_{{\mathcal{R}}}$ & $(2.41\pm 0.11)\times 10^{-9}$& $(2.445\pm 0.096)\times 10^{-9}$\\
& &\\
$h$ & $(71.9^{+ 2.6}_{-2.7})\times 10^{-2}$&$ (70.5\pm 1.3)\times 10^{-2}$\\
\hline
  
  \end{tabular}
  \caption{Current constraints on $\Lambda$CDM cosmology from WMAP $5$-year data and a joint analysis of WMAP, BAO and SN data, \cite{Komatsu}. The quoted intervals indicate $1\sigma$ bounds, i.e. $68\%$ confidence intervals.}
\label{table1}
\end{table}
\par
In addition to these six variables the WMAP team has parametrised the model of the universe with a further seven parameters \cite{0302209,0603449,08030586}. Within their analysis the assumption of a flat universe is dropped, as is the assumption that dark energy is necessarily given by a cosmological constant. This requires constraints on the parameters $(\Omega_k,\Omega_{\Lambda}, \omega)$. Often instead of $\Omega_k$ the total density $\Omega_{tot}$ is given. In addition they allow for the spectral index to run and for a tensor contribution to perturbations. These are given, respectively by the parameters $\alpha_s$ and $r$. The spectral index of the tensor perturbations, $n_t$, is also constrained. These are evaluated at the pivot scale $k=0.002 Mpc^{-1}$. The other variable considered is the neutrino density parameter $\Omega_{\nu}$ (or $\omega_{\nu}=\Omega_{\nu}h^2$). The presence of massive neutrinos may affect the position and size of the Doppler peaks in the CMB at the level of a few percent. Using this fact leads to a constraint on the total mass of neutrinos.
\par
There is a degeneracy in the parameters listed thus far. The Hubble parameter $h$ may be found by evaluating
\begin{align}
h=\sqrt{\frac{\omega_b+\omega_c+\omega_v}{\Omega_{tot}-\Omega_{\Lambda}}}.
\end{align}
In summary we have the twelve parameters $(\Omega_{tot},\Omega_{\Lambda},\omega_b,\omega_c,\omega_{\nu},\omega, \Delta_{{\mathcal{R}}}, r, n_s, n_t, \alpha_s, \tau)$.
\par
In addition to these $12$ variables, another two parameters are also quoted \cite{0608632}. These are the galaxy bias factor, $b$ and the non-linear correction parameter $Q_{nl}$. These constants are necessary to account for the non-linear evolution on small scales. The non-linear correction parameter is defined by writing the galaxy power spectrum in the form
\begin{align}
P_g(k)=P_{\rm{dewiggled}}(k) b^2\frac{1+ Q_{nl}k^2}{1+1.4 k},
\end{align}
where $P_{\rm{dewiggled}}(k)$ accounts for the non-linear suppression of baryon wiggles, and the final factor accounts for non-linear effects and scale-dependent bias of the galaxy distribution. From these $14$ parameters another $45$ may be derived from them as detailed in \cite{0608632}.  Constraints on the extra $9$ parameters beyond the $\Lambda$CDM model are given in Table~\ref{table3}.
\begin{table}
\vskip-5pt
\centering
\begin{tabular}{|c||c||c|}
 \hline

    Parameter  &          WMAP &WMAP+ SDSS    \\

    \hline
    \hline

$\Omega_{tot}$ &$1.054^{+0.064}_{-0.046}$ &$1.003^{+0.010}_{-0.009}$ \\
& &\\
$\Omega_{\Lambda}$ & $0.761^{+0.032}_{-0.037}$&$0.761^{+0.017}_{-0.018}$\\
& &\\
$\omega$ & $-0.82^{+0.23}_{-0.19}$&  $-0.941^{+0.087}_{-0.101}$\\
& &\\
$\alpha_s$ &$ -0.056^{+ 0.031}_{-0.031}$&$  -0.040^{+ 0.027}_{-0.027} $\\
& &\\
$r$ & $<0.65 (95\%)$& $<0.33 (95\%)$\\
& &\\
$n_t+1$ & $ 0.9861^{+ 0.0096}_{-0.0142}$&$0.9861^{+ 0.0096}_{-0.0142}$\\
& &\\
$\omega_{\nu}$ & $<0.024 (95\%)$&$<0.010 (95\%)$\\
& &\\
$b$ & &$ 1.896^{+0.074}_{-0.069}$\\
& &\\
$Q_{nl}$ &&$ 30.3^{+4.4}_{-4.1}$\\
\hline
  
  \end{tabular}
  \caption{Current constraints on the extra $9$ parameters beyond the $\Lambda$CDM model. The constraints are obtained from \cite{0608632} which uses WMAP observations and data from the Sloan Digital Sky Survey. We note that the values of $b$ and $Q_{nl}$ are only obtained from large scale structure and hence from the SDSS data set. The quoted constraints are $68\%$ bounds unless otherwise stated.}
\label{table3}
\end{table}

\section{Non-Gaussianity}\label{sec:nonGaussBgd}
Single field slow roll inflation as we have described is well approximated by a Gaussian random field and so is well described by the power spectrum. To test for alternatives to the standard paradigm it is, therefore, necessary to measure higher order correlators beyond the power spectrum. Of particular interest are the bispectrum and trispectrum which are, respectively, related to the skewness and kurtosis of the underlying primordial distribution. 
\par
\begin{figure}[htp]
\centering 
\includegraphics[width=82mm]{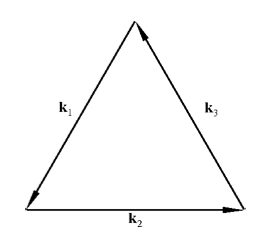}
\caption{Triangle defined by the three wavenumbers $k_1,k_2,k_3$. }
\label{fig:BispTriangle}
\end{figure}
\begin{figure}[htp]
\centering 
\includegraphics[width=102mm]{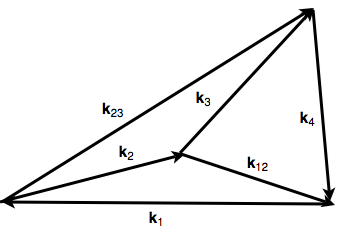}
\caption{Quadrilateral defined by the four wavenumbers $k_i$ and the diagonals $k_{12}=|\bk_1-\bk_2|$ and $k_{2 3}=|\bk_2-\bk_3|$. }
\label{fig:Tquadrilateral}
\end{figure}

The primordial bispectrum is defined by
\begin{align}
\langle \Phi(\mathbf{k}_1)\Phi(\mathbf{k}_2)\Phi(\mathbf{k}_3)\rangle=(2\pi)^3\delta(\mathbf{k}_1+\mathbf{k}_2+\mathbf{k}_3)B_{\Phi}(k_1,k_2,k_3),
\end{align}
where the Dirac delta symbol imposes the closure condition on the wavevectors. The bispectrum, $B_{\Phi}$, defined at a particular scale on a triangle is therefore parametrised by three variables which we may choose to be the wavenumbers $(k_1,k_2,k_3)$ (see Figure~\ref{fig:BispTriangle}). Similarly the primordial trispectrum is given by
\begin{align}\label{trispdefinition}
\langle \Phi(\mathbf{k}_1)\Phi(\mathbf{k}_2)\Phi(\mathbf{k}_3)\Phi(\mathbf{k}_4)\rangle_c=(2\pi)^3\delta(\mathbf{k}_1+\mathbf{k}_2+\mathbf{k}_3+\mathbf{k}_4)T_{\Phi}(\mathbf{k}_1,\mathbf{k}_2,\mathbf{k}_3,\mathbf{k}_4),
\end{align}
where the notation $\langle\dots\rangle_c$ is used to denote the connected component.
The closure condition now implies that the trispectrum, $T_{\Phi}$ is defined on a quadrilateral. Such a quadrilateral may be expressed using the four wavenumbers and its two diagonals (see Figure~\ref{fig:Tquadrilateral}). Thus in the most general case the trispectrum is a function of $6$ variables. The trispectrum may be decomposed according to the symmetries involved in exchanging the fields $\Phi(\mathbf{k}_i)$ and $\Phi(\mathbf{k}_j)$. Firstly we write \eqref{trispdefinition} in the form
\begin{align}\label{eqTTT1}
\langle \Phi(\mathbf{k}_1)\Phi(\mathbf{k}_2)\Phi(\mathbf{k}_3)\Phi(\mathbf{k}_4)\rangle_c=(2\pi)^3\int &d^3 K\delta(\mathbf{k}_1+\mathbf{k}_2 +\mathbf{K})\delta(\mathbf{k}_3+\mathbf{k}_4-\mathbf{K})\nonumber\\
&\times T_{\Phi}(\mathbf{k}_1,\mathbf{k}_2,\mathbf{k}_3,\mathbf{k}_4,\mathbf{K}),
\end{align}
where $\mathbf{K}(=-\mathbf{k}_1-\mathbf{k}_2)$ is a diagonal defined on the quadrilateral.
We now decompose the trispectrum in terms of the three other possible diagonals as
\begin{align}\label{TrispDefinition}
T_{\Phi}(\mathbf{k}_1,\mathbf{k}_2,&\mathbf{k}_3,\mathbf{k}_4,\mathbf{K})=P_{\Phi}(\mathbf{k}_1,\mathbf{k}_2,\mathbf{k}_3,\mathbf{k}_4,\mathbf{K})+\int d^3 K'\Big[ \delta(\mathbf{k}_3-\mathbf{k}_2 -\mathbf{K}+\mathbf{K}')\nonumber\\
&\times P_{\Phi}(\mathbf{k}_1,\mathbf{k}_3,\mathbf{k}_2,\mathbf{k}_4,\mathbf{K}')+\delta(\mathbf{k}_4-\mathbf{k}_2 -\mathbf{K}+\mathbf{K}')P_{\Phi}(\mathbf{k}_1,\mathbf{k}_4,\mathbf{k}_3,\mathbf{k}_2,\mathbf{K}')\Big].
\end{align}
We may further decompose $P_{\Phi}$ in terms of a reduced trispectrum $\mathcal{T}_{\Phi}$, 
\begin{align}\label{eqTTT2}
P_{\Phi}(\mathbf{k}_1,\mathbf{k}_2,\mathbf{k}_3,\mathbf{k}_4,\mathbf{K})=&\mathcal{T}_{\Phi}(\mathbf{k}_1,\mathbf{k}_2,\mathbf{k}_3,\mathbf{k}_4,\mathbf{K})+\mathcal{T}_{\Phi}(\mathbf{k}_2,\mathbf{k}_1,\mathbf{k}_3,\mathbf{k}_4,\mathbf{K})\nonumber\\
&+\mathcal{T}_{\Phi}(\mathbf{k}_1,\mathbf{k}_2,\mathbf{k}_4,\mathbf{k}_3,\mathbf{K})+\mathcal{T}_{\Phi}(\mathbf{k}_2,\mathbf{k}_1,\mathbf{k}_4,\mathbf{k}_3,\mathbf{K}),
\end{align}
with $\mathcal{T}_{\Phi}(\mathbf{k}_1,\mathbf{k}_2,\mathbf{k}_3,\mathbf{k}_4,\mathbf{K})=\mathcal{T}_{\Phi}(\mathbf{k}_3,\mathbf{k}_4,\mathbf{k}_1,\mathbf{k}_2,\mathbf{K})$.
\par
The expression of the trispectrum described by equations \eqref{eqTTT1}, \eqref{TrispDefinition} and \eqref{eqTTT2} often proves useful since, in many cases of interest, the reduced trispectrum, $\mathcal{T}_{\Phi}$, may be written as a function of the magnitude of its five arguments alone, i.e.
\begin{align}
\mathcal{T}_{\Phi}(\mathbf{k}_1,\mathbf{k}_2,\mathbf{k}_3,\mathbf{k}_4,\mathbf{K})=\mathcal{T}_{\Phi}(k_1,k_2,k_3,k_4,K).\nonumber
\end{align}
However, as we shall discuss later, there also exists models which depend only on the magnitude of the wavevectors $\bk_i$ with no diagonal dependence. In the case of such models it is more useful to use equation \eqref{trispdefinition} with $T_{\Phi}(\mathbf{k}_1,\mathbf{k}_2,\mathbf{k}_3,\mathbf{k}_4)=T_{\Phi}(k_1,k_2,k_3,k_4)$. Of course, there is no ambiguity since both expressions are equivalent.
\par
In order to elucidate this discussion we consider the local model of inflation, where the gravitational potential perturbation may be expressed as a Taylor expansion around the Gaussian part \cite{KomatsuSpergel2001},
\begin{align}
\Phi(\mathbf{x})=\Phi_G(\mathbf{x})+f_{NL}\left(\Phi_G^2(\mathbf{x})-\langle \Phi_G(\mathbf{x})\rangle^2\right)+g_{NL}\Phi_G^3(\mathbf{x}).
\end{align}
The parameters $f_{NL}$ and $g_{NL}$ are measures of the level of non-Gaussianity which, as we shall see, are related to the bispectrum and trispectrum respectively. 
In Fourier space we have
\begin{align}
\Phi(\mathbf{k})=\Phi_G(\mathbf{k})+f_{NL}\Phi_{NG,(1)}(\mathbf{k})+g_{NL}\Phi_{NG,(2)}(\mathbf{k}),
\end{align}
and find
\begin{align}
\Phi_{NG,(1)}(\mathbf{k})&=\int \frac{d^3 p}{(2\pi)^3}\Phi_G(\mathbf{k}+\mathbf{p})\Phi_G^*(\mathbf{p})-(2\pi)^3 \delta(\mathbf{k})\langle \Phi_G^2(\mathbf{x})\rangle,\\
\Phi_{NG,(2)}(\mathbf{k})&=\int \frac{d^3 p_1}{(2\pi)^3} \frac{d^3 p_2}{(2\pi)^3}\Phi_G(\mathbf{k}+\mathbf{p}_1+\mathbf{p}_2)\Phi_G^*(\mathbf{p}_1)\Phi_G^*(\mathbf{p}_2),
\end{align}
where
\begin{align}
\langle \Phi_G^2(\mathbf{x})\rangle=\int \frac{d^3 k}{(2\pi)^3}P_{\Phi}(k).
\end{align}
The primordial bispectrum may be evaluated by calculating
\begin{align}
\langle \Phi_G(\mathbf{k}_1)\Phi_G(\mathbf{k}_2)\Phi_{N G}(\mathbf{k}_3)\rangle=&\int \frac{d^3 p}{(2\pi)^3}\langle \Phi_G(\mathbf{k}_1)\Phi_G(\mathbf{k}_2)  \rangle \langle \Phi_G(\mathbf{k}_3+\mathbf{p})\Phi_G^*(\mathbf{p})\rangle\nonumber\\
&+\int \frac{d^3 p}{(2\pi)^3}\langle \Phi_G(\mathbf{k}_1)\Phi_G^*(\mathbf{p})  \rangle \langle \Phi_G(\mathbf{k}_2)\Phi_G(\mathbf{k}_3+\mathbf{p})\rangle\nonumber\\
&+\int \frac{d^3 p}{(2\pi)^3}\langle \Phi_G(\mathbf{k}_1)\Phi_G(\mathbf{k}_3+\mathbf{p})  \rangle \langle \Phi_G(\mathbf{k}_2)\Phi_G^*(\mathbf{p})\rangle\nonumber\\
&-(2\pi)^3\delta(\mathbf{k}_3)\langle \Phi_G(\mathbf{k}_1)\Phi_G(\mathbf{k}_2)\rangle \langle \Phi_G^2(\mathbf{x})\rangle.
\end{align}
This results in the local bispectrum
\begin{align}\label{LocalBisp}
B_{\Phi}(k_1,k_2,k_3)=2f_{NL}(P_{\Phi}(k_1)P_{\Phi}(k_2)+P_{\Phi}(k_1)P_{\Phi}(k_3)+P_{\Phi}(k_2)P_{\Phi}(k_3)).
\end{align}
Measures of the bispectrum are generally expressed using the $f_{NL}$ parameter. Different models will, of course, generally have a different bispectrum. Therefore, a more general definition of $f_{NL}$ is required. This is usually defined as the following ratio
\begin{align}\label{fnldef}
f_{NL}=\frac{B_{\Phi}(k,k,k)}{6 P_{\Phi}(k)^2}.
\end{align}
Clearly this definition agrees with $f_{NL}^{\rm{local}}$ for the local bispectrum. However, since the parameter is defined at a specific value in parameter space- in particular, at the equal $k$ limit - comparison between models may be subject to pathologies in this limit. 
\par
The primordial trispectrum may be evaluated in a similar fashion to the bispectrum. This results in the following local reduced trispectrum
\begin{align}\label{LocalTrisp}
\mathcal{T}_{\Phi}(\mathbf{k}_1,\mathbf{k}_2,\mathbf{k}_3,\mathbf{k}_4,\mathbf{K})=&\frac{25}{9}\tau_{NL} P_{\Phi}(K)P_{\Phi}(k_1)P_{\Phi}(k_3)\nonumber\\
&+g_{NL}\left[	P_{\Phi}(k_2)P_{\Phi}(k_3)P_{\Phi}(k_4)+P_{\Phi}(k_1)P_{\Phi}(k_2)P_{\Phi}(k_4)\right],
\end{align}
where we write $\tau_{NL}=(6 f_{NL}/5 )^2$. Therefore, for single field local inflation the trispectrum offers a measure of the parameters $f_{NL}$ and $g_{NL}$. For multifield local inflation it was shown in \cite{WANDS} that the variable $\tau_{NL}$ is in general independent of $f_{NL}$ except that necessarily $\tau_{NL}\geq (6 f_{NL}/5 )^2$ with equality holding precisely in the single field case.
\par
Again, it is desirable to compare measures of the trispectra of different models. The general measure, $t_{NL}$, is defined in the equilateral limit for which the wavenumbers $k_i$ and the diagonals of the quadrilateral (formed by the wavevectors $\mathbf{k}_i$) have equal values, i.e. $k_1=k_2=k_3=k_4=|\mathbf{k}_1+\mathbf{k}_2|=|\mathbf{k}_1+\mathbf{k}_3|=k$. We define
\begin{align}\label{tnldef}
t_{NL}=\frac{9}{200}\frac{T_{\Phi}(k,k,k,k,k,k)}{P_{\Phi}(k)^3}.
\end{align}
With this definition we find that for the local model $t_{NL}^{\rm{local}}=1.5 \tau_{NL}+1.08 g_{NL}$. As with the bispectrum measure, it is clearly desirable to obtain an integrated measure of the trispectrum.

\section{Inhomogeneities from anisotropies - Collisionless Boltzmann equation}
Given primordial density perturbations we must evolve them to the present day. In this section we will work in the Newtonian gauge, $B=E=0$, such that our metric reads
\begin{align}\label{FLRWpert}
ds^2=-a^2(1+2\Psi)d\tau^2+a^2(1-2\Phi)dx^i dx_{i}.
\end{align}
In general relativity we've a $7$ dimensional $(x^{\mu},\mathbf{p})$ phase space of worldlines satisfying $p^{\mu}=\frac{d x^{\mu}}{d\lambda}$ with $p_{\mu}p^{\mu}=-m^2$ and
\begin{eqnarray}\label{geod}
\frac{dp^{\mu}}{d\lambda}+\Gamma^{\mu}_{\nu \sigma}p^{\nu}p^{\sigma}=0
\end{eqnarray}
For an observer with four velocity $u^{\mu}=a^{-1}(1-\Psi)(1,\mathbf{0})$ orthogonal to a spacelike hypersurface $\Sigma$ the number of worldlines intersecting volume element $dV$ with momenta $p^{\mu}$ in the range $dV_p$ is
\begin{eqnarray}
dN=f(x,p)(-u_{\mu}p^{\mu})dV dV_p, 
\end{eqnarray}
where $f$ represents the distribution function. The distribution function for photons takes the form of the Planck spectrum,
\begin{align}
f(p,T)=\frac{1}{e^{p/T}-1}.
\end{align}
The comoving observer measures the energy of the photon to be,
\begin{align}
E=-g_{\mu\nu}p^{\mu}u^{\nu}=a(1+\Psi)p^0\equiv p,
\end{align}
defining the photon momentum. To zeroth order the temperature redshifts as $1/a$ implying
$f_0(p,T)=f_0(a p)$. Defining $q=ap$ we split the distribution function into a zeroth order part and a first order part,
\begin{align}
f=f_0(q)+f_1(\tau,\mathbf{\hat{n}}^i,\mathbf{x}^i,q),
\end{align}
where $\mathbf{\hat{n}}^i=d\mathbf{x}^i/d\tau$ represents the photon's trajectory.
\par
Liouville's theorem implies that that the phase space volume, $dU=(-u_{\mu}p^{\mu})dV dV_p $, is conserved along the world lines. Thus
\begin{eqnarray}
\frac{d(dN)}{d\lambda}=\frac{df}{d\lambda} dU
\end{eqnarray}
Of course if there are no interactions we get the collisionless Boltzmann equation
\begin{eqnarray}
\frac{df}{d\lambda}=\frac{d x^{\mu}}{d\lambda}\frac{\partial f}{\partial x^{\mu}}+\frac{d p^{i}}{d\lambda}\frac{\partial f}{\partial p^{i}}=0
\end{eqnarray}
If we include interactions with a net collision rate $C[f]dU$ we get the full Boltzmann equation $\dfrac{df}{d\lambda}=C[f]$.
\par
The geodesic equations, \eqref{geod}, imply to first order in the perturbation variables that
\begin{align}
\frac{dq}{d\tau}=&({\Phi}'-\mathbf{\hat{n}}^i\Psi_{,i})q,\nonumber\\
\frac{d\mathbf{\hat{n}}^i}{d\tau}=&\mathcal{O}(\Phi).
\end{align}
Thus in the absence of collisions the CMB photon path is given by
\begin{align}
\mathbf{x}(\tau)=\mathbf{x}(\tau_{\rm{dec}})+\mathbf{\hat{n}}(\tau-\tau_{\rm{dec}})+\mathcal{O}(\Phi).
\end{align}

Substituting these expressions into the Boltzmann equation we find
\begin{align}\label{boltzeq}
\frac{df}{d\tau}\equiv a\frac{df}{dt}=\frac{\partial f_1}{\partial \tau}+\mathbf{\hat{n}}^i\frac{\partial f_1}{\partial \mathbf{{x}}^i}+({\Phi}'-\mathbf{\hat{n}}^i\Psi_{,i})q\frac{\partial f_0}{\partial q}.
\end{align}

Noting that $\dfrac{\partial f_1}{\partial q}$ is absent, we have the ${ansatz}$
\begin{eqnarray}
f(\mathbf{x},\mathbf{p},\tau)&=&f_0 \left(\frac{p}{T(\mathbf{x},\mathbf{\hat{n}},\tau)}\right)=f_0\left( \frac{\overline{T}q}{T}\right)=f_0\left(  \frac{q}{1+\Delta T/T} \right)\nonumber\\\label{boltz}
&\approx& f_0(q)-q\frac{df_0}{dq} \frac{\Delta T}{T}(\mathbf{x},\mathbf{\hat{n}},\tau)=f_0+f_1
\end{eqnarray}
In the remainder of this section we shall work in a locally inertial frame where $g_{\mu \nu}=\eta_{\mu \nu}, p^{\mu}=(q/a)(1,\mathbf{\hat{n}}^i), u^{\mu}=(1,\mathbf{v}^i)$ and $dV_p=d^3p/p^0=d^3 q/(a^2 q)$. We shall also work in Fourier space to simplify the expressions and furthermore we shall consider only scalar modes, in particular assuming that the velocity may be represented as the gradient of a scalar field, $V$. 
\par
The energy-momentum tensor of the photons is given by
\begin{eqnarray}
T^{\mu \nu}=\int p^{\mu}p^{\nu} f(\mathbf{x},\mathbf{p},\tau)dV_p
\end{eqnarray}
Therefore the energy density is $\rho=\overline{\rho}_{\gamma}(1+\delta_{\gamma})=u_0 u_0 T^{00}=\dfrac{1}{a^4}\int q (f_0 + f_1)q^2 dq d\Omega$. Thus
\begin{align}
\overline{\rho}_{\gamma}=&\frac{4\pi}{a^4}\int f_0 q^3 dq,\nonumber\\
\delta_{\gamma}=&\frac{1}{\overline{\rho}_{\gamma}a^4} \int q f_1 q^2 dq d\Omega.
\end{align}
Similarly $T^{0i}=\dfrac{1}{a^2}(\overline{\rho}_{\gamma}+\overline{P}_{\gamma})v^i=\dfrac{1}{a^2}\dfrac{4}{3}\overline{\rho}_{\gamma}ik_i V_{\gamma}=\dfrac{1}{a^6}\int q \hat{n}^i f_1 q^2 dq d\Omega$ (where we define $V_{\gamma}\equiv -i \mathbf{\hat{k}.v}$) which implies that
\begin{eqnarray}
i\mathbf{{k}}.\mathbf{v}=-k V_{\gamma}=\frac{3}{4}\frac{i \hat{k}^i}{\overline{\rho}_{\gamma}a^4}\int q \hat{n}^i f_1 q^2 dq d\Omega
\end{eqnarray}
Using the isotropic integrals we get that
\begin{eqnarray}
T^i_j=\frac{\overline{\rho}_{\gamma}}{3}\delta^i_{j} +\frac{1}{a^4}\int q \hat{n}^i \hat{n}_j f_1 q^2 dq d\Omega
\end{eqnarray}

Multiplying equation \eqref{boltz} by $q^3 dq$ and integrating gives
\begin{eqnarray}
 \int q f_1 q^2 dq &=& -4 \frac{\Delta T}{T}\int q f_0 q^2 dq = -4 \frac{\Delta T}{T}\frac{a^4 \overline{\rho}_{\gamma}}{4 \pi},\nonumber\\
\implies \Theta(\mathbf{x},\mathbf{\hat{n}},\tau)&\equiv&4\frac{\Delta T}{T}= \frac{4\pi}{a^4 \overline{\rho}_{\gamma}}\int q f_1 q^2 dq =4\pi \delta_{\gamma},
\end{eqnarray}
where we have defined the brightness function $\Theta=4\frac{\Delta T}{T}$.
\par
Next, we integrate the Boltzmann equation \eqref{boltzeq} with $q^3 dq$ to obtain the evolution equation for the brightness function
\begin{eqnarray}
\Theta' + ik \mu \Theta = 4( \Phi'-ik\mu\Psi )+a\frac{d\Theta}{dt},
\end{eqnarray}
where $'$ denotes $\partial/(\partial \tau)$ and $\mu=\hat{\mathbf{k}}.\hat{\mathbf{n}}$. 
\par
In the collisionless case, for which $d\Theta/dt=0$, we may solve this equation exactly to find
\begin{align}
\Theta(\tau_0)=(\Theta(\tau_{\rm{dec}})-\Psi(\tau_{\rm{dec}}))e^{-i k\mu(\tau_0-\tau_{\rm{dec}})}+4\int_{\tau_0}^{\tau_{\rm{dec}}}({\Phi}'+{\Psi}')e^{-i k\mu(\tau_0-\tau)}.
\end{align}
As we shall establish in the next section $\Theta(\tau_{\rm{dec}})$ may be written in the form 
\begin{align}\label{decoup}
\Theta(\tau_{\rm{dec}})=\delta_{\gamma}+4\mathbf{\hat{n}.v}_{\rm{dec}}.
\end{align}
Hence we have
\begin{align}
\Theta(\tau_0)=(\delta_{\gamma}+4\mathbf{\hat{n}.v}_{\rm{dec}}-\Psi(\tau_{\rm{dec}}))e^{-i k\mu(\tau_0-\tau_{\rm{dec}})}+4\int_{\tau_0}^{\tau_{\rm{dec}}}({\Phi}'+{\Psi}')e^{-i k\mu(\tau_0-\tau)}.
\end{align}
This formula exhibits the main effects sourcing the CMB anisotropy. Namely, $\delta_{\gamma}$ represents the intrinsic density fluctuation at the surface of last scattering, $\mathbf{\hat{n}.v}$ represents the Doppler shift at the last scattering surface due to photons moving from over-dense regions to underdense ones, $\Psi$ denotes the Sachs-Wolfe term which indicates the difference in gravitational potential between last scattering and today, and the final (integrated) term represents the integrated Sachs-Wolfe effect due to photons passing through time varying gravitational wells between last scattering and today.

\section{Inhomogeneities from anisotropies - Collisional Boltzmann equation}
Before recombination photons and electrons are held in equilibrium by non-relativistic Compton scattering which is almost momentum independent. With a free electron density $n_e$, a photon has a probability $n_e \sigma_T dt$ of scattering in time $dt$. The optical depth 
\begin{align}
\kappa=\int_{t}^{t_0} n_e \sigma_T dt=\int_{\tau}^{\tau_0} a n_e \sigma_T d\tau,
\end{align}
 tells us that the probability of not scattering in $(t,t_0)$ is $e^{-\kappa}$. The mean free path between scatterings is $\lambda_c=1/(n_e \sigma_T)$ which corresponds to a mean free time between collisions of $\tau_c=1/(a n_e \sigma_T)$. For $\tau<\tau_{\rm{dec}}$ equilibrium is maintained because $a'/a\sim \tau^{-1}\ll \tau_c^{-1}$. However, Saha's equation implies that, for $\tau>\tau_{\rm{dec}}$, $n_e\rightarrow 0$ from which we deduce that $\tau_c \gg \tau$.
\newline 
The visibility $|\dot{\kappa}|e^{-\kappa}$ is the probability that a photon last scattered in the time interval $[t,t+dt]$ where $|\dot{\kappa}|=n_e \sigma_T$. We want to incorporate this into the collisional Boltzmann equation\footnote{In this discussion polarisation is ignored and isotropic scattering assumed.}. We should note that for small $k$, i.e. large angular scales, $|\kappa'|e^{-\kappa} \approx \delta( \tau-\tau_{\rm{dec}} )$. Alternatively for large $k$ there is significant damping. In order to include recombination in the Boltzmann equation we include the difference between the photons scattered in $\dfrac{df_{\rm{in}}}{d\lambda}=\delta_{\gamma}+4 i\mu V_b$ and those scattered out $\dfrac{df_{\rm{out}}}{d\lambda}=-{\kappa}'\Theta$, where the gradient of $V_b$ represents the baryon velocity. It should be noted that Thomson scattering will also polarise the incident radiation resulting in an additional term due to polarisation. For simplicity we will neglect this contribution in what follows.
\par
The isotropic collisional term $C[f]=\dfrac{df_{\rm{in}}}{d\lambda}-\dfrac{df_{\rm{out}}}{d\lambda}$ brings the Boltzmann equation to the form
\begin{align}\label{boltz2}
\Theta' + ik \mu \Theta = 4( \Phi'-ik\mu\Psi )+|{\kappa}' |(\delta_{\gamma}+4 i\mu V_b-\Theta).
\end{align}
For small $k$ the Gaussian nature of the visibility function gives the same solution as in the collisionless case. 
\par
In order to solve this equation we adopt a moment expansion of the brightness function. In particular, we write
\begin{align}\label{momentexp}
\Theta(\mathbf{k},\mu,\tau)=\sum_l (-i)^l (2l+1)\Theta_l (k,\tau) P_l (\mu),
\end{align}
where $\Theta_l=\dfrac{i^l}{4\pi}\int d\Omega \Delta(k,\mu,\tau)P_l(\mu)$. In terms of the moments we note that 
\begin{eqnarray}
\Theta_0 &=&\frac{1}{4\pi} \int d\Omega \Theta(k,\mu,\tau) =\frac{1}{a^4 \overline{\rho}_{\gamma}}\int q f_1 q^2 dq = \delta_{\gamma}\nonumber \\
\Theta_1 &=& \frac{i}{4\pi} \int  d\Omega \mu \Theta(k,\mu,\tau) =\frac{i\hat{k}^i}{a^4 \overline{\rho}_{\gamma}}\int q \hat{n}^i f_1 q^2 dq d\Omega = -\frac{4}{3}k V_{\gamma}\nonumber \\
\Theta_2 &=&\frac{1}{a^4 \overline{\rho}_{\gamma}}\int \frac{1}{2} (3\mu^2 -1) q f_1 q^2 dq d\Omega \equiv 2 \sigma_{\gamma} 
\end{eqnarray}
where $\sigma_{\gamma}$ represents the shear. At last scattering we assume a perfect fluid such that
$\Theta_l=0$ $\forall\, l\geq 2$. This implies that
\begin{eqnarray}
\Theta(\mathbf{k},\mu,\tau_{\rm{dec}})&=&\Theta_0 P_0 - 3 i \Theta_1 P_1 \nonumber\\
&=& \delta_{\gamma dec}+4 (\hat{\mathbf{k}}.\mathbf{v}_{\rm{dec}})\mu =\delta_{\gamma dec}+4 \hat{\mathbf{n}}.\mathbf{v}_{\rm{dec}},
\end{eqnarray}
as expressed in equation \eqref{decoup}.
\par
Using the recursion relation for the Legendre polynomials
\begin{align}
(l+1)P_{l+1} (\mu)=(2l+1)\mu P_l(\mu) -lP_{l-1}(\mu),
\end{align}
the Boltzmann equations may be decomposed into equations for the modes $\Theta_l$ reading
\begin{align}
\Theta_0'=&-k\Theta_1+4\Phi' , \qquad\qquad\qquad\qquad \quad\,\,\,\,\,\,\,\Leftrightarrow\,\, \delta_{\gamma}'=-\frac{4}{3}kV_{\gamma}+4\Phi',\nonumber\\
\Theta_1'=&\frac{k}{3}(\Theta_0+2\Theta_2+4\Psi)-\frac{\kappa'}{3}(4   V_b     -3\Theta_1),\,\,\Leftrightarrow \,\,V_{\gamma}'=\frac{k}{4}\left(\delta_{\gamma}+2\Theta_2+4\Psi\right)-\kappa'(V_b-V_{\gamma}),\nonumber\\
\Theta_2'=&\frac{k}{5}(2\Theta_1-3\Theta_3)+\kappa'\Theta_2,\qquad\qquad\quad\,\,\,\,\,\,\,\,\,\Leftrightarrow \, \Theta_2'=\frac{k}{5}(-8V_{\gamma}/3-3\Theta_3)+\kappa'\Theta_2,\nonumber\\
\Theta_{l\geq 3}'=&\frac{k}{2 l+1}\left(l\Theta_{l-1}-(l+1)\Theta_{l+1}\right)+\kappa' \Theta_l.
\end{align}
These equations must be combined with evolution equations for baryons and cold dark matter as well as for the metric perturbations to form a closed set of differential equations which may be solved given a particular set of initial conditions. We assume the universe consists of a perfect fluid composing of different matter contributions, i.e.
\begin{eqnarray}
T^{\mu \nu}=\sum_i\left[  (\rho_i+P_i)u_i^{\mu}  u_i^{\nu} +P_i g^{\mu \nu}   \right].
\end{eqnarray}
Using the Einstein equations we find the following equations for the metric perturbations 
\begin{align}
k^2 \Phi+3\frac{{a}'}{a}\left(\Phi'+\frac{{a}'}{a}\Psi\right)&=-4\pi G a^2 \sum_i \rho_i,\nonumber\\
k^2\left(\Phi'+\frac{{a}'}{a}\Psi\right)&=4\pi G a^2  \sum_i (1+w_i)\overline{\rho}_i V_i,
\end{align}
where the sum runs over the different matter and radiation species and the pressure contribution by species $i$ is given by $P_i=w_i\overline{\rho}_i(1+\delta_i)$.
\par
Approximating cold dark matter as a pressureless fluid we find the equations of motion
\begin{align}
{\delta}_c'=&-kV_c+3{\Phi}' ,\nonumber\\
V_c'=&-\frac{a'}{a}V_c+k\Psi.
\end{align}
Modelling baryons as a fluid with pressure due to Thompon scattering we find
\begin{align}
\delta_b'&=-kV_b+3\Phi',\nonumber\\
V_b'&=-\frac{a'}{a}V_b +k\Psi+c_s^2 k \delta_b+R\kappa'(V_b-V_{\gamma}),
\end{align}
where $R=4\overline{\rho}_{\gamma}/(3\overline{\rho}_b)$ and $c_s^2=R/(3+3R)$ is the effective sound speed which imposes the so-called `baryon drag' on the photon distribution.
\par
Solving this set equations is numerically very challenging since current experiments at high resolution necessitate the solution to $l=\mathcal{O}(10^3)$. Luckily there is an alternative formed by integrating the Boltzmann equations \eqref{boltz2} directly,
\begin{align}
\Theta(\tau_0)=\int_{0}^{\tau_0}\left(4({\Phi}'-ik\mu\Psi) +|\kappa'|\left(\delta_{\gamma}+4i\mu V_b\right) \right)e^{-\kappa}e^{-i k\mu(\tau_0-\tau)}d\tau.
\end{align}
Integrating by parts to remove the dependency on $\mu$, we find
\begin{align}
\Theta(\tau_0)=\int_{0}^{\tau_0}S(k,\tau) e^{-i k\mu(\tau_0-\tau)}d\tau,
\end{align}
where
\begin{align}
S(k,\tau)=4 e^{-\kappa}(\Phi'+\Psi')+(e^{-\kappa})'\left(\delta_{\gamma}+4 \Psi-\frac{4V_b'}{k}\right)-4(e^{-\kappa})''\frac{V_b}{k}.
\end{align}
The source functions may be evaluated using the Boltzmann moment equations described above. Due to backreaction from higher modes it is necessary to solve these equations to $l\approx 10$.
Given the source function, we may again decompose into moments using the following expansion of the exponential function,
\begin{align}\label{exponexp}
e^{i\mathbf{k.x}}=\sum_l i^l (2l+1)j_l(k x) P_l(\mu),
\end{align}
where $\mu=\mathbf{\hat{k}.\hat{x}}$. The multipole moments are then given by
\begin{align}\label{boltzMultipole}
\Theta_l=\int_{0}^{\tau_{0}} S(k,\tau)j_l(k (\tau_0-\tau))d\tau.
\end{align}
The radiation transfer functions, $\Delta_l(k)$ may be defined by setting
\begin{align}\label{transf}
 \Theta_l=4\Delta_l(k)\Phi(\mathbf{k}),
\end{align}
and may be found by setting the initial condition $\Phi(k)=1$ in the above equations. The temperature anisotropies $\Delta T/T =\Theta/4$ may now be represented using the $a_{lm}$ coefficients of a spherical harmonic decomposition of the cosmic microwave sky,
\begin{align}
\frac{\Delta T}{T}(\mathbf{\hat{n}})=\sum_{l m}a_{l m}Y_{l m}(\mathbf{\hat{n}}),
\end{align}
where 
\begin{align}\label{solidangle}
a_{lm}=\int d\Omega \frac{\Delta T}{T}(\mathbf{\hat{n}})Y_{l m}^*(\mathbf{\hat{n}}).
\end{align}
From equation \eqref{momentexp} we have
\begin{align}\label{tempfluc}
\frac{\Delta T}{T}(\mathbf{k},\mathbf{\hat{n}})
&=\frac{1}{4}\sum_l (-i)^l (2l+1)\Theta_l (k,\tau) P_l (\mu)     \nonumber\\
&= \pi \sum_{l m} (-i)^l \Delta_l (k,\tau) \Phi(\mathbf{k})Y_{l m}(\mathbf{\hat{n}})Y_{l m}^*(\mathbf{\hat{k}}),
\end{align}
where in the second line we use the following expansion of the Legendre polynomial,
\begin{align}
P_l(\mu)=\frac{4\pi}{2 l+1}\sum_m Y_{l m}(\mathbf{\hat{n}})Y_{l m}^*(\mathbf{\hat{k}}).
\end{align}
Evaluating the solid angle integral \eqref{solidangle} gives
\begin{align}\label{almphi}
a_{l m}=4\pi (-i)^l \int \frac{d^3 k}{(2\pi)^3}\Delta_l (k) \Phi(\mathbf{k})Y_{l m}^*(\mathbf{\hat{k}}).
\end{align}
Therefore, we interpret the $a_{lm}$ coefficients as the spherical harmonics of the projection of the gravitational potential $\Phi$ from primordial times onto the cosmic microwave sky by the radiation transfer functions, $\Delta_l(k)$. This method for calculating the CMB anisotropy corresponds to the publicly available codes such as CMBFast \cite{9603033} and CAMB \cite{Challinor}.

\section{Signatures of inflation - CMB}


\subsubsection*{CMB Power Spectrum}
Assuming full sky coverage and statistical isotropy the covariance matrix of $a_{lm}$, $C_{l_1 m_1, l_2 m_2}=\langle a_{l_1 m_1}a_{l_2 m_2}^*\rangle$ is rotationally invariant. Therefore, we define the CMB  power spectrum $C_l$ as follows,
\begin{align}
C_{l_1} \delta_{l_1 l_2}\delta_{m_1 m_2}=\langle a_{l_1 m_1}a^*_{l_2 m_2} \rangle.
\end{align}
Using the relationship between $a_{l m}$ and the gravitational potential $\Phi$, we may relate the primordial power spectrum $P_{\Phi}(k)$ to the CMB power spectrum. We find using equation \eqref{almphi}
\begin{align}\label{cleqn}
C_l&=\frac{(4\pi)^2}{(2\pi)^6}\int d^3 k d^3 k' \Delta_l (k) \Delta_l (k') \langle \Phi(\mathbf{k})\Phi(\mathbf{k}')\rangle Y_{l m}^*(\mathbf{\hat{k}})Y_{l m}(\mathbf{\hat{k}}')\nonumber\\
&=\frac{2}{\pi}\int dk  k^2 |\Delta_l(k)|^2 P_{\Phi}(k).
\end{align}
The power spectrum is related to the variance of the temperature map by the following formula,
\begin{align}\label{variance}
\langle\left( \frac{\Delta T}{T}(\mathbf{\hat{n}}\right)^2\rangle=&\sum_{l_i m_i}\langle a_{l_1 m_1}^*a_{l_2 m_2}\rangle \int \frac{d\Omega}{4\pi}Y_{l_1 m_1}(\mathbf{\hat{n}})Y_{l_2 m_2}^*(\mathbf{\hat{n}})\nonumber\\
=&\frac{\sum_l (2 l+1)C_l}{4\pi}.
\end{align}
We may obtain a prediction for the large angle ($>2^o$) CMB power spectrum induced by slow-roll inflation by substituting in the scale invariant power spectrum $P(k)=\tilde{\Delta}_{\phi}/k^3$ (where we define $\tilde{\Delta}_{\phi}=2\pi^2 {\Delta}_{\phi}$) and approximating the radiation transfer function $\Delta_l(k)$. On such scales the perturbations are still on superhorizon scales at last scattering and therefore are frozen in amplitude. Therefore, we assume the gravitational potentials $\Phi$ and $\Psi$ are constant in time, and neglect the collisional terms in equation \eqref{boltz2}. This equation may be rewritten in the form
\begin{align}
\frac{d}{d\tau}(\Theta+4\Psi)=4\frac{\partial}{\partial \tau}(\Phi+\Psi).
\end{align} 
Therefore on superhorizon scales we assume $\Theta+4\Psi=\rm{constant}$.  Therefore, we deduce that
\begin{align}
\Theta(\tau_{0} )-\Theta(\tau_{\rm{dec}} )=4(\Psi( \tau_{\rm{dec}})-\Psi(\tau_{0} ))=4\Psi(\tau_{\rm{dec}}).
\end{align}
Due to gravitational redshift $\Delta T/T=\Theta/4\propto 1/a$. Therefore,
\begin{align}
\Theta(\tau_{\rm{dec}})=-4\frac{\delta a}{a}=-\frac{8}{3}\frac{\delta t}{t}=-\frac{8}{3}\Psi
\end{align}
where the final equality arises from equation \eqref{FLRWpert}. Hence,
\begin{align}
\frac{\Delta T}{T}=\frac{\Psi}{3}.
\end{align}
This is known as the Sachs Wolfe limit. In addition we note that in the absence of anisotropic stress the Einstein equations imply $\Phi=\Psi$. Taking the Fourier transform and expanding the exponential using equation \eqref{exponexp}, we find
\begin{align}
\frac{\Delta T}{T}(\mathbf{\hat{n}})&=\int \frac{d^3 k}{(2\pi)^3}\frac{\Phi(\mathbf{k})}{3} e^{-i \mathbf{k.\hat{n}}(\tau_0-\tau_{\rm{dec}})}\nonumber\\
&=\int \frac{d^3 k}{(2\pi)^3} \sum_{l m}(-i)^l (2l+1) \frac{\Phi(\mathbf{k})}{3} j_l(k (\tau_0-\tau_{\rm{dec}})) P_l(\mu).
\end{align}
By comparison with equation \eqref{tempfluc} we deduce that in the large angle limit
\begin{align}
\Delta_l(k)=\frac{j_l(k (\tau_0-\tau_{\rm{dec}}))}{3}.
\end{align}
Substituting this expression into equation \eqref{cleqn} for the scale invariant power spectrum we find
\begin{align}
C_l^{\rm{SW}}=\tilde{\Delta}_{\Phi}\frac{2}{ 9\pi}\int \frac{d k}{k}(j_l(k))^2=\frac{\tilde{\Delta}_{\Phi}}{9 \pi}\frac{1}{l (l+1)}.
\end{align}
Hence in the Sachs-Wolfe limit $l(l+1) C_l$ is constant for a scale-free primordial power spectrum. This approximation is reasonably accurate for $l\lesssim 60$. Nevertheless, it should be noted that, due to a contribution from the integrated Sachs-Wolfe effect $\int (\Phi'+\Psi' )d\tau$, where the gravitational potential is time varying, the CMB power spectrum is altered somewhat on such scales from the Sachs-Wolfe approximation.
\par
On subhorizon scales, it is necessary to calculate the full radiation transfer function to account for oscillations in the photon-baryon fluid on or before decoupling. After decoupling the photon-baryon fluid no longer oscillates. Soon after, at recombination the photons free-stream from the last scattering surface. Therefore, the CMB is expected to reflect the sound horizon scale of the photon-baryon fluid at last scattering. In particular, the extrema of the modes form a harmonic series based on the sound horizon scale with the first mode representing the first compression mode formed from the photon-baryon fluid moving from an under-dense to over-dense region. In turn the second peak represents the first mode that has compressed and then rarefied to an expansive maximum, etc. This succession of peaks are known as the Doppler peaks. Due to dissipation as the photons make a random walk through the baryons during recombination the acoustic peaks are damped exponentially on scales smaller than this photon diffusion scale. This effect is commonly known as Silk damping.

\subsubsection*{CMB Bispectrum}
As detailed earlier, motivation for the study of the higher order correlators, such as the bispectrum and trispectrum, arises for the prospect of distinguishing more complex models of inflation which can produce non-Gaussianity. In order to measure deviations from Gaussianity, a measure of the skewness and kurtosis must be performed. While the power spectrum corresponds to a measure of the variance \eqref{variance}, the bispectrum and trispectrum can be shown to correspond to measures of the skewness and kurtosis respectively.
\par
The CMB bispectrum is defined to be the three-point correlator of the $a_{l m}$, 
\begin{align}
B^{l_1 l_2 l_3}_{m_1 m_2 m_3}=\langle a_{l_1 m_1}a_{l_2 m_2}a_{l_3 m_3}\rangle.
\end{align} 
Substituting in the expression for the $a_{l m}$ in equation \eqref{almphi} we find
\begin{align}\label{bispLs}
B^{l_1 l_2 l_3}_{m_1 m_2 m_3}=&\frac{(4\pi)^3}{(2\pi)^9}(-i)^{l_1 +l_2 +l_3}\int d^3 k_1 d^3 k_2 d^3 k_3 \Delta_{l_1 }(k_1) \Delta_{l_2 }(k_2)\Delta_{l_3 }(k_3)\nonumber\\
&\times \langle \Phi(\mathbf{k}_1)\Phi(\mathbf{k}_2)\Phi(\mathbf{k}_3)\rangle Y^*_{l_1 m_1}(\mathbf{\hat{k}}_1)Y^*_{l_2 m_2}(\mathbf{\hat{k}}_2)Y^*_{l_3 m_3}(\mathbf{\hat{k}}_3).
\end{align}
We recall that the primordial bispectrum is given by
\begin{align}
 \langle \Phi(\mathbf{k}_1)\Phi(\mathbf{k}_2)\Phi(\mathbf{k}_3)\rangle=(2\pi)^3 \delta(\mathbf{k}_1+\mathbf{k}_2+\mathbf{k}_3)B_{\Phi}(k_1,k_2,k_3).
\end{align}
Due to momentum conservation the wavevectors must form a closed triangle, as imposed by the delta function. This closure condition implies that the bispectrum may be expressed as a function of its wavenumbers only. Note also that the delta function may be written in the following forms,
\begin{align}\label{deltaexp}
\delta(\mathbf{k}_1+\mathbf{k}_2+\mathbf{k}_3)&=\int \frac{d^3 x}{(2\pi)^3}e^{i(\mathbf{k}_1+\mathbf{k}_2+\mathbf{k}_3).\mathbf{x}}\nonumber\\
&=8\sum_{l_i m_i}i^{l_1+l_2+l_3}\left(\int dx x^2 j_{l_1}(k_1 x)j_{l_2}(k_2 x)j_{l_3}(k_3 x)\right)\nonumber\\
&\times Y_{l_1 m_1}(\mathbf{\hat{k}}_1)Y_{l_2 m_2}(\mathbf{\hat{k}}_2)Y_{l_3 m_3}(\mathbf{\hat{k}}_3)\int d\Omega_{\mathbf{\hat{x}}} Y^*_{l_1 m_1}(\mathbf{\hat{x}})Y^*_{l_2 m_2}(\mathbf{\hat{x}})Y^*_{l_3 m_3}(\mathbf{\hat{x}}),
\end{align}
where in the second line we use the expansion of the exponential \eqref{exponexp}.
\par
Substituting these expressions into \eqref{bispLs} reveals
\begin{align}\label{bispLs}
B^{l_1 l_2 l_3}_{m_1 m_2 m_3}=&\left(\frac{2}{\pi}\right)^3 \int  dx d k_1 d k_2 d k_3 (x k_1 k_2 k_3)^2\Delta_{l_1 }(k_1) \Delta_{l_2 }(k_2)\Delta_{l_3 }(k_3)B_{\Phi}(k_1,k_2,k_3)\nonumber\\
&\times j_{l_1}(k_1 x)j_{l_2}(k_2 x)j_{l_3}(k_3 x)\int d\Omega_{\mathbf{\hat{x}}} Y^*_{l_1 m_1}(\mathbf{\hat{x}})Y^*_{l_2 m_2}(\mathbf{\hat{x}})Y^*_{l_3 m_3}(\mathbf{\hat{x}}).
\end{align}
This expression is simplified further by noting that the Gaunt integral is given by
\begin{align}\label{Gauntexp}
\mathcal{G}^{l_1 l_2 l_3}_{m_1 m_2 m_3}=\int d\Omega_{\mathbf{\hat{x}}} Y^*_{l_1 m_1}(\mathbf{\hat{x}})Y^*_{l_2 m_2}(\mathbf{\hat{x}})Y^*_{l_3 m_3}(\mathbf{\hat{x}})=h_{l_1 l_2 l_3}\left( \begin{array}{ccc}
l_1 & l_2 & l_3 \\
m_1 & m_2 & m_3 \end{array} \right),
\end{align}
where
$
\left( \begin{array}{ccc}
l_1 & l_2 & l_3 \\
m_1 & m_2 & m_3 \end{array} \right)
$
denotes the Wigner $3$j symbol and 
\begin{align}
h_{l_1 l_2 l_3}=\sqrt{\dfrac{(2l_1+1)(2l_2+1)(2l_3+1)}{4\pi}}\left( \begin{array}{ccc}
l_1 & l_2 & l_3 \\
0 & 0 & 0 \end{array} \right).
\end{align}
The Gaunt integral is the analog of the Dirac delta symbol in multipole space, imposing constraints on the multipoles $l_i$. Defining the reduced bispectrum, $b_{l_1 l_2 l_3}$, as
\begin{align}
B^{l_1 l_2 l_3}_{m_1 m_2 m_3}=\mathcal{G}^{l_1 l_2 l_3}_{m_1 m_2 m_3}b_{l_1 l_2 l_3},
\end{align}
 we have
 \begin{align}\label{redBisp}
 b_{l_1 l_2 l_3}=&\tilde{\Delta}_{\Phi}^2\left(\frac{2}{\pi}\right)^3 \int  d k_1 d k_2 d k_3  \Delta_{l_1 }(k_1) \Delta_{l_2 }(k_2)\Delta_{l_3 }(k_3)S(k_1,k_2,k_3)\nonumber\\
 &\times \int dx x^2 j_{l_1}(k_1 x) j_{l_2}(k_2 x)j_{l_3}(k_3 x),
 \end{align}
where we define the shape function 
\begin{align}
S(k_1,k_2,k_3)=\frac{(k_1 k_2 k_3)^2}{\tilde{\Delta}_{\Phi}^2} B_{\Phi}(k_1,k_2,k_3).
\end{align}
The shape function is the dimensionless form of the bispectrum. Substituting the primordial local bispectrum into this formula, \eqref{LocalBisp}, we find
\begin{align}
b_{l_1 l_2 l_3}^{\rm{local}}=2 f_{NL}^{\rm{local}}\int x^2 dx \left(\alpha_{l_1}(x)\beta_{l_2}(x)\beta_{l_3}(x)+2\,\rm{permutations}	\right),
\end{align}
where
\begin{align}
\alpha_l(x)&=\frac{2}{\pi}\int dk k^2 \Delta_l(k) j_l(k x), \\
\beta_l(x)&=\frac{2}{\pi}\int dk k^2 P_{\Phi}(k)\Delta_l(k) j_l(k x), 
\end{align}
We may approximate this in the Sachs Wolfe limit for which $\Delta_l(k)\approx j_l(k x_{\rm{dec}})/3$ to find
\begin{align}
b_{l_1 l_2 l_3}^{\rm{local}\, SW}=\frac{2 f_{NL}^{\rm{local}}\tilde{\Delta}_{\Phi}^2}{27 \pi^2} \left(\frac{1}{l_1(l_1+1)l_2 (l_2+1)}+\frac{1}{l_1(l_1+1)l_3 (l_3+1)}+\frac{1}{l_2(l_2+1)l_3 (l_3+1)}	\right).
\end{align}
Similarly for the constant model for which $S(k_1,k_2,k_3)=1$ we find
\begin{align}
b_{l_1 l_2 l_3}^{\rm{constant}\, SW}=\frac{\tilde{\Delta}_{\Phi}^2}{27 } \frac{1}{(2l_1+1) (2l_2+1)(2l_3+1)}\left(\frac{1}{l_1+l_2+l_3+3}+\frac{1}{l_1+l_2+l_3}	\right).
\end{align}
This solution is used commonly as a benchmark with which to compare the reduced CMB bispectra of other models. Many other models have been studied in the literature such as the equilateral model \cite{0405356}, the warm model \cite{MossXiong2007}, etc. (for a review see \cite{LigSef2010}). 
\par
As is clear from equation $\eqref{redBisp}$, analysis of the CMB bispectrum becomes much simpler given a separable shape function, i.e. a shape function of the form $S(k_1,k_2,k_3)=X(k_1)Y(k_2)Z(k_3)+\mathrm{perms}$. In a similar fashion to the above formulae for the local model, such shapes reduce the dimension of integration from four dimensions to two dimensions, thus greatly reducing the computational time required. In \cite{FLS09} a general method for decomposing a shape function into a sum of a products of separable functions has been established. In particular, a (scale-invariant) shape function may be decomposed in the form
\begin{align}
S(k_1,k_2,k_3)=\sum_{p r s}\alpha_{p r s}q_p(k_1)q_r(k_2)q_s(k_3),
\end{align}
where the $q_p$ are basis mode functions spanning the space of all functions on the bispectrum wavenumber domain, $\mathcal{V}_k$, given by
\begin{align}
&k_1\leq k_2+k_3\,\,\rm{for} \,\, k_1\geq k_2,k_3,\quad \rm{or}\,\,k_2\leq k_1+k_3\,\,\rm{for} \,\, k_2\geq k_1,k_3,\nonumber\\
&\rm{or}\,\, k_3\leq k_1+k_2\,\,\rm{for} \,\, k_3\geq k_1,k_2.
\end{align}
 We may introduce a partial ordering on the triples $\{p r s\}$ and write
$S(k_1,k_2,k_3)=\sum_{n(=\{p r s\})}\alpha_{n}q_{\{p}(k_1)q_r(k_2)q_{s\}}(k_3)=\sum_n \alpha_n Q_n(k_1,k_2,k_3)$ where we define the product of the $q$s to be $Q_n$. An inner product may be defined on the wavenumber domain as
\begin{align}
\langle f g \rangle=\int_{\mathcal{V}_k}d\mathcal{V}_k  f(k_1,k_2,k_3)g(k_1,k_2,k_3)\omega(k_1,k_2,k_3),
\end{align}
where $\omega$ is an appropriate weight function. Using this inner product we define the matrix $\Gamma$ such that
\begin{align}
\gamma_{n m}=\langle Q_n Q_m\rangle.
\end{align}
We may evaluate the $\alpha_n$ coefficients using 
\begin{align}
\langle S(k_1,k_2,k_3) Q_m (k_1,k_2,k_3)\rangle&=\sum_n \alpha_n\langle Q_n (k_1,k_2,k_3)Q_m (k_1,k_2,k_3) \rangle=\sum_n \alpha_n \gamma_{n m},\nonumber\\
\implies \alpha_n&=\sum_m\langle S(k_1,k_2,k_3) Q_m (k_1,k_2,k_3)\rangle \gamma_{m n}^{-1}.
\end{align}
Of course the accuracy of the expansion is limited by the number of modes chosen, $n_{max}$. The accuracy of the expansion may be evaluated by forming the correlator
\begin{align}
\mathcal{C}(S,S')=\frac{\langle S S'\rangle}{\sqrt{\langle S S\rangle\langle S' S'\rangle}}
\end{align}
where $S'=\sum_{n=0}^{n_{max}}\alpha_n Q_n$. For the models considered so far only $\mathcal{O}(30)$ modes are required to achieve an accuracy of at least $90-95\%$ \cite{FLS09}.
\par
Given the expansion $S(k_1,k_2,k_3)=\sum_n \alpha_n Q_n(k_1,k_2,k_3)$ the reduced CMB bispectrum becomes
\begin{align}
b_{l_1 l_2 l_3}=\tilde{\Delta}_{\Phi}^2\sum_n \alpha_n \int dx x^2 q^{l_1}_{\{p}(x)q^{l_2}_r(x)q^{l_3}_{s\}}(x),
\end{align}
where
\begin{align}
q_p^l(x)=\frac{2}{\pi}\int dk q_p(k) \Delta_l(k) j_l(k x).
\end{align}
Hence, the separable expansion allows for the reduction in complexity from a four-dimensional integral to a two-dimensional integral given any primordial model. It should be noted that the separable expansion may also be applied to late-time models. For such models the bispectrum does require a line-of-sight integral ($\int dx$). 
\par
Since the CMB bispectrum signal is too weak to measure individual multipoles directly, it is necessary to define an estimator which sums over all multipoles in order to compare theory and observational data. This estimator is given by \cite{Babich}
\begin{align}
\mathcal{E}=\sum_{l_i m_i} b_{l_1 l_2 l_3}\mathcal{G}^{l_1 l_2 l_3}_{m_1 m_2 m_3}\left((C^{-1}a^{\rm{obs}})_{l_1 m_1}(C^{-1}a^{\rm{obs}})_{l_2 m_2}(C^{-1}a^{\rm{obs}})_{l_3 m_3}-3 C^{-1}_{l_1 m_1,l_2 m_2}a_{l_3 m_3}\right),
\end{align}
where $C^{-1}_{l_1 m_1,l_2 m_2}$ denotes the inverse covariance matrix, which is now non-diagaonal because of mode-mode coupling due to the mask and anisotropic noise. For the sake of simplicity we consider the ideal case where the covariance matrix is diagonal and where there is full sky coverage and no anisotropic noise. In particular, we consider
\begin{align}
\mathcal{E}=\sum_{l_i m_i} b_{l_1 l_2 l_3}\mathcal{G}^{l_1 l_2 l_3}_{m_1 m_2 m_3}\frac{a_{l_1 m_1}^{\rm{obs}}a_{l_2 m_2}^{\rm{obs}}a_{l_3 m_3}^{\rm{obs}}}{C_{l_1}C_{l_2} C_{l_3}}.
\end{align}
The expectation value of the estimator is given by
\begin{align} 
\langle \mathcal{E}\rangle=\sum_{l_i m_i}\frac{(b_{l_1 l_2 l_3}\mathcal{G}^{l_1 l_2 l_3}_{m_1 m_2 m_3})^2}{C_{l_1}C_{l_2} C_{l_3}}.
\end{align}
Using the identity
\begin{align}\label{orthogWigner}
\sum_{m_i}\left( \begin{array}{ccc}
l_1 & l_2 & l_3 \\
m_1 & m_2 & m_3 \end{array} \right)^2=1,
\end{align}
and equation \eqref{Gauntexp} we find
\begin{align}
\langle \mathcal{E}\rangle=\sum_{l_i}\frac{(b_{l_1 l_2 l_3}h_{l_1 l_2 l_3})^2}{C_{l_1}C_{l_2} C_{l_3}}.
\end{align}
Using this estimator we may define an integrated measure $F_{NL}$
\begin{align}
F_{NL}=\frac{\mathcal{E}}{\sqrt{\langle \mathcal{E}\rangle \langle \mathcal{E}^{\rm{local}, f_{NL}=1}\rangle}}
\end{align}
where $\mathcal{E}^{\rm{local}, f_{NL}=1}$ is the estimator corresponding to the local model with $f_{NL}=1$. For the local model $F_{NL}$ is precisely $f_{NL}^{\rm{local}}$. However, the advantage of this integrated measure is that it allows for a fair comparison between different models unlike the parameter $f_{NL}$ which is defined at a single point in the wavenumber domain. This is summarised in Table~\ref{table2}.
\begin{table}
\vskip-5pt
\centering
\begin{tabular}{|c||c||c|}
 \hline

    Model  &            $F_{NL}$    &                $f_{NL}$  \\
    \hline
Constant &$35.1\pm 27.4$ &$149.4\pm 116.8$ \\
DBI & $26.7\pm 26.5$&$146.0 \pm 144.5$\\
Equilateral & $25.1\pm 26.4$&  $143.5\pm 151.2$\\
Flat (Smoothed) &$ 35.4\pm 29.2$&$ 18.1 \pm 14.9 $\\
Ghost & $22.0\pm 26.3$& $138.7 \pm 165.4$\\
Local & $54.4\pm 29.4$&$ 54.4\pm 29.4$\\
Orthogonal &$ -16.3\pm 27.3$&$ -79.4\pm 133.3$\\
Single & $28.8 \pm 26.6$& $142.1\pm 131.3$\\
Warm & $24.2\pm 27.3$ & $94.7\pm 106.8$ \\
\hline
  
  \end{tabular}
  \caption{Limits on non-Gaussianity for all known scale-invariant models \cite{FLS10}. The table establishes that the integrated measure $F_{NL}$ allows for a fairer comparison between models than the parameter $f_{NL}$.}
\label{table2}
\end{table}
\par
For completeness we write down the formula for the estimator for a general bispectrum using the separable expansion. In the ideal case the estimator may be written in the form
\begin{align}
\mathcal{E}=\tilde{\Delta}_{\Phi}^2\sum_n \alpha_n \int d\Omega_{\mathbf{\hat{n}}}\int x^2 dx M_{\{p}(\mathbf{\hat{n}},x)M_{r}(\mathbf{\hat{n}},x)M_{s\}}(\mathbf{\hat{n}},x),
\end{align}
where the filtered map $M_{p}(\mathbf{\hat{n}},x)$ is given by
\begin{align}
M_{p}(\mathbf{\hat{n}},x)=\sum_{l m}\frac{a_{l m}Y_{l m}(\mathbf{\hat{n}})}{C_l}q_p^l(x).
\end{align}
Writing the integrated quantity as $\beta_n$ we have succinctly $\mathcal{E}=\tilde{\Delta}_{\Phi}^2\sum_n \alpha_n \beta_n$.
\par
We finally establish the relationship between the bispectrum and the skewness, $g_1$, of the temperature map
\begin{align}
g_1=\frac{\Bigg\langle\left(\frac{\Delta T}{T}(\hat{n}) \right)^3\Bigg\rangle}{\left(\Bigg\langle \left(\frac{\Delta T}{T}(\hat{n}) \right)^2\Bigg\rangle\right)^{3/2}}.
\end{align}
The three-point temperature correlator is given by
\begin{eqnarray}
\Bigg\langle\left(\frac{\Delta T}{T}(\hat{n}) \right)^3\Bigg\rangle&=&\sum_{l_i m_i}\langle a_{l_1 m_1}a_{l_2 m_2}a_{l_3 m_3}\rangle \int \frac{d\Omega_{\hat{n}}}{4\pi} Y_{l_1 m_1}(\hat{\mathbf{n}})Y_{l_2 m_2}(\hat{\mathbf{n}})Y_{l_3 m_3}(\hat{\mathbf{n}})\nonumber\\
&=&\frac{1}{4\pi}\sum_{l_i m_i}\left(\mathcal{G}^{l_1 l_2 l_3}_{m_1 m_2 m_3}\right)^2 b_{l_1 l_2 l_3}.
\end{eqnarray}
Using equation \eqref{Gauntexp} and \eqref{orthogWigner} this may be written as
\begin{eqnarray}
\Bigg\langle\left(\frac{\Delta T}{T}(\hat{n}) \right)^3\Bigg\rangle&=&\frac{1}{4\pi}\sum_{l_i} h_{l_1 l_2 l_3}^2 b_{l_1 l_2 l_3}.
\end{eqnarray}
Therefore from equation \eqref{variance} we have
\begin{align}
g_1=\sqrt{4\pi}\frac{\sum_{l_i} h_{l_1 l_2 l_3}^2 b_{l_1 l_2 l_3}}{\left( \sum_l(2l+1)C_l\right)^{3/2}}.
\end{align}

\subsubsection*{CMB Trispectrum}

The CMB trispectrum is given by the connected four-point correlator of the $a_{lm}$, \cite{Okamoto}
\begin{align}
\langle a_{l_1 m_1}a_{l_2 m_2}a_{l_3 m_3}a_{l_4 m_4}\rangle_c=&\langle a_{l_1 m_1}a_{l_2 m_2}a_{l_3 m_3}a_{l_4 m_4}\rangle-\langle a_{l_1 m_1}a_{l_2 m_2}\rangle\langle a_{l_3 m_3}a_{l_4 m_4}\rangle\nonumber\\
&-\langle a_{l_1 m_1}a_{l_4 m_4}\rangle\langle a_{l_2 m_2}a_{l_3 m_3}\rangle-\langle a_{l_1 m_1}a_{l_3 m_3}\rangle\langle a_{l_2 m_2}a_{l_4 m_4}\rangle\nonumber\\
=&\sum_{L M}(-1)^M \left( \begin{array}{ccc}
l_1 & l_2 & L \\
m_1 & m_2 & -M \end{array} \right)\left( \begin{array}{ccc}
l_3 & l_4 & L \\
m_3 & m_4 & M \end{array} \right)T^{l_1 l_2}_{l_3 l_4}(L),
\end{align}
where isotropy ensures that the dependency on $m_i$ can be separated out. The trispectrum is especially interesting given its use in the study of gravitational lensing in the CMB \cite{0105117,Hanson}.
\par
The connected component indicates the four-point correlator minus the contribution due to two-point correlators. The quantity $T^{l_1 l_2}_{l_3 l_4}(L)$ is generally referred to as the CMB trispectrum. This quantity may be expanded into three pairings according to pairings of the multipoles $l_i$ which form different diagonals, i.e. the pairings $(l_1,l_2), (l_3,l_4)$ or $(l_1,l_3), (l_2,l_4)$ or $(l_1,l_4), (l_2,l_3)$. The four-point function can be expanded into these three pairings as
\begin{align}
\langle a_{l_1 m_1}a_{l_2 m_2}a_{l_3 m_3}a_{l_4 m_4}\rangle_c=&\sum_{L M}(-1)^M \left( \begin{array}{ccc}
l_1 & l_2 & L \\
m_1 & m_2 & -M \end{array} \right)\left( \begin{array}{ccc}
l_3 & l_4 & L \\
m_3 & m_4 & M \end{array} \right)P^{l_1 l_2}_{l_3 l_4}(L)\nonumber\\
&+(l_2\leftrightarrow l_3)+(l_2\leftrightarrow l_4).
\end{align}
There are in total $24$ permutations of the $l_i$. These remaining constraints are  enforced by the reduced function $\mathcal{T}^{l_1 l_2}_{l_3 l_4}(L)$ with
\begin{align}
P^{l_1 l_2}_{l_3 l_4}(L)=\mathcal{T}^{l_1 l_2}_{l_3 l_4}(L)+(-1)^{l_1+l_2+L}\mathcal{T}^{l_2 l_1}_{l_3 l_4}(L)+(-1)^{l_3+l_4+L}\mathcal{T}^{l_1 l_2}_{l_4 l_3}(L)+(-1)^{\sum_i l_i}\mathcal{T}^{l_2 l_1}_{l_4 l_3}(L),
\end{align}
where $\mathcal{T}^{l_1 l_2}_{l_4 l_3}(L)=\mathcal{T}^{l_3 l_4}_{l_1 l_2}(L)$. Defining the 'extra-reduced' trispectrum, $t^{l_1 l_2}_{l_3 l_4}(L)$ as\footnote{It should be noted that there is no ambiguity in the definition of the `extra-reduced' trispectrum in the case $l_1+l_2+L=\mathrm{odd}$. As shall be described further in Chapter~\ref{chapter:methodology}, the `extra-reduced' trispectrum is  defined to factor out terms of the form $h_{l_1 l_2 L}h_{l_3 l_4 L}$ present in the definition of $\mathcal{T}^{l_3 l_4}_{l_1 l_2}(L)$. }
\begin{align}\label{definextrared}
t^{l_1 l_2}_{l_3 l_4}(L)=\frac{\mathcal{T}^{l_1 l_2}_{l_4 l_3}(L)}{h_{l_1 l_2 L}h_{l_3 l_4 L}},
\end{align}
we may finally write
\begin{align}\label{trispectrtotal}
\langle a_{l_1 m_1}a_{l_2 m_2}a_{l_3 m_3}a_{l_4 m_4}\rangle_c=&\sum_{L M}(-1)^M \mathcal{G}^{l_1 l_2 L}_{m_1 m_2 -M}\mathcal{G}^{l_3 l_4 L}_{m_3 m_4 M}t^{l_1 l_2}_{l_3 l_4}(L)+11\rm{\,permutations},
\end{align}
where the Gaunt integral is given in \eqref{Gauntexp}.
\par
The CMB extra-reduced trispectrum $t^{l_1 l_2}_{l_3 l_4}(L)$ may be related to the reduced primordial trispectrum $\mathcal{T}_{\Phi}$, \eqref{TrispDefinition}, where we assume $\mathcal{T}_{\Phi}=\mathcal{T}_{\Phi}(k_1,k_2,k_3,k_4,K)$ using the relation between $a_{ lm}$ and $\Phi$ in equation \eqref{almphi}. We find the contribution to the four-point correlator of the $a_{l m}$ given by the extra-reduced trispectrum $t^{l_1 l_2}_{l_3 l_4}(L)$ is given by
\begin{align}
\sum_{L M}(-1)^M &\mathcal{G}^{l_1 l_2 L}_{m_1 m_2 -M} \mathcal{G}^{l_3 l_4 L}_{m_3 m_4 M}t^{l_1 l_2}_{l_3 l_4}(L)=(4\pi)^4(-i)^{\sum l_i}\int \frac{d^3 k_1}{(2\pi)^3}\frac{d^3 k_2}{(2\pi)^3}\frac{d^3 k_3}{(2\pi)^3}\frac{d^3 k_4}{(2\pi)^3}d^3 K \nonumber\\
&\times Y^*_{l_1 m_1}(\mathbf{\hat{k}}_1)Y^*_{l_2 m_2}(\mathbf{\hat{k}}_2)Y^*_{l_3 m_3}(\mathbf{\hat{k}}_3)Y^*_{l_4 m_4}(\mathbf{\hat{k}}_4)\Delta_{l_1}(k_1)\Delta_{l_2}(k_2)\Delta_{l_3}(k_3)\Delta_{l_4}(k_4)\nonumber\\
&\times (2\pi)^3 \mathcal{T}_{\Phi}(k_1,k_2,k_3,k_4,K)\delta(\mathbf{k}_1+\mathbf{k}_2+\mathbf{K})\delta(\mathbf{k}_3+\mathbf{k}_4-\mathbf{K}).
\end{align}
This implies, using the expansion of the delta function \eqref{deltaexp}, that
\begin{align}
t^{l_1 l_2}_{l_3 l_4}(L)=&\left( \frac{2}{\pi}\right)^5 \int dr_1 dr_2 r_1^2 r_2^2 dk_1 dk_2 dk_3 dk_4 dK(k_1k_2 k_3 k_4 K)^2 j_L(Kr_1) j_L(K r_2)\nonumber\\
&\times [\Delta_{l_1}(k_1) j_{l_1}(k_1 r_1)] [\Delta_{l_2}(k_2) j_{l_2}(k_2 r_1)] [\Delta_{l_3}(k_3) j_{l_3}(k_3 r_2)] [\Delta_{l_4}(k_4) j_{l_4}(k_4 r_2)]\nonumber\\
&\times \mathcal{T}_{\Phi}(k_1,k_2,k_3,k_4,K).
\end{align}
As an example we substitute the primordial local reduced trispectrum into this formula, \eqref{LocalTrisp}, and find
\begin{align}
t^{l_1 l_2}_{l_3 l_4}(L)^{\rm{local}}=&\frac{25}{9}\tau_{NL}^{\rm{local}}\int r_1^2 r_2^2 dr_1 dr_2 F_L(r_1,r_2)\alpha_{l_1}(r_1)\beta_{l_2}(r_1)\alpha_{l_3}(r_2)\beta_{l_4}(r_2)\nonumber\\
&+g_{NL}^{\rm{local}}\int r^2 dr \beta_{l_2}(r)\beta_{l_4}(r)\left[ \alpha_{l_1}(r)\beta_{l_3}(r)+\beta_{l_1}(r)\alpha_{l_3}(r)\right],
\end{align}
where we recall
\begin{align}
\alpha_l(x)&=\frac{2}{\pi}\int dk k^2 \Delta_l(k) j_l(k x), \\
\beta_l(x)&=\frac{2}{\pi}\int dk k^2 P_{\Phi}(k)\Delta_l(k) j_l(k x).
\end{align}
Finally, we relate the CMB trispectrum to the kurtosis, $g_2$, of the temperature map,
\begin{align}
g_2=\frac{\Bigg\langle\left(\frac{\Delta T}{T}(\hat{n}) \right)^4\Bigg\rangle_c}{\left(\Bigg\langle \left(\frac{\Delta T}{T}(\hat{n}) \right)^2\Bigg\rangle\right)^{2}}.
\end{align}
In order to evaluate this quantity we calculate the following,
\begin{align}
K=&\Bigg\langle\left(\frac{\Delta T}{T}(\hat{n}) \right)^4\Bigg\rangle-3\left(\Bigg\langle \left(\frac{\Delta T}{T}(\hat{n}) \right)^2\Bigg\rangle\right)^{2}\nonumber\\
=&\sum_{l_i m_i}\langle a_{l_1 m_1}a_{l_2 m_2}a_{l_3 m_3}a_{l_4 m_4}\rangle \int \frac{d\Omega_{\hat{n}}}{4\pi} Y_{l_1 m_1}(\hat{\mathbf{n}})Y_{l_2 m_2}(\hat{\mathbf{n}})Y_{l_3 m_3}(\hat{\mathbf{n}})Y_{l_4 m_4}(\hat{\mathbf{n}})\nonumber\\
&-3\sum_{l_i m_i}\int \frac{d\Omega_{\hat{n}_1}}{4\pi}  \frac{d\Omega_{\hat{n}_2}}{4\pi}\langle a_{l_1 m_1}a_{l_2 m_2}\rangle \langle a_{l_3 m_3}a_{l_4 m_4}\rangle Y_{l_1 m_1}(\hat{\mathbf{n}}_1)Y_{l_2 m_2}(\hat{\mathbf{n}}_1)Y_{l_3 m_3}(\hat{\mathbf{n}}_2)Y_{l_4 m_4}(\hat{\mathbf{n}}_2).
\end{align}
Using
\begin{align}
\langle a_{l_1 m_1}a_{l_2 m_2}a_{l_3 m_3}a_{l_4 m_4}\rangle=&\langle a_{l_1 m_1}a_{l_2 m_2} a_{l_3 m_3}a_{l_4 m_4}\rangle_c+\langle a_{l_1 m_1}a_{l_2 m_2}\rangle \langle a_{l_3 m_3}a_{l_4 m_4}\rangle\nonumber\\
&+\langle a_{l_1 m_1}a_{l_3 m_3}\rangle \langle a_{l_2 m_2}a_{l_4 m_4}\rangle+\langle a_{l_1 m_1}a_{l_4 m_4}\rangle \langle a_{l_2 m_2}a_{l_3 m_3}\rangle,
\end{align}
it can be shown after some algebra that
\begin{align}
K=\frac{1}{4\pi}\sum_{l_i m_i}\langle a_{l_1 m_1}a_{l_2 m_2}a_{l_3 m_3}a_{l_4 m_4}\rangle_c \int d\Omega_{\hat{n}} Y_{l_1 m_1}(\hat{\mathbf{n}})Y_{l_2 m_2}(\hat{\mathbf{n}})Y_{l_3 m_3}(\hat{\mathbf{n}})Y_{l_4 m_4}(\hat{\mathbf{n}}).
\end{align}
From equation \eqref{trispectrtotal} we find
\begin{align}
K=\frac{12}{4\pi}\sum_{l_i m_i} \sum_{L M}(-1)^M \mathcal{G}^{l_1 l_2 L}_{m_1 m_2 -M}\mathcal{G}^{l_3 l_4 L}_{m_3 m_4 M}t^{l_1 l_2}_{l_3 l_4}(L) \int d\Omega_{\hat{n}} Y_{l_1 m_1}(\hat{\mathbf{n}})Y_{l_2 m_2}(\hat{\mathbf{n}})Y_{l_3 m_3}(\hat{\mathbf{n}})Y_{l_4 m_4}(\hat{\mathbf{n}}).
\end{align}
Next, using
\begin{align}\label{eq:UsefulIdentity}
\int d\Omega_{\hat{n}} Y_{l_1 m_1}(\hat{\mathbf{n}})Y_{l_2 m_2}(\hat{\mathbf{n}})Y_{l_3 m_3}(\hat{\mathbf{n}})Y_{l_4 m_4}(\hat{\mathbf{n}})=\sum_{L' M'}(-1)^{M'}\mathcal{G}^{l_1 l_2 L'}_{m_1 m_2 -M'}\mathcal{G}^{l_3 l_4 L'}_{m_3 m_4 M'}
\end{align}
and the orthogonality of the Wigner 3j symbols
\begin{align}
\sum_{m_1 m_2}(2L+1)\left( \begin{array}{ccc}
l_1 & l_2 & L \\
m_1 & m_2 & M \end{array} \right)\left( \begin{array}{ccc}
l_1 & l_2 & L' \\
m_1 & m_2 & M' \end{array} \right)=\delta_{L L'}\delta_{M M'},
\end{align}
 we obtain the result,
\begin{align}
K=\frac{12}{4\pi}\sum_{l_i L}\frac{h_{l_1 l_2 L}^2h_{l_3 l_4 L}^2}{2L+1}t^{l_1 l_2}_{l_3 l_4}(L).
\end{align}
Hence, the kurtosis is given by
\begin{align}
g_2=\frac{48\pi \sum_{l_i L} h_{l_1 l_2 L}^2h_{l_3 l_4 L}^2t^{l_1 l_2}_{l_3 l_4}(L)/(2L+1)  }{(\sum_l(2l+1)C_l)^2}.
\end{align}
It should be noted that in order to test for non-Gaussianity it is necessary to measure both the bispectrum and trispectrum. Aside from distinguishing models with similar bispectra using their respective trispectra, it is possible that a particular model may possess a symmetry forcing the bispectrum to zero while having a large trispectrum.

\section{Signatures of inflation - Large Scale Structure}
The distribution of matter, as observed via galaxy distributions, is highly non-Gaussian due to the non-linear processes involved in structure formation. This gravitational instability induces a non-Gaussian field even for Gaussian initial conditions. In addition the bias relation between galaxy and matter distributions gives another source of non-Gaussianity. In order to test models of inflation using large scale structure it is, therefore, necessary to distinguish the contribution due to non-Gaussian initial conditions from these other contributions. This necessitates the development of very accurate cosmological perturbation analysis. Despite these difficulties, large scale structure offers a potentially big advantage over the CMB. Galaxy surveys, having a redshift dependence, offer a three-dimensional picture of the density perturbations. The CMB on the other hand, as a two-dimensional object, presents only a snapshot of data. Recent advances in the area have indicated the possibility of reliably testing primordial initial 
conditions using large scale structure (for a review see \cite{DS,LigSef2010}).
\par
The matter density perturbations are related to the primordial fluctuations by the Poisson equation via the expression
\begin{align}
\delta_{\mathbf{k}}(a)=M(k;a)\Phi_{\mathbf{k}},
\end{align}
where $a$ is the scale factor and $M(k;a)$ is given by
\begin{align}
M(k;a)=-\frac{3}{5}\frac{k^2 T(k)}{\Omega_m H_0^2}D_+(a),
\end{align}
where $T(k)$ is the matter transfer function, $D_+(a)$ is the growth factor in linear perturbation theory, $\Omega_m$ is the present value of the dark matter density and $H_0$ is the present value of the Hubble constant. The transfer function is defined as
\begin{align}
T(k)=\frac{\delta_{k}(a_0)}{\delta_{k}(a) D_+(a)}.
\end{align}
Thus, the n-point correlator of the correlation function of matter density perturbations at a given value of the scale factor is given by
\begin{align}
\langle \delta_{\mathbf{k}_1}(a)\delta_{\mathbf{k}_2}(a)\dots \delta_{\mathbf{k}_n}(a)\rangle_c=\left(\Pi_{i=1}^n M(k_i;a)\right)\langle \Phi_{\mathbf{k}_1}\Phi_{\mathbf{k}_2}\dots \Phi_{\mathbf{k}_n}\rangle_c.
\end{align}
The matter power spectrum, bispectrum and trispectrum arising due to the primordial bispectrum are  therefore given by
\begin{align}
P_I(k;a)&=M(k;a)^2 P_{\Phi}(k),\\
B_I(k_1,k_2,k_3;a)&=M(k_1;a)M(k_2;a)M(k_3;a)B_{\Phi}(k_1,k_2,k_3),\\
T_I(\mathbf{k}_1,\mathbf{k}_2,\mathbf{k}_3,\mathbf{k}_4;a)&=M(k_1;a)M(k_2;a)M(k_3;a)M(k_4;a)T_{\Phi}(\mathbf{k}_1,\mathbf{k}_2,\mathbf{k}_3,\mathbf{k}_4).
\end{align}
The contribution to the matter power spectrum and bispectrum due to gravitational collapse are given - at second order with respect to perturbations in $\delta$ - respectively by \cite{FRY,Catelan},
\begin{align}
P_{NG}(k;a)=&2\int \frac{d^3 \mathbf{y}}{(2\pi)^3}B_{I}(\mathbf{k},\mathbf{y},\mathbf{k}-\mathbf{y};a)F_2(\mathbf{y},\mathbf{k}-\mathbf{y}),\\
B_{NG}(k_1,k_2,k_3;a)=&\left[2F_2(\mathbf{k}_1,\mathbf{k}_2)P_{I}(k_1;a)P_{I}(k_2;a)+\rm{cyclic}\right]\nonumber\\
&+\int \frac{d^3 \mathbf{y}}{(2\pi)^3}\left[T_{I}(\mathbf{y},\mathbf{k}-\mathbf{y},\mathbf{k}_1,\mathbf{k}_2;a)F_2(\mathbf{y},\mathbf{k}_3-\mathbf{y})+\rm{cyclic}\right],\label{bbbbi}
\end{align}
where the gravitational kernel $F_2$ is given by
\begin{align}
F_2(\mathbf{y},\mathbf{k}_2)=\frac{17}{21}+P_1(\mu)\left(\frac{y}{k_2}+\frac{k_2}{y} \right)+\frac{4}{21}P_2(\mu),
\end{align}
with $\mu=\mathbf{\hat{y}}.\mathbf{\hat{k}}_2$. We observe from these formulae that the primordial bispectrum gives a contribution to the matter power spectrum, while the primordial trispectrum gives a contribution to the matter bispectrum. It should also be noted from the first term on the right hand side of equation \eqref{bbbbi} that gravitational instability will generate a bispectrum from Gaussian initial conditions. In order to relate these matter spectra to their galaxy counterparts, it is necessary to include the effect of bias discussed in Section 2.5. Writing the Taylor expansion \cite{Gazt}
\begin{align}
\delta_g(\mathbf{x})\approx b_1 \delta(\mathbf{x})+\frac{1}{2}b_2 \delta^2(\mathbf{x})+\dots,
\end{align}
we relate the galaxy and matter perturbations by a series of constant bias parameters, $b_i$. This results in the galaxy power spectrum $P_g(k)=b_1^2 (P_I(k)+P_{NG}(k))$, while the three-point correlator of the galaxy perturbations is given by
\begin{align}
\langle \delta_g(\mathbf{x}_1) \delta_g(\mathbf{x}_2) \delta_g(\mathbf{x}_3)\rangle=b_1^3\langle \delta_g(\mathbf{k}_1) \delta_g(\mathbf{k}_2) \delta_g(\mathbf{k}_3)\rangle+b_1^2 b_2 \langle \delta_g(\mathbf{x}_1) \delta_g(\mathbf{x}_2) \delta_g^2(\mathbf{x}_3)\rangle.
\end{align}
Evaluating, and transforming into Fourier space, gives the galaxy bispectrum in the form
\begin{align}
B_g(k_1,k_2,k_3;a)=b_1^3 \Big( B_I(k_1,k_2,k_3;a)&+B_{NG}(k_1,k_2,k_3;a)\Big)\nonumber\\
&+b_1^2 b_2\Big(P_{I}(k_1;a)P_{I}(k_2;a)+\rm{cyclic}\Big).
\end{align}
\par
When considering luminosity probes of large scale structure the linear peculiar velocities of the galaxies - i.e. that part of the galaxy velocity that is not due to Hubble's expansion law - should be considered. This effect results in a redshift distortion. The relation between redshift-space and real-space density fields is given by
\begin{align}\label{fdefin}
\delta_{m,z}(\mathbf{k})=\delta_{m,r}(\mathbf{k})\left( 1+f(\Omega_{m,0})\mu^2\right),
\end{align}
where $\mathbf{r}$ is the vector along the radial direction, $\mu=\mathbf{\hat{r}.\hat{k}}$ and $f(\Omega_{m,0})\approx \Omega_{m,0}^{0.6}$ is the known as the `velocity suppression factor', \cite{peebles}. Of course, as we have already noted, in galaxy surveys we must account for bias factor, $b_1$, in transforming from matter perturbations to galaxy perturbations. Writing $\delta_{lum}=\delta_m+(b_1-1)\delta_m$, the first term must account for peculiar velocities while the bias accounts for distortions in second term. This results in the expression
\begin{align}
\delta_{lum,z}=\delta_{lum,r}\left(1+\frac{f(\Omega_{m,0})\mu^2}{b_1}\right).
\end{align}
This equation allows us to relate the power spectra in redshift and real space via
\begin{align}
\frac{P_z}{P_r}=(1+\beta \mu^2)^2,
\end{align}
where $\beta=\Omega_{m,0}^{0.6}/b_1$. Similarly the (angle-averaged) bispectrum is given by \cite{0312286, 07050343}
\begin{align}
B_s(k_1,k_2,k_3;z)=\left(1+\frac{2}{3}\beta+\frac{1}{9}\beta^2 \right)B_g(k_1,k_2,k_3;z).
\end{align}
While current measurements of galaxy surveys are not competitive with the CMB for constraining non-Gaussianity, future surveys are expected to be much stronger. Despite the inherent difficulties in analysing galaxy distributions, the area has received much attention of late due to the possibility of greatly improving on CMB non-Gaussianity constraints in the near future \cite{07050343}. 
\par
Alternative tracers of the matter distribution are provided by observations of redshifted $21$-cm radiation  from neutral hydrogen over a wide range of redshifts ($0\leq z \leq 200$) and by the Lyman-$\alpha$ forest in the spectra of distant quasars.  The  Lyman-$\alpha$ forest is the sum of absorption lines arising from the Lyman-$\alpha$ transition of neutral hydrogen in the spectra of these distant galaxies arising from the intergalactic medium through which the light has travelled. Therefore, though these signals both  originate from neutral hydrogen, they arise from different types of astrophysical systems, with the bulk of $21$-cm radiation originating from damped Lyman-$\alpha$ absorbers. The cross-correlation of these tracers has been investigated in \cite{Sarkar}. The major obstacle in detection of these signals is the level of foregrounds from other astronomical sources, which are several orders of magnitude greater. However, much effort has been undertaken recently to detecting the signal \cite{Challinor07,Wyithe}. It should be noted that a major advantage with these data sets is that there is no Silk damping, such that there will be information present up to $l=\mathcal{O}(10^6)$. 
The fluctuations in the $21$-cm radiation are denoted $\delta_T(\mathbf{\hat{n}},z)$. These are related to the underlying matter density perturbations by,
\begin{align}
\delta_{T}(\mathbf{\hat{n}},z)=\overline{T}_b(z) \overline{x}_{HI}(b_1+ f \mu^2)\delta,
\end{align}
where $\overline{x}_{HI}$ is the mean neutral fraction, $f=d \ln D_+/d\ln a$ is the linear growth parameter of density fluctuations (which was given by an approximation under equation \eqref{fdefin}) and $\overline{T}_b$ denotes the mean temperature at redshift $z$. Similarly, the fluctuations in the transmitted flux $\mathcal{F}(\mathbf{\hat{n}},z)$ along the line of sight $\mathbf{\hat{n}}$, denoted $\delta_{\mathcal{F}}(\mathbf{\hat{n}},z)$, may be related to the underlying matter density perturbations by,
\begin{align}
\delta_{\mathcal{F}}(\mathbf{\hat{n}},z)=c_{\mathcal{F}}(1+\beta_{\mathcal{F}} \mu^2)\delta,
\end{align}
where from numerical simulations of the Lyman-$\alpha$ forest $c_{\mathcal{F}}\approx -0.13$ and $\beta_{\mathcal{F}}\approx 1.58$, \cite{McDonald}. Given these expressions we may find the power spectra and higher order correlations as usual. For a review of $21$-cm physics and the Lyman-$\alpha$ forest see \cite{0608032} and \cite{Rauch} respectively.

\newpage
\chapter{Background II: Cosmic Strings}
\markright{Background II}
\label{chapter:bgd2}
\textbf{Summary}\\
\textit{In this chapter we present a brief review of cosmic strings. Our discussion begins with a classification of topological defects and conditions for their stability as provided by Derrick's Theorem. We proceed with a description of the Kibble mechanism and the formation of cosmic strings. The Nambu-Goto action description of cosmic strings is then discussed, along with a sketch of the resultant dynamics. Next, we derive the characteristic conical metric in the vicinity of a straight cosmic string. The properties of a network of cosmic strings are then elucidated with emphasis on the intercommutation of strings as they cross. A network of cosmic strings may be well approximated by using just one length scale. This approximation, known as the one-scale model, and its extension, which incorporates a variable root mean squared (rms) velocity are described briefly. Next, we present a discourse on the potential observational consequences of cosmic strings. We pay particular consideration to the imprint a network of cosmic strings would leave on the cosmic microwave background. Finally, we provide a short  analysis on cosmic superstrings and their potentially observable differences from (field theory) cosmic strings. For a broader review of these concepts the reader is referred to \cite{shellvil,HindKibb,Low,Polchinski,Kibble,Jackson,PolCopeMyers}.}
\vspace{15pt}

\section{Topological Defects}
Topological defects may arise during a phase transition, when the field undergoes spontaneous symmetry breaking leaving the ground state of the theory characterised by a non-zero vacuum expectation value. The formation of topological defects is determined by the topology of the vacuum manifold, $\mathcal{M}_n=G/H$, where the group $G$ is broken down to its subgroup $H$ during the symmetry breaking phase. In particular, let's consider the mappings from a $k$-dimensional sphere $S^k$ into the manifold $\mathcal{M}_n$, as classified by the homotopy group $\pi_k(\mathcal{M}_n)$. If $\mathcal{M}_n$ contains disconnected components, i.e. $\pi_0(\mathcal{M}_n)\neq \mathbf{I}$ then two-dimensional defects, known as domain walls form. Similarly if $\pi_1(\mathcal{M}_n)\neq 0$, one-dimensional defects, called cosmic strings, must form. For $\pi_2(\mathcal{M}_n)\neq 0$, i.e. the manifold contains unshrinkable surfaces, then monopoles form. In general if the space-time is $4$-dimensional and contains a non-trivial $k^{th}$ homotopy group then defects of spacetime dimension $d=3-k$ form. It is also possible to form event-like, $d=0$, defects known as textures characterised by a vacuum manifold with non-contractible three-spheres. 
\par
\begin{figure}[htp]
\begin{center} 
\includegraphics[width=102mm]{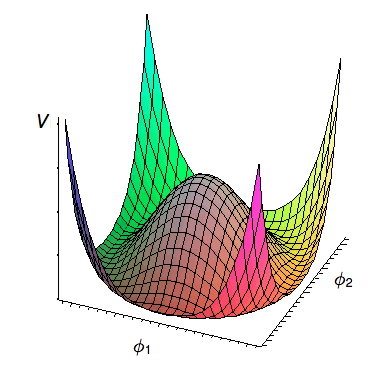}
\end{center}
\caption{`Mexican hat' potential.}\label{fig:mexicanhat}
\end{figure}

In order to elucidate this discussion, consider the simple symmetry breaking Goldstone model of the static complex scalar field $\phi(\bx)$ with
\begin{align}\label{lagran}
\mathcal{L}=\partial_{\mu}\overline{\phi}\partial^{\mu}\phi-V(\phi),
\end{align}
where $V(\phi)$ satisfies the `Mexican-hat' potential (see Figure~\ref{fig:mexicanhat})
\begin{align}\label{potMex}
V(\phi)=(\lambda/4)(|\phi|^2-\eta^2)^2.
\end{align}
This model obeys the global $U(1)$ symmetry $\phi(x)\rightarrow e^{i\alpha}\phi(x)$. It is clear from \eqref{potMex} that the minima lie on a circle $|\phi|=\eta$, i.e. the ground state of the theory satisfies
\begin{align}
\langle 0|\phi |0\rangle=\eta e^{i\theta},
\end{align}
where $\theta$ is an arbitrary phase. Vacua with different values of $\theta$ are equivalent. Thus we set $\theta=0$ in the remainder of this discussion. Perturbing around the vacuum value we write
\begin{align}
\phi=\eta+\frac{1}{\sqrt{2}}(\phi_1+i\phi_2),
\end{align}
where $\phi_1$ and $\phi_2$ are real scalar fields. The Lagrangian \eqref{lagran} now may be written as
\begin{align}
\mathcal{L}=\frac{1}{2}\left((\partial_{\mu}\phi_1)^2+(\partial_{\mu}\phi_2)^2\right)-\frac{\lambda}{2}\eta^2 \phi_1^2 +\mathcal{L}_{int},
\end{align}
where $\mathcal{L}_{int}$ indicates cubic and higher order terms. Thus we may interpret $\phi_1$ as a particle of mass $\sqrt{\lambda}\eta$, and $\phi_2$ a massless field, known as Goldstone bosons. Such symmetry breaking models may be accompanied by topological defects.
\par
The energy of the solution is given by
\begin{align}
E=\int d^{D+1} x \left(\partial_{\mu}\overline{\phi}\partial^{\mu}\phi+V(\phi)\right)=I_1+I_2,
\end{align}
where $I_1$ and $I_2$ represent the gradient and potential terms, respectively. Under a rescaling $\bx \rightarrow \alpha \bx$, the energy becomes
\begin{align}
E_{\alpha}=\alpha^{2-D}I_1+\alpha^{-D}I_2.
\end{align}
Assuming $V(\phi)>0$, then it is clear that for $D\geq 2$, $E_{\alpha}\rightarrow 0$, i.e. time independent defects of spatial dimension two or greater are unstable to collapse. This argument is known as Derrick's Theorem \cite{Derrick}. 
\par
\begin{figure}[htp]
\begin{center} 
\includegraphics[width=102mm]{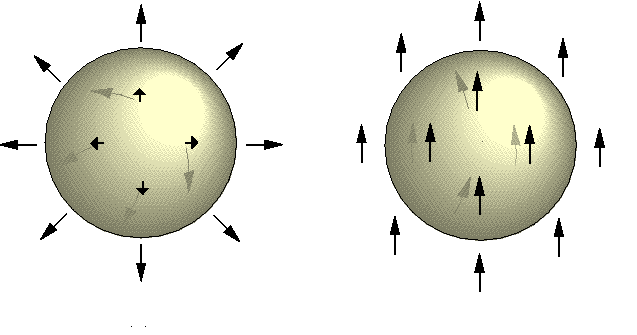}
\end{center}
\caption{Only the three-dimensional `hedgehog' configuration on the left corresponds to a monopole \cite{shellvil}.}\label{fig:monopole}
\end{figure}
\begin{figure}[htp]
\begin{center} 
\includegraphics[width=102mm]{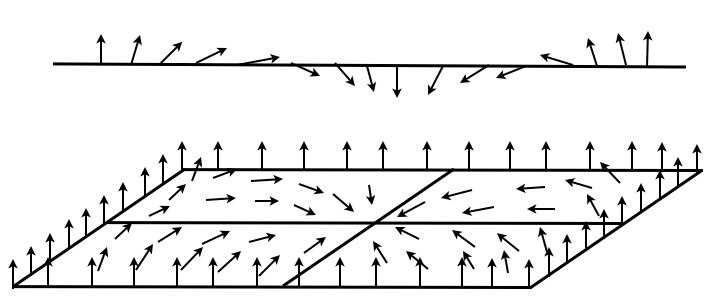}
\end{center}
\caption{Examples of delocalized texture configurations in one and two dimensions \cite{shellvil}.}\label{fig:textures}
\end{figure}

Zero dimensional defects, i.e. point defects, are known as $\bf{monopoles}$ (see Figure \ref{fig:monopole} ).
 These may present a cosmological problem since if they are formed after the inflationary era they will come to dominate the energy density of the universe. This is also a problem true of domain walls (two dimensional defects associated with the breaking of a discrete symmetry). Even if such defects do form they can be diluted away if they are assumed to form before or during an inflationary era. Cosmic strings do not present such a density problem. For global monopoles we know that the force between two global monopoles is independent of distance, and so for a random distribution the global monopoles are attracted to antimonopoles. The pair annihilate to a degree limited by causality and this  is expected to result in a scaling regime with number density $n_M\sim (1-2) d_H^{-3}$ (where the horizon size, $d_H$, is given by $d_H=a(t)\int^t_0 dt'/a(t')$). Global monopoles produce an approximately Harrison- Zeldovich spectrum of fluctuations, but are also expected to form non-Gaussian signatures that should be resolvable on angular scales below $\sim 3.6^o$. 
\newline
\newline
$\bf{Textures}$ (see Figure \ref{fig:textures}) are three dimensional defects, which by Derrick's Theorem are unstable to collapse, have been suggested as a possible explanation for the hot and cold spots in the CMB. 
Since they are nowhere topologically constrained to rise from the minimum of the potential (unlike the other defect considered) then only the massless degrees of freedom are relevant for their descriptions. Therefore they can be effectively reduced to nonlinear sigma models. Local textures emerge at the electroweak phase transition. The field gradients are cancelled by the gauge fields and so local textures have zero energy. Despite this, there are large energy barriers between topologically distinct vacua (corresponding to the different texture solutions). The solitons that interpolate between the vacua are known as sphalerons - excitations of these cause baryon number violating processes since they can change the Chern-Simons number. Since the global textures are unstable, we expect them to shrink and `unwind'. This process is accompanied by the emission of Goldstone boson radiation. Since the unwinding is limited by causality this should lead to a scaling solution. The non-linear collapse should also give rise to non-Gaussian signatures in the density field and also in the CMB which, numerical studies indicate, should be resolvable on angular scales below $1^o$. In particular photons climbing out of a collapsing texture will be redshifted, and blueshifted if they fall into the expanding shell of Goldstone boson radiation.  
\par
 Other defects may be considered if we allow:
\begin{itemize}
\item
 $\bf{extensions}$: for example, models with higher order derivative terms such as $|\nabla \phi|^4$ as in the Skyrme model or by adding gauge fields which allows for instantons in four dimensions. 
 \item
 $\bf{time\, dependence}$: this allows for non-topological non-dissipative solutions such as Q-balls. These are conserved via the Noether charge.
 \item
 $\bf{Non-locality}$: Topological defects are generally assumed to be localised in space. Relaxing this allows for further defects to be considered.
\end{itemize}
Also {\bf{hybrid defects}} may form through a series of phase transitions, e.g. strings corresponding to a particular discrete symmetry will become attached to domain walls when this symmetry is broken. Such a series of phase transitions may form transient defects, for example in the series of transitions
\begin{eqnarray*}
G\rightarrow K \times U(1)\rightarrow K
\end{eqnarray*}
we get monopoles in the first phase transition and strings in the second. The strings are is essence flux tubes connecting monopoles to antimonoples. This connection between the monopoles and antimonopoles leads to their rapid annihilation. Also the strings are not stable in this model since there is a quantum mechanical probability of breakage due to the nucleation of monopole-antimonopole pairs. 
Nonetheless, this may be a solution (aside from assuming that monopoles were formed in a pre-inflationary era) to the monopole problem\footnote{In particular, we assume that $U(1)_{\mbox{em}}$ is broken temporarily and later restored.} (see Section $2.2$).

\section{Cosmic String Formation}\label{sec:Kibblemec}
Since the unification of the electromagnetic force and weak force in a single model known as electroweak theory \cite{Glashow,Salam,Weinberg2}, there has been considerable effort expended in the search for a grand unified theory (GUT) which may unite the strong force with the electroweak force at high energies of order $\mathcal{O}(10^{16})$GeV. Such a GUT theory breaks down in a series of phase transitions to the standard model group that is observed today. Consider again the example of the Goldstone model \eqref{lagran} with the `Mexican-hat' potential \eqref{potMex}, which, as was shown, has a degenerate ground state $\phi=\eta e^{i \theta}$. In particular the ground state does not lie at $\phi=0$. At high temperatures, large random fluctuations of the field, $\delta \phi$, may exceed the ground state value, such that the field does not settle into a a particular ground state. 
\par
\begin{figure}[htp]
\begin{center} 
\includegraphics[width=132mm]{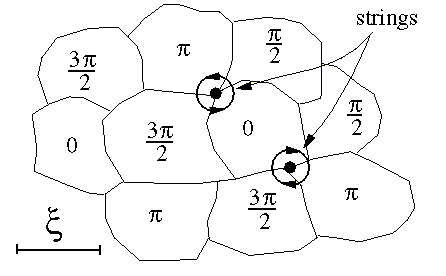}
\end{center}
\caption{The Kibble mechanism for the formation of cosmic strings \cite{shellvil}.}\label{fig:cosmicstringKib}
\end{figure}

Below a critical temperature, $T_c$, set by the symmetry-breaking scale, $\eta$, the energy of the fluctuations is too low to rise above the ground state value. The field must then choose a point on the degenerate circle of minima $|\phi|=\eta$. The choice of the phase, $\theta$, is a random process. As the field settles into its minimum energy configuration the phase angle becomes as uniform as possible. However, in the case of cosmic strings, there is a limit to this process set by the topology of the vacuum manifold, $\mathcal{M}_n$. In particular, if the vacuum manifold has non-shrinkable one-spheres then one part of the system does not `know' the choice of $\theta$ elsewhere. Linear defects, known as cosmic strings form in regions where the phase angle varies by $2\pi$. This mechanism for defect formation is known as the {\textit{Kibble}} mechanism \cite{KibbleMech} (see Figure \ref{fig:cosmicstringKib}). The correlation length, $\xi$, above which the phase values are uncorrelated must satisfy the causality bound,  $\xi<d_H$, where $d_H$ is the causal horizon. Due to continuity of $\phi$, the string must form a closed loop or extend to infinity. It is this stability of cosmic strings that allows them to survive to much later times.

\section{Effective Action and String Dynamics}
An example of a model where strings occur is the abelian Higgs model. This model is described by the Goldstone model with `Mexican-hat' potential, incorporating a charged scalar field. The Lagrangian density now reads
\begin{align}
\textit{L}=\overline{D}_{\mu}\overline{\phi}D^{\mu}\phi -\frac{1}{4}F_{\mu \nu}F^{\mu \nu}-\frac{\lambda}{4}(|\phi|^2-\eta^2)^2,
\end{align}
where $A^{\mu}$ are the associated gauge fields and $D_{\mu}=\partial_{\mu}+ieA_{\mu}$. Of course in the limit $e=0$ we have an abelian Higgs model invariant under a global $U(1)$. In the general case the model here is invariant under the group $U(1)$ of local gauge transformations,
\begin{align}
\phi(x)\rightarrow e^{i \alpha(x)}\phi(x), \quad A_{\mu}(x)\rightarrow A_{\mu}(x)+e^{-1}\partial_{\mu}\alpha(x).
\end{align}
Writing $\phi=\eta+\phi_1/ \sqrt{2}$ we find
\begin{eqnarray*}
L=\frac{1}{2}(\partial_{\mu} \phi_1)^2-\frac{1}{2}{m_{\phi}}^2 \phi_1^2 -\frac{1}{4}F_{\mu\nu} F^{\mu\nu}+\frac{1}{2}m_A^2 A_{\mu}A^{\mu}+L_{\rm{int}},
\end{eqnarray*}
where $m_{\phi}=\sqrt{\lambda}\eta$, $m_A= \sqrt{2}e \eta$. In the global case a (Goldstone) boson accompanies the breaking of the symmetry but this is not present in the local case. This degree of freedom is absorbed, in the local case, into the vector field which now has three independent polarisations instead of two.\footnote{ For {\bf{non-abelian}} gauge theories we similarly find that  Goldstone bosons are present (these correspond to the broken generators) in the global case which are absorbed as additional spin states of the massive vector fields. Fermions can be incorporated by adding the appropriate kinetic and interaction terms to the Lagrangian.}
\par
The action for the abelian-Higgs model with arbitrary potential in a general space-time is given by
\begin{align}
S=\int d^4 y \sqrt{-g}\left(|D_{\mu} \phi|^2 -\frac{1}{4}	F_{\mu\nu} F^{\mu\nu}-V(\phi)\right).
\end{align}
In order to obtain an effective action, a solution to the equations of motion for a curved string, must be used. In the limit of small string curvature (relative to the string length) this is provided by a cylindrically symmetric ansatz known as the Abrikosov-Nielsen-Olsen solution. This solution sweeps out the points labelled by the coordinates $x^{\mu}$ defined on the worldsheet $(\zeta^0,\zeta^1)$. Denoting normal vectors to the worldsheet as $n_{\mu}^{A=0,1}$, with $n_{\mu}^A x^{\mu}_{,a}=0$ we parametrise any point $y^{\mu}$ near the worldsheet as
\begin{align}
y^{\mu}=x^{\mu}+\rho^A n^{\mu}_A.
\end{align}
 Integrating out the massive scalar modes reveals the following action to lowest order,
\begin{align}
S_{\rm{eff}}=-\mu \int d^2 \zeta \sqrt{-\gamma},
\end{align}
where $\gamma_{a b}=g_{\mu \nu}x^{\mu}_{,a}x^{\nu}_{,b}$. This action, known as the {\textit{Nambu-Goto}} action is used widely in studies of cosmic strings. 
\par
As usual the {{equations of motion}} are found by varying the action. Since the action must be invariant under arbitrary reparametrisations of the worldsheet (as well as coordinate transformations) then we are free to choose the worldsheet parametrisations. This is known as gauge fixing. For the Nambu-Goto action a popular choice in flat space is the conformal gauge, $\gamma_{01}=0, \gamma_{00}=-\gamma_{11}$, for which the string equation of motion becomes the two dimensional wave equation,
\begin{align}
\ddot{x}^{\mu}-{x^{\mu}}''=0,
\end{align}
where $'$ denotes partial derivatives with respect to $\zeta_1$.
However there is still a residual gauge left over that allows us to set $\zeta^0=t$ such that the string trajectory reads ${\mathbf{x}}(\zeta^1,\zeta^0=t)$. The general solution is then of the form
\begin{align}
{\mathbf{x}}(\zeta^1,t)=\frac{1}{2}[\mathbf{a}(\zeta^1-t)+\mathbf{b}(\zeta^1+t)],
\end{align}
where $(\partial_{\zeta_1}\mathbf{a})^2=(\partial_{\zeta_1}\mathbf{b})^2=1$. This implies that the energy momentum tensor is
\begin{align}\label{stressEner}
T^{\mu \nu}({\mathbf{x}},t)=\int d\zeta \left( \partial_t x^{\mu}    \partial_t x^{\nu}  -\partial_{\zeta_1} x^{\mu}    \partial_{\zeta_1} x^{\nu}     \right) \delta^{(3)}({\mathbf{x}}-{\mathbf{x}}(\zeta^1,t)).
\end{align}
\par
These results are used to show that for a {\it{closed loop of string at rest}} the motion must be periodic with mean square string velocity $\langle v^2\rangle=1/2$. In curved spacetime a choice of gauge is appropriate such that the spacelike parameter is equal to the invariant length along the string, i.e. $\delta \zeta^1=\delta E/\mu$ (along with choosing $\zeta^0=t$). Of course, {\it{Hubble damping}}, in an expanding universe, is expected, on large scales, to reduce the velocity of a closed loop of string. It is found that, for straight strings on superhorizon scales, perturbations are waves that are conformally stretched by the expansion (i.e. both amplitude and wavelength grow with scale factor) while on subhorizon scales the wavelength grows but the physical amplitude remains constant, i.e. the strings tend to straighten. Numerically this has been shown to hold for strongly curved strings. Due to this effect we expect that loops of string, even if they are irregular initially, will be relatively smooth when they come within the horizon, and  begin to oscillate when they are much smaller than the horizon (where we can ignore expansion effects).
\par
Loops also allow {\bf{cusps}} and {\bf{kinks}}, i.e. points where the string velocity can reach luminal speeds. Kinks are sharp corners where $\partial_{\zeta_1}\mathbf{a}$ ,$-\partial_{\zeta_1}\mathbf{b}$ are discontinuous. Cusps on the other hand are points where the curves described by these two functions intersect. By studying the behaviour of the string near the cusp it is seen that the direction of the cusp is orthogonal to that of the luminal motion. To construct solutions to the equations of motion we can use an expansion in Fourier modes. However, this implicitly assumes smoothness, and to construct kinks all that we must do is to match the ends of straight string segments of possibly different lengths and velocities. Kinks may also be formed by the self-intersection of smooth loops.

\section{Gravitational Effect of Cosmic Strings}\label{sec:metric}
The gravitational interaction strength of cosmic strings is characterised by the dimensionless parameter
\begin{align}
G\mu\sim \left(\frac{\eta}{m_{\rm{pl}}}\right)^2,
\end{align}
where $G$ is Newton's constant, $\mu$ is the string tension, $\eta$ is the symmetry breaking scale and $m_{\rm{pl}}$ is the Planck mass. If formed at the GUT scale $\eta\sim 10^{16}$GeV, then $G\mu\sim 10^{-6}$. Hence, in order to investigate the gravitational effect of cosmic strings we consider a weak gravitational field, resulting in the perturbed Minkowski spacetime
\begin{align}
g_{\mu \nu}=\eta_{\mu \nu}+h_{\mu \nu},\qquad |h_{\mu \nu}|\ll 1.
\end{align}
The linearised Einstein field equations\footnote{We take the convention that the Minkowski metric has mostly positive signature, i.e. $\eta={\rm{diag}}(-1,1,1,1)$.} are 
\begin{align}\label{eomconical}
h_{\mu \nu, \rho \omega} \eta^{\rho \omega} + h_{,\mu \nu}-h^{\rho}_{\mu,\nu\rho}-h^{\rho}_{\nu,\mu\rho}=-16 \pi G S_{\mu \nu},
\end{align}
where $f_{,\mu}={\partial f}/{\partial x^{\mu}}$, $G$ is Newton's constant, $S_{\mu \nu}=T_{\mu \nu}-\dfrac{1}{2}\eta_{\mu \nu}T$ and $h=\eta^{\mu \nu} h_{\mu \nu}$.
In the $\bf{harmonic}$ gauge, i.e. $\partial_{\nu} (h^{\nu}_{\mu}-\frac{1}{2}\delta^{\nu}_{\mu}h)=0$,
equation \eqref{eomconical} may be written in the form
\begin{eqnarray*}
\Box h_{\mu \nu}=-16 \pi G S_{\mu \nu},
\end{eqnarray*}
where $\Box h_{\mu \nu}=h_{\mu \nu, \rho \omega} \eta^{\rho \omega} $. The stress energy tensor \eqref{stressEner} for a straight string reads
\begin{align}
 T_{\mu \nu} &= \mu \delta(x)\delta(y) {\rm{diag}}(1,0,0,1)\\
 \implies S_{\mu \nu} &= -\mu \delta(x)\delta(y)  {\rm{diag}}(0,1,1,0).
\end{align}
Thus, we find the solution 
\begin{align}
h_{\mu \nu}=&0 \quad \forall \, \{i,j \}\, \, \mbox{except}\,\, i=j\in\{1,2\},\nonumber\\
h_{11}=&h_{22}=8G \mu \ln(r/r_0),
\end{align}
where $r^2=x^2+y^2$ and $r_0$ is a constant of integration. In cylindrical coordinates the metric reads
\begin{align}
ds^2=-dt^2+dz^2 +(1-8G\mu \ln(r/r_0))(dr^2+r^2d\theta^2).
\end{align}
Next, we define the coordinates $r',\theta'$ via the expressions
\begin{align}
(1-8G\mu \ln(r/r_0))r^2&=(1-8G\mu){r'}^2,\nonumber\\
\theta'&=(1-4G\mu)\theta.
\end{align}
In terms of these coordinates the metric takes the Minkowskian form
\begin{align}\label{conicalMetric}
ds^2=-dt^2+dz^2+d{r'}^2+{r'}^2 d{\theta'}^2.
\end{align}
Hence, the metric around a straight cosmic string is locally identical to that of flat spacetime. However, noting that the range of $\theta'$ is now $[0,2\pi(1-4G\mu))$, the metric is globally conical, possessing the deficit angle $\Delta=8\pi G \mu$ in the azimuthal direction. Hence, while straight strings do not produce a gravitational effect on nearby matter, their effect may be seen globally via an Aharanov-Bohm type experiment since the conical spacetime of the cosmic string will cause a test particle's own gravitational field to be distorted by the string, thus causing an attractive self-force.

\section{Network Properties}
Strings are produced in the early universe as a tangled network through the Kibble mechanism as described in Section \ref{sec:Kibblemec}. There is around one string segment of (correlation) length $\xi$ per volume $\xi^3$ giving an energy density $\rho=\mu \xi^{-2}$. 
\par
Approximately $20\%$ of the length of the string network is initially in the form of small loops with the remainder forming an infinite network\footnote{Of course, in a closed universe all strings are closed and very large loops take the place of infinite strings.}. In order to determine the $shape$ of the strings we compare the distance between two points on a string $R$ to the average length of string between them, i.e. the string network correlation length $L$. To a good accuracy $L=R^2/\xi$ for the abelian Higgs model. The size distribution of loops is described by $n(R)dR=f(R/\xi)R^{-4}dR$. Network scaling depends on the integral over $f$ approaching a constant, which is observed in all simulations of Nambu-Goto or abelian-Higgs strings.   However the details of loop production remain to be
fully understood. 
	\subsection{Intercommutation and Reconnection}
\begin{figure}[htp]
\begin{center} 
\includegraphics[width=152mm]{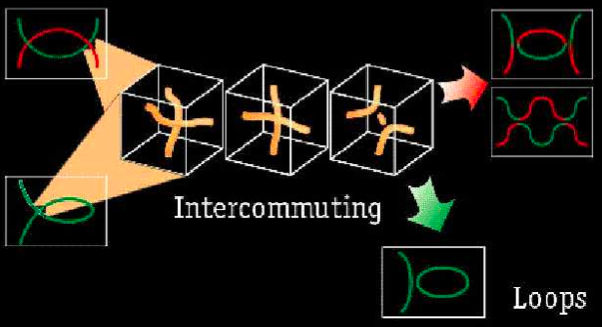}
\end{center}
\caption{Global string interactions leading to loop formation. When two string segments intersect, they reconnect or intercommute (green and red strings - upper part of the figure). Analogously, if a string intersects itself, it can break off a closed loop (green string - bottom part of the figure). In both cases, the interacting string segments first suffer a slight deformation (due to the long-range forces present for global strings), they subsequently fuse and finally exchange partners. An ephemeral unstable amount of energy in the form of a small loop remains in the middle where the energy is high enough to place the Higgs field in the false vacuum. It then quickly collapses, radiating away its energy. The situation is roughly the same for local strings, as simulations have shown \cite{gangui}.}\label{fig:reconnection}
\end{figure}

	In order to understand the evolution of string networks it is important to understand what occurs at string crossing (see for example Figure \ref{fig:reconnection}).
 The effective action breaks down here as well as the zero width approximation. Therefore, to understand the outcome of collisions fully a study of the nonlinear theory must be carried out. However, since the vortex solutions are regular, progress is most easily made by considering the two-dimensional interactions.  In this way reconnection is seen as a result of the underlying field configuration space, i.e. the moduli space. 
\par
For abelian strings, even though in high energy collisions there is often a complex aftermath, the inevitability of reconnection for field theory strings is still a robust result. However this is not a general result. For non-abelian strings there is a topological obstruction to reconnections and so the strings become entangled, with perhaps a third string stretching between them. Upon reconnection the strings coalesce and locally annihilate. The topological flux is then rerouted along the opposite string segments. After this the strings separate and move apart due to their tension \cite{Shellard1987,Matzner1989}. For relativistic interactions there may be enough residual energy to create new zeros of the Higgs field in the form of a closed loop \cite{Battye1993}. 
\par
These results have been shown numerically to hold for parameters of cosmological interest and we can assume that the probability of intercommuting is $p\approx 1$. Numerical studies of high $n$ vortex intercommutations have shown partial intercommutation such that the interconnected segments split apart into lower winding number configurations.

	\subsection{One-Scale Model}\label{sec:scaling}
	The one-scale model \cite{Kibble1985} assumes that the string network may be described by a single length scale, $L$, the inter-string length. In the one-scale model this length scale is assumed to be equal to the correlation length, i.e. $L=\xi$ such that the
energy density is given by $\rho=\mu L^{-2}$. This length scale is assumed to grow in proportion to the horizon, with $L=\gamma(t) t$.  Due to the formation of loops as strings self-intersect the string network will lose energy. This effect is countered by the expansion of the universe. These competing effects result in the crude energy-loss equation
\begin{align}
\dot{E}\approx \frac{\dot{a}}{a}E-\frac{\mu L}{L^4}V,
\end{align}
where $E=\rho V$, with $V\propto a^3$ the physical volume. This implies
\begin{align}
\dot{\rho}=-2\frac{\dot{a}}{a}\rho-\frac{\rho}{L}.
\end{align}
Next, setting $L=\gamma(t)t$ we find
\begin{align}
\frac{\dot{\gamma}}{\gamma}=&\frac{1}{2 t}\left(2(\alpha-1)+\frac{1}{\gamma}\right),\\
\implies \gamma\rightarrow& \frac{1}{2-2\alpha},
\end{align}
where we use $a\propto t^{\alpha}$. In the matter era where $\alpha=2/3$ and in the radiation era where $\alpha=1/2$ this solution clearly approaches a valid attractor solution with $\gamma\approx 3/2$ and $\gamma\approx 1$ in the respective era. Such a solution is known as a `scaling' solution and implies that on large scales the string network follows a self-similar evolution. This scaling behaviour has been observed in numerical simulations of cosmic string networks. Scale invariance on all scales would further imply that the length of small loops produced by the network would be a fixed fraction of the horizon, i.e.
\begin{align}
\langle l \rangle=\alpha t.
\end{align}
It should be noted that since the relative size of small-scale structure must be above the gravitational backreaction scale we necessarily have $\alpha=\Gamma G \mu$, where $\Gamma$ is of order $\mathcal{O}(100)$ and depends in general on the loop's shape and trajectory\footnote{Oscillating loops of string lose their energy by emitting gravitational waves at the rate $\dot{E}=\Gamma G \mu^2$.}.
\par
The one-scale model has been generalised to the `velocity-dependent one scale model' (VOS) \cite{MartShell}. This model incorporates a variable rms velocity $v$. This allows investigation of regimes with frictional damping and the matter-radiation transition. In this model the decay of the string network into small loops is characterised by the `loop chopping efficiency', $\tilde{c}$. The equation of motion for the density parameter becomes
\begin{align}
\dot{\rho}=-2\frac{\dot{a}}{a}(1+v^2)\rho-\frac{\tilde{c}v\rho}{L}.
\end{align}
Noting that the rms velocity is given by
\begin{align}
v^2=\langle \dot{\bx}^2\rangle=\frac{\int \dot{\bx}^2 \epsilon d\zeta}{\int \epsilon d\zeta},
\end{align}
where $\epsilon={\bx'}^2/(1-\dot{\bx}^2)$, we obtain an equation for the acceleration by differentiating this equation and using the equations of motion
\begin{align}
\ddot{\bx}+\left(2\frac{\dot{a}}{a}+\frac{a}{L}\right)(1-\dot{\bx}^2)\dot{\bx}&=\frac{1}{\epsilon}\left(\frac{\bx'}{\epsilon}\right)',\\
\dot{\epsilon}+\left(2\frac{\dot{a}}{a}+\frac{a}{L}\right)\dot{\bx}^2\epsilon=0.
\end{align}
This results in the following expression
\begin{align}
\dot{v}=(1-v^2)\left( \frac{k}{L}-2\frac{\dot{a}}{a}v\right),
\end{align}
where the parameter $k$ encodes information about the small-scale structure on strings and is defined as
\begin{align}
k\equiv \frac{\langle (1-\dot{\bx}^2)(\dot{\bx}.\mathbf{u})\rangle}{v(1-v^2)},
\end{align}
where $\mathbf{u}$ is a unit vector parallel to the curvature radius vector.
An approximation for this parameter is given by
\begin{align}
k=\frac{2\sqrt{2}}{\pi}\left( \frac{1-8v^6}{1+8v^6}\right).
\end{align}
As before it may be shown easily that the VOS model describes a scaling solution with
\begin{align}
\gamma^2=\frac{k(k+\tilde{c})}{4\alpha(1-\alpha)},\qquad v^2=\frac{k(1-\alpha)}{\alpha(k+\tilde{c})}.
\end{align}
This model provides good agreement with results from simulations, especially on large scales. In order to provide a more accurate fit for small scale structure, various extensions of this model have been considered \cite{Martins2000}. 
\section{Observational Consequences}
Despite the width of the core of cosmic strings, $\delta$, being so small, their mass per unit length, $\mu$, may be extremely large in general. Therefore, it is logical to search for signatures of cosmic strings via their gravitational interactions.
\par
Gravitational interactions of cosmic strings are characterised by the dimensionless parameter $G\mu\sim ( \eta/m_{\rm{pl}})^2$. From observational bounds (CMB, pulsars, etc.) we must have $G\mu\lesssim 10^{-6}$.
\paragraph*{Temperature Discontinuities:}
 As described in Section \ref{sec:metric}, for a straight string the spacetime is globally conical. Because of this, when two parallel particles (say, of equal velocities) initially moving along parallel paths pass by the string they will move towards each other at a relative velocity 
 \begin{align}
 \delta u= v \gamma \Delta,
 \end{align}
 where $\Delta$ is the deficit angle, $\bf{v}$ is the initial velocity of both particles and $\gamma$ is the corresponding Lorentz factor. If one of the objects is a source of radiation and the other an observer, then the observer will detect a {{discontinuous change in the radiation frequency}} of magnitude $\delta T/T =\delta u$. This is easily generalised for arbitrary angles between the string, its velocity and the line of sight.  Therefore, with enough resolution, signatures of cosmic strings, if they exist, should be visible in the CMB. The particular pattern would be a superposition of the contribution of strings from different redshifts from the surface of last scattering to today with typical interstring distances of the order $d_H/3$ (see Section \ref{sec:scaling}). It should be clear that near cusps the temperature variation should be quite large.  This signature may be searched for by looking for temperature discontinuities along a line, a few degrees long. 
\par
There may also be {{anisotropies from gravitational waves}} emitted by strings but this does not have a characteristically `stringy' signature.

\paragraph*{Gravitational Lensing:}
As with all massive objects, strings act as {{gravitational lenses}}. However, due to the conical nature of the spacetime this can be more dramatic. In fact, the presence of a cosmic string would result in the formation of a double image of an object, such as a galaxy, located behind the string. The angular separation between such images in the rest frame of the string is given by 
\begin{align}
\delta \phi_0=\Delta \sin (\theta) \frac{l }{d+l}
\end{align}
 for a string at rest, where $l$ is the comoving distance from the object to the string, $d$ the comoving distance from the string to the observer and $\theta$ is the angle between the string and the line of sight. This can be generalised for moving strings by using the fact that since the angular separation, $\delta \phi$ is small (of th order of a few arc seconds) then we have that the frequencies in the frame moving at velocity $\bf{v}$, $\omega$, is related to lowest order in $G\mu$ to the frequencies to the rest frame, $\omega_0$ via $\omega \delta\phi\approx \omega_0 \delta \phi_0$ and, of course, via  $\omega=\gamma(1-\hat{\bf{n}}.\bf{v}) \omega_0$, where $\hat{\bf{n}}$ is the unit vector in the direction from the object to the observer. 
\par
The ideal scenario is for the string to be straight because the images are identical. If the string is not straight, deviations may be caused by unequal time delays, string curvature, etc. Substructure can cause the images to be misaligned. If the string is not perpendicular to the line of sight then the observer sees more distant parts of the string with a retardation. Thus the apparent direction of the string is not the same as its actual direction. The observer sees a projection of the string onto the plane perpendicular to the line of sight. The angle, $\chi$, between the apparent and actual direction is given by $\tan \chi= v \cos \theta$. 
\par
The probability of lensing by infinite strings at redshift $z\sim 1$ is calculated by using the approximate angular separation of images and by noting that numerics tell us the inter-string distance is $\sim d_H/3$. This gives $p_{\infty}(z\sim1)\sim 100G\mu$. As $z$ increases the comoving distances increase as $\sqrt{1+z}$. Putting this in the formula for $\delta \phi$ we get that the angle decreases as $1/\sqrt{1+z}$. Therefore $p_{\infty}$ increases logarithmically with redshift. For $ z \ll 1$ the lensing is primarily by string loops located at redshifts smaller than $z$. Using the loop distribution function we get 
\begin{align}
p_L(z)\sim 9\pi G \mu v z^2 \ln\left(\frac{\alpha}{\Gamma G \mu}\right),
\end{align}
 where it is assumed that the angular size of the loops is greater than $\Delta$ - this should be true for most of the loops, because of the gravitational back reaction cutoff. 

\paragraph*{Gravitational waves}
Gravitational waves are emitted by oscillating loops and so produce a gravitational wave background. The power in gravitational radiation from a weak isolated source is to lowest order in $G$ given by
\begin{align}
P=\sum_n\int d\Omega \frac{d P_n}{d \Omega}=\sum_n\int d\Omega     \frac{  G \omega_n^2}{\pi}  
 \left(        T_{\mu\nu}^*        (\omega_n,\mathbf{k}) T^{\mu\nu}         (\omega_n,\mathbf{k}) - \frac{1}{2}\left| T^{\mu}_{\nu} (\omega_n,\mathbf{k}) \right|^2    \right),
 \end{align}
with $T_{\mu\nu}(\omega_n,\bf{k})$ denoting the Fourier transform of the energy momentum tensor. This gives $P=\Gamma G\mu^2$, with numerical work showing $\left< \Gamma \right>\approx 65$.
\par
To estimate the intensity we use the covariant measure,
\begin{align}
 \Omega_g= \dfrac{\omega}{\rho_c}\dfrac{d\rho_g}{d\omega},
 \end{align}
  where $\rho_c$, $\rho_g$ denote the critical and gravitational wave energy densities respectively and $\omega$ denotes the angular frequency. A simple analysis can be carried out by assuming that the loops radiate only in the lowest few harmonics. The loops emit gravitational waves at frequencies $\omega_n=4\pi n/l$. For loops that decay at time $t_1 < t_{\mbox{eq}}$ then, using the scaling model -which tells us that $l\sim \Gamma G \mu t_1$ and the energy densities of loops in the radiation era- and assuming that almost all the energy is transformed into gravitational waves of frequency $\omega_1\approx 4 \pi/ l$, we get 
 \begin{align}
  \Omega_g (\omega)\sim \rho_g/\rho_c\sim 30\gamma_r^{-2} \sqrt{\frac{\alpha G \mu}{\Gamma}}.
  \end{align}
   Using that the assumption that the early universe was in thermal equilibrium we have $N(T)T^3 \propto a^{-3}$, where $N(T)$ is the effective spin degrees of freedom. Next using the expression $\rho_r =(\pi^2/30)N(T)T^4$ we find $\Omega_r N^{1/3} \propto a^{-4}$. We use this to write
\begin{align}
\Omega_g(\omega)\sim 30 \gamma_r^{-2} \sqrt{\frac{\alpha G \mu}{\Gamma}}\left(\frac{N}{N_{\omega}}\right)^{1/3} \Omega_r,
\end{align}
where $N_{\omega}$ is the value of $N$ when the gravitational waves were emitted. The assumption that the loops radiate only in the lowest harmonics is not valid for loops with cusps or in the matter era. A more general treatment - where we assume the life time of the loop is greater than the Hubble time i.e. $\alpha>\Gamma G \mu$ such that the effect of loop velocities on the spectrum can be neglected - reveals that in the radiation era\footnote{In deriving $\Omega_g$ the further assumptions were made that a) the frequency is so low that waves emitted before positron annihilation do not contribute and factors like $N$ can be ignored, and b) $4\pi n_*\ll \Gamma G \mu \omega t$, where $n_*$ is the mode number beyond which the radiation spectrum of loops can be truncated.}
\begin{align}
\Omega_g(\omega)=\frac{128 \pi}{9 \gamma_r^2}\sqrt{\frac{G\mu \alpha}{\Gamma}}\Omega_r \left[ 1-\left(  1+\frac{\alpha}{\Gamma G \mu}\right)^{-3/2}  \right].
\end{align}
A similar derivation may be carried out for loops radiating in the matter era. It should be noted that infinite strings may also radiate gravity waves. It is found that $\dfrac{\Omega_g(\omega)_{\infty}}{\Omega_g(\omega)_{L}}\sim \dfrac{4\pi^2}{\Gamma}\left( \dfrac{\Gamma G \mu}{\alpha}\right)^{3/2}$ and so if $\alpha\gg \Gamma G \mu$ -i.e. if the loops live much longer than a Hubble time (as is of interest for most applications)- the loops dominate. For $\alpha\sim \Gamma G \mu$ the contributions are comparable.

\paragraph*{Black Holes}
Black Holes may form through the collapse of an oscillating string loop. Since the Schwarzschild radius of a loop of length $l$ is $r_g=2G\mu l$, this would require the loop to shrink by the order of $(G\mu)^{-1}$ during its oscillation and so they are unlikely. However, because the energy density in black holes redshift like matter then, if any are formed in the early universe, they may come to dominate the energy of the universe. This gives the constraint $\Omega_{\rm{bh}}\lesssim 10^{-8}$. 
\par
Using the one-scale model, a loop of mass $m$ is formed at time $t_f\sim m/(\mu \alpha)$. The energy density of the loops at time $t_f$ is given by $\Omega_L=\rho_L/\rho_c$. Using the results in Section \ref{sec:scaling}, we find $\rho_L^{(r)}=\mu \nu_r (t l)^{-3/2}$ in the radiation era. Since $\rho_c=3/(32\pi G t^2)$ we may write $\Omega_L\sim 30 G \mu/\gamma_r^{-2}$. If $P$ is the probability that a loop decays to form a black hole, this gives the constraint $(G\mu)^2 P\lesssim 10^{-29}$ for black holes of mass $\sim 10^{15}g$ where we use the previously quoted numerical estimates for the various parameters.

\paragraph*{Baryon asymmetry}
In the case of grand unified strings, we may consider that the heavy gauge field has baryon number couplings. Thus light fermions that scatter off the string can change their baryon number. This can then obviously cause a {{baryon asymmetry}} if this baryon number scattering is also $CP$-violating. 
\par
This can also occur if a scalar field, that has a Yukawa coupling to fermions or quarks, develops a condensate in the string core. Initial estimates predicted the cross section to be much less than that of particle scattering cross sections. However, further study has shown that the cross section may be enhanced by a resonance phenomenon that is very sensitive to the internal structure of the string.  Other mechanisms suggested to produce baryon asymmetry are: 1) via the decay of heavy particles emitted at cusps; 2) via the disintegration of a loop into heavy particles once it shrinks to a size comparable to its thickness. These heavy particles can then decay in a $B$- and $CP$- violating manner.

\section{Cosmic Strings and the CMB}
\begin{figure}[htp]
\begin{center} 
\includegraphics[width=142mm]{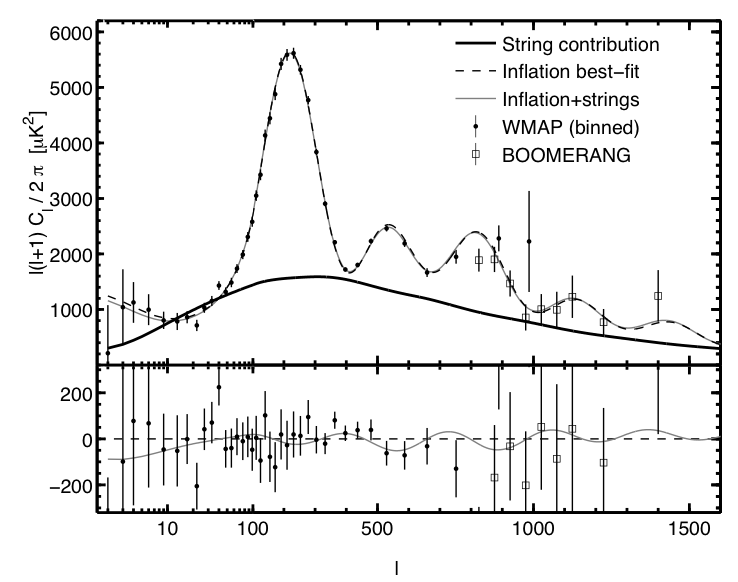}
\end{center}
\caption{  The temperature power spectrum contribution from cosmic strings normalised to match the WMAP data at $l=10$, as well as the best-fit cases from inflation only (model PL) and inflation plus strings (PL+S). These are compared to the WMAP and BOOMERANG data. The lower plot is a repeat but with the best-fit inflation subtracted, highlighting the deviations between the predictions and the data \cite{Bevis}.}\label{fig:bevisstringinfl}
\end{figure}

\begin{figure}[htp]
\begin{center} 
\includegraphics[width=162mm]{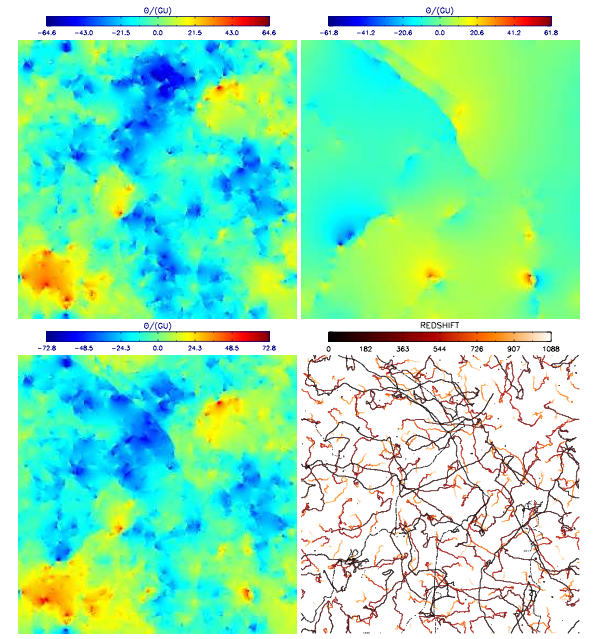}
\end{center}
\caption{  String-induced CMB temperature flutuations on a $7.2^0$ field with a (unrealistic) resolution of $\theta_{res}=0.42'$ ($1024$ pixels). The upper left image shows the fluctuations induced in between the last scattering surface and the redshift $z=36$. The upper right image plots  the fluctuations induced between $z=36$ and $z=0.3$, while the bottom left image plots the overall induced fluctuation. Because of their cosmological scaling, most of the long strings intercept our past light cone close to the last scattering surface. The overall string-induced fluctuations are plotted in the bottom left panel. As can be seen in the bottom right image, the edges in the temperature patterns of the other maps can be identified to strings intercepting our past light cone. Note that active regions corresponding to string intersection and loop formation events lead to the bright spots in these maps. Some of these spots saturate the colour scale \cite{Fraisse}.}\label{fig:ringeval}
\end{figure}

The inflationary paradigm has amassed much strong observational support in recent years, particularly with the observation of the predicted acoustic peaks in the cosmic microwave background (CMB). This has ruled out the possibility of cosmic strings being responsible for the origin of the large scale structure of the universe. However many inflation models predict that a network of cosmic strings should exist after inflation. Current measurements allow the string component of the temperature power spectrum to contribute up to $10\%$ of the overall signal (see \cite{bevis2}). In fact Bevis et al \cite{Bevis} have shown that current CMB data gives moderate preference to the model $n_s=1$ with cosmic strings over the standard zero-strings model with variable tilt (see Figure \ref{fig:bevisstringinfl}).
In this paper the authors have shown that the string tension is currently constrained by observations to the range $G\mu<7\times 10^{-7}$.
Using the null hypothesis that strings should make a zero contribution to the B-mode polarisation the same authors (\cite{hindPol}) have shown that high resolution CMB polarisation experiments should impose the constraint $G\mu<1.2\times 10^{-7}$.
\newline
\newline
Fraisse et al \cite{Fraisse} have used Nambu-Goto simulations to compute the CMB temperature anisotropies induced at arcminute scales by a network of cosmic strings in a FLRW expanding universe. They find that at high multipoles the mean angular power spectrum decays as $l^{-p}$, where $p\approx0.889$. This would suggest that at small angular scale cosmic strings may come to dominate the fluctuations. They discuss how the temperature gradient magnitude operator can trace strings in the context of a typical arcminute diffraction-limited experiment. Allowing for Sunyaev-Zeldovich effects and the inflationary anisotropies they find that strings should be `eye visible' on a $7.2^0$ gradient map down to $G\mu\approx 2\times 10^{-7}$ (see Figure \ref{fig:ringeval}).
\par
In general to find the power spectrum induced by topological defects, it is necessary to solve a system of linear perturbations equations with random sources for each given wavevector $\bk$,
\begin{align}
\mathcal{D}X=\mathcal{S},
\end{align}
where $\mathcal{D}$ is a time-dependent linear differential operator, $X$ a vector containing the matter perturbation variables and $S$ is the source term, containing linear combinations of the energy momentum tensor of the defect. For a given set of initial conditions this equation may be solved using a Green's function $\mathcal{G}(\tau,\tau', \bk)$. The power spectrum $\langle X_i(\tau_0,\bk)X_j^*(\tau_0,\bk')\rangle$ is then found by solving
\begin{align}
\langle X_i(\tau_0,\bk)X_j^*(\tau_0,\bk')\rangle=\int_{\tau_{in}}^{\tau_0}d\tau \mathcal{G}_{im}(\tau_0,\tau,\bk)\int_{\tau_{in}}^{\tau_0}d\tau' \mathcal{G}_{jn}^*(\tau_0,\tau',\bk)\langle {S}_{m}(\tau,\bk){S}_{n}^*(\tau',\bk) \rangle.
\end{align}
The unequal time correlators (UETCs) of the defect stress energy tensor,
\begin{align}
\langle T_{\alpha \beta}(\bk,\tau)T_{\mu \nu }^*(\bk,\tau')\rangle=C_{\alpha \beta, \mu \nu}(\bk,\tau,\tau'),
\end{align}
are highly constrained by causality, scaling and conservation of the stress-energy tensor \cite{TurokPenSeljak}. This greatly reduces the number of UETCs to be calculated and increases the dynamic range greatly. In order to exploit fast Einstein-Boltzmann solvers to calculate cosmological power spectra of topological defects, it is necessary to express the UETC as a sum of eigenvector products,
\begin{align}
C(k,\tau,\tau')=\sum_i \lambda_i v^i(k,\tau) v^i(k,\tau'),
\end{align}
where
\begin{align}
\int d\tau' C(k,\tau,\tau')v^i(k,\tau) w(\tau')=\lambda^i v^i(k,\tau),
\end{align}
where the matrix index summation is replaced by the integral $\int d\tau w(\tau)$. It should be noted that scaling does not hold in the important radiation-matter transition, implying this functional form is no longer valid in such a regime. In such a case the Boltzmann equations must be solved in full on a $3$D grid \cite{shellLan}.

\section{Cosmic Superstrings}
The study of cosmic strings arising from string theory originated with Witten \cite{Witten}. However such cosmic strings correspond to a symmetry breaking scale of order the Planck scale $\mathcal{O}(10^{19})$GeV  were rejected since the tensions of such strings $G\mu\sim 1$ were ruled out by observations. However, recent developments in string theory - such as D-branes and warped dimensions - have warranted further studies. In particular, the discovery by Tye and Sarangi in $2002$ \cite{TyeSarangi}, that cosmic strings are produced in the latter stages of brane inflation, alone justifies their study. In this section we briefly outline a discussion of work presented in \cite{Polchinski,PolCopeMyers,Jackson,Kibble,Sakell,Davis}. 
\par
The basic idea invoked in the brane-inflation idea is the notion of space-time warping. When a spacetime is warped, the metric becomes
\begin{align}
ds^2=e^{2A(y)}\eta_{\mu \nu}dx^{\mu} dx^{\nu}+g_{mn}^{\perp}dy^m dy^n,
\end{align} 
where $y$ are the compact coordinates, $\mu,\nu\in [0,3]$ and $m,n\in [4,9]$. If the string is localised at $y_0$ then $\mu=e^{2A(y_0)}/(2\pi \alpha')$ (where $\alpha'$ determines the fundamental string length scale, $l_s$, via the expression $l_s=4\pi\sqrt{\alpha'}$), thus allowing for a suppression of the string tension. Throats where $e^{2A}\ll 1$ are, therefore, of particular interest. The most detailed model of inflation in string theory is the KKLMMT model which is based on a warped compactification of the IIB string theory \cite{KKLMMT}. Inflation arises from a $D3/\overline{D3}$ pair at the bottom of a throat. F-strings, produced after inflation, lie in the same throat. These have tension $\mu_F\sim 2\times 10^{-10}\sqrt{g_s}$. D-strings, i.e. D-branes which have all but one of their spatial directions wrapped on compact cycles, are also formed by the annihilation of a $D/\overline{D}$ pair of branes.  Bound states of $p$ F-strings and $q$ D-strings have tension $\mu_{(p,q)}=\mu_F \sqrt{p^2 +q^2/g_s^2}$, where $g_s$ is the string coupling. Of course, the only interesting strings are those with non-trivial cycles in the throat, since otherwise the tension is not suppressed and exceeds observational bounds. 
\paragraph*{How the strings form}
The strings form because of the $U(1)\times U(1)$ gauge symmetry on the pair of branes. When the branes annihilate we then get spontaneous symmetry breaking of this gauge. One linear combination is Higgsed so we get a string network via the Kibble mechanism. These are the D1-strings which we can think of as topological defects in the tachyon field describing the annihilation. The other linear combination is confined. Since this can be thought of as dual Higgsing, by a magnetically charged field, then the Kibble mechanism again applies to form F-strings (the fundamental type II strings). In the perturbative theory, at least, it is the quantisation of these F-strings that defines the theory.
\par
 It has also been proposed that F-strings can form in a {{\it{Hagedorn transition}} where strings of infinite length are formed. When $T$ drops below $T_H$ after inflation then (as in the Kibble mechanism) strings form. Such strings would have a lower tension than those formed in the brane scenario due to the inefficiency of the thermal step.  
 \par
 It should be noted that in other models there may be a more complicated geometry that may imply the formation of other types of strings. There may also be other stability issues related to such strings.
\paragraph*{Stability}  While fundamental strings are generally unstable, the existence of a throat allows for cosmic superstrings to grow to macroscopic sizes. However, such strings are still subject to further stability issues such as the fragmentation of an open string into smaller strings and the possibility of breakage on a brane when an F-string encounters a D-string. However, the throat also provides a stabilising potential for such instabilities.
\par
$\bf{Distinguishing\, \,between\, \,cosmic\, \,strings\, \,and\, \,cosmic\, \,superstrings}$         
\par
$\bf{a)\,Intercommutation}$
 In the case of field strings, the reconnection probability is $p\sim 1$. However, for F- strings, reconnection is a quantum process with probability $\mathcal{O}(g_s^2)$. Of course, in general, the potential is a function of the relative angle and velocity in the collision. However, in many models the strings are free to wander over the whole compact space. Thus the strings can miss each other. This causes a large suppression of the reconnection probability. If, as in the case of many realistic models, the strings are localised by a potential in the compact dimensions then this suppression is not so severe. Thermal fluctuations of such a potential will, however, give some suppression of the probability, $p$. Jackson et al found in \cite{Jackson} that for F-F string collisions $10^{-3}\lesssim p \lesssim 1$, for D-D string collisions $10^{-1}\lesssim p \lesssim 1$, while for D-F string collisions the reconnection probability can take any value in $[0,1]$.
\par
 Observations of the overall strength of the signals from cusp events would allow us, in principle, to find $\mu$ and $p$. We note that it is possible for $p\ll 1$ for field theory strings, e.g. electric flux tube strings arising in strongly coupled confining theories have $p\sim 1/N_{\rm{colour}}^2$.
\par
\paragraph*{b) $(p,q)$ strings} When strings of different types collide the intercommutation results in trilinear vertices (except for $(p,q)=\pm (p',q')$). A similar type of scenario occurs for cosmic strings arising in non-abelian field theories. Nonetheless, we note also that $(p,q)$ strings are non-BPS. If a network of such strings is observed it should be distinguishable (to a large confidence) from field theory strings through its particular spectrum. To determine whether such a network would scale or freeze would require simulations. The conjecture is that, if they exist, they don't freeze as otherwise we would expect them to dominate the energy density of the universe today.
\par
\paragraph*{c)  Radiation} As with field theory cosmic strings, we expect gravitational radiation to be emitted from near-cusp regions. This should produce a non-Gaussian signature. Futher, if the string couples strongly to the standard model fields then the string network may decay by the production of cosmic rays, photons and neutrinos.
\newline
\newline
\newline
The study of networks of cosmic superstrings has been developed recently using an adaptation of the VOS model. This adaptation accounts for the junctions formed when strings of different types collide. The interested reader is referred to \cite{AvgousShell,AvgousCope,Pourtsidou}.

\newpage
\thispagestyle{empty}
\mbox{}
\newpage
\chapter{Cosmic Strings Power Spectrum, Bispectrum, Trispectrum}
\label{chapter:stateoftheart}
\textbf{Summary}\\
\textit{We use analytic calculations of the post-recombination gravitational effects of cosmic strings to estimate
the resulting CMB power spectrum, bispectrum and trispectrum.   We place a particular emphasis on
multipole regimes relevant for forthcoming CMB experiments,
notably the Planck satellite.    These calculations use a flat sky approximation, generalising
previous work by integrating string contributions from last scattering to the present day, finding
the dominant contributions to the correlators for multipoles $l> 50$.   We find a well-behaved shape
for the string bispectrum (without divergences) which is easily distinguishable from the
inflationary bispectra which possess significant acoustic peaks.   We estimate that the
nonlinearity parameter characterising the bispectrum is approximately $0\gtrsim f_{NL} \gtrsim -40$ (given
present string constraints from the CMB power spectrum).
We also apply these unequal time correlator methods to calculate the trispectrum for
parrallelogram configurations, again valid over a large range of angular scales relevant
for WMAP and Planck, as well as on very small angular scales.   We find that, unlike the
bispectrum which is suppressed by symmetry considerations, the trispectrum for cosmic
strings is large.    Our current estimate
for the trispectrum parameter is $\tau_{NL} \sim 10^5$, which may provide one of the strongest
constraints on the string model in future analysis.}
\vspace{15pt}
\vspace{35pt}


\section{Introduction}

Cosmic strings are a common feature in fundamental cosmology scenarios, such as brane and hybrid inflation \cite{Tye_or_review}, and
they appear to be generic in realistic grand unified theories \cite{Jeannerot}.  Cosmic strings
leave a distinct `line-like' signature in the cosmic microwave background (CMB) \cite{Gott,KAISERSTEBBINS} which
offers one of the best prospects for their detection.   Strings create this imprint after recombination by perturbing photons through relativistic gravitational effects.  While inflationary fluctuations are believed
to dominate the overall CMB power spectrum, current constraints suggest that up to 10\% of the
signal could be contributed by cosmic strings \cite{Battye, Kunzetal}. A higher proportion
is incompatible with WMAP because the cosmic string power spectrum around
multipoles $l \approx 200$ is dominated by relatively featureless metric fluctuations, rather than the pre-recombination acoustic peaks characterising inflation.  The gravitational strength of cosmic strings
is determined by the parameter $G\mu = (\eta/m_{Pl})^2$ where $\eta$ is the symmetry breaking
scale at which strings form; $G\mu$ is essentially the ratio of the string tension $\mu$ to
the square of the Planck mass $m_{Pl}$.     The present WMAP limit on the string contribution to
the CMB translates into a strong constraint on the parameter $G\mu \leq 2.5 \times 10^{-7}$
 \cite{Battye} (or from field theory simulations $G\mu\leq 7\times 10^{-7}$ \cite{Bevis}).
\par
Inflationary CMB fluctuations in the standard picture are very nearly Gaussian, so higher order correlators offer the prospect of  further differentiation between these and competing signals from
cosmic strings.
The spherical harmonic transform of the three-point CMB correlator is the bispectrum $B_{l_1 l_2 l_3}$.   Usually discussions of the bispectrum are simplified further by focusing on the nonlinearity parameter $f_{NL}$ which is roughly the ratio of the three-point correlator to the square of the two-point correlator,
that is, $f_{NL}\sim \langle \zeta\zeta\zeta \rangle /\langle\zeta\zeta\rangle^2$ where the $\zeta$ are the primordial curvature fluctuations (dimensionless) which seed the CMB anisotropies. For a perturbative theory like inflation, we expect the second-order fluctuations $\zeta^{(2)}$ to be constructed from convolutions of the linear perturbations $\zeta^{(2)} \sim \zeta^{(1)}* \zeta^{(1)}$.  This implies that
the leading order term in the three-point correlator can be expected to behave as
$\langle \zeta^{(1)}\zeta^{(1)}\zeta^{(2)}\rangle \sim \langle \zeta^{(1)}\zeta^{(1)} \rangle^2$, that is, $f_{NL} \sim 1$.   In fact, for standard single field inflation there is considerable further suppression from slow-roll and the primordial signal $f_{NL} \approx \mathcal{O}(0.01)$ \cite{Maldacena}.  One can only obtain a significantly larger $f_{NL}$ in more exotic inflationary models with multiple scalar fields or non-canonical kinetic terms (see, for example, the reviews in \cite{Chen,Fergusson}).  Similar arguments
apply to the trispectrum or four-point correlator characterised by $\tau_{NL} \sim \langle \zeta\zeta\zeta\zeta\rangle/\langle\zeta\zeta\rangle^3$, with $\tau_{NL} \lesssim  1$ expected for standard single field inflation.
\par
In contrast, the cosmic string signature has an inherently non-perturbative origin so we do not expect
$f_{NL}$ to be small nor, more particularly, $\tau_{NL}$.    For the higher order correlators from cosmic strings (measured relative
to the string power spectrum $\langle \zeta\zeta\rangle_{cs}$), the relevant scaling with the parameter $G\mu$ is $f_{NL} \sim {\cal A} (G\mu)^{-1}$ \cite{Hind09} and, as we shall discuss here, $\tau_{NL} \sim {\cal B}(G\mu)^{-2}$
 with coefficients ${\cal A}$ and ${\cal B}$ determined by geometric and dynamical considerations.
There is a suppressed amplitude ${\cal A}$ for the bispectrum (due to symmetry
cancellations \cite{Hind09}), and we contrast this here with one of the key results
of this chapter which is the much larger relative value of
${\cal B}$ for the trispectrum.  Here, we shall compare the higher order correlators from strings we
calculate with  the dominant inflationary power spectrum $\langle \zeta\zeta\rangle_{\rm{inf}}$ and we will show
that the trispectrum with $\tau_{NL}\gg 1$ is expected to produce stronger
 constraints on cosmic strings than the bispectrum.
\par
In this chapter, we use analytic calculations of the post-recombination string signature to estimate
the power spectrum, bispectrum and trispectrum relevant for forthcoming CMB experiments,
particularly the Planck satellite.    We generalise previous analytic calculations of Hindmarsh and
collaborators for the power spectrum \cite{Hindmarsh} and bispectrum \cite {Hind09} based on a flat sky approximation.  While this
was an important first step, the summation employed was essentially confined only to strings close
to the surface of last scattering and it is only relevant for very small angular scales, i.e.
multipoles $l_1,\,l_2,\,l_3 > 2000$.   Here,
we note that the dominant contribution on large angular scales (as well as very small angles)
comes from the same gravitational GKS effects but from strings at late times well after last scattering.
By integrating in time over the unequal time correlators, we obtain an approximate bispectrum
valid for all $l \gtrsim 50$ (where the flat sky approximation breaks down), i.e. a result which is useful
for both WMAP and Planck.   We also apply these
unequal time correlator methods to the first calculation of the trispectrum (for parralelogram
configurations), which is again valid over a large range of angular scales.   Our calculation
also differs from ref.~\cite{Hind09} by eliminating the divergences they find in their Gaussian integrals for flattened triangles.  We note that these are actually cut-off by the behaviour of other terms in the integrand, yielding a finite result in this regime.   The cross-sectional
shape of the bispectrum we present is relatively
featureless, except for a finite rise near the edges (for flattened triangles) and
suppression towards the corners because of causality contraints (squeezed triangle limit).  This
well-behaved shape is suitable for the bispectrum estimation methods being developed for Planck
and is easily distinguishable  from bispectra predicted
by inflation because of the absence of acoustic peaks \cite{Fergusson}. Of course, these analytic calculations neglect important recombination
effects which also provide significant contributions to the string bispectrum for $500\lesssim l
\lesssim 2000$.
However, determining the extent to which these contributions reinforce or confuse the
non-Gaussian `line-like' signature of cosmic strings will be the subject of future study \cite{FergLandetal}.
\par
These results for the string bispectrum and trispectrum are important for CMB experiments and should be
valid and dominant at both
large and very small angular scales.    While the Planck satellite does not have the resolution
to see individual string signatures, it should be possible to obtain statistically significant
constraints on cosmic strings that compete with limits on $G\mu$ from the power spectrum,
especially for the trispectrum when these techniques are fully developed. On very small
angular scales the string power spectrum begins to dominate over the inflationary signature
($l>3000$) because it is not influenced by exponentially decaying transfer functions.  Here
direct detection of `line-like' signatures may prove possible provided that experiments can
achieve $\mu K$ sensitivities, as anticipated by AMI and ACT, for example. Again, searching
in larger data sets for higher order correlators may provide statistically more significant
results.

\section{Gott-Kaiser-Stebbins effect}
In this section we calculate the Gott-Kaiser-Stebbins (GKS) effect which is expected to give the principal contribution to cosmic strings on subhorizon scales where we may use the flat sky approximation and can ignore the response of cosmological fluids. With these approximations the integrated Sachs-Wolfe effect is equal to the temperature discontinuity in the CMB. In fact we shall show in this chapter that it is the `late time small angular effect' induced by the GKS effect that also dominates large angular scales for cosmic strings. The discussion here follows that of \cite{Hindmarsh}.
\par
 Consider a photon with $4$-momentum $p_{\mu}=(E,0,0,E)$ and a string of coordinates $X^{\mu}(\sigma,t)$, where $(\sigma, t)$ are the worldsheet coordinates of the string (with the gauge chosen such that $t$ corresponds to the (conformal) time coordinate). The unperturbed geodesics can be written as
\begin{eqnarray*}
Z^{\mu}=x^{\mu}+\lambda p^{\mu}.
\end{eqnarray*}
The perturbation to the $4$-momentum along the photon path (which gives the ISW effect) is
\begin{eqnarray}
\delta p_{\mu}=-\frac{1}{2}\int_{\lambda_1}^{\lambda_0}h_{\nu\rho,\mu}(Z(\lambda))p^{\nu}p^{\rho} d\lambda,
\end{eqnarray}
where $h_{\mu\nu}$ is the metric perturbation which in general must be calculated. However, in
refs.~\cite{Stebbins,Hindmarsh} it was pointed out that we can simplify the discussion by considering
\begin{eqnarray*}
\mathbf{\nabla}_{\perp}^2 \delta p_{\mu},
\end{eqnarray*}
where $\mathbf{\nabla}_{\perp}$ represents partial derivatives with respect to the transverse coordinates to the unperturbed geodesics of the photons.
\newline Now suppose that the photon motion is in the $z$-direction. This implies that $(\partial_t-\partial_z)f(Z)=E^{-1}\frac{df}{d\lambda} $. Also we note the following
\begin{eqnarray*}
\mathbf{\nabla}_{\perp}^2 \delta p_{\mu}&=&(\partial_t^2-\partial_z^2-\partial^2)\delta p_{\mu}\\
&=&-\frac{1}{2}\int_{\lambda_0}^{\lambda_1}d\lambda \left[      (\partial_t+\partial_z)(\partial_t-\partial_z)     -\partial^2\right]h_{\nu\rho,\mu}(Z(\lambda))p^{\nu}p^{\rho}.
\end{eqnarray*}
Now suppose we use the harmonic gauge for the metric perturbations. Then we have $\partial^2 h_{\mu \nu}=16\pi G(T_{\mu\nu}-1/2 \eta_{\mu\nu}T)$ which implies that
\begin{eqnarray}
\mathbf{\nabla}_{\perp}^2 \delta p_{\mu}&=&\left[       \frac{1}{2E}(\partial_t+\partial_z)h_{\nu\rho,\mu}p^{\nu}p^{\rho}       \right]^{\lambda_0}_{\lambda_1}+8\pi G\int_{\lambda_0}^{\lambda_1}d\lambda  \partial_{\mu}T_{\nu\rho}p^{\nu}p^{\rho} \,.
\end{eqnarray}
Next we define $\hat{p}^{\mu}=p^{\mu}/E$. Since we are interested in the energy we focus on the $\mu=0$ component. We then obtain\footnote{Note that the formula  $(\partial_t-\partial_z)f(Z)=E^{-1}\frac{df}{d\lambda}$ generalises for a general photon path to $(\partial_t+\hat{p}^i\partial_i)=E^{-1}\frac{d}{d\lambda}$. }
\begin{eqnarray*}
T_{\rho i,0}\hat{p}^i&=&\frac{1}{E}\frac{dT_{\rho i}\hat{p}^i}{d\lambda} -T_{\rho i,j}\hat{p}^i\hat{p}^j\\
\implies T_{\rho\nu,0}\hat{p}^{\nu}&=&\frac{1}{E}\frac{dT_{\rho i}\hat{p}^i}{d\lambda} -T_{\rho i,j}\hat{p}^i\hat{p}^j +T_{\rho j,j}\\
&=&\nabla_{\perp}^i T_{\rho i}+\frac{1}{E}\frac{dT_{\rho i}\hat{p}^i}{d\lambda}\,,
\end{eqnarray*}
where we have used  $T_{\rho j,j}=T_{\rho 0,0}$ in the second line and we define $\nabla_{\perp}^i =\partial^i-\hat{p}^i\hat{p}^j\partial_j$.
This leads to
\begin{eqnarray}
\mathbf{\nabla}_{\perp}^2 \frac{\delta E}{E}=8\pi G\int d\lambda E \nabla_{\perp}^i T_{i \rho}\hat{p}^{\rho}+\frac{1}{2}\left[    (\partial_t+\partial_z)\partial_t\hat{h}-16\pi GT_{\rho i}\hat{p}^i\hat{p}^{\rho}\right]_{\lambda_0}^{\lambda_1}\,.
\end{eqnarray}
The terms in square brackets are boundary terms. These may be important at decoupling. However, due to the finite thickness of the last scattering surface, we expect these fluctuations to be smeared out on the scales of interest. Therefore we neglect these terms. For a string source\footnote{The energy momentum tensor in the conformal gauge is $T^{\mu\nu}=\int dt d\sigma( \dot{X}^{\mu}\dot{X}^{\nu}-{X'}^{\mu}{X'}^{\mu})\delta^{(4)}(x-X)$.} in the light cone gauge ($X^0+X^3=t$) this yields
\begin{eqnarray}\label{mainequation}
\mathbf{\nabla}_{\perp}^2\delta=-8\pi G \mu \int d\sigma \dot{X}.\mathbf{\nabla}_{\perp}\delta^{(2)}(\mathbf{x-X}),
\end{eqnarray}
where the quantities are evaluated at $t_r=t+z-X^3(\sigma,t_r)$,  $\delta^{(2)}({{\mathbf{X}}})$ represents the two dimensional Dirac delta function, and we denote $\delta=\delta E/E$. The wavenumber ${\mathbf{k}}$ is related to the multipole moment, $l$, for $l\gtrsim 60$ via $k^2\approx l(l+1)/(t_0-t_r)^2\approx l(l+1)/t_0^2$, where $t_0$ is the conformal time today.
\newline
In Fourier space we have
\begin{eqnarray}\label{densitypert}
-k^2 \delta_k(t_r)=i8\pi G \mu k^A \int d\sigma \dot{X}^A(\sigma,t_r) e^{i\mathbf{k.X}(\sigma,t_r)},
\end{eqnarray}
where $A=1,2$ runs over the transverse coordinates. In order to compute the spectra of cosmic strings we will assume the string network is in the scaling regime in the matter era. For clarity we will use the conformal time $\eta$ in what follows.

\section{Power Spectrum}
\subsection{Power Spectrum on small angular scales $l\gtrsim 500$}
In order to calculate the power spectrum, we find the  unequal (conformal) time correlator for density perturbations formed at different light cone crossing times and then sum the contributions between the last scattering surface and today to get the total power spectrum. 

Defining the unequal time correlator as
\begin{eqnarray}
\langle\delta_{\mathbf{k_1}}(\eta_1)\delta_{\mathbf{k_2}}(\eta_2)\rangle=(2\pi)^2\delta^{(2)}(\mathbf{k_1+k_2})P(k_1,\eta_1,\eta_2),
\end{eqnarray}
 we can find it by integrating over the string source terms \eqref{densitypert} at the given times,
\begin{eqnarray}\label{PowerSpecUnequal}
P(k,\eta_1,\eta_2)=(8\pi G \mu )^2\frac{k^A k^B}{\mathcal{A}k^4}\int d\sigma d\sigma'\langle \dot{X}^A(\sigma,\eta_1)\dot{X}^B(\sigma',\eta_2) e^{i\mathbf{k.(X(\sigma,\eta_1)-X(\sigma',\eta_2))}} \rangle
\end{eqnarray}
where $\mathcal{A}=(2\pi)^2 \delta(0)$ is a formal area factor. 
\par
Now we use the following three assumptions (for the network at a fixed time as in ref.\cite{Hindmarsh}): (i)
The string ensemble is a Gaussian process, i.e. we can find all the correlation functions in terms of two-point correlators.  (ii)
We have reflection and translation invariance of the transverse coordinates.  (iii)
We have reflection and translation invariance of the worldsheet coordinates.   This means
that for equal time correlators we can write
\begin{eqnarray}\label{ETCs}
 \langle   \dot{X}^A(\sigma,\eta)\dot{X}^B(\sigma',\eta) \rangle &=&\frac{\delta^{AB}}{2}V(\sigma-\sigma',\eta),\\
 \langle  \dot{X}^A(\sigma,\eta){X'}^B(\sigma',\eta) \rangle &=&   \frac{\delta^{AB}}{2}M_1(\sigma-\sigma',\eta),\\
 \langle {X'}^A(\sigma,\eta){X'}^B(\sigma',\eta) \rangle &=&\frac{\delta^{AB}}{2}T(\sigma-\sigma',\eta),
\end{eqnarray}
and hence we have
\begin{align}\label{ETCs2}
 \langle  (X^A(\sigma,\eta)-X^A(\sigma',\eta))^2 \rangle &=&\int_{\sigma'}^{\sigma} \int_{\sigma'}^{\sigma} d\sigma_1 d\sigma_2 T(\sigma_1-\sigma_2,\eta)\equiv\Gamma(\sigma-\sigma',\eta),\\
 \langle  (X^A(\sigma,\eta)-X^A(\sigma',\eta))\dot{X}^A(\sigma',\eta) \rangle &=&\int_{\sigma'}^{\sigma} d\sigma_1 M_1(\sigma_1-\sigma,\eta)\equiv \Pi(\sigma-\sigma',\eta)\,.
\end{align}
This is an extremely powerful relation since we can use the fact that at small angles, i.e. small distances for correlators we have that $\Gamma(\sigma,\eta)\propto \sigma^2, V(\sigma,\eta)\propto \sigma^0, \Pi(\sigma,\eta)\propto \sigma^2$. In particular, we write (for small angles)
\begin{eqnarray}\label{SmallAngle}
\Gamma(\sigma,\eta)\approx s^2\sigma^2,\qquad V(\sigma,\eta)\approx \overline{v}^2,\qquad \Pi(\sigma,\eta)\approx \frac{c_0}{2\hat{\xi}}\sigma^2,
\end{eqnarray}
where $s^2$ is a constant we will derive later, $c_0=\hat{\xi}\langle {X^A}'' .\dot{X}^A\rangle$ and $\hat{\xi}=\tilde{\tilde{\xi}} \eta$ is the (comoving) correlation length of the network which scales in (conformal) time, i.e. $\tilde{\tilde{\xi}}=\rm{constant}$. The values of the parameters $(\overline{v}, s, c_0)$ as indicated by simulations are discussed in Appendix A (Section~\ref{sec:appendixA}). We observe that in \cite{Hindmarsh} $s$ was denoted $\overline{t}$. We have changed notation to avoid confusion with the time parameter, $t$. In what follows, we define $\tilde{\xi}=\tilde{\tilde{\xi}}\eta_{\rm{lss}}/\eta_0\approx 1/500$ to simplify notation. At time $\eta\gtrsim \eta_{\rm{lss}}$ we have  the following relation between the wavenumber $k$ and the multipole $l$
\begin{eqnarray}\label{Notation}
k\hat{\xi}&\approx& l\left(\tilde{\tilde{\xi}}\frac{\eta_{\rm{lss}}}{\eta_0} \right)\frac{\eta}{\eta_{\rm{lss}}} = l\tilde{\xi} \frac{\eta}{\eta_{\rm{lss}}}\nonumber \\
&\approx&\frac{l}{500}\frac{\eta}{\eta_{\rm{lss}}},
\end{eqnarray}
that is, the correlation length at time $\eta\gtrsim \eta_{\rm{lss}} $ corresponds to a multipole $500 (\eta_{\rm{lss}}/\eta)$ since the correlation length at last scattering corresponds to a multipole of approximately $l\approx500$. Therefore, for angular scales below $500$ we cannot integrate back to last scattering and instead to $\eta_{\rm{start}}=\eta_{\rm{lss}}(500/l)>\eta_{\rm{lss}}$.
\par
For unequal times, we have on small angular scales
\begin{eqnarray}
\tilde{\Gamma}(\sigma,\sigma',\eta_1,\eta_2)&=& \langle  (X^A(\sigma,\eta_1)-X^A(\sigma',\eta_2))^2 \rangle \approx s^2 (\sigma-\sigma')^2+\overline{v}^2(\eta_1-\eta_2)^2.
 \end{eqnarray}
Hence, on small angular scales,
\begin{eqnarray}
P(k,\eta_1,\eta_2)\propto e^{-k^2 \overline{v}^2(\eta_1-\eta_2)^2/4},
\end{eqnarray}
and so we have that $P(k,\eta_1,\eta_2)\approx P(k,\eta_1,\eta_1)  $. Thus, to an accurate approximation, we need only consider equal time correlators. In order to obtain the total power spectrum today it is necessary to include the contributions from the equal time correlators between last scattering and today, i.e.
\begin{eqnarray}
P(k)\approx \int_{1}^{\eta_0/\eta_{\rm{lss}}} d\eta_1 P(k,\eta_1,\eta_1),
\end{eqnarray}
where we use the normalised time $\eta/\eta_{\rm{lss}}$.
\par
Now using
\begin{eqnarray*}
P(k,\eta,\eta)=(8\pi G \mu )^2\frac{k^A k^B}{4\mathcal{A}k^4}\int d\sigma_+ \int d\sigma_- \langle   \dot{X}^A(\sigma,\eta)\dot{X}^B(\sigma',\eta) e^{i\mathbf{k.(X}(\sigma,\eta)-{{\mathbf{X}}}(\sigma',\eta))} \rangle, 
\end{eqnarray*}
where $\sigma_+=\sigma+\sigma'$ and  $\sigma_-=\sigma-\sigma'$, we approximate the integral as
\begin{align}\label{UEPowerSpec}
P(k,\eta,\eta)=(8\pi G \mu )^2\frac{\mathcal{L}}{\mathcal{A}}\frac{1}{2 k^2}\int d\sigma_-\left(V(\sigma_-,\eta)+\frac{k^2}{2}\Pi(\sigma_-,\eta)  \right)\exp\left(-\frac{k^2}{4}\Gamma(\sigma_-,\eta)\right),
\end{align}
where we note that the range of integration (in $\sigma$) is limited by the length of string in the network. For large wavenumbers, $k/\eta_0\gtrsim 5000$, we may use the small angle approximations as outlined in \eqref{SmallAngle}. However for smaller wavenumbers we must constrain the range of $\sigma_-$ to $(-\hat{\xi},\hat{\xi})$, since simulations show that in this range the approximations made above are reasonably accurate for $k/\eta_0\gtrsim 500$. The second term in \eqref{UEPowerSpec} gives a subdominant contribution to the power spectrum at all angular scales and therefore we discuss only the first term below. We can make a better estimation motivated by the velocity correlator as found numerically in ref.~\cite{MartShell}\footnote{The velocity correlator here is in the light cone gauge. However the behaviour in any other temporal gauge should follow a similar behaviour.}. In particular, we approximate the velocity correlator (see Figure~\ref{fig:velocity})
 to be\footnote{This approximation obeys the constraint $\int d\sigma V(\sigma)=0$ which is a reflection of conservation of momentum in the network.}
\begin{eqnarray}\label{VelocImprov}
V(\sigma,\eta)\approx \overline{v}^2 \left(1-\frac{|\sigma|}{\hat{\xi}}\right)\exp\left(-|\sigma|/\hat{\xi}\right).
\end{eqnarray}

\begin{figure}[htp]
\centering 
\includegraphics[width=152mm]{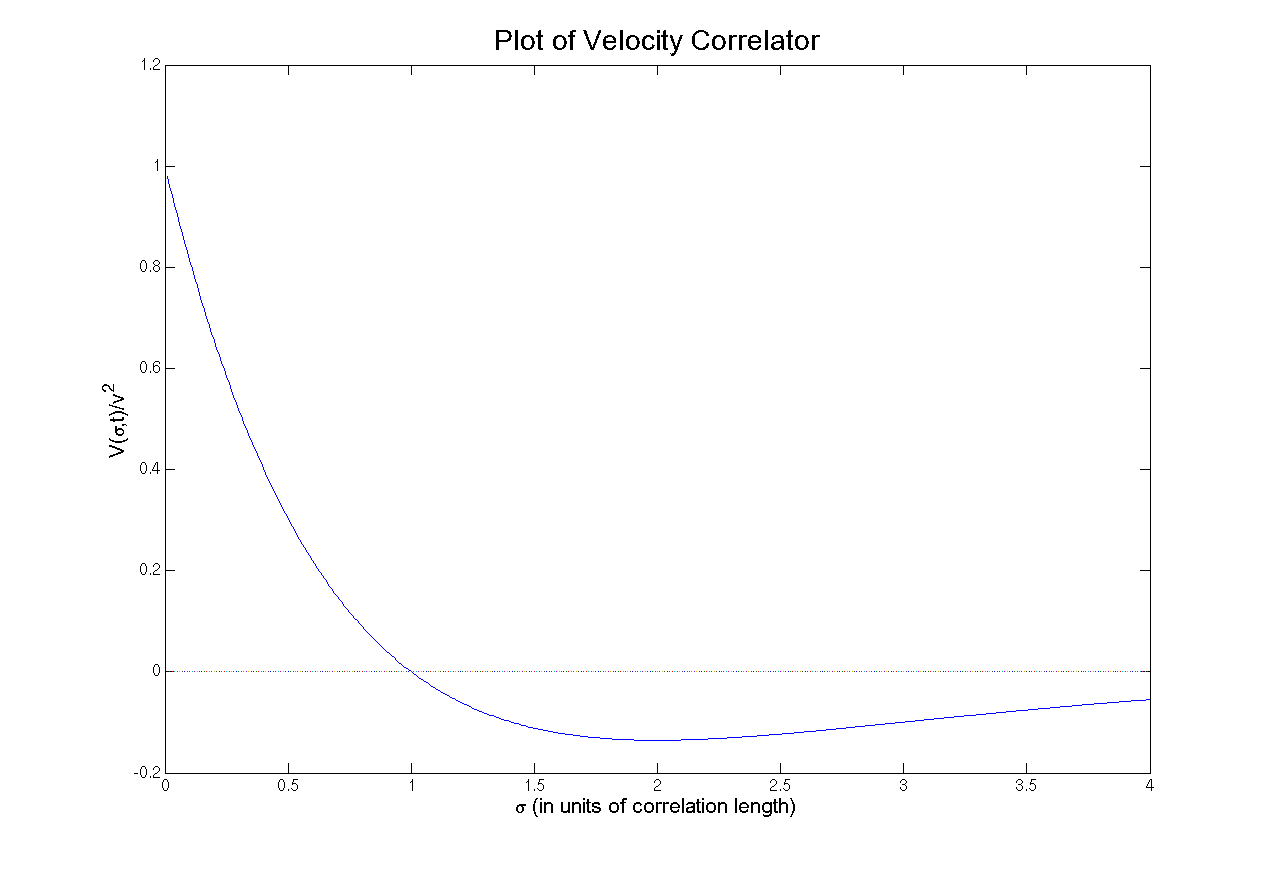}
\caption{Plot of the improved approximation of $V(\sigma,\eta)/(\overline{v}^2)$ (see equation \eqref{VelocImprov}). The velocity approaches zero at the correlation length at each time $t$ and displays an anti-correlation which is an consequence of conservation of momentum in the network.}
\label{fig:velocity}
\end{figure}
Using this we may safely push the range of integration for $V(\sigma,\eta)$ to $\sigma\in(-\infty,\infty)$. This improvement makes little difference to the total power spectrum but becomes very important for the bispectrum and trispectrum. With this approximation for $V$ we find
\begin{align}
&\int d\sigma V(\sigma)\exp\left(-\frac{s^2 k^2 \sigma^2}{4}\right)\approx 2\overline{v}^2 \int_0^{\infty} d\sigma \left(1-\frac{\sigma}{\hat{\xi}}\right)\exp\left(-\sigma/\hat{\xi}\right)\exp\left(-\frac{s^2 k^2 \sigma^2}{4}\right)\\
&\approx 2\overline{v}^2 \frac{\sqrt{\pi}(2+\hat{\xi}^2k^2s^2) \rm{erfc}(1/(\hat{\xi}ks)) \exp(1/(\hat{\xi}ks))^2-2\hat{\xi}ks}{k^3 \hat{\xi}^2s^3}\equiv\frac{2\overline{v}^2}{k^3 \hat{\xi}^2 s^3}f_2(k\hat{\xi}).
\end{align}
\par
 The power spectrum is therefore given by
 \begin{eqnarray*}
k^2P(k)\approx l^2 C_l\approx &&(8\pi G \mu )^2\int_1^{\eta_0/\eta_{\rm{lss}}} \frac{d\eta}{\tilde{\xi}\eta} \frac{\mathcal{L}\hat{\xi}}{\mathcal{A}} \overline{v}^2 \frac{f_2(k\hat{\xi}) }{k^3 \hat{\xi}^2 s^3}\\
\approx &&(8\pi G \mu )^2   \frac{c \overline{v}^2}{s^3} \frac{ 1}{l^3 \tilde{\xi}^3}  \int_1^{\eta_0/\eta_{\rm{lss}}} \frac{d\eta}{\eta^3} \frac{1}{1+\eta} f_2(l\tilde{\xi}\eta).
\end{eqnarray*}
 where we use $\frac{\mathcal{L}\hat{\xi}}{\mathcal{A}}=c/(1+\eta)$ (see Appendix A) with $c$ a constant expected to be of order unity. Using the approximate values for the string network parameters, given in Appendix A, we find the power spectrum is well approximated by - see Figure \ref{fig:powerspec} -
 \begin{eqnarray}
 k^2P(k)\approx l^2 C_l\approx &&(8\pi G \mu )^2 \frac{c \overline{v}^2}{s^3}\times\sqrt{\pi} s^2 \ln\left( \frac{2\eta_0/\eta_{lss}}{1+\eta_0/\eta_{lss}}\right)\frac{1}{1.9+l\tilde{\xi}}\nonumber\\
 \approx&&(8\pi G \mu )^2 \frac{c \overline{v}^2}{s}\sqrt{\pi} \ln\left( \frac{2\eta_0/\eta_{lss}}{1+\eta_0/\eta_{lss}}\right)\frac{1}{1.9+l\tilde{\xi}},
 \end{eqnarray}
 where we recall that $\tilde{\xi}\approx 1/500$ and $\eta_0/\eta_{lss}\approx 50$. Hence, we obtain the result that the small angle power spectrum for multipoles $l \gg 500$ obeys $l^2 C_l \sim l^{-1}$.

\subsection{Power Spectrum on large angular scales, $l\lesssim 500$}
In ref.~\cite{Hindmarsh}, it was shown that if we compute the large angle power spectrum using the contribution of strings at the last scattering surface then $k^2 P(k)\propto k^2$ or $l^2C_l\propto l^2$
on length scales above the correlation length ($l\lesssim 500$). However, we will show in this section the well-known result that the late time small angle contribution of strings, i.e. the sub-horizon effect of cosmic strings as the network evolves between last scattering and today, is the dominant contribution to the low $l$ part of the spectrum. We find this by integrating the unequal time correlator for all times between $\eta>\eta_{\rm{lss}}$ and $\eta_0$ and finding where the peak of the spectrum from such times is located. Such contributions are, in effect, on superhorizon scales at last scattering, and thus are only seeded at later times. Essentially we are tracking the evolution of the peak of the spectrum (located at $l\approx 500$ at last scattering) in time.  Here we simply note that the contributions on superhorizon scales at a time $\eta=\eta_H$ fall off
sufficiently rapidly for $k<aH$ that they can be neglected relative to the late time contributions
$ \eta>\eta_H$ from subhorizon strings.

For large angular scales with $l\lesssim 500$ we no longer integrate back to the last scattering surface since the small angle approximation is no longer valid at that point. Since the correlation length grows with time the small angle approximation will be valid from some time $\eta_{\rm{start}}>\eta_{lss}$. In particular the small angle approximation is assumed to be valid on scales below the correlation length, i.e. $k\hat{\xi}\gtrsim1$. This is given by times $\eta>\eta_{\rm{start}}$ such that
\begin{eqnarray}\label{tstartCalc}
k\hat{\xi}&\approx& l\left(\frac{\tilde{\tilde{\xi}}\eta_{\rm{start}}}{\eta_0} \right)\frac{\eta}{\eta_{\rm{start}}}=l\tilde{\xi} \frac{\eta}{\eta_{\rm{start}}},
\end{eqnarray}
where $\tilde{\xi}$ is now given by $1/\min(500,l)=1/l_m$ and we note that $\eta_0/\eta_{\rm{start}}\approx(\eta_0/\eta_{lss})l_m/500\approx l_m/10$. For $l<500$ $k\hat{\xi}\approx\eta/\eta_{\rm{start}}$.

Integrating the string sources over all times since last scattering we obtain the following approximation to the power spectrum relevant
on large angular scales $l\lesssim 500$ (which obey $l\tilde{\xi}=1$ and $l_m=l$),
 \begin{align}\label{PowerSpec2}
k^2P(k)\approx l^2 C_l\approx &(8\pi G \mu )^2\int_1^{\eta_0/\eta_{\rm{start}}} \frac{d\eta}{\tilde{\xi}\eta} \frac{\mathcal{L}\hat{\xi}}{\mathcal{A}} \overline{v}^2 \frac{f_2(k\hat{\xi}) }{k^3 \hat{\xi}^2 s^3}\nonumber\\
\approx &(8\pi G \mu )^2   \frac{c \overline{v}^2}{s^3} \frac{ 1}{l^3 \tilde{\xi}^3}  \int_1^{\eta_0/\eta_{\rm{start}}} \frac{d\eta}{\eta^3} \frac{1}{1+\frac{500}{l_m}\eta} f_2(l\tilde{\xi}\eta)\nonumber\\
\approx & (8\pi G \mu )^2 \frac{c \overline{v}^2}{s}\sqrt{\pi} \ln\left( \frac{2\eta_0/\eta_{lss}}{1+\eta_0/\eta_{lss}}\right)\frac{1}{2.9}\frac{\ln(1+l/1000)}{\ln(1.5)}.
\end{align}
In the second line of this expression we make note that in general $\frac{\mathcal{L}\hat{\xi}}{\mathcal{A}} =c/(1+ (500/l_m)\eta)$ as outlined in Appendix A.
In figure \ref{fig:powerspec} this analytic approximation to the power spectrum is verified. In summary we have the following approximation to the power spectrum, valid for all $l$,
 \begin{align}\label{PowerSpecAll}
k^2P(k)\approx l^2 C_l\approx & (8\pi G \mu )^2 \frac{c \overline{v}^2}{s}\sqrt{\pi} \ln\left( \frac{2\eta_0/\eta_{lss}}{1+\eta_0/\eta_{lss}}\right)\frac{1}{1.9+l\tilde{\xi}}\frac{\ln(1+l_m/1000)}{\ln(1.5)},
\end{align}
where $l_m=\min(500,l)$ and $\tilde{\xi}=1/l_m$.

This is only strictly valid, as mentioned above, for $l\gtrsim 60$ where the flat sky approximation remains good. However, it can be extrapolated to lower $l$ to provide useful estimates as can our subsequent
results for the bispectrum and trispectrum in these regimes.   From  this expression and the small angle result, we see that the peak of the power spectrum lies near $l\approx 500$ with a logarithmic deviation from scale-invariance on large angular scales and a steeper $l^{-1}$ fall-off on smaller scales (see Figure~\ref{fig:powerspec}). For realistic networks we note that the unequal time correlator source terms will have a broader distribution than our analytic approximation here, which, together with a smoother transition between regimes, will tend to ``wash out'' the sharp central feature.

\begin{figure}[htp]
\centering 
\includegraphics[width=102mm]{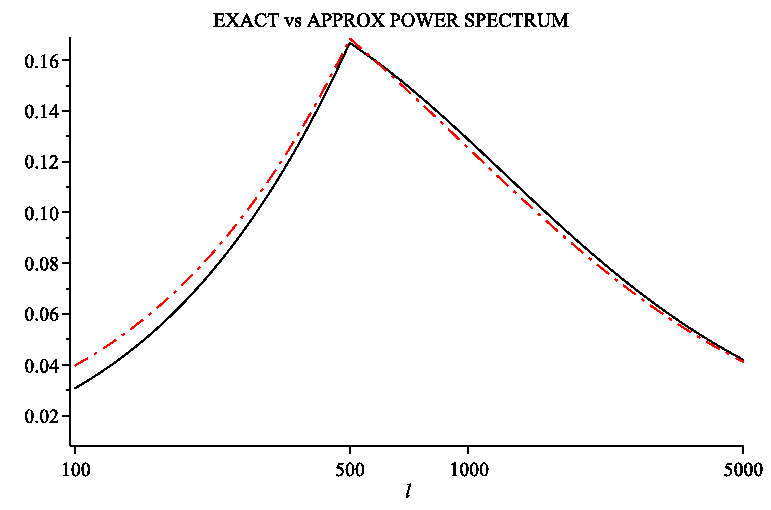}
\caption{Plot of the angular power spectrum $l(l+1)C_l$ against angular multipole for $l\gtrsim 60$. The power spectrum is given in units of $(8\pi G\mu)^2$. The signal peaks at $l\approx 500$ which corresponds to the correlation length at the last scattering surface. The approximation to the power spectrum~\eqref{PowerSpecAll} is given by the solid (black) line while the `exact' result given by the first line of equation~\eqref{PowerSpec2} is given by the dot dash (red) line. The approximation is shown to agree well with the exact result for all multipoles.}
\label{fig:powerspec}
\end{figure}


\subsection{Comparison with simulations}
The proportionality constant used to define $\mathcal{L/A}$ is set by comparison to the normalisation in the Fraisse et al simulations \cite{Fraisse} who found $l^2 C_l/(2\pi)|_{l=1000}\approx 14 (G\mu)^2$.  We choose to compare to these simulations since, as with the approximations used to find the analytic results in this chapter, they also neglect recombination effects.

Substituting the numerical values for $\overline{v}, s$ into the approximation for the power spectrum and also setting $\eta_0/\eta_{lss}=50$ we find
\begin{eqnarray}
\frac{l^2 C_l}{2\pi}^{\rm{cosmic\,\, strings}}|_{l=1000}\approx 15 c (G\mu)^2.
\end{eqnarray}
Therefore we set $c=1$. 
The large angle analytic result is qualitatively in agreement with more sophisticated numerical work on this problem in ref.~\cite{Pogosian} (see more recent results in \cite{Battye2}), though we note that quantitatively the rise towards a peak at $l\approx 500$ is much
steeper for two reasons: (i) The matter-radiation transition breaks scale-invariance because the
density of strings is substantially higher at early times and is therefore dropping through recombination
and afterwards.  (ii) We do not take into account  the pre-recombination perturbations seeded in the
cosmological fluid by cosmic strings which create additional intrinsic CMB anisotropies.  Since the
effect of recombination is not our focus here in this analytic work, we leave these quantitative issues for
further study elsewhere \cite{FergLandetal}.

\section{Bispectrum}

\subsection{Bispectrum on small angular scales}

The calculation for the power spectrum can be easily extended to the bispectrum using
\begin{eqnarray}\label{bispecunequal}
 \langle \delta_{{\mathbf{k}}_1}(\eta_1)\delta_{{\mathbf{k}}_2}(\eta_2)\delta_{{\mathbf{k}}_3}(\eta_3) \rangle =(2\pi)^2\delta(\mathbf{k_1+k_2+k_3})B^T({\mathbf{k}}_1,{\mathbf{k}}_2,{\mathbf{k}}_3,\eta_1,\eta_2,\eta_3),
\end{eqnarray}
yielding \footnote{Again we note that with our conventions for the dimensionless constant our range of $\eta$ is $[1,\eta_0/\eta_{\rm{lss}}]$.}
\begin{align}\label{bispecunequal2}
&B^T({\mathbf{k}}_1,{\mathbf{k}}_2,{\mathbf{k}}_3,\eta_1,\eta_2,\eta_3)=\nonumber\\
&i(8\pi G \mu )^3\frac{1}{\mathcal{A}}\frac{k_1^A k_2^B k_3^C}{k_1^2 k_2^2 k_3^2}\int d\sigma_1d\sigma_2d\sigma_3 \langle 	\dot{X}^A(\sigma_1,\eta_1)\dot{X}^B(\sigma_2,\eta_2)\dot{X}^C(\sigma_3,\eta_3)	e^{i\mathbf{k_a.X_a}     } \rangle, 
\end{align}
where $a=1,2,3$. Similarly to the case for the power spectrum we find that the bispectrum reduces to the integration of the equal time bispectrum between last scattering and today. Therefore, we temporarily drop the subscript referring to time. A similar calculation to this was carried out in \cite{Hind09}. However, in that work the contribution is considered to be dominated by the strings near the last scattering surface. General studies of the bispectrum focus on regular configurations, namely the collapsed, equilateral and squeezed shapes (see Figures~\ref{fig:collapsed},\ref{fig:equil},\ref{fig:squeezed} respectively). The work here is valid for all configurations but, as we will note later we can estimate the signal to noise quite well using only equilateral shapes.

\begin{figure}[htp]
\centering 
\includegraphics[width=72mm]{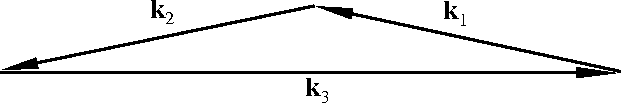}
\caption{Plot of the collapsed (or flattened) triangle configuration with $k_3\approx k_1+k_2$.}
\label{fig:collapsed}
\end{figure}

\begin{figure}[htp]
\centering 
\includegraphics[width=52mm]{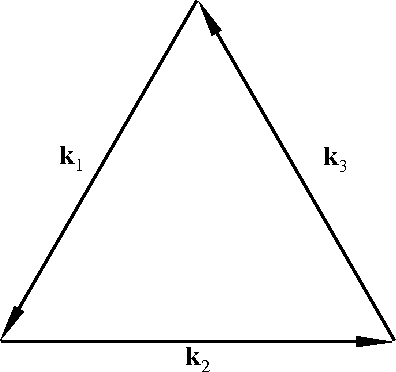}
\caption{Plot of the equilateral triangle configuration with $k_1\approx k_2 \approx k_3$.}
\label{fig:equil}
\end{figure}

\begin{figure}[htp]
\centering 
\includegraphics[width=72mm]{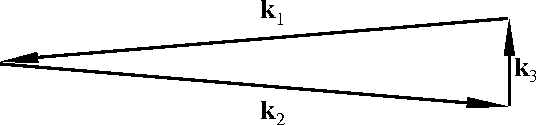}
\caption{Plot of the squeezed triangle configuration with $k_3\ll k_1+k_2$.}
\label{fig:squeezed}
\end{figure}

Using the assumption of a Gaussian process we get that
\begin{eqnarray}\label{bispecExpan}
 \langle 	C^{ABC}e^{iD}	 \rangle =i\left( \langle 	C^{ABC}D	 \rangle     - \langle \dot{X}^A(\sigma_1)D \rangle   \langle \dot{X}^B(\sigma_2)D \rangle  \langle \dot{X}^C(\sigma_3)D \rangle    \right)e^{-\frac{D^2}{2}},
\end{eqnarray}
where we write $C^{ABC}=\dot{X}^A(\sigma_1)\dot{X}^B(\sigma_2)\dot{X}^C(\sigma_3)	$ and $D=k_a.X_a$. Next, setting $\dot{X}_i=\dot{X}(\sigma_i)$, we have
\begin{align*}
 \langle 	C^{ABC}D	 \rangle =& \langle  	\dot{X}_1^A	\dot{X}_2^B \rangle  \langle  	\dot{X}_3^C	\mathbf{k_a.X_a} \rangle + \langle  	\dot{X}_1^A	\dot{X}_3^C \rangle  \langle  	\dot{X}_2^B	\mathbf{k_a.X_a} \rangle + \langle  	\dot{X}_2^B	\dot{X}_3^C \rangle  \langle  	\dot{X}_1^A	\mathbf{k_a.X_a} \rangle \\
=&\frac{\delta^{AB}}{4}V(\sigma_1-\sigma_2)\left[k_1^C\Pi(\sigma_1-\sigma_3)+k_2^C\Pi(\sigma_2-\sigma_3)	\right]\\
&+\frac{\delta^{AC}}{4}V(\sigma_1-\sigma_3)\left[k_1^B\Pi(\sigma_1-\sigma_2)+k_3^B\Pi(\sigma_3-\sigma_2)	\right]\\
&+\frac{\delta^{BC}}{4}V(\sigma_2-\sigma_3)\left[k_2^A\Pi(\sigma_2-\sigma_1)+k_3^A\Pi(\sigma_3-\sigma_1)	\right]
\end{align*}
and
\begin{align*}
 \langle 	D^2	 \rangle =& \langle \mathbf{k_a.X_a}	\mathbf{k_b.X_b} \rangle \\
 =& \langle (k_1^A(X_1^A-X_3^A)+k_2^A(X_2^A-X_3^A))(k_2^B(X_2^B-X_1^B)+k_3^B(X_3^B-X_1^B)) \rangle \\
=&-\frac{1}{2}\left[	\kappa_{13}\Gamma(\sigma_1-\sigma_3)+\kappa_{23}\Gamma(\sigma_2-\sigma_3)	+\kappa_{12}\Gamma(\sigma_1-\sigma_2)		\right],
\end{align*}
where $\kappa_{ij}=\mathbf{k_i . k_j}$. We also find that
\begin{align*}
& \langle \dot{X}_i^A D \rangle =\sum_{j\neq i}\frac{k_j^A}{2} \Pi_{ji}\\
\implies & k_1^A k_2^B k_3^C  \langle \dot{X}^A_1D \rangle   \langle \dot{X}^B_2 D \rangle  \langle \dot{X}^C_3 D \rangle \\
&\qquad\qquad\qquad\qquad=\frac{1}{8}(\kappa_{12}\Pi_{12}+\kappa_{13}\Pi_{13})(\kappa_{12}\Pi_{12}+\kappa_{23}\Pi_{23})(\kappa_{13}\Pi_{13}+\kappa_{23}\Pi_{23})
\end{align*}
(where we use the notation $\Pi_{ij}=\Pi(\sigma_i-\sigma_j)$). The second term in \eqref{bispecExpan} gives a subdominant (as in the case of the power spectrum) contribution to the bispectrum and so we neglect it in the following.
\par
In order to calculate the bispectrum we make a transformation from $\sigma_1,\sigma_2,\sigma_3$ to the variables  $\sigma_{12},\sigma_{13},\sigma_{23}$, where $\sigma_{ij}=\sigma_i-\sigma_j$. It should be noted that only two of $\sigma_{12},\sigma_{13},\sigma_{23}$ are independent. The choice of independent variables depends on the quantities under consideration, e.g. for the term $V_{13}\Pi_{12}$ we use $\sigma_{12}, \sigma_{13}$ and any term involving $\sigma_{23}$ is then written in terms of these variables. Clearly one of the integrations is then independent of the two independent variables chosen and integrates to give $\mathcal{L}$, as in the case of the power spectrum. To elucidate this discussion we consider the term $V_{13}\Pi_{12}$, i.e.
\begin{eqnarray*}
\int d\sigma_1 d\sigma_2 d\sigma _3 V_{13}\Pi_{12}\exp(-D^2/2)=\mathcal{L}\int d\sigma_{12}d\sigma_{13} V_{13}\Pi_{12}\exp(-D^2/2).
\end{eqnarray*}
We can assume the small angle approximations for $\Gamma$ and $\Pi$ but as we will outline we need to use the better approximation in the case of $V$, i.e. equation \eqref{VelocImprov}. In order to simplify the calculation we note that
\begin{align*}
\frac{-D^2}{2}\approx&\frac{s^2}{4}\left(	\kappa_{12}\sigma_{12}^2+ \kappa_{13}\sigma_{13}^2 +\kappa_{23}	(\sigma_{12}-\sigma_{13})^2	\right)\\
\approx&-\frac{s^2}{4}\left((k_2\sigma_{12}-\frac{\kappa_{23}}{k_2}\sigma_{13})^2  +\frac{k_2^2 k_3^2 -\kappa_{23}^2}{k_2^2}\sigma_{13}^2	\right)\\
\approx&-\frac{s^2}{4}\left((k_2\sigma_{12}-\frac{\kappa_{23}}{k_2}\sigma_{13})^2  +K_2^2\sigma_{13}^2	\right),
\end{align*}
where $K_2 =\sqrt{k_2^2 k_3^2 -\kappa_{23}^2}/k_2$.
This indicates that we can make the integration separable using
\begin{align*}
\int d\sigma_{12}d\sigma_{13} &V_{13}\Pi_{12}\exp(-D^2/2)\\
&\approx\int d\sigma_{12} \Pi_{12}\exp\left(-\frac{s^2 k_2^2 \sigma_{12}^2}{4}\right)\int d\sigma_{13}V_{13}\exp\left(-s^2 \frac{ {{\rm{Area}}}_{\triangle}^2}{k_2^2} \sigma_{13}^2\right)\\
&\approx\int d\sigma_{12} \Pi_{12}\exp\left(-\frac{s^2 k_2^2 \sigma_{12}^2}{4}\right)\int d\sigma_{13}V_{13}\exp\left(-\frac{s^2 K_2^2 \sigma_{13}^2}{4}\right),
\end{align*}
where ${{\rm{Area}}}_{\triangle}=\sqrt{k_2^2 k_3^2 -\kappa_{23}^2}/2$ and (for simplicity of notation) we denote $K_i= 2 {{\rm{Area}}}_{\triangle}/k_i $. The separable approximation is clearly valid for $\kappa_{23}/k_2\ll k_2$ and only possibly breaks down for $\kappa_{23}\approx k_3^2$. Assuming $k_2>k_3$ the approximation is therefore valid except for quite degenerate configurations. Also for large wavenumbers the (Gaussian) integrals become quite sharp. In such a case, the separability assumption becomes exact. 
\par
Since $K_2$ (divided by $\eta_0$) may be less than $500$, i.e. correspond to an angular scale below that of the correlation length, we should use a better approximation to $V_{12}$. From earlier considerations we find that
\begin{eqnarray*}
&&\int d\sigma \Pi(\sigma)\exp\left(-\frac{s^2 k^2 \sigma^2}{4}\right)=\frac{c_0}{\hat{\xi}}\int_{0}^{\hat{\xi}} d\sigma \sigma^2 \exp\left(-\frac{s^2 k^2 \sigma^2}{4}\right)\\
&&\approx \frac{c_0}{\hat{\xi}}\frac{2\sqrt{\pi}{\rm{erf}}(k\hat{\xi}s/2)-2 k\hat{\xi}s\exp(-(k\hat{\xi}s/2)^2)}{k^3 s^3}\equiv \frac{c_0}{\hat{\xi}}\frac{1}{k^3 s^3}f_1(k\hat{\xi})
\end{eqnarray*}
and we recall that
\begin{align*}
\int d\sigma V(\sigma)\exp\left(-\frac{s^2 K^2 \sigma^2}{4}\right)
&\approx 2\overline{v}^2 \frac{\sqrt{\pi}(2+\hat{\xi}^2K^2s^2) \rm{erfc}(1/(\hat{\xi}Ks)) \exp(1/(\hat{\xi}Ks))^2-2\hat{\xi}Ks}{K^3 \hat{\xi}^2s^3}\\
&\equiv\frac{2\overline{v}^2}{K^3 \hat{\xi}^2 s^3}f_2(K\hat{\xi}).
\end{align*}
Using this notation (i.e. the above definitions of $f_1$ and $f_2$) the total bispectrum on small angular scales (for which all multipoles $l_i=k_i/\eta_0\gtrsim 500$), after some algebra, reads
\begin{align}\label{bispecAllL}
&B({k_1,k_2,k_3})=\int_1^{\eta_0/\eta_{\rm{lss}}} d\eta(8\pi G\mu)^3 \frac{\mathcal{L}\hat{\xi}}{\mathcal{A}}\frac{c_0 \overline{v}^2}{16 s^6 \hat{\xi}^4}\frac{1}{k_1^2 k_2^2 k_3^2 {{\rm{Area}}}_{\triangle}^3}\nonumber\\
&\times \left( \kappa_{12}k_3^2 f_1(k_3\hat{\xi})f_2(K_3 \hat{\xi}) +\kappa_{13}k_2^2f_1(k_2\hat{\xi})f_2(K_2 \hat{\xi}) +\kappa_{23}k_1^2 f_1(k_1\hat{\xi})f_2(K_1 \hat{\xi}) \right)\\
\implies &(l_1 l_2 l_3)^{4/3}b_{l_1 l_2 l_3}= (8\pi G\mu)^3 c \frac{c_0 \overline{v}^2}{16 s^6 \tilde{\xi}^4}\frac{1}{(l_1 l_2 l_3)^{2/3}{{\rm{Area}}}_{\triangle l}^3 }\int_{1}^{\eta_0/\eta_{lss}}\frac{1}{\eta^4 (1+\eta)}\nonumber\\
&\times\left( l_{12}l_3^2 f_1(l_3 \tilde{\xi}\eta)f_2(L_3 \tilde{\xi}\eta)+l_{13}l_2^2 f_1(l_2 \tilde{\xi}\eta)f_2(L_2 \tilde{\xi}\eta)+l_{23}l_1^2 f_1(l_1 \tilde{\xi}\eta)f_2(L_1 \tilde{\xi}\eta)\right),
\end{align}
where we use $ \frac{\mathcal{L}\hat{\xi}}{\mathcal{A}}=c/(1+\eta)$, ${{\rm{Area}}}_{\triangle l}=\sqrt{l_1^2 l_2^2-l_{12}^2}/2$ and $(k_1 k_2 k_3)^{4/3}B({k_1,k_2,k_3})= (l_1 l_2 l_3)^{4/3}b_{l_1 l_2 l_3}$, with $b_{l_1 l_2 l_3}$ denoted the reduced bispectrum. We denote the angular wavenumbers $l_i=k_i/\eta_0, L_i=K_i/\eta_0, l_{ij}=\kappa_{ij}/\eta_0^2$, and note that
\begin{eqnarray*}
L_i=K_i/\eta_0=2{{\rm{Area}}}_{\triangle}/(k_i\eta_0)=\sqrt{l_i^2 l_j^2 -l_{ij}^2}/l_i= l_{j}\sqrt{1-\cos^2(\theta_{ij})},
\end{eqnarray*}
where $j\neq i$ and $\theta_{ij}$ is the angle between ${\mathbf{k}}_i$ and ${\mathbf{k}}_j$.
We have calculated the bispectrum, \eqref{bispecAllL}, numerically over the full range of multipoles for which the flat sky approximation is valid and the results are shown in Figure~\ref{fig:bispec1}. The end result is a fairly featureless flat bispectrum with very localised and modest upturns for flattened triangles and suppression of squeezed triangles because of causality. This is quite unlike the inflationary bispectra found in ref.~\cite{Fergusson} and should be easily distinguishable given a sufficiently significant signal. In the analytic spirit of this chapter, however, we press on to give some simple analytic approximations to the full bispectrum in different regimes.

Firstly, (for the purpose of comparison to \cite{Hind09}) we consider the range $l_i\tilde{\xi}, L_i\tilde{\xi}\gtrsim 10 \,\, \forall \,i$. Using $\tilde{\xi}=1/500$ this corresponds to the angular multipole range $l_i,L_i\gtrsim 5000$, for which we have $f_1(l_i\tilde{\xi}\eta)\approx 2\sqrt{\pi}$ and $f_2(L_i \tilde{\xi}\eta)\approx \sqrt{\pi}\tilde{\xi}^2 L_i^2 s^2 \eta^2$. This then implies that
\begin{align}\label{Hindapprox1}
 ((l_1^2 l_2^2 l_3^2)^{2/3}b_{l_1 l_2 l_3})^{l \gg 500}&\approx-(8\pi G\mu)^3 c \frac{c_0 \overline{v}^2\pi}{4 s^4\eta_0^2 \tilde{\xi}^2}\frac{l_1^2+l_2^2+l_3^2}{(l_1^2 l_2^2 l_3^2)^{1/3} {{\rm{Area}}}_{\triangle l}} \int_{1}^{\eta_0/\eta_{lss}}d\eta \frac{1}{\eta^2 (1+\eta)}\nonumber\\
 &\approx-(8\pi G\mu)^3 c \frac{c_0 \overline{v}^2\pi}{2 s^4  \tilde{\xi}^2}\frac{l_1^2+l_2^2+l_3^2}{(l_1^2 l_2^2 l_3^2)^{1/3} \sqrt{l_1^2l_2^2-l_{12}^2}}(1-\ln(2))\propto \frac{1}{\tilde{\xi}^2 l^2}.
\end{align}
The difference between this result and that of Hindmarsh \cite{Hind09} is due to the approximation made to make the integration over the $\sigma$ coordinates separable. This makes little difference to the quantitative result. 
\begin{figure}[htp]
\centering 
\includegraphics[width=102mm]{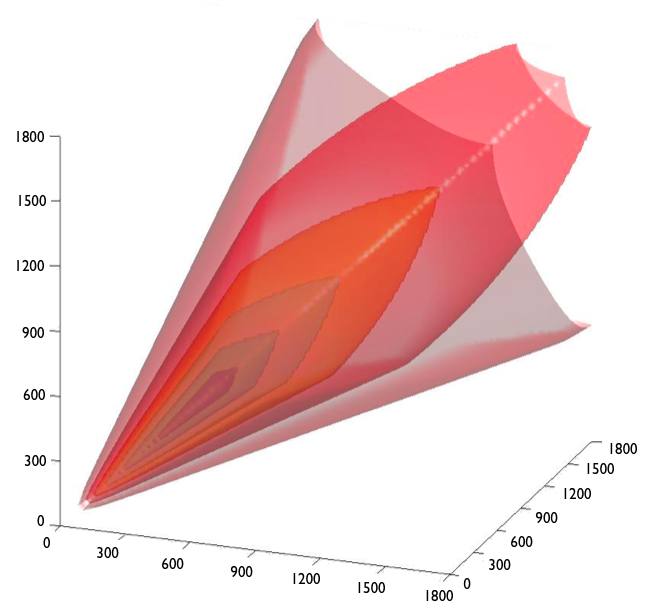}
\caption{Plot of the 3D bispectrum  $(l_1 l_2 l_3)^{4/3}b_{l_1 l_2 l_3}$. The signal is seen to peak at near the correlation length at last scattering for which all $l_i\approx 500$. Due to the resolution of the data points we do not pick up the rise towards the edge. The lighter shading for lower multipoles ($l<500$) indicates the expected slow (logarithmic) fall off indicated in equation~\eqref{bispecFunction2}. It is evident that the bispectrum from cosmic strings is quite flat unlike the local or equilateral bispectra (figure~\ref{fig:local3d}) which are quite `bumpy' due to the acoustic peaks from the transfer functions.}
\label{fig:bispec1}
\end{figure}

\subsection{Bispectrum including large angular scales}
If $\min k_i /\eta_0=l_m\lesssim 500$, then we must restrict the range of times over which we sum to $(\eta_{\rm{start}},\eta_0)$. 
From our numerical considerations we find that, as in the case for the power spectrum, the bispectrum decreases approximately as a logarithm as we restrict the range, i.e. the bispectrum $\propto \ln(1+(l_m/1000))$. Therefore, we can include this effect by multiplying the approximations by $\ln(1+(l_m/1000))/\ln(1.5)$ for $l_m\lesssim 500$. The extension to large angular scales allows us to estimate the level of non-Gaussianity in the parameter range of WMAP and Planck. Taking this into consideration we consider the integral \eqref{bispecAllL}. From this equation it is clear that we must compute the following integral:
\begin{eqnarray}\label{keyApprox1}
I(l_i \tilde{\xi}, L_i \tilde{\xi})=\int_{1}^{\eta_0/\eta_{\rm{start}}}d\eta \frac{1}{\eta^4(1+(500/l_m)\eta)} f_1(l_i\tilde{\xi}\eta)f_2(L_i \tilde{\xi}\eta),
\end{eqnarray}
where we note that in general $\frac{\mathcal{L}\hat{\xi}}{\mathcal{A}}=c/(1+(500/l_m)\eta)$.
As for the large angle power spectrum we set $\tilde{\xi}=\tilde{\tilde{\xi}}\eta_{\rm{start}}/\eta_0=1/\min(l_i,500)=1/l_m$. We note here that in the integral we use the normalised time $\eta/\eta_{\rm{start}}$. 
We now proceed to analytically estimate the integral \eqref{keyApprox1} across the full range of multipoles. In particular we express our estimate over three different ranges of the multipoles (which will also encompass equation~\eqref{Hindapprox1}) . These approximations are derived in more detail in Appendix B. The range of validity of these respective expressions is shown in figure~\ref{fig:BispecRange}.
\begin{figure}[htp]
\centering 
\includegraphics[width=102mm]{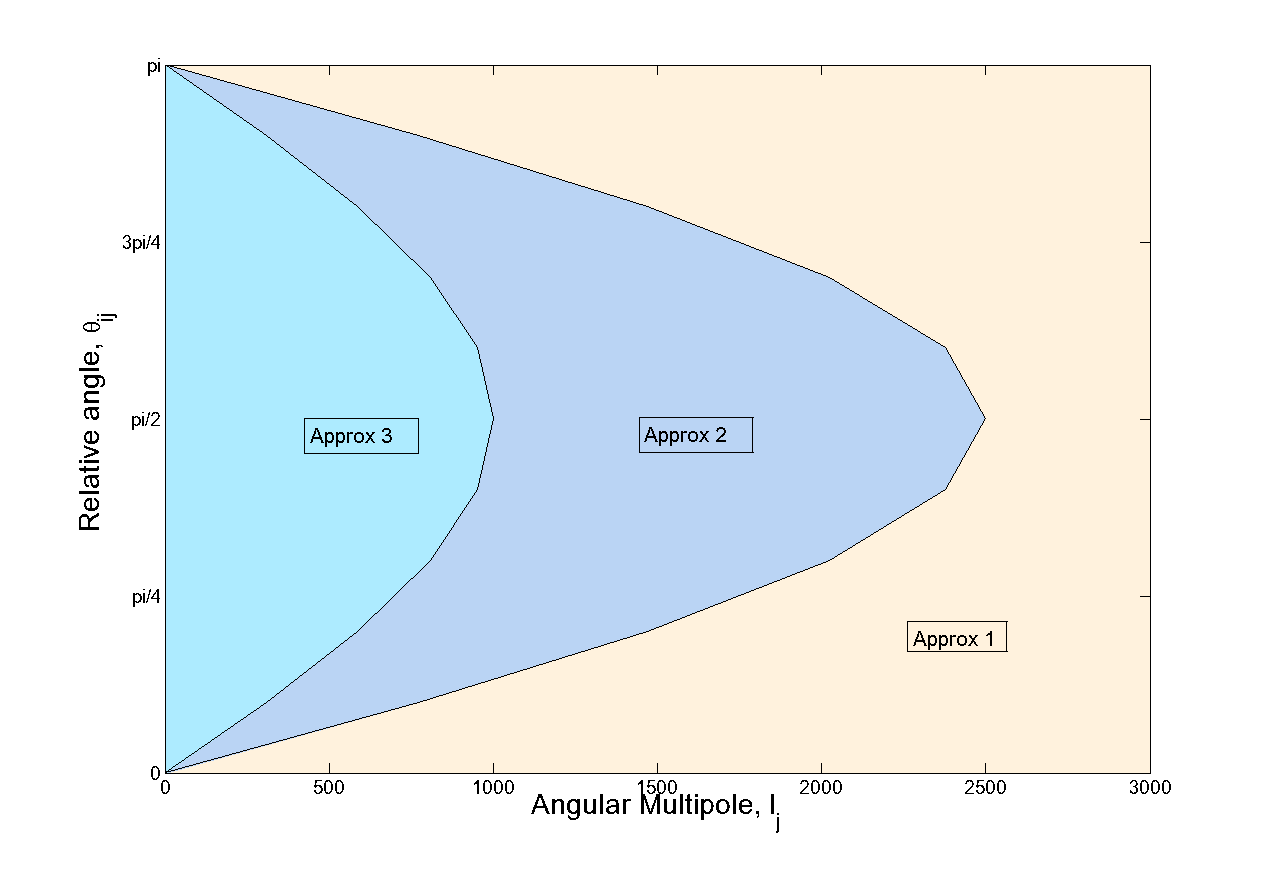}
\caption{Plot of the domain of validity of Approximations $1,2,3$ described in equations \eqref{fullapprox2Aa}-\eqref{fullapprox4Aa}. The angular multipoles, $l_j$, are expressed in units of $l_m/500$ where $l_m=\rm{min}(500,l_i)$. In particular, Approximation 1, valid for $L_i\tilde{\xi}>5$ or  $l_{j}\sqrt{1-\cos^2(\theta_{ij})}>2500(l_m/500)$, is shown in the rightmost region of the plot. Approximation 3, valid for $L_i\tilde{\xi}<5$ or  $l_{j}\sqrt{1-\cos^2(\theta_{ij})}<1000(l_m/500)$ is shown in the leftmost region of the plot. Approximation 2 covers the remaining region.}
\label{fig:BispecRange}
\end{figure}

\paragraph*{Approximation (1):}
For $L_i \tilde{\xi}\gtrsim 5$ (or $l_{j}\sqrt{1-\cos^2(\theta_{ij})}\gtrsim 2500(l_m/500)$ for $j\neq i$) and any value of $l_i$ we find that 
\begin{align}\label{fullapprox2Aa}
I(l_i \tilde{\xi}, & L_i \tilde{\xi})_{\rm{approx-}1} \nonumber\\
&\approx 2\pi s^2 (1-\ln(2))(L_i\tilde{\xi})^2 \left(1-\frac{2}{\sqrt{\pi}s L_i \tilde{\xi}}\right){\rm{erf}}(s l_i\tilde{\xi}/2)\exp\left(-\frac{0.6}{l_i\tilde{\xi}} \right)\frac{\ln(1+l_m/1000)}{\ln(1.5)}
\end{align}
The error function appears since the Gaussian is not sharp enough for small values of $l$.
 
\paragraph*{Approximation (2):}
For $L_i\tilde{\xi}\in (2,5)$ (or $1000(l_m/500)\lesssim l_{j}\sqrt{1-\cos^2(\theta_{ij})}\lesssim 2500(l_m/500)$ for $j\neq i$) we find 
\begin{align}\label{fullapprox3Aa}
I(l_i \tilde{\xi},& L_i \tilde{\xi})_{\rm{approx-}2}\nonumber\\
&\approx 0.3 \frac{(L_i\tilde{\xi})^3}{1+L_i\tilde{\xi}/3}\exp\left(-\frac{0.15}{L_i\tilde{\xi}} \right){\rm{erf}}(s l_i\tilde{\xi}/2)\exp\left(-\frac{0.45}{l_i\tilde{\xi}} \right)\frac{\ln(1+l_m/1000)}{\ln(1.5)}.
\end{align}
The choice of constants assure that we match (approximately) with \eqref{fullapprox2Aa} at $L\tilde{\xi}=5$ or $L=2500$. 

\paragraph*{Approximation (3):}
For $L_i\tilde{\xi}\lesssim 2$  (or $l_{j}\sqrt{1-\cos^2(\theta_{ij})}\lesssim 1000(l_m/500)$) we find that 
\begin{align}\label{fullapprox4Aa}
I(l_i \tilde{\xi}, & L_i \tilde{\xi})_{\rm{approx-}3}\nonumber\\ 
&\approx 0.18 (L_i\tilde{\xi})^3   \exp\left(-\frac{0.15}{L_i\tilde{\xi}} \right){\rm{erf}}(s l_i\tilde{\xi}/2)\exp\left(-\frac{0.15}{l_i\tilde{\xi}} \right)\frac{\ln(1+l_m/1000)}{\ln(1.5)},
\end{align}
where the choice of constant $0.18$ again assures approximate matching with  \eqref{fullapprox3Aa} at $L\tilde{\xi}=2$ or $L=1000$.

 It is important from these approximations to note that, although the bispectrum grows as we approach collapsed configurations -for which ${{\rm{Area}}}_{\triangle l}\propto L\rightarrow 0$ -, it is not divergent (see equations \eqref{bispecAllL} and \eqref{fullapprox4Aa}) and attains a finite value on the boundary.
  \par 
To summarise, let us provide a final simple expression which is valid over all multipoles. Expressions \eqref{fullapprox2Aa}, \eqref{fullapprox3Aa} and \eqref{fullapprox4Aa} imply
\begin{align}\label{bispecFunction2}
&(l_1 l_2 l_3)^{4/3}b_{l_1 l_2 l_3}\nonumber\\
&= (8\pi G\mu)^3 c \frac{c_0 \overline{v}^2}{2 s^6 \tilde{\xi}^4}\frac{1}{(l_1 l_2 l_3)^{2/3}}\left( l_{12}l_3^2 \frac{I(l_3 \tilde{\xi},L_3 \tilde{\xi})}{l_3^3 L_3^3}+l_{13}l_2^2  \frac{I(l_2 \tilde{\xi},L_2 \tilde{\xi})}{l_2^3 L_2^3}+l_{23}l_1^2  \frac{I(l_1 \tilde{\xi},L_1 \tilde{\xi})}{l_1^3 L_1^3}\right),\nonumber\\
&= (8\pi G\mu)^3 c \frac{c_0 \overline{v}^2}{2 s^6 }\frac{1}{(l_1 l_2 l_3)^{2/3}} \frac{1}{\tilde{\xi}^2 \sqrt{l_1^2 l_2^2 -l_{12}^2}}    \Big( l_{12}{\rm{erf}}(sl_3\tilde{\xi}/2)  D(l_3 \tilde{\xi},L_3 \tilde{\xi})  \nonumber \\
&\,\,+l_{13}{\rm{erf}}(sl_2\tilde{\xi}/2)  D(l_2 \tilde{\xi},L_2 \tilde{\xi}) + l_{23}{\rm{erf}}(sl_1\tilde{\xi}/2)  D(l_1 \tilde{\xi},L_1 \tilde{\xi})   		 \Big)\frac{\ln(1+l_m/1000)}{\ln(1.5)},
\end{align}
where we use ${{\rm{Area}}}_{\triangle l}=\sqrt{l_1^2 l_2^2-l_{12}^2}=2l_i L_i$, and where
\begin{align}
D(l_i\tilde{\xi},L_i\tilde{\xi})&\approx2\pi s^2 (1-\ln(2))\left(1-\frac{2 }{\sqrt{\pi}s \tilde{\xi}L_i}\right) {\rm{exp}}\left(-\frac{0.6}{l_i\tilde{\xi}}\right) \quad {\rm{if}}\,\, L_i\tilde{\xi}> 5 \quad (L_i>2500)\nonumber  \\
&\approx0.3 \frac{L\tilde{\xi}}{1+L\tilde{\xi}/3}{\rm{exp}}\left(-\frac{0.15}{L\tilde{\xi}}\right){\rm{exp}}\left(-\frac{0.45}{l_i\tilde{\xi}}\right)\,\,\quad\quad\, {\rm{if}}\,\, L_i\tilde{\xi}\in (2,5) \quad \nonumber\\
&\approx0.18 L\tilde{\xi}{\rm{exp}}\left(-\frac{0.15}{L\tilde{\xi}}\right){\rm{exp}}\left(-\frac{0.15}{l_i\tilde{\xi}}\right) \quad\qquad\qquad {\rm{if}}\,\, L_i\tilde{\xi}< 2 \quad (L_i<1000)).
\end{align}
It should be noted here that for ${{\rm{Area}}}_{\triangle l}\propto L_i<1000$ that $D(l_i\tilde{\xi},L_i\tilde{\xi})\propto \tilde{\xi}\sqrt{l_1^2 l_2^2 -l_{12}^2}/l_i$. Therefore in this regime $b_{l_1 l_2 l_3}$ is independent of ${{\rm{Area}}}_{\triangle l}$ and therefore remains finite. This is contrary to the expectation of equation \eqref{Hindapprox1} which shows that for large multipoles $b_{l_1 l_2 l_3}$ increases as ${{\rm{Area}}}_{\triangle l}$ decreases. Thus, reassuringly, the bispectrum is shown to remain finite over all configurations.
This approximation is valid for multipoles $l\gtrsim 50$, i.e. including the range of multipoles relevant for Planck and WMAP ($l\lesssim 2000$). As can be observed from equation \eqref{bispecFunction2}, if $c_0=0$, as would be the case for a network of strings in a non-expanding universe with no small scale structure, the bispectrum is zero. In particular, $c_0$ indicates the level of velocity-curvature correlation. However, as with the case of the power spectrum we caution that this analysis neglects contributions due to recombination effects.


\subsection{Skewness}
As shown in \cite{RSF10} the skewness $g_1=\langle (\Delta T/T)^3\rangle/\langle (\Delta T/T)^2\rangle^{3/2}$ can be written in terms of multipoles as
\begin{eqnarray}\label{skewform}
g_1=   \sqrt{4\pi}    \frac{      \sum_{l_i} h_{l_1 l_2 l_3}^2 b_{l_1 l_2 l_3}     }{(\sum_l (2 l+1)C_l)^{3/2}       },
\end{eqnarray}
where $b_{l_1 l_2 l_3}$ is the reduced bispectrum and $h_{l_1 l_2 l_3}$ is given in terms of the Wigner $3j$ symbol by
\begin{eqnarray*}
h_{l_1 l_2 l_3}=\sqrt{\frac{(2l_1 +1)(2l_2 +1)(2l_3 +1)}{4\pi}} \left( \begin{array} {ccc}
l_1 & l_2 &l_3 \\
0 & 0  &0\end{array} \right).
\end{eqnarray*}
Since, in the case of cosmic strings, the dominant contribution is given for $l\gtrsim 100$, we can approximate the skewness formula using $(2l+1)\approx 2l$. We have also seen that the bispectrum shape is quite flat for multipoles in the range of Planck. The skewness formula~\eqref{skewform} is clearly dominated by these multipoles since $l^3h_{lll}^2 b_{lll}\sim l^4 b_{l l l}\sim l^{-2}$ for large $l$. Therefore we may approximate the skewness using the equilateral values of the bispectrum. In particular,
\begin{eqnarray}
g_1\approx \sqrt{\frac{\pi}{2}} \frac{\sum_l l^2 h_{l l l}^2 b_{l l l}}{(\sum_l l C_l)^{3/2}}.
\end{eqnarray}
In \cite{Fergusson} it was shown that $h_{l l l}^2\approx (4/\sqrt{3}\pi^2)l$. This implies
\begin{eqnarray} 
g_1\approx 0.3 \frac{\sum_l l^3 b_{l l l}}{(\sum_l l C_l)^{3/2}}.
\end{eqnarray}
Substituting in the equilateral values of the reduced bispectrum for cosmic string we find
\begin{eqnarray}
g_1^{\rm{cosmic\, \,strings}}\approx -\mathcal{O}(0.1).
\end{eqnarray}
This agrees well with the numerical simulations of Fraisse et al \cite{Fraisse,Ringeval} who found the mean sample skewness from a set of $300$ independent CMB maps to be $g_1\approx -0.22\pm0.12$.

\subsection{Estimate of $f_{NL}$}
Current constraints for non-Gaussianity are generally expressed in terms of the parameter $f_{NL}$. Due to calculational difficulties in finding bispectrum estimators in general, the constraints are given in terms of three (separable) primordial models \cite{Fergusson}. In particular, for local and equilateral (a separable approximation to DBI inflation) inflation \cite{Zaldarr2, Moss, Komatsu}
\begin{eqnarray}
-7<f_{NL}^{\rm{local}}<59\\
-151<f_{NL}^{\rm{equil}}<253.
\end{eqnarray}
However, note the much more generic set of bispectrum constraints presented recently in \cite{FLS10}.
\par
In this section we will attempt to estimate the level of non-Gaussianity by comparing the cosmic string bispectrum with that of the theoretical local model for which $l^4b_{l l l}\approx 2\times 10^{-17}f_{NL}$. In order to remove the dependency on a particular value of $l$ we will, in particular, normalise the signal to noise ratio of cosmic strings to that of the local model. However, as we shall note again later, cosmic strings are not expected to follow a local model. This is highlighted by comparing the bispectrum of cosmic strings in figure~\ref{fig:bispec1} with that of the local model and the equilateral model in figure~\ref{fig:local3d}.
\begin{figure}[htp]
\centering 
\includegraphics[width=102mm]{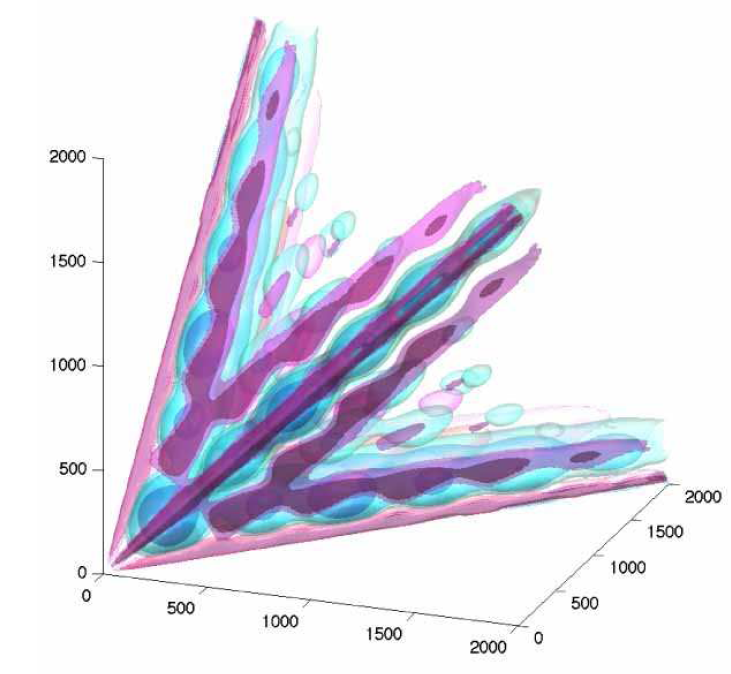}
\includegraphics[width=102mm]{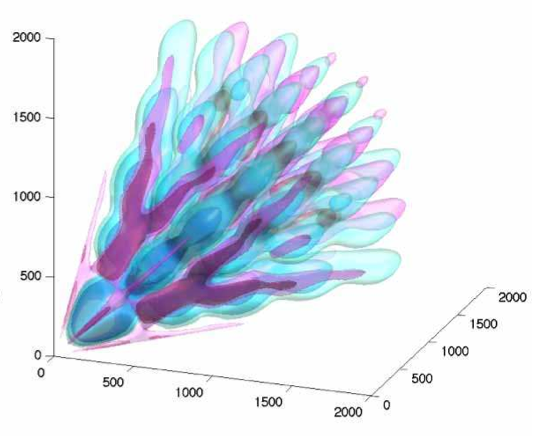}
\caption{(Top) 3D plot of the reduced bispectrum for the local model of inflation~\cite{Fergusson}. The primordial signal is peaked for squeezed configurations. (Bottom) 3D plot of the reduced bispectrum for the equilateral model of inflation~\cite{Fergusson}. The model is so-called because the primordial signal is peaked for equilateral configurations.}
\label{fig:local3d}
\end{figure}
\newline
As detailed in \cite{Babich}, the signal to noise ratio of the bispectrum is given by
\begin{eqnarray}
(S/N)^2=\frac{2}{\pi}\sum_{l_1, l_2 ,l_3}l_1 l_2 l_3\left( \begin{array} {ccc}
l_1 & l_2 &l_3 \\
0 & 0  &0\end{array} \right)^2 \frac{b_{l_1 l_2 l_3}^2}{C_{l_1}C_{l_2}C_{l_3}},
\end{eqnarray}
where $b_{l_1 l_2 l_3}$ corresponds to the reduced bispectrum, which is precisely the bispectrum as calculated in the flat sky limit and where we assume that the noise is cosmic variance dominated.
As shall be established later in this section (see Figures \ref{fig:bispec1},\ref{fig:slice750} - \ref{fig:slice1500}), within the range of interest for Planck we can find a reasonably accurate estimate for $f_{NL}$ by summing over the equal $l$ values. This gives
\begin{eqnarray}
(S/N)^2&\approx&\frac{2}{\pi}\sum_{l}l^5 \left( \begin{array}{ccc}
l & l &l \\
0 & 0  &0\end{array} \right)^2 \frac{b_{l l l}^2}{C_{l}C_{l}C_{l}}\nonumber\\
&\approx&\frac{1}{2\sqrt{3}\pi^5}\sum_{l} \frac{l (l^4 b_{lll})^2}{(l^2 C_l/(2\pi))^3}.
\end{eqnarray}
For the local model of inflation the signal to noise ratio is proportional to $f_{NL}$. Since the local bispectrum is proportional to $l^{-4}$ for $l\lesssim 1000$ \cite{Fergusson}  we normalise the signal to noise ratio of cosmic strings to that of local models of inflation, for which $l^4 b_{lll}\approx 2\times 10^{-17}f_{NL}$\footnote{It should be noted that, by parametrising the signal to noise with the term $f_{NL}$, we are, in essence, expressing the amplitude of the cosmic string bispectrum by comparison to a (primordial) local model which would give the same signal to noise. Of course the cosmic string bispectrum and the local model bispectrum have different shape dependencies and so may be otherwise distinguished.}. We have computed $l^2 C_l/(2\pi)$ using CMBFAST with WMAP5 values for the parameters.
Therefore, defining,
\begin{eqnarray}
f_{NL}^2=(S/N)^{2}_{\rm{cs}}/(S/N)^{2}_{\rm{inf}},
\end{eqnarray}
we plot $f_{NL}$ against $l$ for multipoles summed between 60 and $l$ in Figure~\ref{fig:fnl}. The fluctuations in the plot arise since cosmic strings do not strictly follow a local model and because of the transfer functions appearing in the $C_l$s. 
\newline


Using these values of the cosmic string model parameters and $G\mu=7\times 10^{-7}$ we estimate $f_{NL}\sim -40$ (see Figure ~\ref{fig:fnl}). On the other hand, taking $G\mu=2.5\times 10^{-7}$ (the current constraint from Nambu-Goto simulations) we would estimate $f_{NL}\sim -2$.  Compare this with the most stringent estimate of $f_{NL}$ for local models $f_{NL}=38\pm 21$ (\cite{Zaldarr2}). It should also be stressed that cosmic strings produce a very different shape to  local non-Gaussianity, so analysis of compatibility with WMAP or Planck needs to be specifically investigated (\cite{FergLandetal}).

We have plotted the three dimensional bispectrum in the multipole range of Planck $l_i< 2000$ by numerically integrating the analytic result (see Figure~\ref{fig:bispec1}). In particular we plot the quantity $(l_1 l_2 l_3)^{4/3}b_{l_1 l_2 l_3}$. We find that the signal peaks at around $(500,500,500)$ and drops towards the corners. Therefore, our assumption that using the equal $l$ values to estimate the level of non-Gaussianity seems valid.
However, we do expect that the signal rises very near the edge (before it levels out) and so we use our approximations as detailed earlier to track this behaviour on slices across the bispectrum. In particular we show the contour slices for  $l_1+l_2+l_3=750$ (Figure~\ref{fig:slice750}), $l_1+l_2+l_3=2000$ (Figure~\ref{fig:slice2000})  and $l_1+l_2+l_3=3000$ (Figure~\ref{fig:slice3000}). In order to see the behaviour more clearly we have multplied the bispectrum by a negative sign. The slices are plotted in units of $\epsilon^3=(8\pi G\mu)^3$. There is a peak towards the edge but only a factor of a few larger than at the centre point. To illustrate this further we plot the slice through the bispectrum for $l_1+l_2+l_3=1500$ (Figure~\ref{fig:slice1500}) so that the edge is highlighted. This shows that the growth of the peak only begins very near the edge for these slices. Therefore, we feel justified in neglecting the edges in our estimator for $f_{NL}$.


\begin{figure}[htp]
\centering 
\includegraphics[width=102mm]{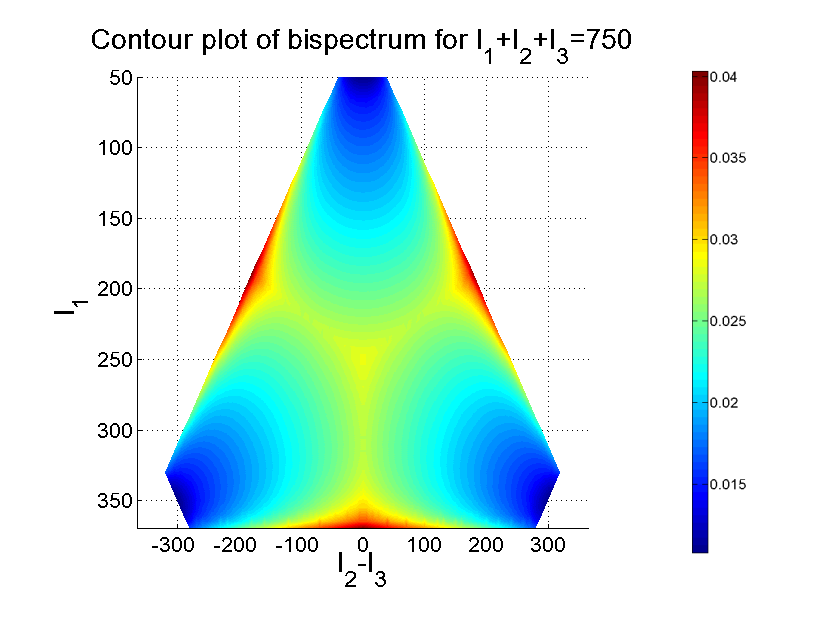}
\caption{Contour plot of the slice through the string bispectrum at $\sum_i l_i=750$.}
\label{fig:slice750}
\end{figure}
\begin{figure}[htp]
\centering 
\includegraphics[width=102mm]{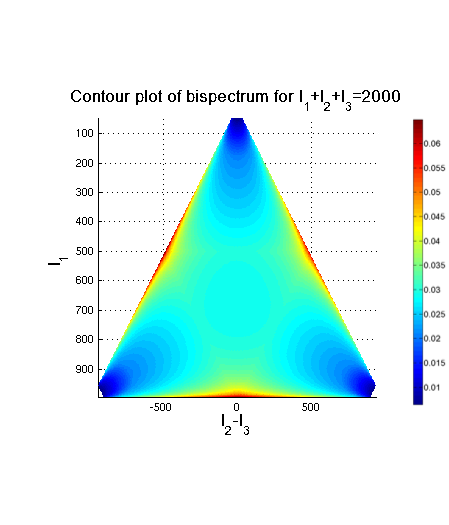}
\caption{Contour plot of the slice through the string bispectrum at $\sum_i l_i=2000$.}
\label{fig:slice2000}
\end{figure}
\begin{figure}[htp]
\centering 
\includegraphics[width=102mm]{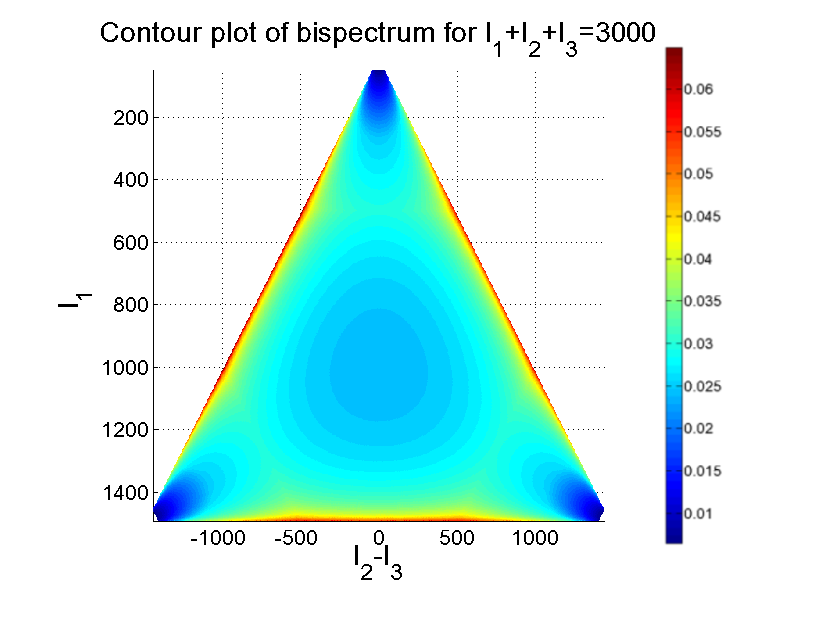}
\caption{Contour plot of the slice through the string bispectrum at $\sum_i l_i=3000$.}
\label{fig:slice3000}
\end{figure}
\begin{figure}[htp]
\centering 
\includegraphics[width=102mm]{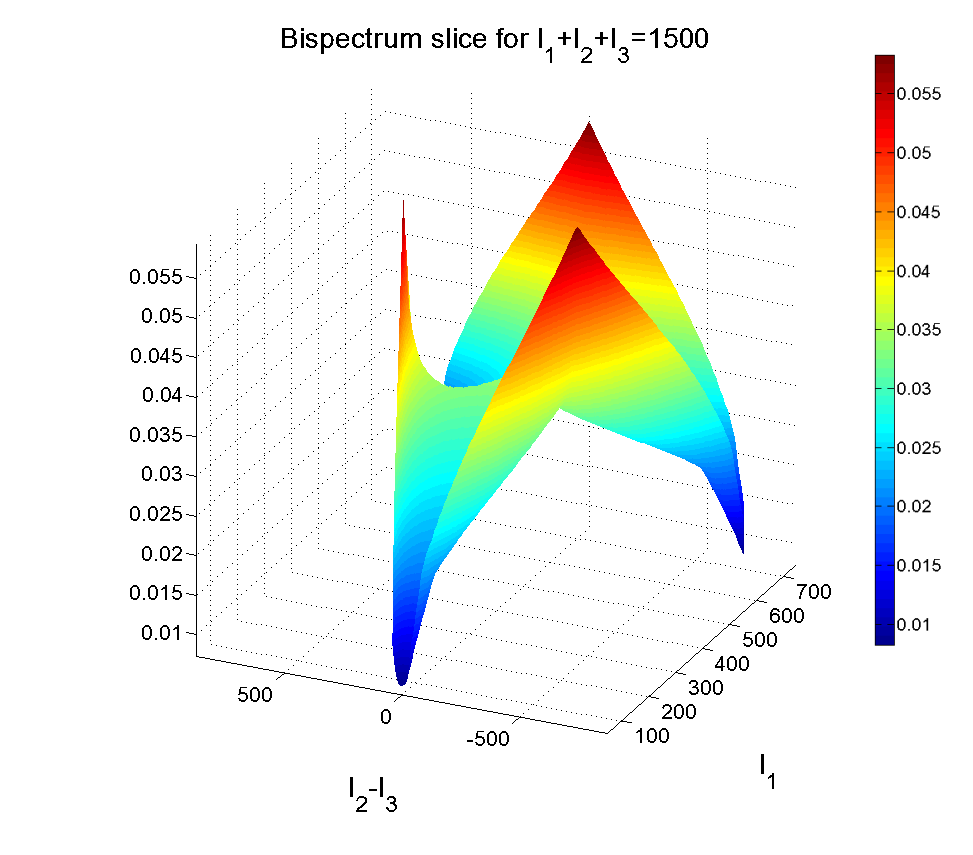}
\caption{Plot of the slice through the string bispectrum at $\sum_i l_i=1500$.}
\label{fig:slice1500}
\end{figure}

\begin{figure}[htp]
\centering 
\includegraphics[width=152mm]{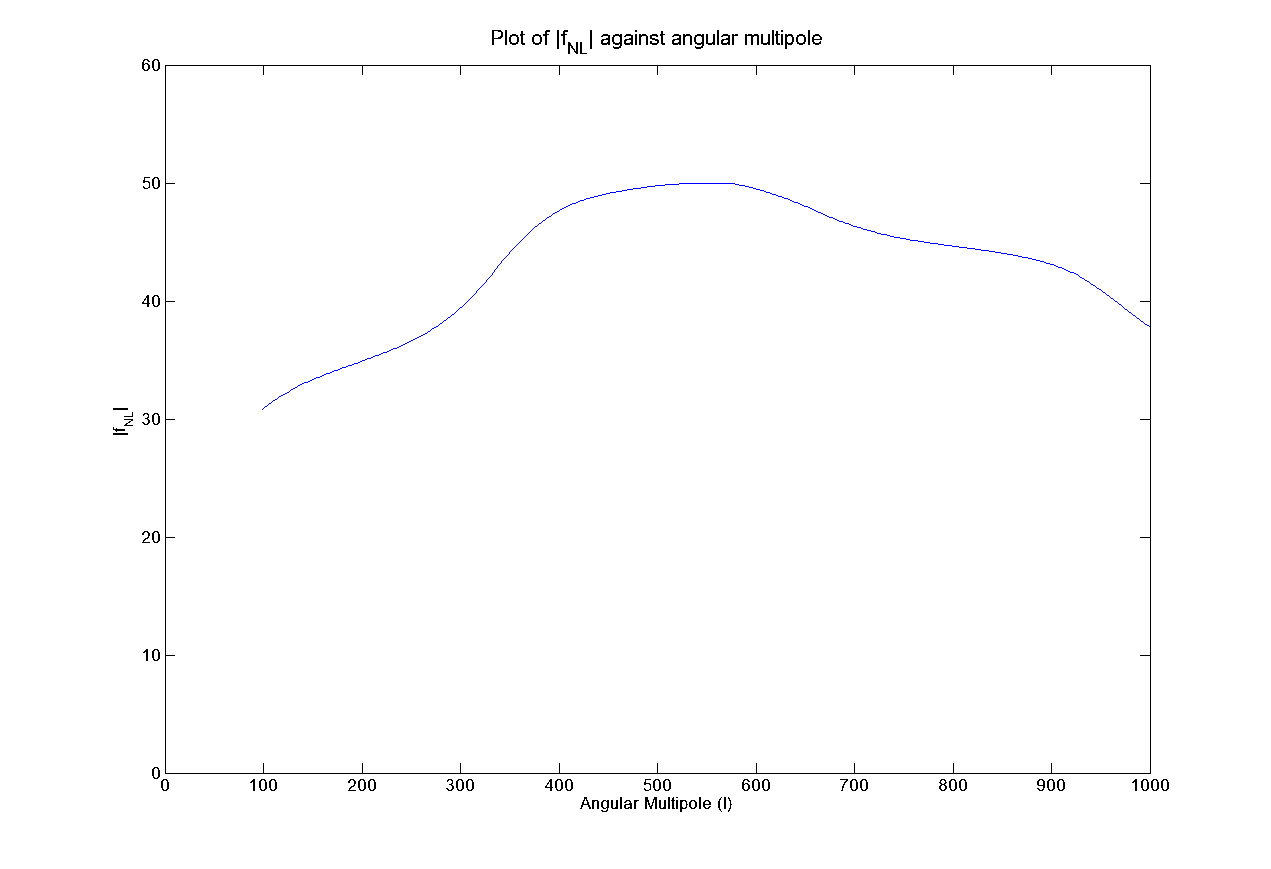}
\caption{Measure of the level of non-Gaussianity using $f_{NL}^{\rm{local}}$ as a function of $l_{\rm{max}}$, the maximum observational resolution. The string tension is assumed to be $G\mu=7\times 10^{-7}$. It should be noted that the string prediction has $f_{NL}<0$.}
\label{fig:fnl}
\end{figure}

\section{Trispectrum}
The calculation for the trispectrum follows a similar path to that of the power spectrum and bispectrum. Again we find that the total trispectrum reduces to the sum of the equal time trispectra between $\eta_{\rm{start}}\gtrsim\eta_{\rm{lss}}$ and $\eta_0$.
\par
The trispectrum, in the flat sky approximation, is given (at any given time) by
\begin{eqnarray}\label{Trispec}
 \langle \delta_{{\mathbf{k}}_1}\delta_{{\mathbf{k}}_2}\delta_{{\mathbf{k}}_3}\delta_{{\mathbf{k}}_4} \rangle _c=(2\pi)^2 \delta^{(2)}(\mathbf{k_1+k_2+k_3+k_4})T_c(\mathbf{k_1,k_2,k_3,k_4}),
\end{eqnarray}
where the subscript $c$ denotes the connected part. The total trispectrum sums the connected part for all times between last scattering and today. The unconnected part, denoted hereafter by the subscript $uc$, is given by
\begin{align}\label{Unconnected}
 \langle \delta_{{\mathbf{k}}_1}\delta_{{\mathbf{k}}_2}\delta_{{\mathbf{k}}_3}\delta_{{\mathbf{k}}_4} \rangle _{uc}=&(2\pi)^4  \delta^{(2)}(\mathbf{k_1+k_2})\delta^{(2)}(\mathbf{k_3+k_4})P(k_1)P(k_3)\nonumber\\
 &+(2\pi)^4  \delta^{(2)}(\mathbf{k_1+k_3})\delta^{(2)}(\mathbf{k_2+k_4})P(k_1)P(k_2)\nonumber\\
&+(2\pi)^4  \delta^{(2)}(\mathbf{k_1+k_4})\delta^{(2)}(\mathbf{k_2+k_3})P(k_1)P(k_2)\nonumber\\
=&\mathcal{A}^2\Big( \delta_{k_1,k_2}\delta_{k_3,k_4}P(k_1)P(k_3)+\delta_{k_1,k_3}\delta_{k_2,k_4}P(k_1)P(k_2)\nonumber\\
&\qquad\qquad\qquad\qquad\qquad\qquad\qquad+\delta_{k_1,k_3}\delta_{k_2,k_4}P(k_1)P(k_2)\Big).
\end{align}

We will calculate the four-point correlator in this section which includes the connected and unconnected components. In particular, we denote
\begin{eqnarray}
\langle \delta_{{\mathbf{k}}_1}\delta_{{\mathbf{k}}_2}\delta_{{\mathbf{k}}_3}\delta_{{\mathbf{k}}_4}\rangle=(2\pi)^2 \delta^{(2)}({\mathbf{k}}_1+{\mathbf{k}}_2+{\mathbf{k}}_3+{\mathbf{k}}_4) T^{{\rm{whole}}} ({\mathbf{k}}_1,{\mathbf{k}}_2,{\mathbf{k}}_3,{\mathbf{k}}_4).
\end{eqnarray}
In the remainder of this section we will refer to $T^{{\rm{whole}}}$ as the trispectrum (and drop the superscript `${\rm{whole}}$') unless otherwise stated.
At first we shall only consider parallelogram configurations. In this regard we are motivated by the earlier section on the bispectrum which shows that although as the triangle collapses it grows (though does not become infinite) the equilateral configuration is adequate to approximately describe the bispectrum behaviour in the multipole range of Planck.  Clearly the trispectrum is expected to replicate this behaviour. We will return to deal with this issue later in the section. 

The trispectrum may be further divided into contributions from different orderings of the wavevectors ${\mathbf{k}}_i$. In particular, writing the trispectrum in the form 
\begin{align}
\langle \delta_{{\mathbf{k}}_1}\delta_{{\mathbf{k}}_2}\delta_{{\mathbf{k}}_3}\delta_{{\mathbf{k}}_4}\rangle=(2\pi)^2\int d^2 K \delta^{(2)}({\mathbf{k}}_1+{\mathbf{k}}_2-{\mathbf{k}})    \delta^{(2)}({\mathbf{k}}_3+{\mathbf{k}}_4+{\mathbf{k}}) T ({\mathbf{k}}_1,{\mathbf{k}}_2,{\mathbf{k}}_3,{\mathbf{k}}_4;{\mathbf{k}}),
\end{align}
we can decompose the trispectrum into the following sum
\begin{align}
T ({\mathbf{k}}_1,{\mathbf{k}}_2,{\mathbf{k}}_3,{\mathbf{k}}_4;{\mathbf{k}})=&p({\mathbf{k}}_1,{\mathbf{k}}_2,{\mathbf{k}}_3,{\mathbf{k}}_4;{\mathbf{k}})\nonumber\\
&+\int d^2 K' \delta^{(2)}({\mathbf{k}}_3-{\mathbf{k}}_2
-{\mathbf{k}}+{\mathbf{k}}') p({\mathbf{k}}_1,{\mathbf{k}}_3,{\mathbf{k}}_2,{\mathbf{k}}_4;{\mathbf{k}}')\nonumber\\
&+\int d^2 K' \delta^{(2)}({\mathbf{k}}_4-{\mathbf{k}}_2-{\mathbf{k}}+{\mathbf{k}}') p({\mathbf{k}}_1,{\mathbf{k}}_4,{\mathbf{k}}_3,{\mathbf{k}}_2;{\mathbf{k}}').
\end{align}
Alternatively, we may write
\begin{eqnarray}\label{pdecomp}
 T({\mathbf{k}}_1,{\mathbf{k}}_2,{\mathbf{k}}_3,{\mathbf{k}}_4)= p({\mathbf{k}}_1,{\mathbf{k}}_2,{\mathbf{k}}_3,{\mathbf{k}}_4)+p({\mathbf{k}}_1,{\mathbf{k}}_3,{\mathbf{k}}_2,{\mathbf{k}}_4)+p({\mathbf{k}}_1,{\mathbf{k}}_4,{\mathbf{k}}_3,{\mathbf{k}}_2).
\end{eqnarray}
\par
In a similar fashion to the calculation of the power spectrum and bispectrum, we find that
\begin{align}
T({\mathbf{k}}_1,{\mathbf{k}}_2,{\mathbf{k}}_3,{\mathbf{k}}_4)=\frac{(8\pi G \mu)^4}{\mathcal{A}}&\frac{k_1^A k_2^B k_3^C k_4^D}{k_1^2 k_2^2 k_3^2 k_4^2}\int d\sigma_1 d\sigma_2 d\sigma_3 d\sigma_4 \nonumber\\
&\times\langle \dot{X}^A(\sigma_1)\dot{X}^B(\sigma_2)\dot{X}^C(\sigma_3)	\dot{X}^D(\sigma_4)	e^{i\mathbf{k_a.X_a}}   \rangle.
\end{align}
Setting $C^{ABCD}=\dot{X}^A(\sigma_1)\dot{X}^B(\sigma_2)\dot{X}^C(\sigma_3)\dot{X}^D(\sigma_4)	$ and $E=k_a.X_a$ and using the assumption of a Gaussian process we get that
\begin{eqnarray*}
\langle	C^{ABCD}e^{iE}	\rangle=\langle C^{ABCD}- \frac{1}{2} C^{ABCD} E^2+\frac{1}{24}C^{ABCD}E^4+\dots		\rangle.
\end{eqnarray*}
Now
\begin{align*}
\langle	C^{ABCD}	\rangle	=&\langle	\dot{X}^A_1	\dot{X}^B_2	\dot{X}^C_3	\dot{X}^D_4	\rangle
\nonumber\\
=&\frac{\delta^{AB}}{2}V_{12}\frac{\delta^{CD}}{2}V_{34}+\frac{\delta^{AC}}{2}V_{13}\frac{\delta^{BD}}{2}V_{24}+\frac{\delta^{AD}}{2}V_{14}\frac{\delta^{BC}}{2}V_{23}\\
\implies k_1^A k_2^B k_3^C k_4^D \langle	C^{ABCD}	\rangle=&\frac{1}{4}\left(\kappa_{12}\kappa_{34}V_{12}V_{34}+\kappa_{13}\kappa_{24}V_{13}V_{24}+\kappa_{14}\kappa_{23}V_{14}V_{23}	\right).
\end{align*}
Next, we consider (writing $C^{ABCD}$ as $ABCD$)
\begin{align*}
\langle	ABCD E^2\rangle=&\langle	C^{ABCD}	\rangle \langle	E^2	\rangle +
2\langle AB \rangle \langle CE \rangle \langle DE \rangle+\rm{perms},
\end{align*}
where $\rm{perms}$ denotes the contribution from terms found by permutating the symbols $A,B,C,D$ in the second term on the right hand side of the equality\footnote{We note that the factor of two appears since $ \langle AB \rangle = \langle BA \rangle $.}.
Now we note that (using $\sum_i {\mathbf{k}}_i=0$)
\begin{eqnarray*}
 \langle AB \rangle  \langle CE \rangle  \langle DE \rangle &=&\frac{\delta^{AB} V_{12}}{2} \langle \dot{X}^C_3 \sum_{j\neq 3} k_j^E X_{j3}^E \rangle  \langle \dot{X}^D_4 \sum_{j\neq 4} k_j^E X_{j4}^E \rangle \\
&=&\frac{\delta^{AB} V_{12}}{8}\sum_{j\neq 3}k_j^C \Pi_{3j} \sum_{j\neq 4}k_j^D \Pi_{4j},
\end{eqnarray*}
where we use the fact that $\Pi_{ij}= \langle  X^A_{ij} \dot{X}_j^A \rangle =- \langle \dot{X}_j^AX^A_{ji} \rangle $.
Suppressing the summation notation we find that
\begin{eqnarray*}
&&k_1^A k_2^B k_3^C k_4^D \langle 	ABCD E^2 \rangle = \langle 	C^{ABCD}	 \rangle  \langle 	E^2	 \rangle  +\\
&& \frac{1}{4}\left( V_{12}\kappa_{12}	\kappa_{3j}\Pi_{3j} \kappa_{4j}\Pi_{4j}	+V_{13}\kappa_{13}	\kappa_{2j}\Pi_{2j} \kappa_{4j}\Pi_{4j}	+V_{14}\kappa_{14}	\kappa_{2j}\Pi_{2j} \kappa_{3j}\Pi_{3j}	\right)+\\
&&\frac{1}{4}\left( V_{23}\kappa_{23}	\kappa_{1j}\Pi_{1j} \kappa_{4j}\Pi_{4j}	+V_{24}\kappa_{24}	\kappa_{1j}\Pi_{1j} \kappa_{3j}\Pi_{3j}	+V_{34}\kappa_{34}	\kappa_{1j}\Pi_{1j} \kappa_{3j}\Pi_{3j}	\right).
\end{eqnarray*}

Now we proceed to investigate
\begin{align*}
 \langle 	ABCD E^4 \rangle =& \langle ABCDE^2 \rangle  \langle E^2 \rangle +24 \langle AE \rangle  \langle BE \rangle  \langle CE \rangle  \langle DE \rangle .
\end{align*}
We have seen above that $ \langle AE \rangle =-\dfrac{1}{2}\sum_{j\neq 1}k_j^A \Pi_{1j}$ so (again suppressing the summation notation) we have
\begin{eqnarray*}
k_1^A k_2^B k_3^C k_4^D \langle AE \rangle  \langle BE \rangle  \langle CE \rangle  \langle DE \rangle =\frac{24}{16} \kappa_{1j}\Pi_{1j} \kappa_{2j}\Pi_{2j} \kappa_{3j}\Pi_{3j} \kappa_{4j}\Pi_{4j}. 
\end{eqnarray*}
Finally, we collect the terms to find that
\begin{align*}
k_1^A k_2^B k_3^C k_4^D \langle 	C^{ABCD}e^{iE}	 \rangle =&[ \frac{1}{4}\left(\kappa_{12}\kappa_{34}V_{12}V_{34}+\kappa_{13}\kappa_{24}V_{13}V_{24}+\kappa_{14}\kappa_{23}V_{14}V_{23}	\right)\\
&- \frac{1}{8}\left( V_{12}\kappa_{12}	\kappa_{3j}\Pi_{3j} \kappa_{4j}\Pi_{4j}	+\rm{perms}	\right)\\&+\frac{1}{16} 
\kappa_{1j}\Pi_{1j} \kappa_{2j}\Pi_{2j} \kappa_{3j}\Pi_{3j} \kappa_{4j}\Pi_{4j} ]e^{- \langle E^2 \rangle 2},
\end{align*}
where we have 
\begin{eqnarray*}
 \langle E^2 \rangle &=& \langle (k_1^A X_{14}^A+k_2^A X_{24}^A+k_3^A X_{34}^A  )(k_2^B X_{21}^B+k_3^B X_{31}^B+k_4^B X_{41}^B) \rangle \\
&=&-\frac{1}{2}\left(\kappa_{12}\Gamma_{12}+\kappa_{13}\Gamma_{13}+\kappa_{14}\Gamma_{14}+\kappa_{23}\Gamma_{23}+\kappa_{24}\Gamma_{24}+\kappa_{34}\Gamma_{34}\right),
\end{eqnarray*}
and where we use the notation $\rm{perms}$ to represent the non-equivalent contributions by permuting the symbols $(1,2,3,4)$ of the term in the brackets.
The dominant term on all angular scales is found to be given by the first term ($\sim C^{ABCD}e^{-E^2/2}$) and, therefore, in what follows we neglect the remaining terms.
\par
Clearly the cosmic string trispectrum may be readily decomposed as in equation~\eqref{pdecomp}. We have
\begin{eqnarray} 
p({\mathbf{k}}_1,{\mathbf{k}}_2,{\mathbf{k}}_3,{\mathbf{k}}_4)=\frac{(8\pi G \mu)^4}{\mathcal{A}}\frac{1}{4 k_1^2 k_2^2 k_3^2 k_4^2}\int d\sigma_1 d\sigma_2 d\sigma_3 d\sigma_4 \kappa_{13}\kappa_{24} V_{13}\Pi_{24}.
\end{eqnarray}
\par
For a parallelogram configuration $\kappa_{13}=-k_1^2,\, \kappa_{24}=-k_2^2,\, \kappa_{23}=\kappa_{14}=-\kappa_{34}=-\kappa_{12}$. This implies that, in the small angle approximation,
\begin{eqnarray*}
 \langle E^2 \rangle &=&\frac{-s^2}{2}\left(\kappa_{12}\sigma_{12}^2-k_{1}^2\sigma_{13}^2-\kappa_{12}\sigma_{14}^2-\kappa_{12}\sigma_{23}^2-k_2^2\sigma_{24}^2+\kappa_{12}\sigma_{34}^2\right).
\end{eqnarray*}
 Again the choice of independent variables is dependent on the integrand under consideration. To elucidate this we consider the term
 \begin{eqnarray*}
\int d\sigma_1 d\sigma_2 d\sigma _3 d\sigma_4 V_{13}V_{24}\exp(-E^2/2)=\mathcal{L}\int d\sigma_{12}d\sigma_{13}d\sigma_{24} V_{13}\Pi_{12}\exp(-E^2/2).
\end{eqnarray*}
 In a similar manner to the case of the bispectrum we can simplify the calculation by making a separable approximation. In particular, using
 \begin{align*}
\exp(- \langle E^2 \rangle /2)=&\exp\left(\frac{s^2}{4}\left(\kappa_{12}\sigma_{12}^2-k_{1}^2\sigma_{13}^2-\kappa_{12}\sigma_{14}^2-\kappa_{12}\sigma_{23}^2-k_2^2\sigma_{24}^2+\kappa_{12}\sigma_{34}^2\right)\right)\\
=&\exp\left(-\frac{s^2}{4}\left((k_1\sigma_{13}-\frac{\kappa_{12}}{k_1}\sigma_{24})^2+\frac{k_1^2 k_2^2 -\kappa_{12}^2}{k_1^2}\sigma_{24}^2)\right)\right)\nonumber\\
\approx&  \exp\left(-\frac{k_1^2s^2\sigma_{13}^2}{4}\right) \exp\left(-\frac{K_1^2s^2\sigma_{24}^2}{4}\right),
 \end{align*}
 we find that
\begin{align}\label{separabletrisp}
\int d\sigma_1 d\sigma_2 d\sigma _3 d\sigma_4 V_{13}V_{24}\exp(-E^2/2)=2\mathcal{L}\hat{\xi}\int d\sigma_{13} & V_{13} \exp\left(-\frac{k_1^2s^2\sigma_{13}^2}{4}\right)\nonumber\\
&\times \int d\sigma_{24} V_{24} \exp\left(-\frac{K_1^2s^2\sigma_{24}^2}{4}\right),
\end{align}
where $K_i= 2 {{\rm{Area}}}_{\triangle}/k_i ={\rm{Area}}/k_i $ as defined in the section on the bispectrum (note that the area of the triangle is half that of the parallelogram). It should be noted that in the evaluation of this integral we must restrict the domain of $\sigma_{12}$ to $(-\hat{\xi},\hat{\xi})$ as the small angle approximation used to evaluate $\langle E^2\rangle$ is only valid in this range. This is unlike the integral that gives $\mathcal{L}$ which is independent of the approximations. We also observe that, similarly to the case of the bispectrum, this separable approximation holds for $k_1>k_2$. In the case $k_2>k_1$ we must make the replacement $k_1\rightarrow k_2$ and $K_1\rightarrow K_2$. Since $K_i$ (divided by $\eta_0$) may be less than $500$, i.e. correspond to an angular scale below that of the correlation length then we should use a better approximation to $V_{24}$. Following these approximations for all terms and using the results of the previous sections we find that
\begin{eqnarray}
p({\mathbf{k}}_1,{\mathbf{k}}_2,{\mathbf{k}}_3,{\mathbf{k}}_4)\approx \frac{\mathcal{L}\hat{\xi}}{2\mathcal{A}}    \frac{(8\pi G\mu)^4}{ k_1^2 k_2^2 k_3^2 k_4^2} \int d\sigma_{13}d\sigma_{24}k_1^2 k_2^2V_{13} V_{24}e^{-E^2/2}.
\end{eqnarray}
In order to compute the integral we use the more accurate approximation for $V$, i.e. equation \eqref{VelocImprov}. Using the previous expressions for the separate integrals we find
\begin{eqnarray}
p({\mathbf{k}}_1,{\mathbf{k}}_2,{\mathbf{k}}_3,{\mathbf{k}}_4)\approx \frac{\mathcal{L}\hat{\xi}}{\mathcal{A}}    \frac{(8\pi G\mu)^4}{ k_1^2 k_2^2 k_3^2 k_4^2} \frac{2\overline{v}^4}{s^6\hat{\xi}^4 {\rm{Area}}^3 }\left(k_1^2 k_2^2	f_2(K_1\hat{\xi})f_2(k_1\hat{\xi})\right),
\end{eqnarray}
where ${\rm{Area}}=2{{\rm{Area}}}_{\triangle}=\sqrt{k_1^2 k_2^2 -\kappa_{12}^2}$ and $K_i={\rm{Area}}/k_i$.
\par
This result must be integrated as usual to give the total trispectrum due to cosmic strings between last scattering (or if $l_m<500$ between $\eta_{\rm{start}}=\eta_{\rm{lss}}500/l_m$) and today. Returning the time dependence using the renormalised time $\eta/\eta_{\rm{start}}$ and understanding the multipoles to refer to angular multipoles we have that the total trispectrum contribution by the parallelogram considered here is
 \begin{align}\label{TrispecTotal}
p^{{\rm{total}}}({\mathbf{k}}_1,{\mathbf{k}}_2,{\mathbf{k}}_3,{\mathbf{k}}_4)\approx& \frac{(8\pi G\mu)^4}{ k_1^2 k_2^2 k_3^2 k_4^2}\frac{2\overline{v}^4 }{s^6  {\rm{Area}}^3 }\int_{1}^{\eta_0/\eta_{\rm{lss}}} d\eta\frac{1}{\hat{\xi}^4 (1+\eta)}\left(k_1^2 k_2^2	f_2(K_1\hat{\xi} )f_2(k_1\hat{\xi} )\right).
\end{align}
In order to understand the behaviour for the trispectrum we must numerically estimate the integral given by
\begin{align}\label{IntegralTrisp}
\int_{1}^{\eta_0/\eta_{\rm{lss}}} d\eta \frac{1}{\hat{\xi}^4 (1+\eta)}\left(k_1^2 k_2^2	f_2(K_1\hat{\xi} )f_2(k_1\hat{\xi} )\right)\propto\int_{1}^{\eta_0/\eta_{\rm{lss}}} d\eta\frac{1}{ \eta^4(1+\eta)}	f_2(L_1\tilde{\xi}\eta )f_2(l_1\tilde{\xi}\eta )\equiv I.
\end{align}
We find, using the numerical values of the parameters as given earlier, that
\begin{eqnarray}\label{trispapprox11}
I\approx \ln\left(\frac{1+\eta_0/\eta_{lss}}{2}\right)\pi s^4  \frac{\left( L_1\tilde{\xi}\right)^3}{(0.63+L_1\tilde{\xi})} \frac{\left( l_1\tilde{\xi}\right)^3}{(0.63+l_1\tilde{\xi})}
\end{eqnarray}
is an accurate approximation to the integral (see figure \ref{fig:TrispHigh}). Therefore, 
\begin{align}\label{TrispecApprox}
(k_1 k_2 k_3 k_4)^{3/2}& p^{{\rm{total}}}({\mathbf{k}}_1,{\mathbf{k}}_2,{\mathbf{k}}_3,{\mathbf{k}}_4)\nonumber\\&\approx (8\pi G\mu)^4\frac{2\overline{v}^4 \pi }{s^2 }\frac{ l_2^2}{(l_1 l_2 l_3 l_4)^{1/2}}   \frac{1}{(0.63+L_1\tilde{\xi})} \frac{\left( l_1\tilde{\xi}\right)^2}{(0.63+l_1\tilde{\xi})} \ln\left(\frac{1+\eta_0/\eta_{lss}}{2}\right),
\end{align}
where we use ${\rm{Area}}=l_1 L_1/\eta_0^2$. We should note that in the common notation used to describe the trispectrum in terms of multipoles, this quantity may be expressed as
\begin{eqnarray}
(k_1 k_2 k_3 k_4)^{3/2}p^{{\rm{total}}}({\mathbf{k}}_1,{\mathbf{k}}_2,{\mathbf{k}}_3,{\mathbf{k}}_4)=(l_1 l_2 l_3 l_4)^{3/2}{p^{{\rm{total}}}}(l_1,l_2, l_3,l_4,L),
\end{eqnarray}
where $L$ is the diagonal $|{\mathbf{k}}_1+{\mathbf{k}}_2|/\eta_0=\sqrt{l_1^2+l_2^2+2 l_{12}}$. It should be noted that in the equilateral limit $l_1=l_2=l_3=l_4=L_1$ we find $L_1=L(\sqrt{l^2-L^2/4}/l)$. Thus for $L\lesssim l$ we may use $L_1\approx L$. 
\par
It should be noted that as $L_1\rightarrow 0$ the formula for $p^{{\rm{total}}}$ is independent of $L_1$, i.e. of ${\rm{Area}}$. Therefore, as for the bispectrum, the trispectrum grows towards the edge but is not divergent (i.e. it reaches a cutoff). For large wavenumbers this growth towards the edge becomes more and more pronounced. However, in the multipole range of Planck this growth is suppressed as in the case of the bispectrum.
\begin{figure}[htp]
\centering 
\includegraphics[width=79mm]{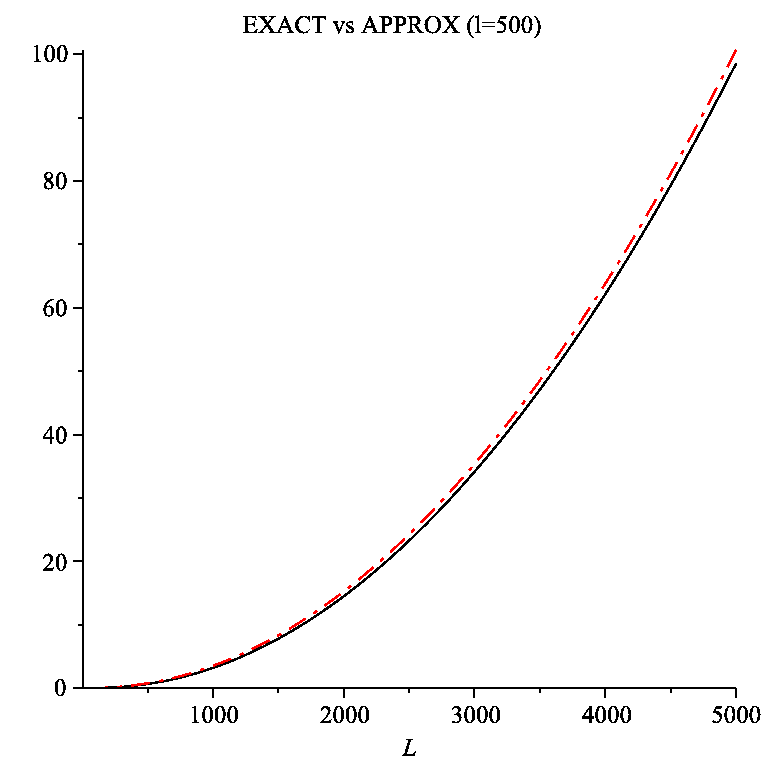}
\includegraphics[width=79mm]{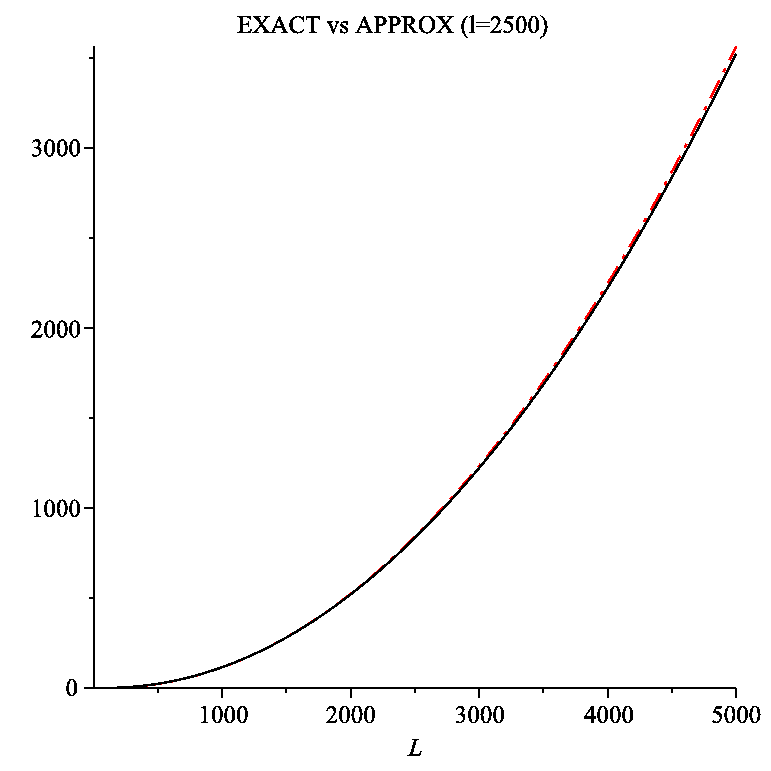}
\includegraphics[width=79mm]{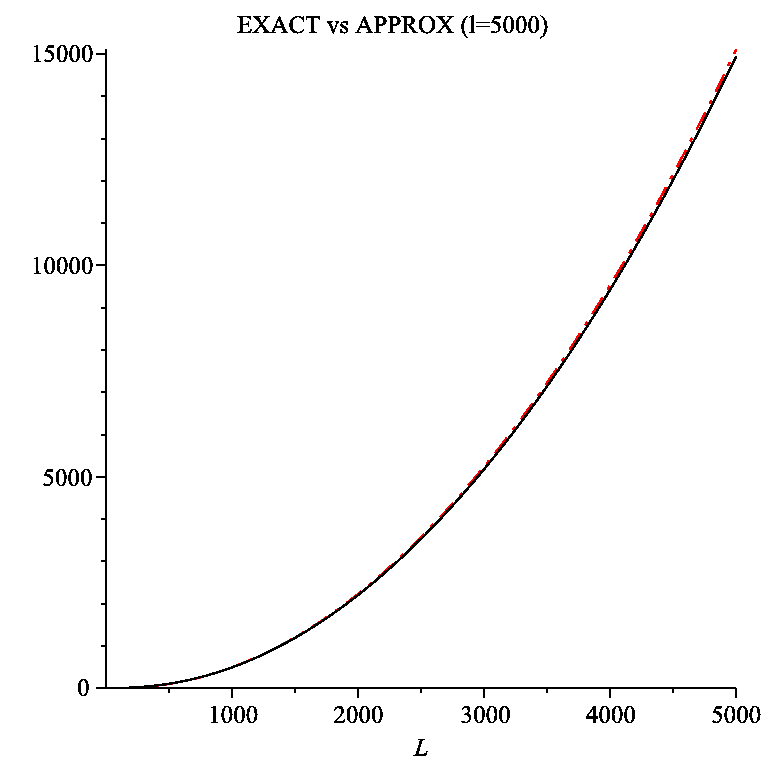}
\caption{Comparison of approximation of \eqref{IntegralTrisp} given by \eqref{trispapprox11} for $l=500$, $2500$ and $l=5000$ respectively. The exact result is given by the solid (black) line, while the approximation is given by the dot dash (red) line. The plots verify the accuracy of the trispectrum approximation.}
\label{fig:TrispHigh}
\end{figure}

\subsection{Trispectrum including large angular scales}
If $\min k_i/\eta_0 =l_m\lesssim 500$, then we must restrict the range of times over which we sum to $(\eta_{\rm{start}},\eta_0)$. The integral $I$ in \eqref{IntegralTrisp} becomes
\begin{eqnarray}\label{IntegralTrisp2}
I=\int_{1}^{\eta_0/\eta_{\rm{start}}} d\eta\frac{1}{ \eta^4(1+(500/l_m)\eta)}	f_2(L_1\tilde{\xi}\eta )f_2(l_1\tilde{\xi}\eta ).
\end{eqnarray}
For  $l_m\lesssim 500$ we recall that $\tilde{\xi}$ becomes $1/l_m$. We find the following fit to the integral
\begin{eqnarray}\label{trispeapprox}
I\approx  \ln\left(\frac{1+\eta_0/\eta_{lss}}{2}\right)\pi s^4  \frac{\left( L_1\tilde{\xi}\right)^3}{(0.63+L_1\tilde{\xi})} \frac{\left( l_1\tilde{\xi}\right)^3}{(0.63+l_1\tilde{\xi})}\left( \frac{2}{1+500/l_m}\right)^{2.3}.
\end{eqnarray}
In figure \ref{fig:TrispLow} this approximation is verified.
This is the same effect as was evident in the case of the bispectrum and the power spectrum. That we can extend the range of multipoles to less than $500$ means that we can use this formalism to make a prediction for $\tau_{NL}$ generated by cosmic strings in the multipole range of Planck. Therefore, we have the following approximation to equation \eqref{TrispecTotal} which is valid for angular multipoles that satisfy the flat sky approximation, i.e. $l\gtrsim 60$:
\begin{align}\label{TrispecApprox}
&(k_1 k_2 k_3 k_4)^{3/2}p^{{\rm{total}}}({\mathbf{k}}_1,{\mathbf{k}}_2,{\mathbf{k}}_3,{\mathbf{k}}_4)\nonumber\\ &\approx (8\pi G\mu)^4\frac{2\overline{v}^4 \pi }{s^2 }\frac{ l_2^2}{(l_1 l_2 l_3 l_4)^{1/2}}   \frac{1}{(0.63+L_1\tilde{\xi})} \frac{\left( l_1\tilde{\xi}\right)^2}{(0.63+l_1\tilde{\xi})} \ln\left(\frac{1+\eta_0/\eta_{lss}}{2}\right)\left( \frac{2}{1+500/l_m}\right)^{2.3},
\end{align}
where we note again that $L_1=\sqrt{l_1^2 l_2^2 -l_{12}^2 }/l_1$, $l_m=\min(500,l_i)$ and $\tilde{\xi}=1/l_m$.
\begin{figure}[htp]
\centering 
\includegraphics[width=82mm]{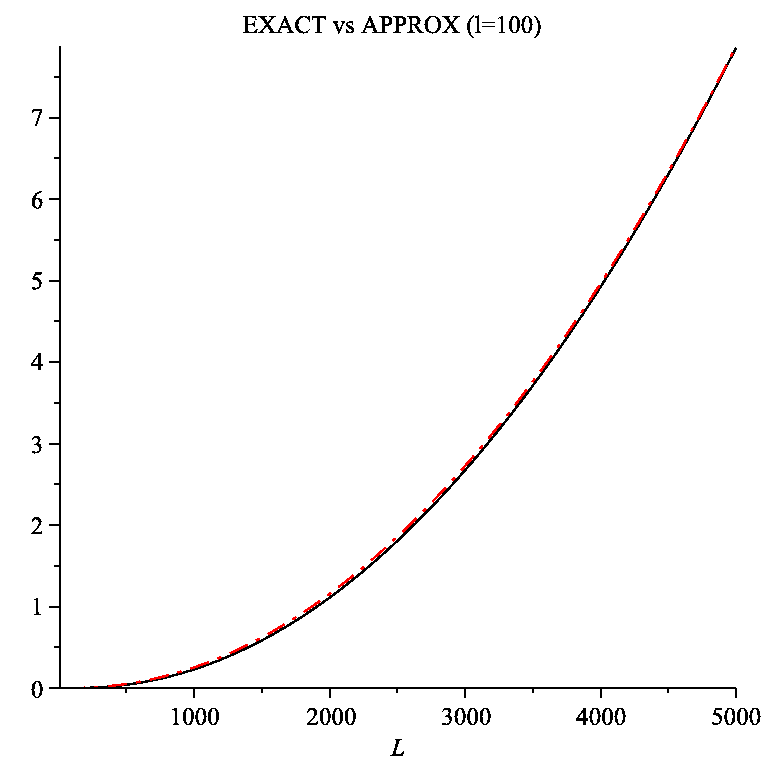}
\includegraphics[width=82mm]{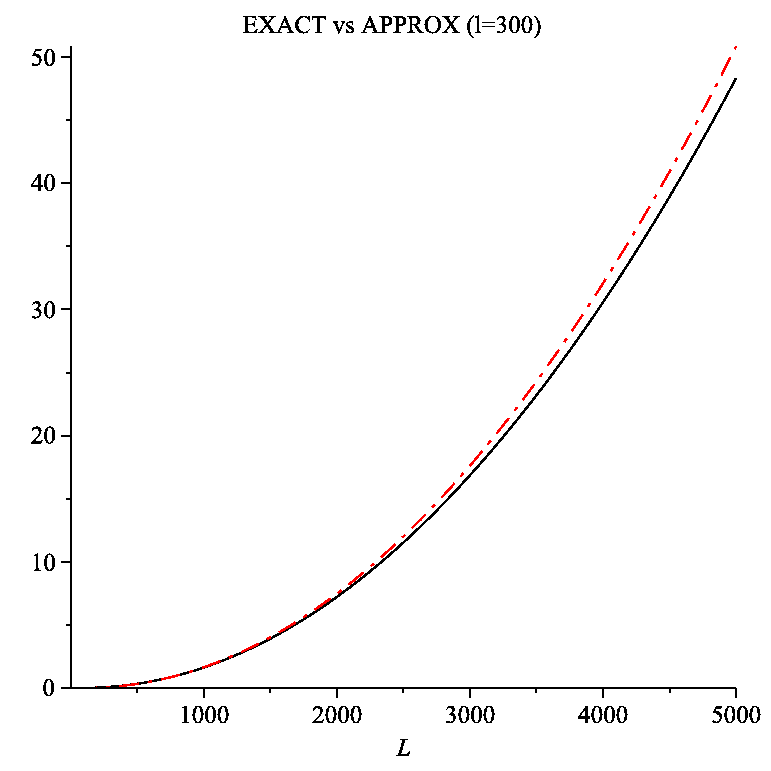}
\caption{Comparison of approximation of \eqref{IntegralTrisp2} given by \eqref{trispeapprox} for $l=100$ and $l=300$ respectively. The exact result is given by the solid (black) line, while the approximation is given by the dot dash (red) line. The plots verify the accuracy of the trispectrum approximation and in particular the logarithmic behaviour for low $l$ of the integral.}
\label{fig:TrispLow}
\end{figure}

\subsection{Non-parallelogram configurations}
For completeness we include the case of non-parallelogram configurations in this section. However, we will find for much of the multipole range of Planck a reasonable approximation to integrated measures of the size of the trispectrum may be given by considering the parallelogram case only. 
\par
The quantity $\langle E^2 \rangle$ in the leading order small angle approximation may be expressed in the form $\langle E^2\rangle=-(s^2/2) \sum_{i j}\kappa_{i j}\sigma_{i 4}\sigma_{j 4}$. This is a quadratic form and thus may be expressed as the sum of two squares. In the case of the parallelogram $\sigma_{12},\sigma_{13}, \sigma_{24}$ were chosen to be our independent variables and it was found that $\sigma_{12}$ does not appear in the exponent. Although in general $\sigma_{12}, \sigma_{13}$ and $\sigma_{24}$ may appear in the exponent, we may exploit the fact that $\langle E^2 \rangle$ is a quadratic form to transform to a different set of variables such that just two appear in the quadratic form and the third drops out (similarly to $\sigma_{12}$ in the parallelogram case). Thus, to leading order we may neglect $\sigma_{12}$ in the exponent for the calculation of the trispectrum term $p({\mathbf{k}}_1,{\mathbf{k}}_2,{\mathbf{k}}_3,{\mathbf{k}}_4)$. It is therefore necessary to include the next to leading order contribution in the exponent in order to calculate the integral over $\sigma_{12}$. We will find that in the case of parallelogram configurations the calculation above is valid. However, for non-parallelogram configurations it is necessary to include the next to leading order effect.
This contribution is found by noting that the next to leading order contribution to $T(\sigma)$ is given by
\begin{eqnarray}
T(\sigma)\approx\langle {X'}^{A}(\sigma){X'}^A(0)\rangle&\approx& s^2-\frac{A}{3}\left( \frac{\sigma}{\eta}\right)^{2\chi}
\approx s^2-\frac{A \tilde{\tilde{\xi}}^{2\chi} }{3}        \left( \frac{\sigma}{\hat{\xi}} \right)^{2\chi},
\end{eqnarray}
where we use $\hat{\xi}=\tilde{\tilde{\xi}}\eta$ in the second line. This expression was derived by Polchinski and Rocha \cite{PolchRocha} who found that in the matter era $A\approx 0.61$ and $2\chi\approx 0.5$. This implies that
\begin{eqnarray}
\Gamma(\sigma)&\approx& s^2\sigma^2-\tilde{c}_1    \left( \frac{\sigma^{2\chi+2}}{\hat{\xi}^{2\chi}} \right),
\end{eqnarray}
where $\tilde{c}_1=A \tilde{\tilde{\xi}}^{2\chi}/(3(1+\chi)(1+2\chi))$. Using the values of these parameters we find $\tilde{c}_1\approx 0.05$. In order to compute the trispectrum we assume that the next to leading order contribution gives a negligible contribution to the integrals over $\sigma_{13}$ and $\sigma_{24}$. In addition we assume that these (Gaussian) integrals are sufficiently sharp so that we may set $\sigma_{13}=0=\sigma_{24}$ in the next to leading order contribution. This is certainly true for large wavenumbers and we will find that the case of low wavenumbers differs little from the parallelogram calculation. With this approximation we find that the contribution dependent on $\sigma_{12}$ is given by
\begin{eqnarray}
\langle E^2\rangle_{\sigma_{12}}=\tilde{c}_1|{\mathbf{k}}_1+{\mathbf{k}}_3|^2 \frac{|\sigma_{12}|^{2\chi+2}}{\hat{\xi}^{2\chi}}.
\end{eqnarray}
We note that, in the case of a parallelogram, $|{\mathbf{k}}_1+{\mathbf{k}}_3|=0$ thus verifying the validity of the earlier calculation. The $\sigma_{12}$ integral becomes, in general
\begin{align}
\int_{-\hat{\xi}}^{\hat{\xi}} d\sigma_{12} \exp\left(-\frac{\tilde{c}_1}{2}|{\mathbf{k}}_1+{\mathbf{k}}_3|^2 \frac{|\sigma_{12}|^{2\chi+2}}{\hat{\xi}^{2\chi}}\right)=&2\int_{0}^{\hat{\xi}} d\sigma_{12} \exp\left(-\frac{\tilde{c}_1}{2}|{\mathbf{k}}_1+{\mathbf{k}}_3|^2 \frac{\sigma_{12}^{2\chi+2}}{\hat{\xi}^{2\chi}}\right)\nonumber\\
=&2\hat{\xi}  M\left(\frac{1}{2\chi+2};1+\frac{1}{2\chi+2};-\frac{\tilde{c}_1}{2}|{\mathbf{k}}_1+{\mathbf{k}}_3|^2 \hat{\xi}^2\right),
\end{align}
where $M$ is the confluent hypergeometric function. Using the inequality $|{\mathbf{k}}_1+{\mathbf{k}}_3|\leq |{\mathbf{k}}_1|+|{\mathbf{k}}_3|$ we have $\frac{\tilde{c}_1}{2}|{\mathbf{k}}_1+{\mathbf{k}}_3|^2 \hat{\xi}^2\leq 2\tilde{c}_1 \max(k_1,k_3)^2 \hat{\xi}^2= 2\tilde{c}_1 \max(l_1,l_3)^2 \tilde{\xi}^2 \eta^2.$

Substituting into the confluent hypergeometric function we find that, assuming $2\tilde{c}_1 \times\\
\max(l_1,l_3)^2 \tilde{\xi}^2\lesssim 0.4$, the hypergeometric function stays within around $10\%$ of unity for $\eta\lesssim2$ (where $\eta$ is the conformal time normalised by the conformal time at last scattering). This is shown in figure \ref{fig:HyperGeom}. This condition is satisfied for $l\lesssim 1000$. Since the overall time dependence of the trispectrum scales as $1/(1+\eta)$ (c.f. equation \eqref{TrispecTotal}) then we may reasonably take the integral over $\sigma_{12}$ as $2\hat{\xi}$ in this range and use the approximations indicated in the parallelogram section. Beyond this range the trispectrum drops off faster with multipole. Hence in estimating the integrated size of the trispectrum using the kurtosis or in estimating the non-linearity parameter $\tau_{NL}$ we will restrict to this range.

\begin{figure}[htp]
\centering 
\includegraphics[width=102mm]{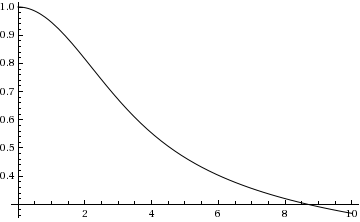}
\caption{Plot of the confluent hypergeometric function $M\left(\frac{1}{2\chi+2};1+\frac{1}{2\chi+2};- 0.4\hat{\xi}^2\right)$ against the normalised time $\eta$. We take $2\chi=0.5$. For $\eta\lesssim 2$ the function stays within approximately $10\%$ of unity.}
\label{fig:HyperGeom}
\end{figure}

\par
We also note that for shapes that satisfy $\frac{\tilde{c}_1}{2}|{\mathbf{k}}_1+{\mathbf{k}}_3|^2 \hat{\xi}^2\gg1$ we have 
\begin{align}
\int_{-\hat{\xi}}^{\hat{\xi}} d\sigma_{12} \exp\left(-\frac{\tilde{c}_1}{2}|{\mathbf{k}}_1+{\mathbf{k}}_3|^2 \frac{|\sigma_{12}|^{2\chi+2}}{\hat{\xi}^{2\chi}}\right)\approx& \frac{\Gamma(1/(2+2\chi))}{2+2\chi}\frac{2^{1/(2+2\chi)}}{{\tilde{c}_1}^{1/(2+2\chi)}} \nonumber\\
&\times \frac{2\hat{\xi}}{|{\mathbf{k}}_1+{\mathbf{k}}_3|^{1/(1+\chi)}  \hat{\xi}^{1/(1+\chi)} }.
\end{align}
For such shapes we may approximate the trispectrum using the value at last scattering (which is valid for large enough multipoles). In particular, we have
\begin{align}\label{TrispecApproxNonParallel}
(k_1 k_2 k_3 k_4)^{3/2}p^{{\rm{whole}}}({\mathbf{k}}_1,{\mathbf{k}}_2,{\mathbf{k}}_3,{\mathbf{k}}_4)\approx&
(8\pi G\mu)^4\frac{2\overline{v}^4 \pi }{s^2 }\frac{ l_2^2}{(l_1 l_2 l_3 l_4)^{1/2}}   \frac{ l_1^2}{\sqrt{l_1^2 l_2^2 -l_{12}^2 }}  \frac{\Gamma(1/(2+2\chi))}{2+2\chi}\nonumber\\
 &   \times   \frac{2^{1/(2+2\chi)}}{{\tilde{c}_1}^{1/(2+2\chi)}}   \frac{1}{\left( \tilde{\xi}\sqrt{l_1^2+l_3^2+2l_{13}}\right)^{1/(1+\chi)}},
\end{align}
where we use the identity $L_1=\sqrt{l_1^2 l_2^2 -l_{12}^2 }/l_1$, assume $l_1>l_2$ (see paragraph below equation \eqref{separabletrisp}) and use the superscript `${\rm{whole}}$' to indicate that the expression includes contributions from the unconnected components of the four-point correlator.
\par
In the remainder of this section we will calculate integrated measures of the trispectrum valid in the multipole range of Planck. As described above we may approximate such measures using the expression for the parallelogram trispectrum assuming that the multipoles satisfy $l_i\lesssim 1000$.

\subsection{Kurtosis}
The connected four-point function of the multipoles $a_{lm}$ may be written in the form
\begin{align}
\langle a_{l_1 m_1}a_{l_2 m_2}a_{l_3 m_3}a_{l_4 m_4}\rangle_c=&\sum_{L M}(-1)^M \left( \begin{array} {ccc}
l_1 & l_2 &L \\
m_1 & m_2  &-M\end{array} \right)\left( \begin{array} {ccc}
l_3 & l_4 &L \\
m_3 & m_4  &M\end{array} \right){P}^{l_1 l_2}_{l_3 l_4}(L)\nonumber\\
&+(l_2\leftrightarrow l_3)+(l_2\leftrightarrow l_4).
\end{align}
In \cite{RSF10} it was demonstrated that the kurtosis $g_2=\langle (\Delta T/T)^4\rangle_c/\langle (\Delta T/T)^2\rangle^{2}$ may be written in terms of multipoles as
\begin{align}\label{kurt1}
g_2=\frac{48\pi  \sum_{l_i,L}h_{l_1 l_2 L}^2 h_{l_3 l_4 L}^2 t^{l_1 l_2}_{l_3 l_4}(L)/(2L+1)      }{(\sum_l (2 l+1)C_l)^{2}}=\frac{12\pi \sum_{l_i,L}h_{l_1 l_2 L}^2 h_{l_3 l_4 L}^2 p^{l_1 l_2}_{l_3 l_4}(L)/(2L+1)    }{(\sum_l (2 l+1)C_l)^{2}},
\end{align}
where $t^{l_1 l_2}_{l_3 l_4}(L)$ is the so called extra-reduced trispectrum, $p^{l_1 l_2}_{l_3 l_4}(L)=t^{l_1 l_2}_{l_3 l_4}(L)+t^{l_2 l_1}_{l_3 l_4}(L)+t^{l_1 l_2}_{l_4 l_3}(L)+t^{l_2 l_1}_{l_4 l_3}(L)$ and we note ${P}^{l_1 l_2}_{l_3 l_4}(L)=h_{l_1 l_2 L}h_{l_3 l_4 L}{p}^{l_1 l_2}_{l_3 l_4}(L)$. Alternatively, since we may write $g_2=\langle (\Delta T/T)^4\rangle/\langle (\Delta T/T)^2\rangle^{2}-3$ we instead decompose the multipole four-point function as
\begin{align}
\langle a_{l_1 m_1}a_{l_2 m_2}a_{l_3 m_3}a_{l_4 m_4}\rangle=&\sum_{L M}(-1)^M \left( \begin{array} {ccc}
l_1 & l_2 &L \\
m_1 & m_2  &-M\end{array} \right)\left( \begin{array} {ccc}
l_3 & l_4 &L \\
m_3 & m_4  &M\end{array} \right){P^{{\rm{whole}}}}^{l_1 l_2}_{l_3 l_4}(L)\nonumber\\
&+(l_2\leftrightarrow l_3)+(l_2\leftrightarrow l_4).
\end{align}
It should be noted that $P^{{\rm{whole}}}$ contains a contribution due to the unconnected components of the four-point function. Now setting ${{p}^{{\rm{whole}}}}^{l_1 l_2}_{l_3 l_4}(L)={P^{{\rm{whole}}}}^{l_1 l_2}_{l_3 l_4}(L)/(h_{l_1 l_2 L}h_{l_3 l_4 L})$ we may write the kurtosis as
\begin{eqnarray}
g_2=\frac{12\pi \sum_{l_i,L}h_{l_1 l_2 L}^2 h_{l_3 l_4 L}^2 {p^{{\rm{whole}}}}^{l_1 l_2}_{l_3 l_4}(L)/(2L+1)    }{(\sum_l (2l+1) C_l)^{2}}-3.
\end{eqnarray}
As we have shown for multipoles  $l\lesssim1000$ the trispectrum is quite flat. Beyond this the trispectrum is expected to fall off more rapidly. Hence we approximate the kurtosis using
\begin{eqnarray}
g_2&\approx&\frac{3\pi \sum_l l^4 h_{l l l}^4  {p^{{\rm{whole}}}}^{l l}_{l l}(l)/(2 l)}{(\sum_l l C_l)^{2}}-3\nonumber\\
&\approx& \frac{8}{\pi^3}\frac{\sum_{l=100}^{1000} l^5 {p^{{\rm{whole}}}}^{l l}_{l l}(l)     }{   (\sum_l l C_l)^{2}         }-3.
\end{eqnarray}
It should be noted that the flat-sky reduced trispectrum combination $p(l_1,l_2,l_3,l_4,L)$ corresponds to the extra-reduced trispectrum combination $p^{l_1 l_2}_{l_3 l_4}(L)$ as shown in \cite{Hu}. Similarly ${p^{{\rm{whole}}}}(l_1,l_2,l_3,l_4,L)$ as described earlier in the section corresponds to ${p^{{\rm{whole}}}}^{l_1 l_2}_{l_3 l_4}(L)$. Substituting in the values for the cosmic string trispectrum and power spectrum we find
\begin{eqnarray}   
g_2^{\rm{cosmic\, \,strings}}\approx \mathcal{O}(1).
\end{eqnarray}
Again this compares favourably to the simulations of Fraisse et al \cite{Fraisse,Ringeval} who found $g_2=0.69\pm0.29$.

\subsection{Estimate of $\tau_{NL}$}
In order to estimate the magnitude of the cosmic string trispectrum we compare to the local model of inflation. In \cite{Kogo} it was shown that in the Sachs-Wolfe limit the local model trispectrum with negligible cubic self-interaction term gives
\begin{eqnarray}
l^6 {p_{\rm{loc}}}^{l l}_{l l}(l)\approx 100 \tau_{NL}(l^2 C_l^{\rm{SW}})^3.
\end{eqnarray}
We note that $l^2 C_l^{\rm{SW}}\approx 6\times 10^{-10}$. In order to estimate a representative value for $l^6 {p}^{l l}_{l l}(l)$ for cosmic strings\footnote{Note that $p^{{\rm{whole}}}$ includes $p$ and a contribution due to the unconnected component of the four-point function.} we use our approximation for the kurtosis and equation \eqref{kurt1} to write
\begin{eqnarray}
l^6 {p_{\rm{c.s.} } }^{l l}_{l l}(l)|_{l=l^*}\approx \frac{\pi^3}{8}g_2 (\sum_l l C_l^{\rm{c.s.}} )^2.
\end{eqnarray}
Comparing with the Sachs-Wolfe limit of the local model we obtain
\begin{eqnarray}
\tau_{NL}\approx 120000 g_2 \left( \frac{G\mu}{2.5\times 10^{-7}}\right)^4 \approx \mathcal{O}(10^5)\left( \frac{G\mu}{2.5\times 10^{-7}}\right)^4.
\end{eqnarray}
\par
Observationally the trispectrum has not received much attention with the rather
weak constraint $\tau_{NL} < 10^8$ \cite{Lyth}.   Recently there has been
a significant improvement claimed using N-point probability distributions \cite{vielva2}
with $-5.6\times 10^5<\tau_{NL}<6.4\times 10^5$. Therefore, for  $G\mu=7\times10^{-7}$ (the bound indicated by Abelian-Higgs simulations) the result is marginally in conflict with observations. On the other hand, taking $G\mu=2.5\times 10^{-7}$ (the current constraint from Nambu-Goto simulations) we would estimate $\tau_{NL}\sim 10^5$. As it is expected that the Planck satellite will be sensitive to a value of $|\tau_{NL}|\sim 560$ (see \cite{Kogo}) it is clear that the trispectrum may provide the strongest constraint yet on cosmic strings in the near future.
Again we note that these results are for a
local-type non-Gaussianity which is unlike the less peaked cosmic string trispectrum,
so a more specific analysis will be necessary to achieve more accurate quantitative constraints\footnote{Of course since the local model is a primordial source of non-Gaussianity, while cosmic strings are an active source, the relation between the cosmic string bispectrum/trispectrum and that of the local model is not straightforward. Furthermore, the shape of the respective spectra are quite different over the multipole range of interest. Therefore the predictions for the level of $\tau_{NL}$ in this context should be interpreted as order of magnitude estimates of the expected signal to noise.}.

Methods have recently been developed to measure the level of non-Gaussianity from the trispectrum efficiently. It is hoped to apply such methods to cosmic strings in order to estimate the expected deviation from Gaussianity due to the four-point function \cite{FergLandetal}.


\section{Conclusions}

In this chapter, we have endeavoured to analytically calculate the cosmic string
power spectrum, bispectrum and trispectrum on both large and small scales.    We have been
particularly focused on extending previous work to multipole ranges ($l\lesssim2000$)
applicable for the Planck satellite,  an experiment which has the potential to
dramatically improve constraints on all these correlators.   We have presented
a relatively featureless shape for the bispectrum over the relevant range which should be easily
distinguishable from the oscillatory peaks of inflationary bispectra \cite{Fergusson} if there
is a significant signal discovered (see figures~\ref{fig:local3d} and \ref{fig:bispec1}).    Our preliminary estimates of $f_{NL}$ from cosmic
strings indicate that Planck constraints from the bispectrum should be competitive with
those from the power spectrum.  In particular we find $f_{NL}\simeq-40$ for string tension $G\mu=7\times 10^{-7}$.   We note that we obtain a much smaller estimate than
the $f_{NL} \simeq -1000$ in ref.~\cite{Hindmarsh} (which was also calculated for same value of string tension) for several reasons, including the extension
of  our analysis over lower multipole ranges relevant for WMAP, a more careful comparison
with  $f_{NL}$ estimators used in the literature, and a lower normalisation of the string
spectrum derived for Nambu strings \cite{Battye2}, rather than that obtained from field
theory simulations \cite{Kunzetal}.   Nevertheless, as we shall discuss, our estimate
contains many uncertainties and more detailed and accurate forecasts are the subject
of ongoing work \cite{FergLandetal}.

We have also evaluated the CMB trispectrum (for
parallelogram configurations) on the relevant multipole scales for WMAP and Planck.
Again we find a relatively constant trispectrum, finite in all regimes and for which
we do not expect dramatic features for other configurations.    Our results indicate a
relatively larger signature  for the trispectrum,  in
contrast to the amplitude of the bispectrum which is suppressed because it depends on
the poor correlation between the string velocity and curvature.   Our preliminary estimate
of $\tau_{NL} $ a few times $10^5$ needs a more detailed analysis to characterise key
uncertainties more carefully.   Nevertheless, the trispectrum deserves much closer scrutiny observationally and the prospect of constraining cosmic strings should motivate the
development of suitable estimators.

These analytic calculations of the post-recombination gravitational effects of cosmic
strings offer important physical insights into the CMB correlations they induce.
 However, we note that there are many directions in which they
can be substantially improved.   Detailed numerical investigations of the CMB power spectrum
created by cosmic strings and other topological defects already takes into account
a far wider range of physical effects, notably recombination physics around decoupling
and a better description of the evolving string network.    It should be possible to
similarly develop the unequal time correlation methods presented for the late
time GKS signatures here to calculate both the bispectrum and the trispectrum to
high accuracy numerically.   In the meantime, however, it is a matter of comparing
the present analytic results with CMB maps induced by cosmic string networks on both
large (full sky) and small scales in order to get a more accurate normalisation and
characterisation of the bispectrum and trispectrum.   Given the stark contrast
between the string shapes with those predicted by inflation, there needs also to be
a specific search for these signatures in present and forthcoming CMB data,
a project which is being actively investigated \cite{FergLandetal}.   There seem
to be  good prospects of using forthcoming CMB data to obtain new insight into
cosmic string scenarios using higher order correlators.

\section{Appendix A: String network parameters}\label{sec:appendixA}
In order to calculate $\frac{\mathcal{L}\hat{\xi}}{\mathcal{A}}$ we use that the energy density of long strings is given by $\rho_{\infty}=\mu/\hat{\xi}^2$. Therefore the comoving length of string per comoving volume $\mathcal{A}d\eta$ is given by
\begin{eqnarray}
\frac{d\mathcal{L}}{\mathcal{A}d\eta}\propto\frac{1}{\eta^2}.
\end{eqnarray}
Dividing the time period between $\eta_{\rm{lss}}$ and $\eta_{0}$ into discrete timesteps, i.e. $(\eta_{\rm{lss}},2\eta_{\rm{lss}},\dots, n \eta_{\rm{lss}})$ where we use $\eta_0/\eta_{\rm{lss}}=n$. Integrating the above expression between times $\eta_i$ and $\eta_{i+1}$ in this series we find
\begin{eqnarray}
\frac{\mathcal{L}\eta_i}{\mathcal{A}}\propto\frac{1}{1+i}.
\end{eqnarray}
Next, noting that $\hat{\xi}\propto \eta$ we set
\begin{eqnarray}
\frac{\mathcal{L}\hat{\xi}_i}{\mathcal{A}}=\frac{c}{1+i}.
\end{eqnarray}
Making this continuous we have at time $\eta$ that $\dfrac{\mathcal{L}\hat{\xi}}{\mathcal{A}}=\dfrac{c}{1+\eta/\eta_{lss}}$. More generally if we are integrating between $\eta_{\rm{start}}$ and $\eta_0$ we use
\begin{eqnarray}
\frac{\mathcal{L}\hat{\xi}}{\mathcal{A}}=\frac{c}{1+(\eta/\eta_{\rm{start}})(\eta_{\rm{start}}/\eta_{\rm{lss}})}=\frac{c}{1+(\eta/\eta_{\rm{start}})(500/l_m)},
\end{eqnarray}
where $l_m=\min(500,l_i)$ and where we use equation~\eqref{tstartCalc} to write $\eta_{\rm{start}}/\eta_{\rm{lss}}=500/l_m$. We observe here that the constant $c$ is expected to be of order unity since the  string length per unit volume is of the order the correlation length (see also \cite{Hind09}). Upon comparison of the analytic power spectrum to simulations we will find that we may set $c\approx 1$.
\par
In order to estimate the parameters $(\overline{v}^2,s^2, c_0)$ we use representative values from simulations \cite{BBS, MartShell, AllShel}. However, these (flat-sky) simulations are generally computed using the temporal gauge for which
\begin{eqnarray*}
X^0=\eta;\quad \dot{X}^i \dot{X}^i +X'^{i}X'^{i}=1;\quad \dot{X}^i X'^i =0.
\end{eqnarray*}
In this gauge equation~\eqref{mainequation} becomes
\begin{align}
\mathbf{\nabla}_{\perp}^2\delta=&-8\pi G \mu \int d\sigma \left(\dot{X}-\frac{(X'.\hat{p})}{(\dot{X}.\hat{p})}X'\right).\mathbf{\nabla}_{\perp}\delta^{(2)}(\mathbf{x-X})\nonumber\\
=&-8\pi G \mu \int d\sigma {\mathbf{u}}.\mathbf{\nabla}_{\perp}\delta^{(2)}(\mathbf{x-X}).
\end{align}
Therefore, to relate the results of simulations in the temporal gauge to the lightcone gauge we make the replacement $\dot{X}_{\rm{lightcone}}^A\rightarrow u^A$ and ${X'}_{\rm{lightcone}}^A(\sigma)\rightarrow {X'}^A$.
Using the prescription outlined in \cite{Hindmarsh} and the value $V^2=\langle \dot{X}^i\dot{X}^i \rangle=0.37$ (indicated from the simulations) we find that
\begin{eqnarray}
s^2&=& \langle X'^A X'^A \rangle=\frac{2}{3} \langle X'^i X'^i \rangle=\frac{2}{3}(1-V^2)\\
\overline{v}^2= \langle u^A u^A \rangle&=& \langle \dot{X}^A\dot{X}^A  \rangle
-2 \langle \frac{(X'.\hat{p})}{(\dot{X}.\hat{p})}\dot{X}^A X'^A \rangle+ \langle \frac{(X'.\hat{p})^2}{(\dot{X}.\hat{p})^2}X'^A X'^A  \rangle\\
&\approx& \frac{2}{3}V^2 +0 +  \frac{2}{9}(1-V^2)^2 (1+V^2)\approx 0.365.
\end{eqnarray}
We may approximate $c_0$ using the following expression described in \cite{MartShell},
\begin{eqnarray}
\frac{k}{R}V(1-V^2)=\frac{1}{R}\langle \dot{X}^i.{X^i}''\rangle,
\end{eqnarray}
where we may identify $R$ as $\hat{\xi}$. It was shown that $k$ may be approximated by
\begin{eqnarray}
k \approx \frac{2\sqrt{2}}{\pi}\frac{1-8V^6}{1+8v^6}.
\end{eqnarray} 
Therefore,
\begin{eqnarray}
c_0&=&\hat{\xi}\langle u^A.{X^A}''\rangle=\hat{\xi}\langle \dot{X}^A.{X^A}''\rangle=\frac{2}{3}\hat{\xi}\langle \dot{X}^i.{X^i}''\rangle\nonumber\\
&\approx& \frac{2}{3}\frac{2\sqrt{2}}{\pi}\frac{1-8V^6}{1+8V^6}V(1-V^2)\approx 0.1.
\end{eqnarray}
This value for $c_0$ is an adequate match to simulations \cite{MartShell} with an accuracy sufficient for our purposes here. Greater quantitative precision requires investigation with high resolution simulations in which small-scale structure is stabilised.

\section{Appendix B: Bispectrum integral approximations}
In this appendix we derive analytic estimates for equation~\eqref{keyApprox1}. Evaluating this integral allows us to construct an analytic approximation to the bispectrum relevant for the Planck satellite, i.e. for angular multipoles, $l\lesssim 2000$. Firstly, we derive an estimate for $L\tilde{\xi}\gtrsim 5$ (equivalently $L\gtrsim 2500 (l_m/500)$). Next, we derive an expression valid for $L\tilde{\xi}\lesssim 2$ (equivalently $L\lesssim 1000 (l_m/500)$) and, finally, motivated by these two expressions, we find an approximation valid in the range $2\lesssim L\tilde{\xi}\lesssim 5$ (equivalently $1000 (l_m/500)\lesssim L\lesssim 2500 (l_m/500)$).

\subsection{Approximation 1}
For $L\tilde{\xi}\gtrsim 5$ or $L\gtrsim 2500$ we may approximate the functions $f_1, f_2$ using
\begin{eqnarray}
f_1&\approx& 2\sqrt{\pi}{\rm{erf}}\left(\frac{k\hat{\xi}s}{2}\right)= 2\sqrt{\pi}{\rm{erf}}\left(\frac{l\tilde{\xi}s \eta}{2}\right)\\
f_2&\approx&\sqrt{\pi}  K^2 \hat{\xi}^2 s^2-2 K \hat{\xi} s =\sqrt{\pi}  L^2 \tilde{\xi}^2 s^2 \eta^2\left(1-\frac{2}{\sqrt{\pi}L\tilde{\xi}s\eta}\right),
\end{eqnarray}
where $\eta\in(1,\eta_0/\eta_{\rm{start}})$ is the renormalised time.
In fact using these approximations for the functions we can find an accurate approximation to the integral. In particular, for $L\gtrsim 2500$ and $l\gg 500$ we find the following approximation to the integral,
\begin{eqnarray}\label{approx1}
I_{L\gg2500}&\approx&\left( \int_1^{\eta_0/\eta_{\rm{lss}}} \frac{d\eta}{\eta^2 (1+\eta)}\right){\rm{erf}}\left(\frac{l\tilde{\xi}s}{2}\right) 
2\pi s^2 L^2\tilde{\xi}^2 \left(1-\frac{2}{\sqrt{\pi}L\tilde{\xi}s}\right)\nonumber\\
 &\approx& 2\pi s^2 (L\tilde{\xi})^2\left(1-\frac{2}{\sqrt{\pi}L\tilde{\xi}s}\right){\rm{erf}}\left(\frac{l\tilde{\xi}s}{2}\right)(1-\ln(2)). 
\end{eqnarray}
For multipoles $l\approx 500$ this approximation must be corrected slightly to agree with the exact integral. Multiplying the above estimate by $\exp(-0.6/(l\tilde{\xi}))$ gives an accurate fit as established
in figure~\ref{fig:BispecApprox1} where we compare the exact integral against this approximation for $L\tilde{\xi}\gtrsim 5$ (or $L=\gtrsim 2500$) for various values of $l$. 
\newline
For $l\lesssim 500$ it is found that multiplying equation \eqref{approx1} by $\ln(1+l_m/1000)/\ln(1.5)$ gives an accurate quantitative fit. In figure~\ref{fig:BispecApprox1} the accuracy of this estimate is also established. Therefore, for $L\tilde{\xi}\gtrsim 5$ and any value of $l$, an accurate quantitative fit to the integral is given by
\begin{align}\label{fullapprox1}
I_{\rm{approx-}1} \approx2\pi s^2 (1-\ln(2))(L\tilde{\xi})^2&\left(1-\frac{2}{\sqrt{\pi}L\tilde{\xi}s}\right){\rm{erf}}\left(\frac{l\tilde{\xi}s}{2}\right)\nonumber\\
&\times \exp\left(-\frac{0.6}{l\tilde{\xi}}\right)\frac{\ln(1+l_m/1000)}{\ln(1.5)},
\end{align}
where we recall that $l_m=\min(500,l)$ and $\tilde{\xi}=1/l_m$.



\begin{figure}[htp]
\centering 
\includegraphics[width=78mm]{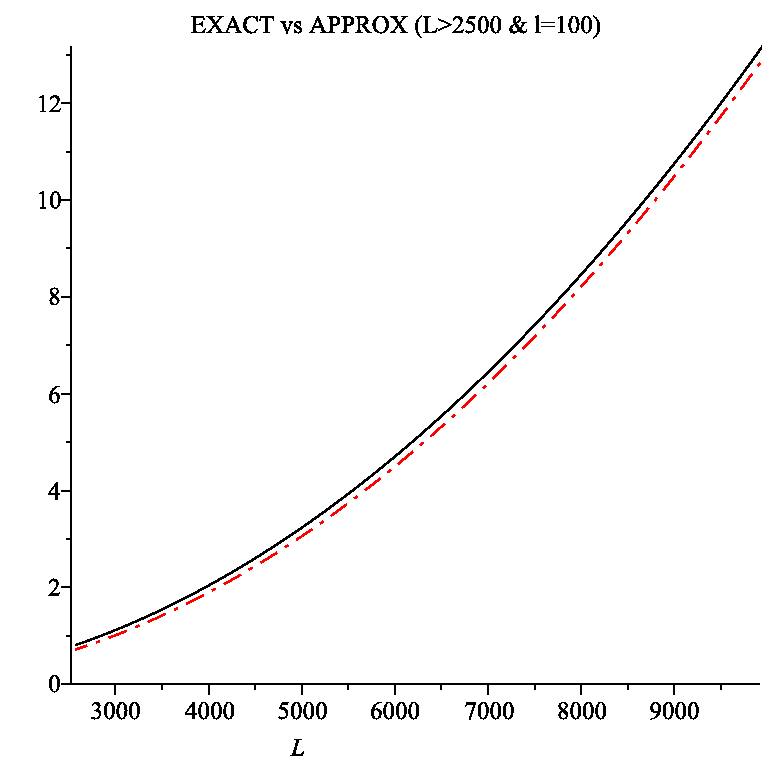}
\includegraphics[width=78mm]{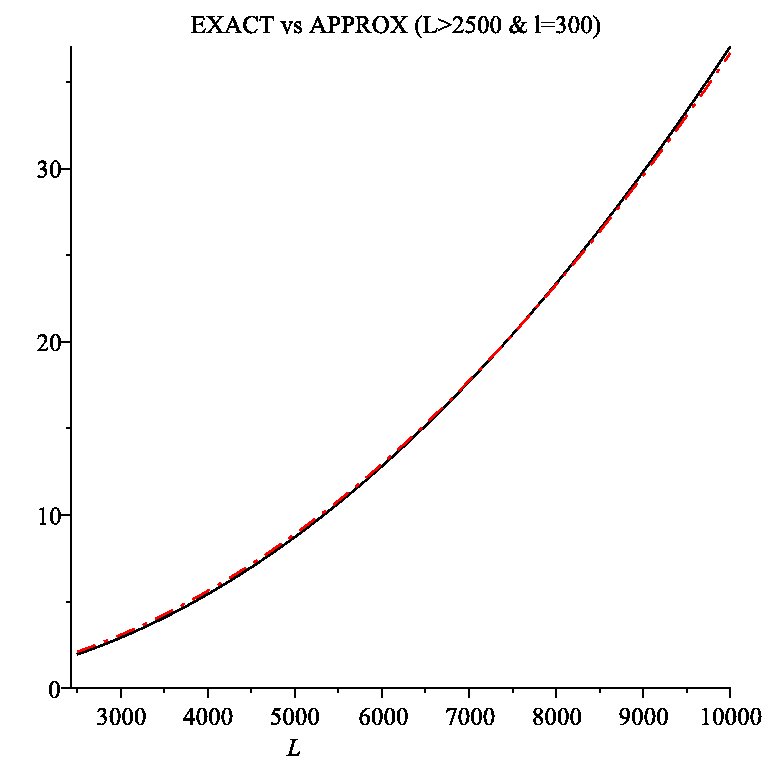}
\includegraphics[width=78mm]{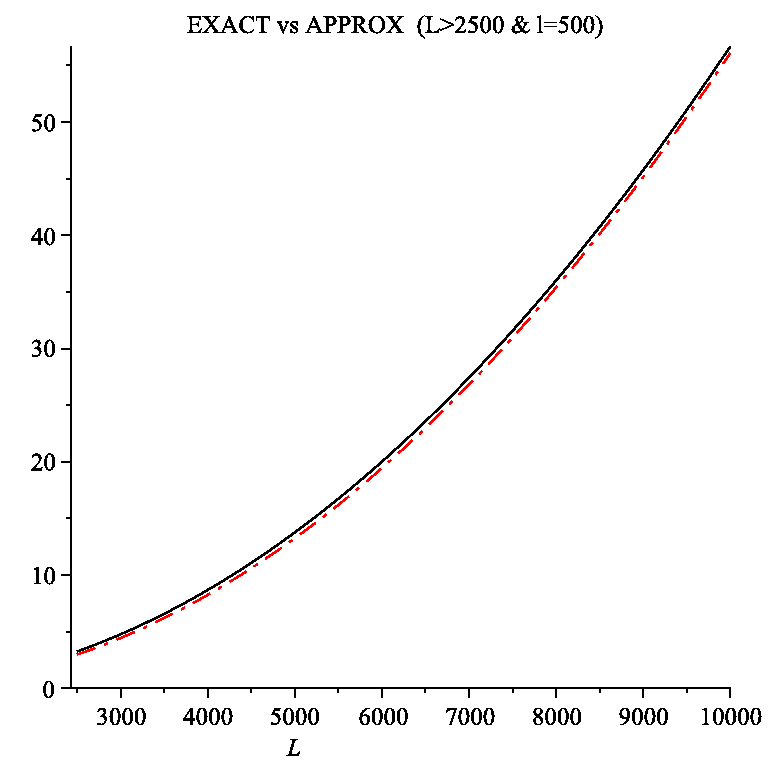}
\includegraphics[width=78mm]{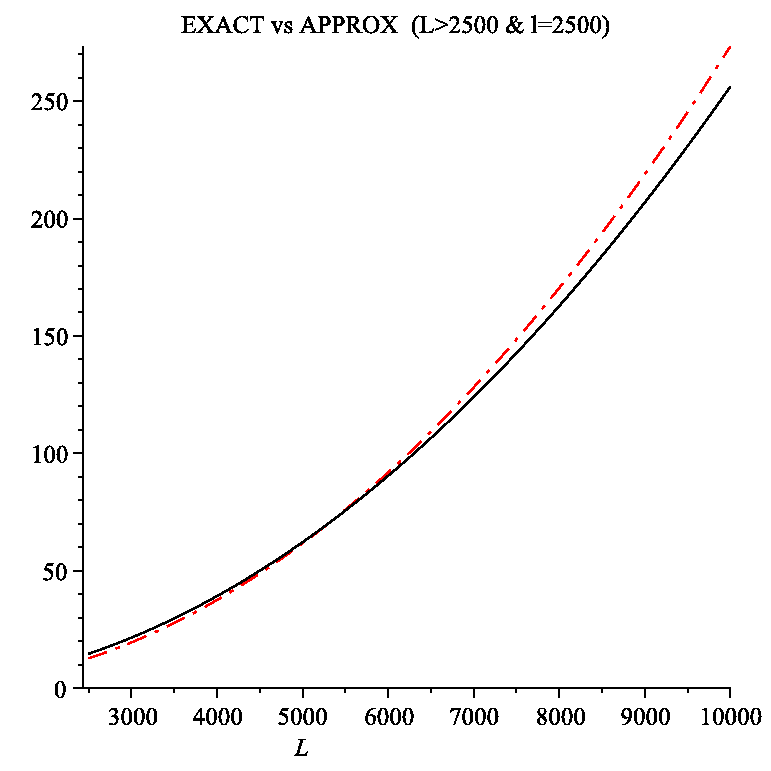}
\caption{Comparison of approximation of \eqref{keyApprox1} for $L>2500$ given by \eqref{fullapprox1} for $l=100,300,500,2500$ respectively. The exact result is given by the solid (black) line, while the approximation is given by the dot dash (red) line. The comparison clearly indicates the accuracy of the analytic approximation to the integral.}
\label{fig:BispecApprox1}
\end{figure}


\subsection{Approximation 3}
For $x$ large we can expand $\rm{erfc}(x)$ in the form,
\begin{eqnarray}
\rm{erfc}(x)=\frac{\exp(-x^2)}{\sqrt{\pi} x}\left(1+\sum_{n=1}^{\infty} (-1)^n	\frac{1.3.5\dots (2n-1)}{(2 x^2)^n}	\right).
\end{eqnarray}
This implies that, for large $x$,
\begin{eqnarray}
\sqrt{\pi}(2+x^{-2})\rm{erfc}(x)\exp(x^2)-2 x^{-1}\approx x^{-5} +\mathcal{O}(x^{-7}).
\end{eqnarray}
Clearly identifying $x^{-1}$ as $K\hat{\xi} s$ we have that, for $K\hat{\xi}$ small,
\begin{eqnarray} 
f_2\approx (K\hat{\xi}s)^5=(L\tilde{\xi}s)^5 \eta^5.
\end{eqnarray}
In particular, in assuming that, for $K\hat{\xi}\lesssim 2$, $f_2$ follows this behaviour we will derive an approximation that follows the exact result quite accurately. We note that this condition limits the range of time for our integral to $\eta \in [1,2/(L\tilde{\xi})]$ where $\tilde{\xi}=1/l_m$. This implies that, for $l\gg 500$ and $K\hat{\xi}\lesssim 2$ at last scattering,
\begin{eqnarray}
I_{\rm{approx-}3}&\sim&\int_{1}^{2/(L\tilde{\xi})} \frac{2\sqrt{\pi}{\rm{erf}}(\frac{l\tilde{\xi}s}{2}\eta)}{ \eta^4 (1+\eta)}(L\tilde{\xi}s)^5 \eta^5 d\eta \nonumber\\
&\sim& 2\sqrt{\pi}{\rm{erf}}\left(\frac{l\tilde{\xi}s}{2}\right)(L\tilde{\xi}s)^5 \int_{1}^{2/(L\tilde{\xi})} \frac{\eta}{1+\eta} d\eta \nonumber\\
&\sim&2\sqrt{\pi}{\rm{erf}}\left(\frac{l\tilde{\xi}s}{2}\right)(L\tilde{\xi}s)^5 \left(\frac{2}{L\tilde{\xi}}-\ln\left(1+\frac{2}{L\tilde{\xi}}\right)\right)\nonumber\\
&\sim&2\sqrt{\pi}{\rm{erf}}\left(\frac{l\tilde{\xi}s}{2}\right)(L\tilde{\xi}s)^5 \frac{2}{(L\tilde{\xi})^2}\sim {\rm{erf}}\left(\frac{l\tilde{\xi}s}{2}\right)(L\tilde{\xi})^3
\end{eqnarray}
where we extract the leading order term in the expansion of the logarithm in the last line. This shows that for $L\lesssim1000$ we can expect that $I\propto (L\tilde{\xi})^3$. Similarly to Approximation 2 we expect corrections to this formula for smaller multipoles $l$ and, in particular a logarithmic correction for $l\lesssim 500$. We find the following fit to the integral valid for all multipoles satisfying $L\lesssim 1000$
\begin{align}\label{fullapprox3}
I_{\rm{approx-}3}\approx 0.18 (L_i\tilde{\xi})^3   \exp\left(-\frac{0.15}{L_i\tilde{\xi}} \right){\rm{erf}}(s l_i\tilde{\xi}/2)\exp\left(-\frac{0.15}{l_i\tilde{\xi}} \right)\frac{\ln(1+l_m/1000)}{\ln(1.5)}.
\end{align}
 Using figure~\ref{fig:BispecApprox3} we compare the exact integral against this approximation for $L\tilde{\xi}\lesssim 2$ (or $L\lesssim 1000$) for various values of $l$. It is clear from this figure that the approximation gives a good quantitative fit to the exact expression.


\begin{figure}[htp]
\centering 
\includegraphics[width=78mm]{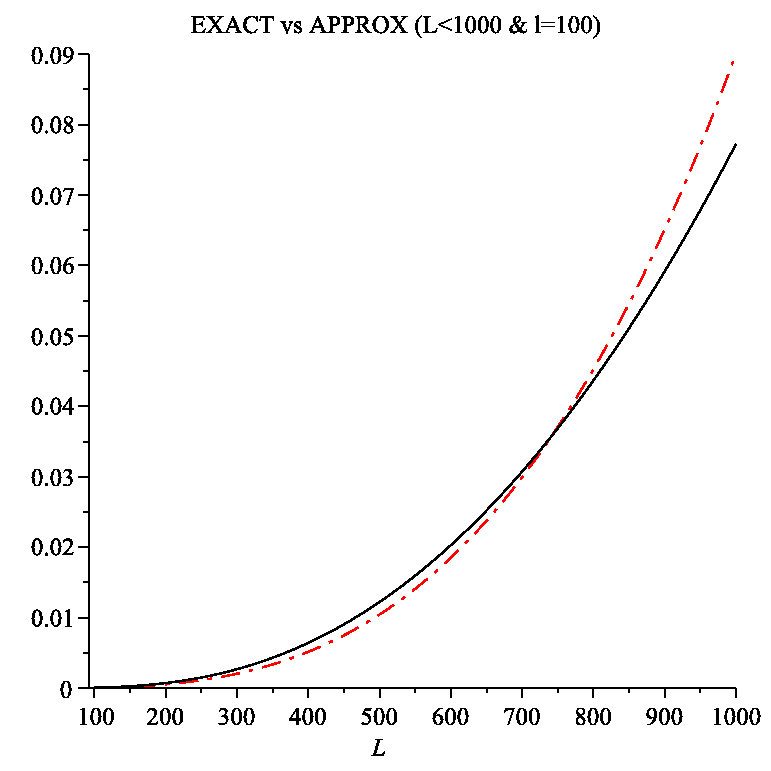}
\includegraphics[width=78mm]{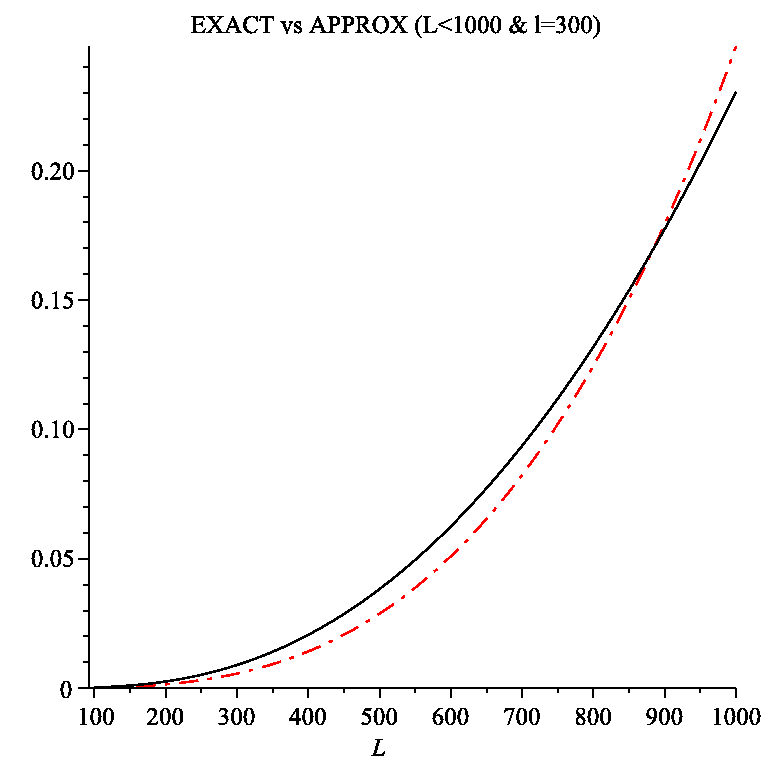}
\includegraphics[width=78mm]{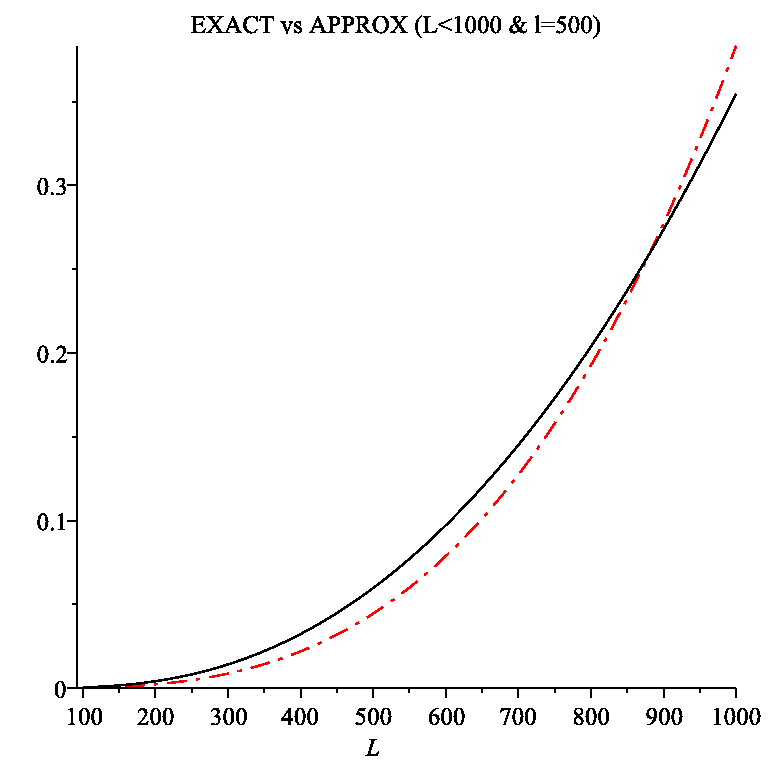}
\includegraphics[width=78mm]{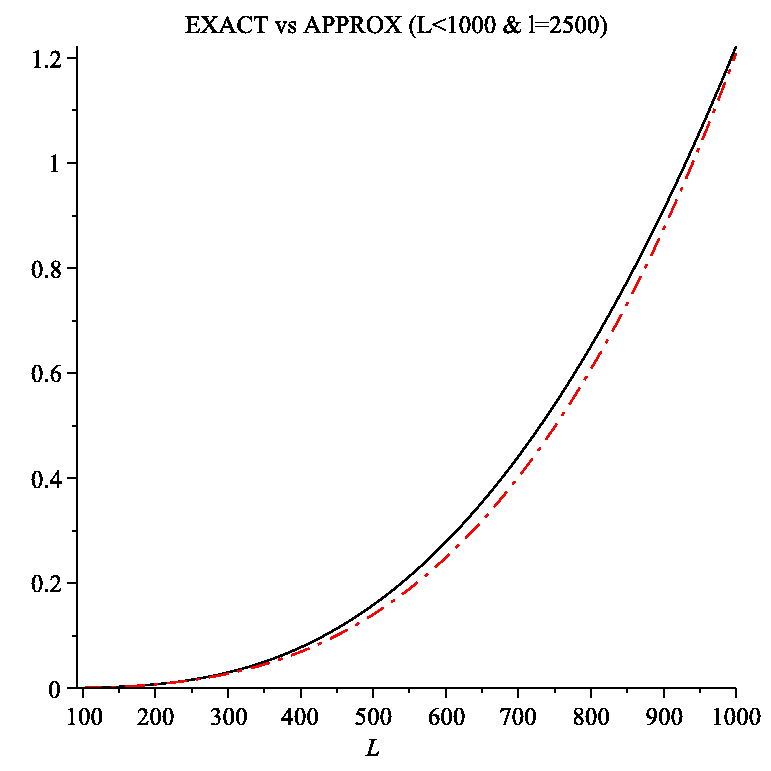}
\caption{Comparison of approximation of \eqref{keyApprox1} for $L<1000$ given by \eqref{fullapprox3} for $l=100,300,500,2500$ respectively. The exact result is given by the solid (black) line, while the approximation is given by the dot dash (red) line. The comparison shows the analytic approximation gives an accurate fit to the integral.}
\label{fig:BispecApprox3}
\end{figure}





\subsection{Approximation 2}
For $2 \lesssim L\tilde{\xi}\lesssim 5$ we make a fit to the exact expression motivated by expressions \eqref{fullapprox1} and \eqref{fullapprox3}. In particular, we will approximate the integral in this range with the assumption $I_{\rm{approx-}2} \propto (L\tilde{\xi})^{3}/(1+aL\tilde{\xi})$ (where $a$ is a constant to be determined) to interpolate between the previous expressions. We expect similar logarithmic behaviour for $l\lesssim 500$. This leads to the following estimate
\begin{align}\label{fullapprox2}
I(l_i \tilde{\xi}, L_i \tilde{\xi})_{\rm{approx-}2}\approx & 0.3 \frac{(L_i\tilde{\xi})^3}{1+L_i\tilde{\xi}/3}\exp\left(-\frac{0.15}{L_i\tilde{\xi}} \right){\rm{erf}}(s l_i\tilde{\xi}/2)\exp\left(-\frac{0.45}{l_i\tilde{\xi}} \right)\nonumber\\
&\times\frac{\ln(1+l_m/1000)}{\ln(1.5)}.
\end{align}
In figure~\ref{fig:ApproxALL} we plot the approximations to~$\eqref{keyApprox1}$ for all $L\lesssim 5000$ using various values of $l$. It is evident that the approximations \eqref{fullapprox1}, \eqref{fullapprox3} and \eqref{fullapprox2} fit the exact result quite accurately for any value of $l$.

\begin{figure}[htp]
\centering 
\includegraphics[width=78mm]{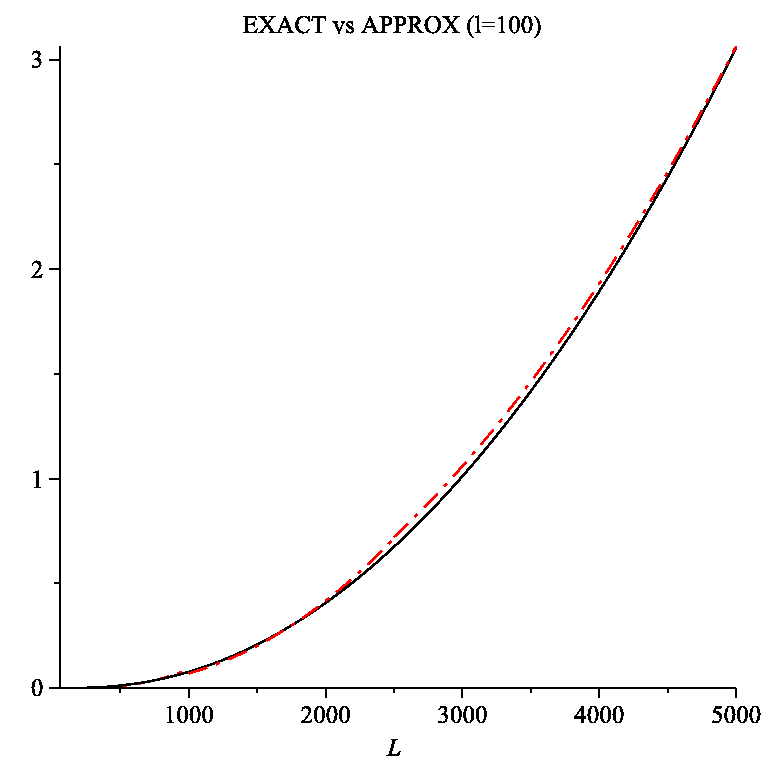}
\includegraphics[width=78mm]{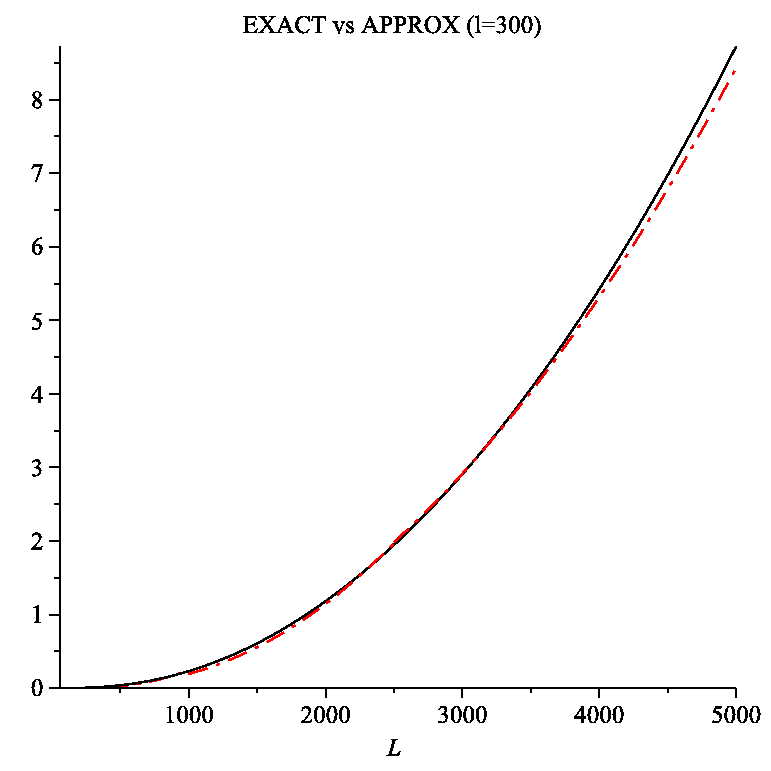}
\includegraphics[width=78mm]{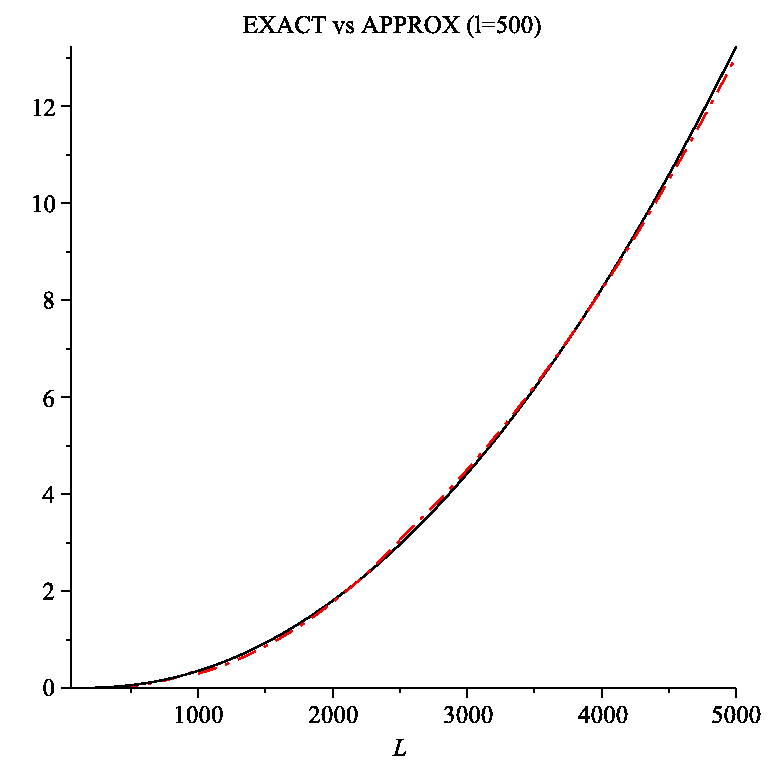}
\includegraphics[width=78mm]{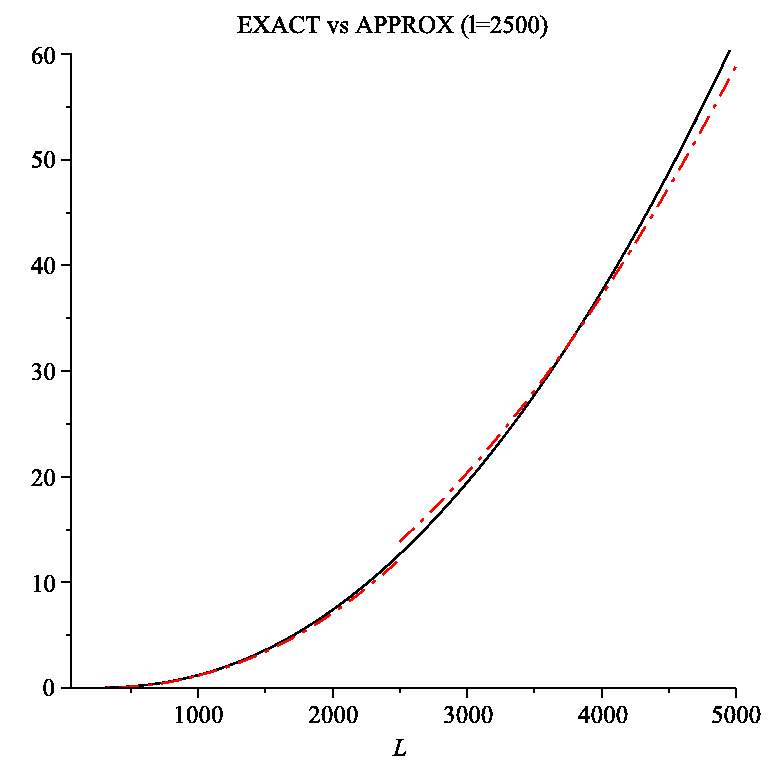}
\caption{Comparison of the approximations to \eqref{keyApprox1} given by  \eqref{fullapprox1}, \eqref{fullapprox3} and \eqref{fullapprox2} for $l=100,300,500,2500$ respectively. The exact result is given by the solid (black) line, while the approximation is given by the dot dash (red) line. The comparison shows the analytic approximations give an accurate fit to the integral.}
\label{fig:ApproxALL}
\end{figure}

\newpage

\thispagestyle{empty}
\mbox{}
\newpage
\chapter{Analysis of the CMB Trispectrum}
\label{chapter:methodology}

\textbf{Summary}\\
\textit{In this chapter we present trispectrum estimation methods which can be applied
to general non-separable primordial and CMB trispectra.  We review the relationship
between the reduced CMB trispectrum and  the reduced primordial trispectrum.
We present a general optimal  estimator for the connected part of the trispectrum,
for which we derive a quadratic term to incorporate the effects of inhomogeneous
noise and  masking.  We describe a general algorithm for creating simulated maps
with given arbitrary (and independent) power spectra, bispectra and trispectra.
We propose a universal definition of the trispectrum parameter $T_{NL}$,
 so that the integrated trispectrum on the observational domain can be consistently
 compared between theoretical models.   We define a shape function for the
 primordial trispectrum, together with a shape correlator and a useful parametrisation
 for visualizing the trispectrum; these methods might also be applied to the late-time trispectrum for large scale structure. We derive separable analytic CMB solutions in the
 large-angle limit for constant and local models.
 We present separable mode decompositions which can be used to describe any primordial
 or CMB trispectra on their respective wavenumber or multipole domains.   By extracting
 coefficients of these separable basis functions from an observational map, we are able
 to present an efficient estimator for any given theoretical model with a nonseparable
 trispectrum.   The estimator  has two manifestations,  comparing the theoretical and
 observed coefficients at either primordial or late times, thus encompassing a wider
 range of models, such as secondary anisotropies, lensing and cosmic strings.
 We show that these mode decomposition methods are numerically tractable with
 order $l^5$ operations for the CMB estimator and approximately order $l^6$ for the
 general primordial estimator (reducing to order $l^3$ in both cases for a special class of models). We also demonstrate how the trispectrum can be reconstructed
 from observational maps using these methods.}
\vspace{50pt}

\section{Introduction}
Single field slow-roll inflationary fluctuations in the standard picture of cosmology predict a nearly scale invariant spectrum of adiabatic perturbations with a nearly Gaussian distribution. Hence it can be described very accurately by its angular power spectrum. These predictions agree well with measurements of the cosmic microwave background (CMB) and large scale structure, such as those provided by WMAP and SDSS. However, it remains possible that there exists a mechanism for generating large non-Gaussianities in the early Universe. Measurements of such non-Gaussianities open up the opportunity of investigating the physics of the early universe including different inflationary models and competing alternative scenarios. In order to study such observations, higher order correlators, beyond the two-point function, offer possibly the best prospects. General methods for comparing the three-point correlator, dubbed the bispectrum, were developed in \cite{Fergusson,Ferg2,Ferg3}. In those papers an integrated measure of the bispectrum was defined, as well as a set of formalisms for comparing, evolving and constraining the bispectrum in the case of both the primordial and CMB three-point correlators. In this chapter we will generalise many of these methods to the four-point correlator which is denoted the trispectrum. We will emphasise the application of these methods to the primordial and CMB trispectra. The primary motivation for this chapter is to develop formalisms to bring observations to bear on this broader class of cosmological models. We will demonstrate that despite the complexity of trispectrum estimation, these methods are numerically tractable given 
present resources, even at Planck satellite resolution.  
\par
In order to get large non-Gaussianity we must move away from the standard single field slow-roll inflation~\cite{Chen3}. Multifield inflation allows the possibility for superhorizon evolution. Non-Gaussianities are generated when this evolution is nonlinear. We can consider superhorizon behaviour as occurring in patches separated by horizons which evolve independently of each other. This locality in position space translates to non-locality in momentum space and indicates that for such models we expect the signal to peak for $k_4\ll k_1,k_2,k_3$. This forms the so-called local model. Such models have been investigated in the context of the trispectrum in \cite{Seery1,Seery2,Seery3,Adshead,Bartolo,Valen,Rodriguez,Huang1,Lehners}. Since subhorizon modes oscillate and so average out, the only chance to have large non-Gaussianity in single field inflationary models is when all modes have similar wavelengths and exit at the same time. A non-standard kinetic term allows for such a possibility. Since the signal peaks when the modes have similar wavelengths this class of forms are known as equilateral models and have been investigated using the trispectrum in \cite{aChen,bArroja,cChen4,dArroja2,eArroja3,fSenatZal,gGao}. It should be noted that this amplification of nonlinear effects around the time the modes exit the horizon is not possible for slow-roll single field inflation. It has also been shown in \cite{Huang2,Izumi} that a large trispectrum may be generated in the ghost inflation model. These models are so-called as they are based on the idea of a ghost condensate, i.e. a kind of fluid with equation of state $p=-\rho$, that can fill the universe, and which provides an alternative method of realising de Sitter phases in the early universe. Of course there are other methods to generate non-Gaussianity such as having sharp features in the potential or a non-Bunch-Davies vacuum. Also there are models which have features that resemble the aforementioned forms in different regimes, e.g. quasi-single field inflation~\cite{Chen2}, or have mixed contributions, e.g. in multifield DBI inflation~\cite{RenauxDBI}.
\par
One of the motivations for studying the four-point correlator is that it may be possible that the bispectrum is suppressed but still have a large trispectrum. In particular, this behaviour may be realised in quasi-single field inflation~\cite{Chen2} or in the curvaton model~\cite{Sasaki}. It also occurs in the case of cosmic strings where the bispectrum is suppressed by symmetry considerations \cite{Hindmarsh2009,Regan:2009hv}. The effects of non-Gaussianity  could also be detectable in a wide range of astrophysical measurements, such as cluster abundances and the large scale clustering of highly biased tracers. In \cite{Sefusatti} the possibility of using the galaxy bispectrum to constrain the local form of the trispectrum has been reviewed. 
\par
The trispectrum, $T(\bk_1,\bk_2,\bk_3,\bk_4)$, is generally parametrised using the variable $\tau_{NL}$ which schematically is given by the ratio $\tau_{NL}\approx T(k,k,k,k;k,k)/P(k)^3$ (where the $6$ parameters denote the four sides and two of the diagonals of the quadrilateral created by the $\bk_i$). Standard slow-roll inflation predicts $\tau_{NL}\lesssim r/50$ where $r<1$ is the tensor to scalar ratio~\cite{Seery1}. Such a low signal would be undetectable since it is below the level of non-Gaussian contamination that would be expected from secondary anisotropies $\tau_{NL}\approx \mathcal{O}(1)$. Using the analysis of N-point probability distribution of the CMB anisotropies~\cite{vielva}, where a local non-linear perturbative model $\Phi=\Phi_L+f_{NL}(\Phi^2_L-\langle \Phi_L^2\rangle)+g_{NL}\Phi_L^3+\mathcal{O}(\Phi_L^4)$ is used to characterise the large scale anistropies, the constraint $-5.6\times 10^5<g_{NL}<6.4\times 10^5$ was obtained\footnote{It should be noted that for single field local inflation $\tau_{NL}^{\rm{loc}}=\left(\frac{5}{6}f_{NL}\right)^2$. Since $f_{NL}$ is constrained by the bispectrum, $g_{NL}$ is the quantity that is constrained by the trispectrum directly in this case.}. For the more general case, there is only a weak experimental bound imposed on non-Gaussianity by the trispectrum, which is roughly $|\tau_{NL}|\lesssim 10^8$~\cite{Lyth}. In~\cite{Cooray,Cooray2} an improved constraint on $\tau_{NL}$ was presented using estimators to allow a joint fit of $f_{NL}$ and $g_{NL}$ using the trispectrum of WMAP5 data. However, the analysis therein included an incomplete formula for the CMB trispectrum due to local non-Gaussianity\footnote{The formula for the reduced local CMB trispectrum has been used in place of the full local CMB trispectrum,
which appears to simplify the analysis.}. Nonetheless, the approach indicates that vast improvements to trispectrum constraints should be achievable in the near future. In fact, it is expected that the Planck satellite will be sensitive to a value of $|\tau_{NL}|\sim 560$~\cite{Kogo}.
\par
The analysis of the trispectrum is a computationally intensive operation. In fact only the trispectrum induced by the local shape has been constrained so far by CMB data. The local form is an example of a separable shape - a notion which we will define more concretely in this chapter. Essentially, since the primordial trispectrum is a six dimensional quantity, separability means the trispectrum is the product of one dimensional functions of each of these variables. Exploiting this separability reduces the problem from one of $\mathcal{O}(l_{\rm{max}}^7)$ operations to a more manageable $\mathcal{O}(l_{\rm{max}}^5)$. In special cases we get a further reduction to $\mathcal{O}(l_{\rm{max}}^4)$. 
\par
In the next section we shall describe the CMB trispectrum and its relation to the primordial equivalent. We will make use of a particular parametrisation of the reduced primordial trispectrum and exploit a Legendre series expansion in terms of one of these parameters to write an expression for the reduced CMB trispectrum. We will also outline a general correlation method for comparing different trispectra. In this section we will also give a formula for the kurtosis in terms of the multipoles. In Section \ref{sec:III} we define a shape function which is a scale invariant form of the trispectrum. Using this function we define a shape correlator that is expected to predict closely the correlation between the respective trispectra. Next, we show how to decompose this shape in order to provide a method for visualising trispectra. We apply this visualisation to the case of the local and equilateral models which we describe in Section \ref{sec:IV}. We also present the Sachs Wolfe limit ($l<100$) for the local and constant models. In Section \ref{sec:V} we describe how to form a mode expansion for general non-separable shapes. This provides a rigorous method to find a separable approximation to any shape and therefore makes analysis of the trispectrum far more tractable. This expansion can be performed for both the primordial and CMB trispectra. Of immediate relevance in terms of Planck is to find a general measure for the size of the trisectrum. This is addressed in Section \ref{sec:VI} in both the primordial and CMB cases. We also discuss, in this section, the reduction in computational complexity for the class of trispectra which are diagonal-free. It is clearly desirable to be able to reconstruct the underlying trispectrum given the data. As we shall describe in Section \ref{sec:VII} this is a computationally intensive task, but it is tractable. We will observe here that there is a degeneracy in reconstruction of the primordial trispectrum.   Finally in Section \ref{sec:VIII} we outline a method for performing CMB map simulations for given general bispectra and trispectra.

\section{The CMB Trispectrum}\label{sec:II}
\subsection{Definition of the primordial and CMB trispectra}
We are concerned with the analysis of the four-point function induced by a non-Gaussian primordial gravitational potential $\Phi(\mathbf{k})$ in the CMB temperature fluctuation field. The temperature anisotropies
may be represented using the $a_{lm}$ coefficients of a spherical harmonic decomposition of the cosmic microwave sky,
\begin{eqnarray}\label{DeltaT}
\frac{\Delta T}{T}(\hat{\mathbf{n}})=\sum_{l,m} a_{lm}Y_{lm}(\hat{\mathbf{n}}).
\end{eqnarray}
The primordial potential $\Phi$ induces the multipoles $a_{lm}$ via a convolution with the transfer functions $\Delta_l(k)$ through the relation
\begin{eqnarray}
a_{lm}=4\pi (-i)^l \int \frac{d^3k}{(2\pi)^3} \Delta_l(k) \Phi(\mathbf{k}) Y_{lm}(\hat{\mathbf{k}}).
\end{eqnarray}
The connected part of the four-point correlator of the $a_{lm}$ gives us the trispectrum. In particular, 
\begin{eqnarray}\label{eq:Tconnls}
T_{l_1 m_1 l_2 m_2 l_3 m_3 l_4 m_4}&=&\langle a_{l_1 m_1}	 a_{l_2 m_2} a_{l_3 m_3} a_{l_4 m_4}	\rangle_c \nonumber\\
&=& (4\pi)^4 (-i)^{\sum_i l_i}\int  \frac{d^3 k_1 d^3 k_2 d^3 k_3 d^3 k_4}{(2\pi)^{12}}\Delta_{l_1}(k_1)\Delta_{l_2}(k_2)\Delta_{l_3}(k_3)\Delta_{l_4}(k_4)\times \nonumber\\
&&\langle \Phi(\mathbf{k_1})\Phi(\mathbf{k_2})\Phi(\mathbf{k_3})\Phi(\mathbf{k_4}) \rangle_c Y_{l_1 m_1}(\hat{\mathbf{k_1}}) Y_{l_2 m_2}(\hat{\mathbf{k_2}}) Y_{l_3 m_3}(\hat{\mathbf{k_3}}) Y_{l_4 m_4}(\hat{\mathbf{k_4}}),\nonumber\\
&&
\end{eqnarray}
where $k_i=|\mathbf{k}_i|$ and the subscript $c$ is used to denote the connected component. Naively, we would define the primordial trispectrum as 
\begin{eqnarray*}
\langle \Phi(\mathbf{k_1})\Phi(\mathbf{k_2})\Phi(\mathbf{k_3})\Phi(\mathbf{k_4}) \rangle_c=(2\pi)^3 \delta(\mathbf{k_1+k_2+k_3+k_4}) T'_{\Phi}(\mathbf{k}_1,\mathbf{k}_2,\mathbf{k}_3,\mathbf{k}_4).
\end{eqnarray*}
Here, the four wave-vectors form a quadrilateral as shown in Figure~\ref{fig:Tquad}.  
\begin{figure}[htp]
\centering 
\includegraphics[width=102mm]{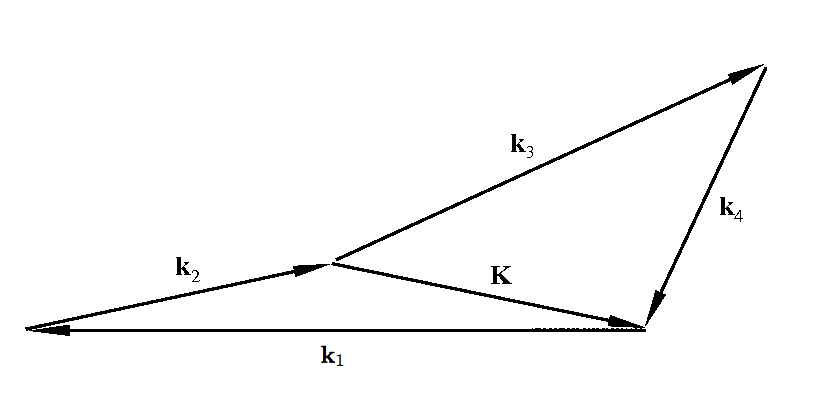}
\caption{Quadrilateral defined by the four wave-vectors ${\bf k}_i$.  The diagonal is represented
by ${\bf K}$.}
\label{fig:Tquad}
\end{figure}
However, a more useful definition is to write
\begin{eqnarray}\label{eq:Tconn}
\langle \Phi(\mathbf{k_1})\Phi(\mathbf{k_2})\Phi(\mathbf{k_3})\Phi(\mathbf{k_4}) \rangle_c&=&(2\pi)^3 \int d^3 K \delta (\mathbf{k_1+k_2+K})\delta (\mathbf{k_3+k_4-K})\nonumber\\
&&\times T_{\Phi}(\mathbf{k}_1,\mathbf{k}_2,\mathbf{k}_3,\mathbf{k}_4;\mathbf{K}).
\end{eqnarray}
Here the delta function indicates that the diagonal $\mathbf{K}$ makes triangles with $(\mathbf{k_1, k_2})$ and $(\mathbf{k_3, k_4})$, respectively. Of course there are symmetries implicit in this definition of $T_{\Phi}$ - namely, that we may form triangles with different combinations of the vectors. In particular,
\begin{align}
T_{\Phi}(\mathbf{k}_1,\mathbf{k}_2,\mathbf{k}_3,\mathbf{k}_4;\mathbf{K})=&P_{\Phi}(\mathbf{k}_1,\mathbf{k}_2,\mathbf{k}_3,\mathbf{k}_4;\mathbf{K}) \nonumber \\
&+\int d^3 K'[\delta (\mathbf{k_3-k_2-K+K'})P_{\Phi}(\mathbf{k}_1,\mathbf{k}_3,\mathbf{k}_2,\mathbf{k}_4;\mathbf{K'})\nonumber \\
&\qquad\qquad+\delta (\mathbf{k_4-k_2-K+K'})P_{\Phi}(\mathbf{k}_1,\mathbf{k}_4,\mathbf{k}_3,\mathbf{k}_2;\mathbf{K'}) ],
\end{align}
where $P_{\Phi}$ are constructed using a reduced trispectrum $\mathcal{T}_{\Phi}$ via
\begin{align}
P_{\Phi}(\mathbf{k}_1,\mathbf{k}_2,\mathbf{k}_3,\mathbf{k}_4;\mathbf{K})=&\mathcal{T}_{\Phi}(\mathbf{k}_1,\mathbf{k}_2,\mathbf{k}_3,\mathbf{k}_4;\mathbf{K})+\mathcal{T}_{\Phi}(\mathbf{k}_2,\mathbf{k}_1,\mathbf{k}_3,\mathbf{k}_4;\mathbf{K})\nonumber\\
&+\mathcal{T}_{\Phi}(\mathbf{k}_1,\mathbf{k}_2,\mathbf{k}_4,\mathbf{k}_3;\mathbf{K})+\mathcal{T}_{\Phi}(\mathbf{k}_2,\mathbf{k}_1,\mathbf{k}_4,\mathbf{k}_3;\mathbf{K}),
\end{align}
with $\mathcal{T}_{\Phi}(\mathbf{k}_1,\mathbf{k}_2,\mathbf{k}_3,\mathbf{k}_4;\mathbf{K})=\mathcal{T}_{\Phi}(\mathbf{k}_3,\mathbf{k}_4,\mathbf{k}_1,\mathbf{k}_2;-\mathbf{K})$.
Therefore, we need only consider the reduced trispectrum $\mathcal{T}$ from one particular arrangement of the vectors and form the other contributions by permuting the symbols. 

The CMB trispectrum may also be written in a rotationally invariant way as
\begin{eqnarray}\label{Ttot}
T_{l_1 m_1 l_2 m_2 l_3 m_3 l_4 m_4}=\sum_{LM} (-1)^M \left( \begin{array}{ccc}
l_1 & l_2 & L \\
m_1 & m_2 & -M \end{array} \right) \left( \begin{array}{ccc}
l_3 & l_4 & L \\
m_3 & m_4 & M \end{array} \right) T^{l_1 l_2}_{l_3 l_4}(L).
\end{eqnarray}
The Wigner $3j$ symbols impose the triangle conditions on the multipole combinations $(l_1, l_2 ,L)$ and $(l_3,l_4 ,L)$.
As in the case of the primordial trispectrum, there are symmetries implicit in this definition. In a similar manner to the primordial case we can write
\begin{align}\label{Prel1}
T_{l_1 m_1 l_2 m_2 l_3 m_3 l_4 m_4}=&\sum_{LM} (-1)^M \left( \begin{array}{ccc}
l_1 & l_2 & L \\
m_1 & m_2 & -M \end{array} \right) \left( \begin{array}{ccc}
l_3 & l_4 & L \\
m_3 & m_4 & M \end{array} \right) P^{l_1 l_2}_{l_3 l_4}(L) \nonumber\\
&+ (l_2\leftrightarrow l_3)+ (l_2\leftrightarrow l_4),
\end{align}
with
\begin{align}\label{Prel2}
P^{l_1 l_2}_{l_3 l_4}(L) =&\mathcal{T}^{l_1 l_2}_{l_3 l_4}(L)+(-1)^{l_1+l_2+L} \mathcal{T}^{l_2 l_1}_{l_3 l_4}(L)+(-1)^{l_3+l_4+L} \mathcal{T}^{l_1 l_2}_{l_4 l_3}(L)\nonumber\\
&+(-1)^{l_1+l_2+l_3+l_4} \mathcal{T}^{l_2 l_1}_{l_4 l_3}(L),
\end{align}
where the factors of powers of $(-1)$ are induced by identities of the Wigner $3j$ symbol. Therefore, we again need only consider the reduced trispectrum $\mathcal{T}$ from one particular arrangement of the multipoles. Indeed we need only find the reduced CMB trispectrum induced by the reduced primordial trispectrum. In particular, we denote
\begin{eqnarray}\label{RedTrisp}
\mathcal{T}_{l_1 m_1 l_2 m_2 l_3 m_3 l_4 m_4}=\sum_{LM} (-1)^M \left( \begin{array}{ccc}
l_1 & l_2 & L \\
m_1 & m_2 & -M \end{array} \right) \left( \begin{array}{ccc}
l_3 & l_4 & L \\
m_3 & m_4 & M \end{array} \right) \mathcal{T}^{l_1 l_2}_{l_3 l_4}(L),
\end{eqnarray}
and observe that
\begin{align}\label{TotalRedTrisp}
&{T}_{l_1 m_1 l_2 m_2 l_3 m_3 l_4 m_4}=\nonumber\\
&\qquad\qquad\mathcal{T}_{l_1 m_1 l_2 m_2 l_3 m_3 l_4 m_4}+\mathcal{T}_{l_2 m_2 l_1 m_1 l_3 m_3 l_4 m_4}+\mathcal{T}_{l_1 m_1 l_2 m_2 l_4 m_4 l_3 m_3 }+\mathcal{T}_{ l_2 m_2 l_1 m_1 l_4 m_4 l_3 m_3 }\nonumber\\
&\qquad\quad+\mathcal{T}_{l_1 m_1 l_3 m_3  l_2 m_2 l_4 m_4}+\mathcal{T}_{ l_3 m_3 l_1 m_1 l_2 m_2 l_4 m_4}+\mathcal{T}_{l_1 m_1  l_3 m_3 l_4 m_4 l_2 m_2}+\mathcal{T}_{l_3 m_3 l_1 m_1   l_4 m_4 l_2 m_2}\nonumber\\
&\qquad\quad+\mathcal{T}_{l_1 m_1  l_4 m_4 l_2 m_2 l_3 m_3}+\mathcal{T}_{ l_4 m_4 l_1 m_1 l_2 m_2 l_3 m_3}+\mathcal{T}_{l_1 m_1  l_4 m_4  l_3 m_3 l_2 m_2}+\mathcal{T}_{ l_4 m_4 l_1 m_1 l_3 m_3  l_2 m_2}.
\end{align}

\subsection{Relation between the primordial and CMB trispectra}
In order to relate the above definitions for the  primordial and CMB trispectra we use the following identities
\begin{eqnarray}\label{Identities}
\delta(\mathbf{k})&=&\frac{1}{(2\pi)^3}\int e^{i \mathbf{r.k}}d^3 r,\nonumber\\
e^{i \mathbf{r.k}}&=& 4\pi \sum_{l,m} i^l j_l(k r) Y_{l m}(\hat{\mathbf{k}}) Y_{lm}^*(\hat{\mathbf{r}}),\nonumber\\
Y_{l -m}&=&(-1)^m Y_{lm}^*.
\end{eqnarray}
We find using these identities with equations~\eqref{eq:Tconnls} and~\eqref{eq:Tconn}
\begin{align}
\mathcal{T}_{l_1 m_1 l_2 m_2 l_3 m_3 l_4 m_4}&=\left(\frac{2}{\pi}\right)^5 (-i)^{\sum l_i}\int \left(\Pi_{i=1}^4d^3 k_i \Delta_{l_i}(k_i)  Y_{l_i m_i}(\hat{\mathbf{k}}_i) \right)d^3 K  \mathcal{T}_{\Phi}(\mathbf{k}_1,\mathbf{k}_2,\mathbf{k}_3,\mathbf{k}_4;\mathbf{K})\nonumber \\
&\times \sum_{l_i',L' ,L''} \sum_{m_i',M',M''}\int d^3 r_1 d^3 r_2 i^{\sum_{i'} l_i' +L'-L''} [j_{l_1'}(k_1 r_1) Y_{l_1' m_1'}(\hat{\mathbf{k}}_1) Y^*_{l_1' m_1'}(\hat{\mathbf{r}}_1)  ] \nonumber \\
&\times [j_{l_2'}(k_2 r_1) Y_{l_2' m_2'}(\hat{\mathbf{k}}_2) Y^*_{l_2' m_2'}(\hat{\mathbf{r}}_1)  ] [j_{l_3'}(k_3 r_2) Y_{l_3' m_3'}(\hat{\mathbf{k}}_3) Y^*_{l_3' m_3'}(\hat{\mathbf{r}}_2)  ] \nonumber\\
&\times [j_{l_4'}(k_4 r_2) Y_{l_4' m_4'}(\hat{\mathbf{k}}_4) Y^*_{l_4' m_4'}(\hat{\mathbf{r}}_2)  ]  [j_{L'}(K r_1) Y_{L' M'}(\hat{\mathbf{K}}) Y^*_{L' M'}(\hat{\mathbf{r}}_1)  ]\nonumber\\
&\times [j_{L''}(K r_2) Y^*_{L'' M''}(\hat{\mathbf{K}}) Y_{L'' M''}(\hat{\mathbf{r}}_2)  ]\,, 
\end{align}
where $\hat{\mathbf{k}}_i$ represents the unit vector in the direction $\mathbf{k}_i$.
\begin{figure}[htp]
\centering 
\includegraphics[width=102mm]{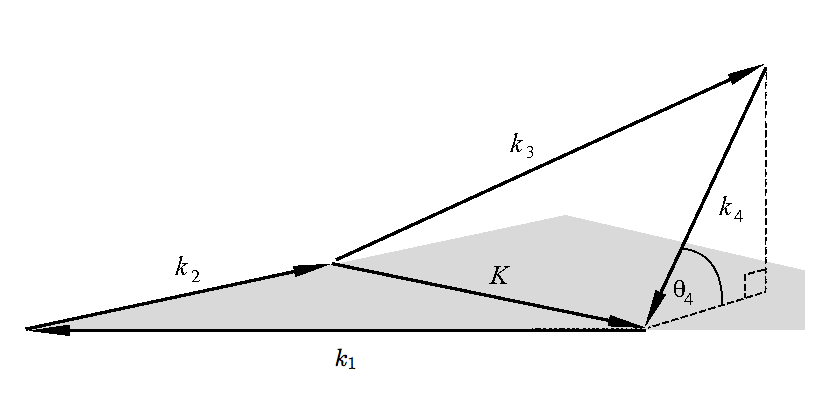}
\caption{Quadrilateral defined by the four wavenumbers $ k_i$, the diagonal $K$, and the 
angle $\theta_4$ out of the plane of the first triangle.}
\label{fig:Tquad2}
\end{figure}

To calculate further, we must choose an appropriate parametrisation for $\mathcal{T}_{\Phi}$. 
We note that the primordial trispectrum shape has $6$ degrees of freedom. We could define the quadrilateral uniquely by the lengths of the four sides $k_i = |{\bf k}_i|$,
together with the two diagonals $K = |{\bf K}|$ and $\tilde K = |\tilde{\bf K}|$.  However, we find it 
more convenient to represent the sixth degree of freedom with the angle $\theta_4$ which represents
the deviation of the quadrilateral from planarity (as illustrated in Figure~\ref{fig:Tquad2})\footnote{This angle $\theta_4$ is in effect the angle between the plane defined by the vectors $\bk_1$ and $\bk_2$ and the plane defined by the vectors $\bk_3$ and $\bk_4$.}.  Many well-motivated primordial models, such as
the local and equilateral cases we shall discuss, are planar (i.e. $\theta_4=0$). So we choose the independent parameters to identify the shape to be $(k_1,k_2, k_3,k_4, K, \theta_4)$, that is, 
 $\mathcal{T}_{\Phi}=\mathcal{T}_{\Phi}(k_1,k_2, k_3,k_4; K, \theta_4)$. We can decompose this expression by expanding the primordial trispectrum as a Legendre series. In particular, we write
\begin{eqnarray}\label{expansion}
\mathcal{T}_{\Phi}(k_1,k_2, k_3,k_4; K, \theta_4)=\sum_{n=0}^{\infty}\mathcal{T}_{\Phi,n}(k_1,k_2, k_3,k_4; K) P_n(\cos\theta_4).
\end{eqnarray}
This is an expansion about the $n=0$ planar mode which, as we have noted, is sufficient for describing many well-motivated models. In what follows we shall, therefore, consider only  trispectra which are independent of $\theta_4$, i.e. $\mathcal{T}_{\Phi}(k_1,k_2, k_3,k_4; K, \theta_4)=\mathcal{T}_{\Phi,0}(k_1,k_2, k_3,k_4; K) $, unless otherwise stated.
 \par
 With this parametrisation, and using the following identities,
\begin{align}\label{Gaunt}
\int d\Omega_{\hat{r}} Y_{l_1 m_1}(\hat{\mathbf{r}})Y_{l_2 m_2}(\hat{\mathbf{r}})Y_{l_3 m_3}(\hat{\mathbf{r}})&=\sqrt{\frac{(2l_1+1)(2l_2+1)(2l_3+1)}{4\pi}}\left( \begin{array}{ccc}
l_1 & l_2 & l_3 \\
0 & 0 & 0 \end{array} \right)\nonumber\\
&\,\,\times\left( \begin{array}{ccc}
l_1 & l_2 & l_3 \\
m_1 & m_2 & m_3 \end{array} \right),\nonumber\\
\int d \Omega_{\hat{r}}Y_{lm}(\hat{\mathbf{r}})Y_{l' m'}^*(\hat{\mathbf{r}})&=\delta_{l l'}\delta_{m m'}
\end{align}
and~\eqref{Identities} we find
\begin{align}
&\mathcal{T}_{l_1 m_1 l_2 m_2 l_3 m_3 l_4 m_4}=\left(\frac{2}{\pi}\right)^5 \sum_{L',M'}\sum_{l_4',m_4'} (-1)^{M'}\int (k_1 k_2 k_3 k_4 K)^2 dk_1 dk_2 dk_3 dk_4 dK\nonumber\\
&\times  r_1^2 dr_1 r_2^2 dr_2 j_L(K r_1) j_L(K r_2) [j_{l_1}(k_1 r_1)\Delta_{l_1}(k_1)][j_{l_2}(k_2 r_1)\Delta_{l_2}(k_2)][j_{l_3}(k_3 r_2)\Delta_{l_3}(k_3)]\nonumber\\
&\times[j_{l_4}(k_4 r_2)\Delta_{l_4}(k_4)] h_{l_1 l_2 L'} h_{l_3 l_4' L'} (-1)^{m_4'}\int d\Omega_{\hat{\mathbf{k}}_4} \mathcal{T}_{\Phi,0}(k_1,k_2, k_3,k_4; K)Y_{l_4 m_4}(\hat{\mathbf{k}}_4)Y_{l_4' m_4'}(\hat{\mathbf{k}}_4)\nonumber\\
&\times \left( \begin{array}{ccc}
l_1 & l_2 & L' \\
m_1 & m_2 & -M' \end{array} \right)\left( \begin{array}{ccc}
l_3 & l_4' & L' \\
m_3 & -m_4' & M' \end{array} \right),
\end{align}
where we write
\begin{eqnarray}
h_{l_1 l_2 L'}=\sqrt{\frac{(2l_1+1)(2l_2+1)(2L'+1)}{4\pi}}\left( \begin{array}{ccc}
l_1 & l_2 & L' \\
0 & 0 & 0 \end{array} \right). 
\end{eqnarray}
Next, we note that inverting equation~\eqref{RedTrisp} gives the expression
\begin{align}
\mathcal{T}^{l_1 l_2}_{l_3 l_4}(L)&=\sum_{m_i,M}(2L+1)(-1)^M \left( \begin{array}{ccc}
l_1 & l_2 & L \\
m_1 & m_2 & M \end{array} \right)\left( \begin{array}{ccc}
l_3 & l_4 & L \\
m_3 & m_4 & -M \end{array} \right) \mathcal{T}_{l_1 m_1 l_2 m_2 l_3 m_3 l_4 m_4}.\nonumber\\
&
\end{align}
The sum over $m_1, m_2$ is proportional to
\begin{eqnarray}\label{Orthog1}
\sum_{m_1 ,m_2} (2L+1) \left( \begin{array}{ccc}
l_1 & l_2 & L \\
m_1 & m_2 & M \end{array} \right)\left( \begin{array}{ccc}
l_1 & l_2 & L' \\
m_1 & m_2 & -M' \end{array} \right)=\delta_{L,L'} \delta_{M,-M'}
\end{eqnarray}
and therefore the sum over $L', M'$ implies $L'=L$ and $M'=-M$. The sum over $m_3, M$ is then proportional to
\begin{eqnarray}\label{Orthog2}
\sum_{m_3 ,M} \left( \begin{array}{ccc}
l_3 & l_4 & L \\
m_3 & m_4 & -M \end{array} \right)\left( \begin{array}{ccc}
l_3 & l_4' & L \\
m_3 &- m_4' & -M \end{array} \right)=\frac{1}{2l_4'+1}\delta_{l_4,l_4'} \delta_{m_4,-m_4'}.
\end{eqnarray}
Combining these we find that our expression for the CMB trispectrum becomes
\begin{align}\label{TrispRed2}
\mathcal{T}^{l_1 l_2}_{l_3 l_4}(L)=& h_{l_1 l_2 L} h_{l_3 l_4 L} \left(\frac{2}{\pi}\right)^5 \int (k_1 k_2 k_3 k_4 K)^2 dk_1 dk_2 dk_3 dk_4 dK r_1^2 dr_1 r_2^2 dr_2 \nonumber\\
&\times j_L(K r_1) j_L(K r_2)[j_{l_1}(k_1 r_1)\Delta_{l_1}(k_1)][j_{l_2}(k_2 r_1)\Delta_{l_2}(k_2)][j_{l_3}(k_3 r_2)\Delta_{l_3}(k_3)]\nonumber\\
&\times[j_{l_4}(k_4 r_2)\Delta_{l_4}(k_4)] \mathcal{T}_{\Phi,0}(k_1,k_2, k_3,k_4; K). 
\end{align}
\par
From equation~\eqref{TrispRed2} it is clear that the definition of the reduced trispectrum \cite{Hu} includes an unnecessary geometrical factor $h_{l_1 l_2 L} h_{l_3 l_4 L}$ and we therefore advocate the use of the true reduced trispectrum,
\begin{eqnarray}\label{ExtraTrispRed}
t^{l_1 l_2}_{l_3 l_4}(L)=\frac{\mathcal{T}^{l_1 l_2}_{l_3 l_4}(L)}{h_{l_1 l_2 L}h_{l_3 l_4 L}},
\end{eqnarray}
by analogy with the reduced bispectrum $b_{l_1 l_2 l_3}=B_{l_1 l_2 l_3}/h_{l_1 l_2 l_3}$, where $B_{l_1 l_2 l_3}$ represents the angle-averaged bispectrum. To prevent confusion, however, we refer to $t^{l_1 l_2}_{l_3 l_4}(L)$ as the `extra'-reduced trispectrum.

\subsection{Relationship between the primordial trispectrum and other probes}
As has been discussed in \cite{SefZal} the matter density perturbations are related to the primordial fluctuations by the Poisson equation via the expression
\begin{eqnarray}
\delta_{\mathbf{k}}(a)=M(k;a) \Phi_{\mathbf{k}},
\end{eqnarray}
where $a$ is the scale factor and $M(k;a)$ is given by
\begin{eqnarray}
M(k;a)=-\frac{3}{5}\frac{k^2 T(k)}{\Omega_m H_0^2}D_+(a),
\end{eqnarray}
where $T(k)$ is the transfer function, $D_+(a)$ is the growth factor in linear perturbation theory, $\Omega_m$ is the present value of the dark matter density and $H_0$ is the present value of the Hubble constant.
Therefore, the primordial contribution to the $n$- point connected correlation function of matter density perturbations at a given value of the scale factor is given by
\begin{eqnarray}
\langle \delta_{\mathbf{k}_1}(a) \delta_{\mathbf{k}_2}(a)\dots\delta_{\mathbf{k}_n} (a) \rangle_c =M(k_1;a)M(k_2;a)\dots M(k_n;a)\langle \Phi_{\mathbf{k}_1} \Phi_{\mathbf{k}_2}\dots\Phi_{\mathbf{k}_n}  \rangle_c.
\end{eqnarray}
 Possible probes of the matter density perturbations include galaxy surveys and the Lyman alpha forest, i.e. the sum of absorption lines from the Ly-$\alpha$ transition of the neutral hydrogen in the spectra of distant galaxies and quasars. There are three sources of non-Gaussianity in such surveys \cite{ReviewLFSS}: one primordial, one due to gravitational instability and the last due to nonlinear bias. $21$cm observations offer another probe of non-Gaussianity which are less subject to the unknown galaxy bias, especially at high redshift. However, uncertainties in the neutral fraction replaces the uncertainties in the bias in this case. There are also complications due to redshift space distortions arising from peculiar velocities. Despite these drawbacks, recent advances in this area suggest that probes of the matter density perturbations potentially represent a powerful tool to detect non-Gaussianity and possibly break the degeneracy implicit in trispectrum measurements using the CMB. The study of such data may involve using the full Legendre expansion of the primordial trispectrum as in equation~\eqref{expansion}. In the remainder of this chapter we proceed to investigate the CMB trispectrum. However, many of the results presented here are straightforwardly extended to alternative probes of non-Gaussianity as discussed here.

\subsection{Ideal Estimator}\label{sec:idealestimator}
Unfortunately the trispectrum signal, like the bispectrum, is too weak for us to measure individual multipoles directly. Therefore, in order to compare theory with observations it is necessary to use an estimator that sums over all multipoles. Estimators can be thought of as performing a least squares fit of the trispectrum predicted by theory, $\langle a_{l_1m_1}a_{l_2m_2}a_{l_3m_3}a_{l_4m_4}\rangle_c$, to the trispectrum obtained from observations. The trispectrum from observations is given by $(a_{l_1 m_1}^{\rm{obs}}a_{l_2 m_2}^{\rm{obs}}a_{l_3 m_3}^{\rm{obs}}a_{l_4 m_4}^{\rm{obs}})_c$ where we subtract the unconnected or Gaussian part, denoted $\rm{uc}$, from the four-point function, $a_{l_1 m_1}^{\rm{obs}}a_{l_2 m_2}^{\rm{obs}}a_{l_3 m_3}^{\rm{obs}}a_{l_4 m_4}^{\rm{obs}}$. This unconnected part is related to the observed
angular power spectrum $C_{l}^{\rm{obs}}$ by
\begin{align}
(a_{l_1 m_1}^{\rm{obs}}a_{l_2 m_2}^{\rm{obs}}a_{l_3 m_3}^{\rm{obs}}a_{l_4 m_4}^{\rm{obs}})_{\rm{uc}}=&(-1)^{m_1+m_3}C_{l_1}^{\rm{obs}}C_{l_3}^{\rm{obs}}\delta_{l_1, l_2}\delta_{m_1, -m_2}\delta_{l_3, l_4}\delta_{m_3, -m_4}\nonumber\\
+(-1)^{m_1+m_2}C_{l_1}^{\rm{obs}}C_{l_2}^{\rm{obs}}&\Big(\delta_{l_1, l_3}\delta_{m_1, -m_3}\delta_{l_2, l_4}\delta_{m_2, -m_4}
+\delta_{l_1, l_4}\delta_{m_1, -m_4}\delta_{l_2, l_3}\delta_{m_2, -m_3}\Big).
\end{align}
We define the estimator to be
\begin{eqnarray}\label{Estimator}
\mathcal{E}=\frac{1}{N_T}\sum_{l_i m_i}\frac{\langle a_{l_1m_1}a_{l_2m_2}a_{l_3m_3}a_{l_4m_4}\rangle_c \left(a_{l_1 m_1}^{\rm{obs}}a_{l_2 m_2}^{\rm{obs}}a_{l_3 m_3}^{\rm{obs}}a_{l_4 m_4}^{\rm{obs}}\right)_c}{C_{l_1}C_{l_2}C_{l_3}C_{l_4}}
\end{eqnarray}
where the normalisation factor $N_T$ is given by 
\begin{align}
N_{T}=&\sum_{l_i m_i}\frac{\langle a_{l_1m_1}a_{l_2m_2}a_{l_3m_3}a_{l_4m_4}\rangle_c \langle a_{l_1m_1}a_{l_2m_2}a_{l_3m_3}a_{l_4m_4}\rangle_c}{C_{l_1}C_{l_2}C_{l_3}C_{l_4}}. 
\end{align}
We may simplify this expression by using equation~\eqref{Ttot} to expand $N_{T}$ in the form
\begin{align}
N_{T}=&\sum_{l_i L L'}\sum_{m_i M M'} \frac{ 1      }{C_{l_1}C_{l_2}C_{l_3}C_{l_4}}(-1)^M \left( \begin{array}{ccc}
l_1 & l_2 & L \\
m_1 & m_2 & -M \end{array} \right) \left( \begin{array}{ccc}
l_3 & l_4 & L \\
m_3 & m_4 & M \end{array} \right)\nonumber\\
&\times  T^{l_1 l_2}_{l_3 l_4}(L)   (-1)^{M'} 
\left( \begin{array}{ccc}
l_1 & l_2 & L' \\
m_1 & m_2 & -M' \end{array} \right) \left( \begin{array}{ccc}
l_3 & l_4 & L' \\
m_3 & m_4 & M' \end{array} \right) T^{l_1 l_2}_{l_3 l_4}(L').\nonumber    
\end{align}
Now, using
\begin{eqnarray}\label{Orthog3}
\sum_{m_1 m_2} \left( \begin{array}{ccc}
l_1 & l_2 & L \\
m_1 & m_2 & -M \end{array} \right)  \left( \begin{array}{ccc}
l_1 & l_2 & L' \\
m_1 & m_2 & -M' \end{array} \right)&=&\frac{\delta_{L,L'} \delta_{M,M'} }{2L+1},\nonumber\\
\sum_M \sum_{m_3 m_4} \left( \begin{array}{ccc}
l_3 & l_4 & L \\
m_3 & m_4 & M \end{array} \right)  \left( \begin{array}{ccc}
l_3 & l_4 & L \\
m_3 & m_4 & M \end{array} \right)&=&1,
\end{eqnarray}
we find that
\begin{align}\label{eq:NormalT}
N_{T}=\sum_{l_i, L}\frac{T^{l_1 l_2}_{l_3 l_4}(L)T^{l_1 l_2}_{l_3 l_4}(L)}{(2L+1)C_{l_1}C_{l_2}C_{l_3}C_{l_4}}.
\end{align}
Thus, we may use equation~\eqref{eq:NormalT} to normalise the estimator~\eqref{Estimator}.
\par
We can expand $N_T$ in terms of the reduced trispectrum using
\begin{eqnarray}
N_T=12\sum_{l_i m_i}\frac{\mathcal{T}_{l_1 m_1 l_2 m_2 l_3 m_3 l_4 m_4}T_{l_1 m_1 l_2 m_2 l_3 m_3 l_4 m_4}}{C_{l_1}C_{l_2}C_{l_3}C_{l_4}}.
\end{eqnarray}
Then with the identity for the Wigner $6j$ symbol (see Appendix in \cite{Hu})
\begin{align}
\left\{ \begin{array}{ccc}
a & b & e \\
c & d & f \end{array}\right\}=\sum_{\alpha \beta \gamma}\sum_{\delta \epsilon \phi}(-1)^{e+f+\epsilon+\phi}& \left( \begin{array}{ccc}
a & b & e \\
\alpha & \beta & \epsilon \end{array} \right) \left( \begin{array}{ccc}
c & d & e \\
\gamma & \delta & -\epsilon \end{array} \right)\nonumber\\
&\times \left( \begin{array}{ccc}
a & d & f \\
\alpha & \delta & -\phi \end{array} \right) \left( \begin{array}{ccc}
c & b & f \\
\gamma & \beta & \phi \end{array} \right),
\end{align}
the identities \eqref{Orthog3} and relations for $P$ in \eqref{Prel1} and \eqref{Prel2} we find that
\begin{align}
N_T=12\sum_{l_i,L}\frac{\mathcal{T}^{l_1 l_2}_{l_3 l_4}(L)}{C_{l_1}C_{l_2}C_{l_3}C_{l_4}}\Big(\frac{P^{l_1 l_2}_{l_3 l_4}(L)}{2L+1}+\sum_{L'}&(-1)^{l_2+l_3}\left\{ \begin{array}{ccc}
l_1 & l_2 & L \\
l_4 & l_3 & L' \end{array}\right\} P^{l_1 l_3}_{l_2 l_4}(L')\nonumber\\
&+\sum_{L'}(-1)^{L+L'}\left\{ \begin{array}{ccc}
l_1 & l_2 & L \\
l_3 & l_4 & L' \end{array}\right\} P^{l_1 l_4}_{l_3 l_2}(L')\Big).\nonumber\\
\end{align}
Due to the presence of the $6$j symbols the calculation of $N_T$ is computationally very expensive in general.
\par
As is clear from the earlier discussion, assuming isotropy for a given theoretical model, we need only calculate the reduced trispectrum, $\mathcal{T}^{l_1 l_2}_{l_3 l_4}(L)$,  rather than the more challenging full trispectrum $\langle a_{l_1m_1}a_{l_2m_2}a_{l_3m_3}a_{l_4m_4}\rangle_c$. 
\par
The estimator in~\eqref{Estimator} naturally defines a correlator for testing whether two competing trispectra could be differentiated by an ideal experiment. Replacing the observed trispectrum with one calculated from a competing theory we have,
\begin{eqnarray}
\mathcal{C}(T,T')&=&\frac{1}{N_T}\sum_{l_i, m_i}\frac{\langle a_{l_1m_1}a_{l_2m_2}a_{l_3m_3}a_{l_4m_4}\rangle_c \langle a_{l_1m_1}'a_{l_2m_2}' a_{l_3m_3}' a_{l_4m_4}'\rangle_c }{C_{l_1}C_{l_2}C_{l_3}C_{l_4}}\nonumber\\
&=&\frac{1}{N_T}\sum_{l_i, L}\frac{T^{l_1 l_2}_{l_3 l_4}(L)T'^{l_1 l_2}_{l_3 l_4}(L)}{(2L+1)C_{l_1}C_{l_2}C_{l_3}C_{l_4}},
\end{eqnarray}
where now the normalisation $N_T$ is defined as follows,
\begin{eqnarray}
N_T=\sqrt{\sum_{l_i, L}\frac{T^{l_1 l_2}_{l_3 l_4}(L)T^{l_1 l_2}_{l_3 l_4}(L)}{(2L+1)C_{l_1}C_{l_2}C_{l_3}C_{l_4}}}\sqrt{\sum_{l_i, L}\frac{T'^{l_1 l_2}_{l_3 l_4}(L)T'^{l_1 l_2}_{l_3 l_4}(L)}{(2L+1)C'_{l_1}C'_{l_2}C'_{l_3}C'_{l_4}}}.
\end{eqnarray}
An alternative correlator between two trispectra, which is easier to solve numerically, is found by replacing the trispectra by the respective reduced trispectra in the above definitions. Therefore, when comparing two trispectra we shall use this latter definition, $\mathcal{C(T,T')}$. The exact relation between the two correlators can be deduced from Appendix B in ref.~\cite{Hu}.

\subsection{General Estimator}
The above estimator is applicable for general trispectra in the limit where non-Gaussianity is small and the observed map is free of instrument noise and foreground contamination. Of course, this is an idealised case and we need to consider taking into account the effect of sky cuts and inhomogeneous noise. Here we follow the approach of~\cite{Amendola} (an approach that is further elucidated in~\cite{Babichopt} and~\cite{Komatsu2}). 
\par
When non-Gaussianity is weak we can exploit the multivariate Edgeworth expansion~\cite{Amendola} around the Gaussian probability distribution function (PDF), $P^G(a)$, i.e.
\begin{align}
P(a)=&\Bigg[1-\sum_{l_i m_i}\langle a_{l_1 m_1}a_{l_2 m_2}a_{l_3 m_3}\rangle \frac{\partial}{\partial a_{l_1 m_1}}	\frac{\partial}{\partial a_{l_2 m_2}}	\frac{\partial}{\partial a_{l_3 m_3}}	\nonumber\\
&+\sum_{l_i m_i}\langle a_{l_1 m_1}a_{l_2 m_2}a_{l_3 m_3}a_{l_4 m_4}\rangle_c \frac{\partial}{\partial a_{l_1 m_1}}	\frac{\partial}{\partial a_{l_2 m_2}}	\frac{\partial}{\partial a_{l_3 m_3}}\frac{\partial}{\partial a_{l_4 m_4}}+\dots	\Bigg]P^G(a),
\end{align}
where the Gaussian PDF is given by
\begin{eqnarray}
P^G(a)=\frac{e^{-\frac{1}{2}\sum_{l m}\sum_{l' m'}a_{l m} (C^{-1})_{l m, l' m'} a_{l' m'}
 }}{(2\pi)^{N/2}|C|^{1/2}},
\end{eqnarray}
with $C_{l m, l' m'}=\langle a_{l m} a_{l' m'}\rangle$ and $N$ the number of $l$ and $m$. Maximising over the three-point correlator results in the optimal bispectrum estimator. Here we will ignore this term (setting it to zero for convenience) and concentrate on the four-point correlator. We find
\begin{align}
\frac{P(a)}{P^G(a)}=&\Bigg[1+\sum_{l_i m_i}\langle a_{l_1 m_1}a_{l_2 m_2}a_{l_3 m_3}a_{l_4 m_4}\rangle_c\Bigg((C^{-1} a^{ })_{l_1 m_1}(C^{-1} a^{ })_{l_2 m_2} (C^{-1} a^{ })_{l_3 m_3} (C^{-1} a^{ })_{l_4 m_4}\nonumber\\
&-6(C^{-1})_{l_2 m_2,l_1 m_1} (C^{-1} a^{ })_{l_3 m_3} (C^{-1} a^{ })_{l_4 m_4}+3 (C^{-1})_{l_1 m_1,l_2 m_2}(C^{-1})_{l_3 m_3,l_4 m_4}		\Bigg)	\Bigg],
\end{align}
where $(C^{-1} a)_{l m}=\sum_{l' m'}C^{-1}_{l' m', l m}a_{l' m'}$.
Parametrising the size of the trispectrum by $\mathcal{E}$ we wish to maximise the PDF with respect to this. We assume that $ (a_{l_1 m_1}a_{l_2 m_2}a_{l_3 m_3}a_{l_4 m_4})_c \propto \mathcal{E}$ such that the second term is proportional to $\mathcal{E}$.\footnote{The notation $(a_{l_1 m_1}a_{l_2 m_2}a_{l_3 m_3}a_{l_4 m_4})_c$ is used to denote $a_{l_1 m_1}a_{l_2 m_2}a_{l_3 m_3}a_{l_4 m_4}-6C_{l_1 m_1, l_2 m_2}a_{l_3 m_3}a_{l_4 m_4}+3C_{l_1 m_1, l_2 m_2}C_{l_3 m_3, l_4 m_4}$.} Maximising the PDF means that we wish to set $(d P/d\mathcal{E})=0$, such that the Taylor expansion about $\mathcal{E}=0$ reads
\begin{align}
P(a)= \Bigg[1+\frac{d (P/P^G)}{d\mathcal{E}}\mathcal{E} +\frac{1}{2}\frac{d^2  (P/P^G)}{d\mathcal{E}^2} \mathcal{E}^2+\dots\Bigg]P^G(a) \approx  \Bigg[1+\frac{1}{2}\frac{d^2  (P/P^G)}{d\mathcal{E}^2} \mathcal{E}^2 \Bigg]P^G(a),
\end{align}
Since 
\begin{align*}
\frac{d^2 P}{d\mathcal{E}^2}\propto &2 \sum_{l_i m_i}  \langle a_{l_1m_1}a_{l_2m_2}a_{l_3m_3}a_{l_4m_4}\rangle_c  (C^{-1})_{l_1 m_1,l_1' m_1'} (C^{-1})_{l_2 m_2,l_2' m_2'}  \nonumber\\
&\times (C^{-1})_{l_3 m_3,l_3' m_3'} (C^{-1})_{l_4 m_4,l_4' m_4'}( a_{l_1' m_1'}a_{l_2' m_2'}a_{l_3' m_3'}a_{l_4' m_4'})_c,
\end{align*}
we find that the estimator is maximised by setting (with appropriate choice of proportionality constant)
\begin{align}\label{eq:Estimator2}
\mathcal{E}=&\frac{1}{\tilde{N}}\sum_{l_i m_i}\langle a_{l_1 m_1}a_{l_2 m_2}a_{l_3 m_3}a_{l_4 m_4}\rangle_c\Bigg((C^{-1} a^{ })_{l_1 m_1}(C^{-1} a^{ })_{l_2 m_2} (C^{-1} a^{ })_{l_3 m_3} (C^{-1} a^{ })_{l_4 m_4}\nonumber\\
&-6(C^{-1})_{l_1 m_1,l_2 m_2} (C^{-1} a)_{l_3 m_3} (C^{-1} a^{ })_{l_4 m_4}+3  (C^{-1})_{l_1 m_1,l_2 m_2}(C^{-1})_{l_3 m_3,l_4 m_4}		\Bigg),
\end{align}
where 
\begin{align*}
\tilde{N}=& \sum_{l_i m_i}  \langle a_{l_1m_1}a_{l_2m_2}a_{l_3m_3}a_{l_4m_4}\rangle_c  (C^{-1})_{l_1 m_1,l_1' m_1'} (C^{-1})_{l_2 m_2,l_2' m_2'} \nonumber\\
&\times(C^{-1})_{l_3 m_3,l_3' m_3'} (C^{-1})_{l_4 m_4,l_4' m_4'} ( a_{l_1' m_1'}a_{l_2' m_2'}a_{l_3' m_3'}a_{l_4' m_4'})_c.
\end{align*}
We note that, in general, the covariance matrix $C$ is now non-diagonal due to mode-mode coupling introduced by the mask and anisotropic noise. Due to the breaking of isotropy extra terms have been added in order to maintain the optimality of the estimator. The optimal estimator, in the case that the covariance matrix is diagonal, reads
\begin{eqnarray}
\mathcal{E}&=&\frac{1}{N_T}\sum_{l_i m_i}  \frac{\langle a_{l_1m_1}a_{l_2m_2}a_{l_3m_3}a_{l_4m_4}\rangle_c }{C_{l_1}C_{l_2}C_{l_3}C_{l_4}}\Big[a^{\rm{obs}}_{l_1 m_1}a^{\rm{obs}}_{l_2 m_2}a^{\rm{obs}}_{l_3 m_3}a^{\rm{obs}}_{l_4 m_4}\\
&&-6 (-1)^{m_1}C_{l_1}\delta_{l_1 l_2}\delta_{m_1 -m_2}a^{\rm{obs}}_{l_3 m_3}a^{\rm{obs}}_{l_4 m_4}+3(-1)^{m_1+m_2}\delta_{l_1 l_2}\delta_{m_1 -m_2} \delta_{l_3 l_4}\delta_{m_3 -m_4}C_{l_1}C_{l_3}\Big],\nonumber
\end{eqnarray}
where $N_T$ is given by equation~\eqref{eq:NormalT}. We note also that the average of this estimator is
\begin{eqnarray}
\langle\mathcal{E}\rangle&=&\frac{1}{N_T}\sum_{l_i m_i}  \frac{\langle a_{l_1m_1}a_{l_2m_2}a_{l_3m_3}a_{l_4m_4}\rangle_c \langle a^{\rm{obs}}_{l_1 m_1}a^{\rm{obs}}_{l_2 m_2}a^{\rm{obs}}_{l_3 m_3}a^{\rm{obs}}_{l_4 m_4}\rangle_c}{C_{l_1}C_{l_2}C_{l_3}C_{l_4}},
\end{eqnarray}
as expected.

In the remainder of this chapter we shall refer to the ideal estimator unless otherwise stated. However, this formula is important for the general implementation of the formalisms introduced here.

\subsection{Kurtosis as a measure of non-Gaussianity}
As an aside,  we note that the use of non-optimal estimators may also provide useful information, e.g. as a reality check on these complex calculations. The kurtosis of the one point temperature distribution offers such an estimator. The kurtosis, $g_2$, is defined as
\begin{eqnarray}\label{Kurtosis1} 
g_2=\frac{\Bigg\langle\left(\frac{\Delta T}{T}(\hat{n}) \right)^4\Bigg\rangle}{\left(\Bigg\langle \left(\frac{\Delta T}{T}(\hat{n}) \right)^2\Bigg\rangle\right)^2}-3.
\end{eqnarray}
As was shown in Section~\ref{sec:nonGaussBgd} the kurtosis may be written in the following form
\begin{eqnarray}\label{Kurtosis2}
g_2=\frac{48\pi\sum_{l_i,L} h_{l_1 l_2 L}^2 h_{l_3 l_4 L}^2t^{l_1 l_2}_{l_3 l_4}(L)/(2L+1)}{\left(\sum_l (2 l+1)C_l\right)^2}.
\end{eqnarray}
The calculation of this quantity is relatively straightforward compared to the full estimator due to the absence of Wigner $6j$ symbols in the expression.

\section{The Shape of Primordial Trispectra}\label{sec:III}

\subsection{Shape function}
It is known from CMB observations that the power spectrum is nearly scale-invariant. Analysis of the bispectrum is performed using the shape function, which is a scale invariant form of the bispectrum. To parallel this analysis we wish to write a scale invariant form of the trispectrum (or in particular the trispectrum modes). Therefore, we need to eliminate a $k^9$ scaling. Motivated by~\eqref{TrispRed2}, we define this shape by
\begin{eqnarray}\label{eq:Shape}
S_T(k_1,k_2,k_3,k_4,K)&=&\frac{(k_1 k_2 k_3 k_4)^2 K}{\tilde{\Delta}_{\Phi}^3 N} T_{\Phi,0}(k_1,k_2,k_3,k_4;K),
\end{eqnarray}
where $N$ is an appropriate normalisation factor. For clarity in what follows we note that we shall use the symbol $S_{\mathcal{T}}$ when referring to the shape induced by the reduced primordial trispectrum. Of course, this choice of the shape function is not unique. Another choice of shape function is
\begin{eqnarray}\label{Shape2}
\tilde{S}_T(k_1,k_2,k_3,k_4,K)&=&\frac{(k_1 k_2 k_3 k_4)^{9/4}}{\tilde{\Delta}_{\Phi}^3 N} T_{\Phi,0}(k_1,k_2,k_3,k_4;K),
\end{eqnarray}
which has the advantage of remaining independent of the diagonal $K$ if the underlying trispectrum has this property. Such a class of models are discussed further in Section~\ref{sec:SpecialClass}. Nonetheless we proceed with $S_{T}$ as our choice of shape function in this chapter, leaving further investigation of this issue to a future publication~\cite{Regan2}. We should also notice that we have only included the zeroth mode of the Legendre expansion as indicated by~\eqref{TrispRed2}. This zeroth mode incorporates almost all of the trispectrum shapes studied in the literature. However, for more general shapes of non-Gaussianity, as discussed in Section \ref{sec:II}, the full Legendre expansion described by equation~\eqref{expansion} may be required. In such a case the analysis outlined here can be applied mode-by-mode. Due to orthogonality of the Legendre modes, extending the study is a trivial task.
\par
 If we rewrite the reduced CMB trispectrum in terms of the shape function, $S_{\mathcal{T}}$, we have
\begin{eqnarray} 
\mathcal{T}^{l_1 l_2}_{l_3 l_4}(L)&=&N  h_{l_1 l_2 L} h_{l_3 l_4 L} \left(\frac{2}{\pi}\right)^5 \int d\mathcal{V}_k  S_{\mathcal{T}} (k_1,k_2,k_3,k_4,K)K\nonumber\\
&&\times \Delta_{l_1}(k_1)\Delta_{l_2}(k_2)\Delta_{l_3}(k_3)\Delta_{l_4}(k_4) I^G_{l_1 l_2 l_3 l_4 L}(k_1,k_2,k_3,k_4,K),
\end{eqnarray}
where the integral $I^G$ is given by
\begin{align}
 I^G_{l_1 l_2 l_3 l_4 L}(k_1,k_2,k_3,k_4,K)&=\nonumber\\
 \int r_1^2 r_2^2& dr_1 dr_2 j_L(K r_1) j_L (K r_2) j_{l_1}(k_1 r_1) j_{l_2}(k_2 r_1) j_{l_3}(k_3 r_2) j_{l_4}(k_4 r_2)
 \end{align}
and $d \mathcal{V}_k$ corresponds to the area inside the region $k_i,K/2\in[0,k_{\rm{max}}]$ allowed by the triangle conditions. Therefore the shape function is the signal that is evolved via the transfer functions to give the trispectrum today. Essentially, $I^G$ acts like a window function on all the shapes as it projects from $k$ to $l-$space, that is, it will tend to smear out their sharper distinguishing features. This means that the shape function $S_{\mathcal{T}}$, especially in the scale invariant case, can be thought of as the primordial counterpart of the reduced CMB trispectrum $\mathcal{T}^{l_1 l_2}_{l_3 l_4}(L)$ before projection.

\subsection{Shape correlators}
We wish to construct a primordial shape correlator that predicts the value of the CMB correlator $\mathcal{C(T,T')}$. To this end we should consider something of the form
\begin{align}
F(S_{\mathcal{T}},S_{\mathcal{T}'})=\int d\mathcal{V}_k  S_{\mathcal{T}}(k_1,k_2,k_3,k_4,K) S_{\mathcal{T}'}(k_1,k_2,k_3,k_4,K) \omega(k_1,k_2,k_3,k_4,K),
\end{align}
where $\omega$ is an appropriate weight function. With this choice of weight the primordial shape correlator then takes the form
\begin{eqnarray}
\overline{\mathcal{C}}(S_{\mathcal{T}},S_{\mathcal{T}'})=\frac{F(S_{\mathcal{T}},S_{\mathcal{T}'})}{\sqrt{F(S_{\mathcal{T}},S_{\mathcal{T}})F(S_{\mathcal{T}'},S_{\mathcal{T}'})}}.
\end{eqnarray}
\par
The question now is what weight function should we choose? Our goal is to choose $S^2 \omega$ in $k$- space such that it produces the same scaling as the estimator $T^2/((2L+1)C^4)$ in $l$- space. Let's consider the simplest case where $k=k_1=k_2=k_3=k_4=K$ and $l=l_1=l_2=l_3=l_4=L$. For primordial trispectra which are scale invariant, then
\begin{eqnarray}
S_{\mathcal{T}}^2(k,k,k,k,k)\,\omega(k,k,k,k,k)\propto \omega(k,k,k,k,k).
\end{eqnarray}
If we work in the large angle approximation, and assume $l+1\approx l$, then we know $C_l\propto l^{-2}$ and from the analytic solution for the local model which we will describe below (see equations~\eqref{LocA} and~\eqref{LocB}) we have
\begin{eqnarray}
T^{l l}_{l l}(l)\propto h_{l l l}^2 \frac{1}{l^6}.
\end{eqnarray}
Now $h_{ l l l}\propto l^{3/2} \left( \begin{array}{ccc}
l & l & l \\
0 & 0 & 0 \end{array} \right)$ and the Wigner $3j$ symbol has an exact solution for which
\begin{eqnarray}
\left( \begin{array}{ccc}
l & l & l \\
0 & 0 & 0 \end{array} \right)\approx (-1)^{3l/2}\frac{1}{\sqrt{3 l+1}}\sqrt{\frac{l!^3}{3l!}}\frac{(3l/2)!}{(l/2)!^3}\approx (-1)^{3l/2}\sqrt{\frac{2}{\sqrt{3}\pi}}\frac{1}{l}.
\end{eqnarray}
Therefore $T^{l l}_{l l}(l)\propto  l^{-5}$ and so
\begin{eqnarray}
\frac{T^{l l}_{l l}(l)^2}{(2l+1)C_l^4 }\propto l^{-3}.
\end{eqnarray}
Hence we find that we should choose a weight function $\omega(k,k,k,k,k)\propto k^{-3}$. The particular choice of $\omega$ may significantly improve forecasting accuracy - by, for instance, using a phenomenological window function to incorporate damping due to photon diffusion or smoothing due to projection from $k$- to $l$- space, but it does not impact important qualitative insights. A specific choice of weight function, motivated by the choice of weight function for the bispectrum, is the following
\begin{eqnarray}\label{Weightk}
w(k_1,k_2,k_3,k_4,K)=\frac{K}{(k_1+k_2+K)^2(k_3+k_4+K)^2}.
\end{eqnarray}

\subsection{Shape Decomposition}
Given the strong observational limits on the scalar tilt we expect all shape functions to exhibit behaviour close to scale-invariance, so that $S_{\mathcal{T}}(k_1,k_2,k_3,k_4,K)$ will depend only weakly on the overall magnitude of the summed wavenumbers. Here we choose to parametrise the magnitude of the wavenumbers with the quantity
\begin{eqnarray} 
k=\frac{1}{2}(k_1+k_2+K).
\end{eqnarray}
$k$ is the semi-perimeter of the triangle formed by the vectors $\mathbf{k}_1,\mathbf{k}_2,\mathbf{K}$.
Due to the scaling behaviour the form of the shape function on a cross-section is essentially independent of $k$, so that 
\begin{eqnarray}
S_{\mathcal{T}}(k_1,k_2,k_3,k_4,K)=f(k)\overline{S}_{\mathcal{T}}(\hat{k}_1,\hat{k}_2,\hat{k}_3,\hat{k}_4,\hat{K}),
\end{eqnarray}
where $\hat{k}_i=k_i/k$ and $\hat{K}=K/k$. Since we are restricted to the region where the wavenumbers $(k_1,k_2,K)$ and $(k_3,k_4,K)$ form triangles by momentum conservation, we will reparametrise the allowed region to separate out the overall scale $k$ from the behaviour on a cross-sectional slice. This four-dimensional slice is spanned by the remaining coordinates. Concentrating on each triangle individually, we reparametrise in a similar fashion to the analysis done in~\cite{Ferg2}. For triangle $(k_1,k_2,K)$ we have
\begin{eqnarray}
K&=&k(1-\beta),\nonumber\\
k_1&=&\frac{k}{2}(1+\alpha+\beta),\nonumber\\
k_2&=&\frac{k}{2}(1-\alpha+\beta),
\end{eqnarray}
while for triangle $(k_3,k_4,K)$ we have
\begin{eqnarray}
K&=&\epsilon k(1-\delta),\nonumber\\
k_3&=&\frac{\epsilon k}{2}(1+\gamma+\delta),\nonumber\\
k_4&=&\frac{\epsilon k}{2}(1-\gamma+\delta),
\end{eqnarray}
where $\epsilon$ parametrises the ratio of the perimeters of the two triangles, i.e. $\epsilon=\frac{k_3+k_4+K}{k_1+k_2+K}$. We consider $1\leq\epsilon<\infty$. The different expressions for $K$ imply that
\begin{eqnarray}
1-\beta=\epsilon (1-\delta).
\end{eqnarray}
The conditions for triangle $(k_1,k_2,K)$ that $0\leq k_1,k_2,K\leq k$ imply that $0\leq \beta \leq 1$ and $-(1-\beta)\leq \alpha \leq 1-\beta$, while the conditions for triangle $(k_3,k_4,K)$ that $0\leq k_1,k_2,K\leq \epsilon k$, along with the relationship between $\delta$ and $\beta$ and the requirement that $\epsilon\geq 1$, imply that 
$-(1-\beta)/\epsilon \leq\gamma\leq (1-\beta)/\epsilon$. In summary, we have the following domains,
\begin{align}
&0\leq k<\infty,\quad  1\leq\epsilon<\infty,\quad 0\leq \beta \leq 1,\quad -(1-\beta)\leq \alpha \leq 1-\beta,\nonumber\\
& -\frac{1-\beta}{\epsilon} \leq\gamma\leq \frac{1-\beta}{\epsilon}.
\end{align}
With this parametrisation we can re-write the shape function and the volume element respectively as
\begin{align}
S_{\mathcal{T}}(k_1,k_2,k_3,k_4,K)=f(k)\overline{S}_{\mathcal{T}}(\alpha,\beta,\gamma,\epsilon),\quad d\mathcal{V}_{k}=d k_1 d k_2 dk_3 dk_4 dK= \epsilon k^4 dk d\alpha d\beta d\gamma d\epsilon.
\end{align}
In order to represent the shape function graphically we can choose fixed values of $\epsilon$ in which case the shape $\overline{S}$ becomes three dimensional. The particular three dimensional domain is shown in Figure~\ref{fig:shape}. From the image we see how the particular triangles created by the wavenumbers generate the three dimensional slice for each $\epsilon$. We can envisage  the four-dimensional shape by
imagining an orthogonal direction for $\epsilon$ out of the page, along which are located increasingly squeezed rectangular pyramids. 

\begin{figure}[htp]
\centering 
\includegraphics[width=172mm]{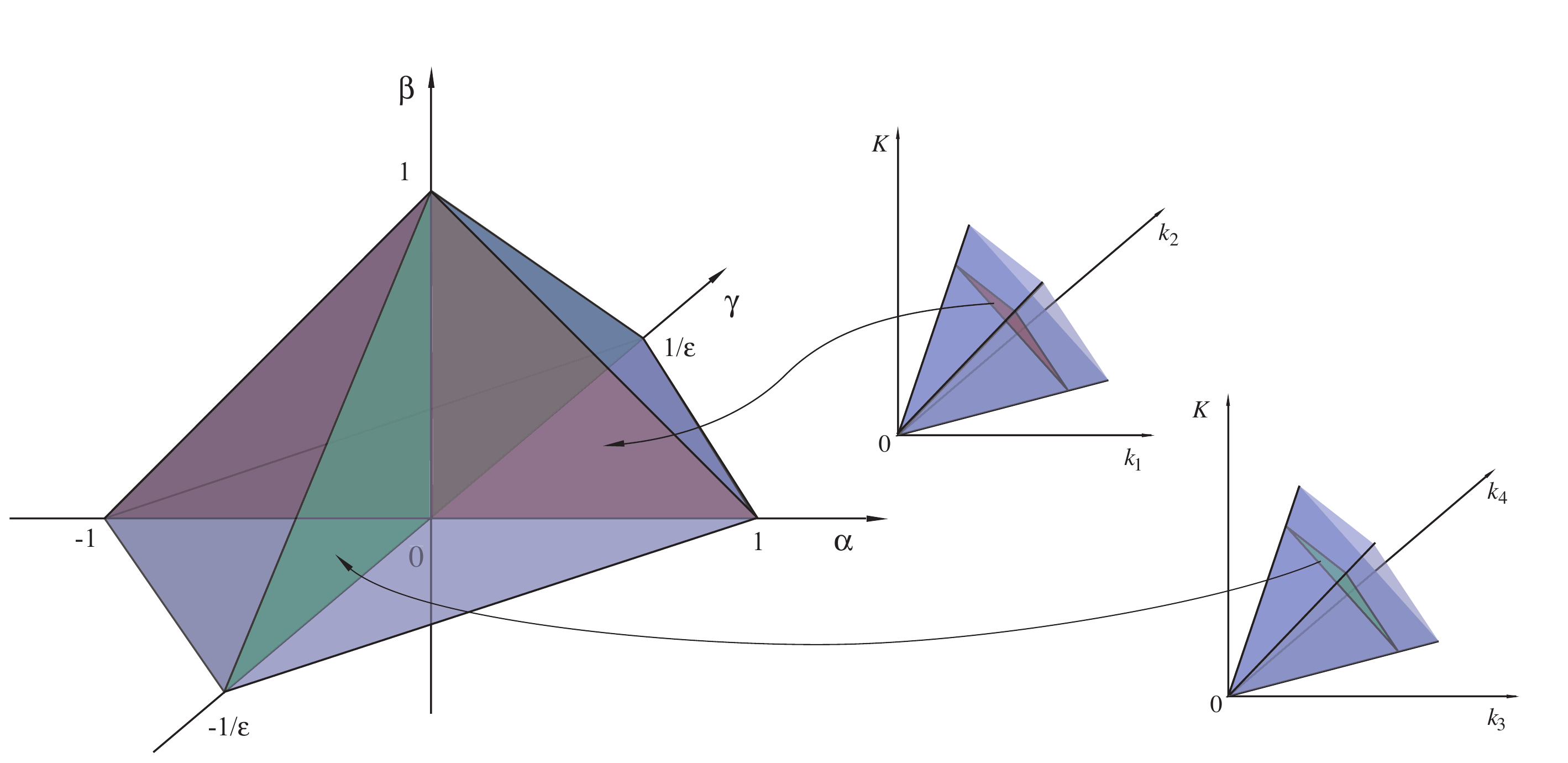}
\caption{Three-dimensional shape function domain for a fixed value of $\epsilon$, i.e. for a particular ratio of the perimeters of the triangles created by the wavenumbers $(k_1,k_2,K)$ and $(k_3,k_4,K)$ respectively. Note that the triangle conditions on these two wavenumber sets restrict them to the two tetrahedral domains illustrated (right), slices through which are mapped as shown into the full domain, a rectangular pyramid (left).}
\label{fig:shape}
\end{figure}

\section{Separable Shapes}\label{sec:IV}

\subsection{Examples: Local, equilateral and constant models}
The local model is given by the reduced primordial trispectrum 
\begin{eqnarray}
\mathcal{T}^{\rm{loc}}_{\Phi}(\mathbf{k}_1,\mathbf{k}_2,\mathbf{k}_3,\mathbf{k}_4;\mathbf{K})=\mathcal{T}^{\rm{loc}}_{\Phi A}(\mathbf{k}_1,\mathbf{k}_2,\mathbf{k}_3,\mathbf{k}_4;\mathbf{K})+\mathcal{T}^{\rm{loc}}_{ \Phi B} (\mathbf{k}_1,\mathbf{k}_2,\mathbf{k}_3,\mathbf{k}_4;\mathbf{K}),
\end{eqnarray}
where
\begin{eqnarray}
\mathcal{T}^{\rm{loc}}_{\Phi A}(\mathbf{k}_1,\mathbf{k}_2,\mathbf{k}_3,\mathbf{k}_4;\mathbf{K})&=&\frac{25}{9}\tau_{NL} P_{\Phi}(K)P_{\Phi}(k_1)P_{\Phi}(k_3),\\
\mathcal{T}^{\rm{loc}}_{ \Phi B} (\mathbf{k}_1,\mathbf{k}_2,\mathbf{k}_3,\mathbf{k}_4;\mathbf{K})&=&g_{NL}\left[P_{\Phi}(k_2)P_{\Phi}(k_3)P_{\Phi}(k_4)+P_{\Phi}(k_1)P_{\Phi}(k_2)P_{\Phi}(k_4)\right].
\end{eqnarray}
For single field inflation we have $\tau_{NL}=(\frac{6}{5}f_{NL})^2$. This relationship breaks down for multifield inflation (see \cite{Byrnes}). 
We can see clearly here that the local trispectrum is independent of the angle $\theta_4$, i.e. the zeroth mode of the local trispectrum is exactly the full local trispectrum. The primordial shapes for each of these expressions may be shown visually using the prescription described in the previous section and they are shown in Figures~\ref{fig:LocA} and~\ref{fig:LocB}. As expected for the local model the signal peaks in the corners. However, as is easily observable the `peaking' behaviour is somewhat orthogonal between the two models. Working in the Sachs-Wolfe approximation, where we replace the transfer function with a Bessel function,
\begin{eqnarray}
\Delta_l(k)=\frac{1}{3}j_l((\tau_0-\tau_{\rm{dec}})k),
\end{eqnarray}
the integral for the reduced trispectrum can be expressed in closed form. Setting $P_{\Phi}(k)=\tilde{\Delta}_{\Phi}k^{-3}$ we find
\begin{align}\label{LocA}
\mathcal{T}^{l_1 l_2\rm{loc}}_{l_3 l_4, A}(L)&=\frac{25\tau_{NL}}{9} \frac{\tilde{\Delta}_{\Phi}^3}{3^4} \left(\frac{2}{\pi}\right)^5 h_{l_1 l_2 L}h_{l_3 l_4 L}\int r_1^2 d r_1 r_2^2 d r_2 (K k_1 k_3)^{-1}dK dk_1
dk_3 I_{l_2}(2,r_1) I_{l_4}(2,r_1)\nonumber\\
&=\frac{25\tau_{NL}}{9}  \frac{\tilde{\Delta}_{\Phi}^3}{3^4} \left(\frac{2}{\pi}\right)^5 h_{l_1 l_2 L}h_{l_3 l_4 L} \left( \frac{\pi}{2}\right)^2 I_L(-1,1) I_{l_1}(-1,1) I_{l_3}(-1,1)\nonumber\\
&=\frac{25\tau_{NL}}{36} \pi^2\frac{\tilde{\Delta}_{\Phi}^3}{3^4} \left(\frac{2}{\pi}\right)^5 h_{l_1 l_2 L}h_{l_3 l_4 L} \left( \frac{1}{2 L(L+1)2 l_1(l_1+1)2 l_3(l_3+1)} \right),
\end{align}
where
\begin{eqnarray}
I_l (p,x)=\int k^p dk j_l(k)j_l(x k),
\end{eqnarray}
and we have used
\begin{eqnarray}\label{IdentitySpecial}
I_l(2,r)&=& \frac{\pi}{2 r^2}\delta(r-1) \\
I_l(-1,1)&=&\frac{1}{2l(l+1)}.
\end{eqnarray}
Similarly,
\begin{align}\label{LocB}
\mathcal{T}^{l_1 l_2 \rm{loc}}_{l_3 l_4, B}(L)=&
 g_{NL}\frac{\pi^2}{4}\frac{\tilde{\Delta}_{\Phi}^3}{3^4} \left(\frac{2}{\pi}\right)^5 h_{l_1 l_2 L}h_{l_3 l_4 L} \nonumber\\
 &\times\left( \frac{1}{2 l_2(l_2+1)2 l_3(l_3+1)2 l_4(l_4+1)}+(l_1\leftrightarrow l_3)\right).
\end{align}
\begin{figure}[htp]
\centering 
\includegraphics[width=120mm]{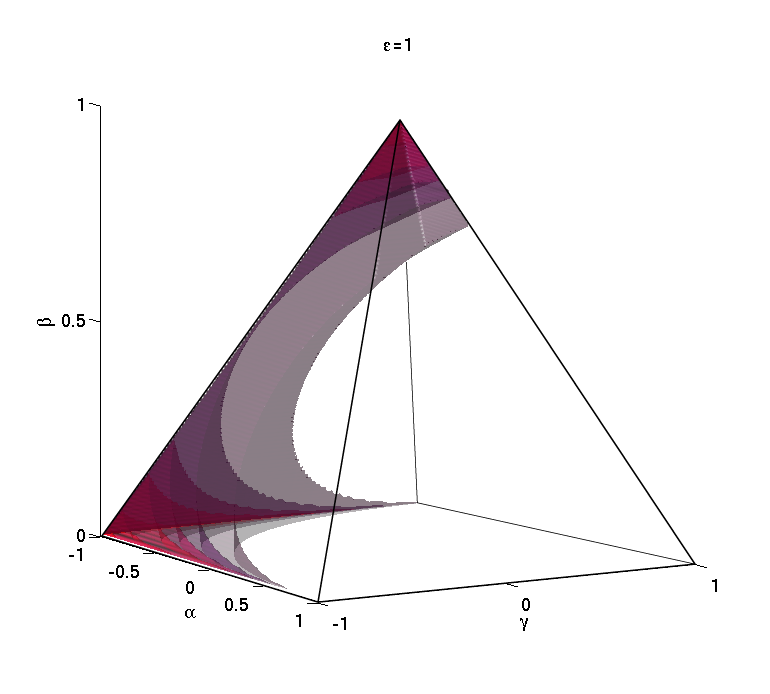}
\includegraphics[width=80mm]{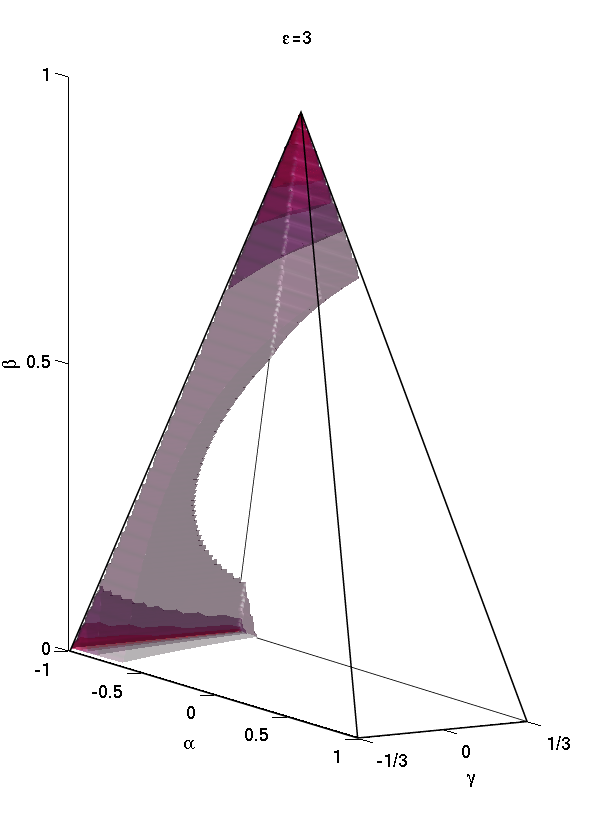}
\includegraphics[width=62mm]{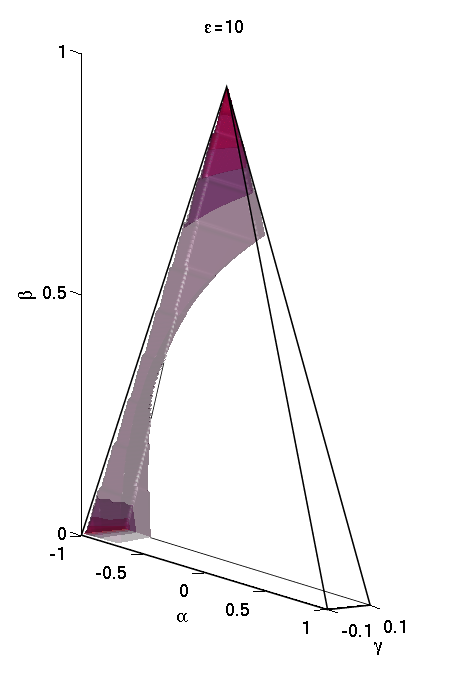}
\caption{Local $A$ model~\eqref{LocA}. The peak as $\beta\rightarrow 1$ corresponds to $K\rightarrow 0$, i.e. the `doubly-squeezed' limit. The other peak corresponds to $k_1\rightarrow 0, k_3\rightarrow (\epsilon-1)k$. As $\epsilon$ rises above unity (i.e. for triangle $(k_3,k_4,K)$ bigger than triangle $(k_1,k_2,K)$) we expect this peak to become suppressed as observed.}
\label{fig:LocA}
\end{figure}

\begin{figure}[htp]
\centering 
\includegraphics[width=120mm]{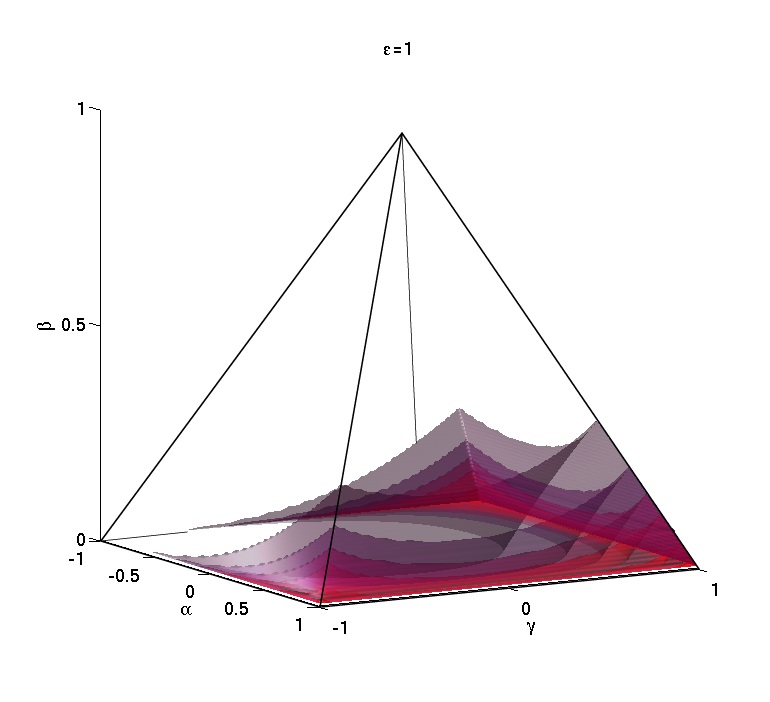}
\includegraphics[width=80mm]{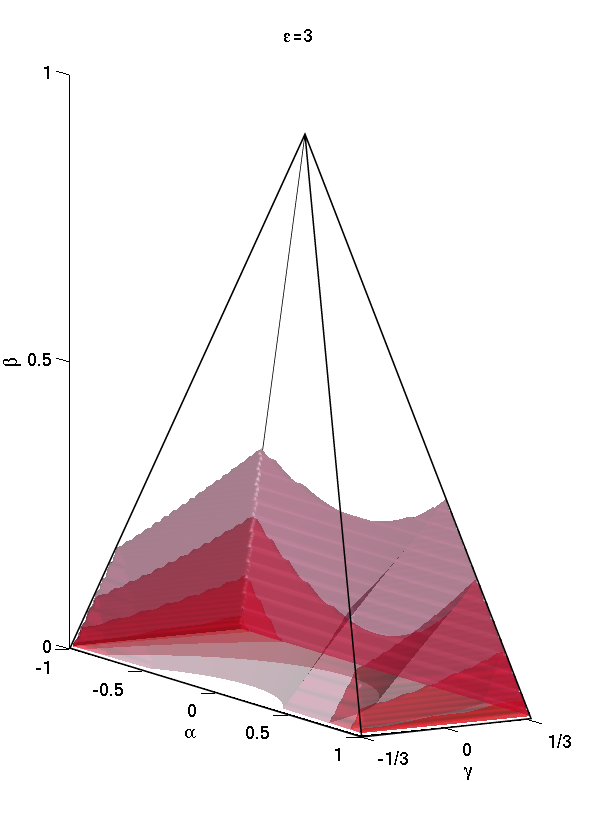}
\includegraphics[width=70mm]{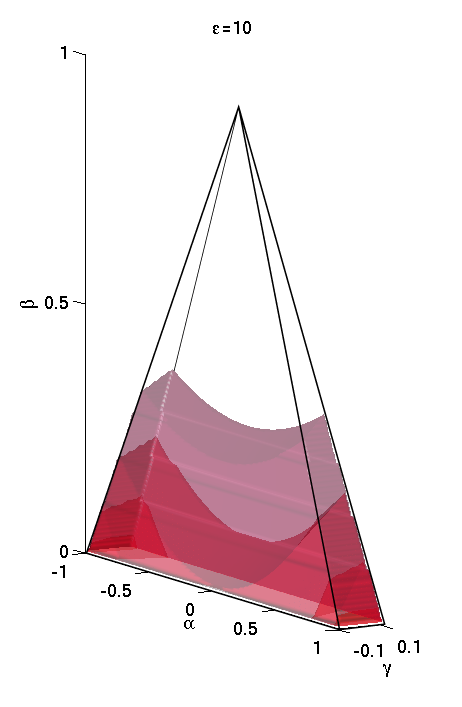}
\caption{Local $B$ model~\eqref{LocB}. The peaks at $\alpha=1$ correspond to $k_2\rightarrow 0$, while the peaks at $\alpha=-1$ correspond to $k_1\rightarrow 0$. For $\epsilon=1$ (i.e. equal triangle sizes $(k_1,k_2,K)$ and $(k_3,k_4,K)$) we see a more confined peaking at $\gamma=-1,\gamma=1$, i.e. $k_3\rightarrow 0, k_4\rightarrow 0$ respectively. We observe the peaking of the local $B$ model to be somewhat orthogonal to that of the local $A$ model.}
\label{fig:LocB}
\end{figure}

Next, we propose a $constant$ model for the primordial trispectrum, analogous to the 
simplest model for the bispectrum. This is given by
\begin{eqnarray}\label{Constant}
\frac{1}{\tilde{\Delta}_{\Phi}^3 N}(k_1 k_2 k_3 k_4)^2 K\mathcal{T}_{\Phi,0}(k_1,k_2,k_3,k_4;{K})=S_{\mathcal{T}}(k_1,k_2,k_3,k_4,K)=1,
\end{eqnarray}
with $N$ the normalisation factor of equation~\eqref{Constant} and the choice $\tilde{\Delta}_{\Phi}^3$ motivated by comparison to the local model.
Again, as for the local model, the primordial trispectrum is already a zero mode quantity with respect to angle $\theta_4$, i.e. $T=T_{,0}$. Using the Sachs Wolfe approximation, the integral~\eqref{TrispRed2} can be written as
\begin{align}
\mathcal{T}^{l_1 l_2\rm{const}}_{l_3 l_4}(L)=&\frac{\tilde{\Delta}_{\Phi}^3 N}{3^4 }\left(\frac{2}{\pi}\right)^5 h_{l_1 l_2 L}h_{l_3 l_4 L}\nonumber\\
&\times\int dx x^2  dr_1 r_1^3 I_{l_1}(0,r_1) I_{l_2}(0,r_1) I_{l_3}(0,r_1 x) I_{l_4}(0,r_1 x) I_L(1,x),
\end{align}
where we write $r_2/r_1=x$. Now we can evaluate
\begin{eqnarray}
I_l(0,x)&=&\frac{\pi}{2(2l+1)} x^{-(l+1)}\qquad \mbox{for}\,\, x>1 \nonumber \\
&=&\frac{\pi}{2(2l+1)} x^{l}\qquad\qquad \,\mbox{for}\,\, x<1, \\
I_L(1,x)&=&\frac{\pi \Gamma(L+1)}{2\Gamma(L+3/2)} x^{-(L+2)}   {}_2F_1(\frac{1}{2},L+1;L+\frac{3}{2};x^{-2}) \qquad \, \mbox{for}\,\, x>1 \nonumber \\
&=&\frac{\pi \Gamma(L+1)}{2\Gamma(L+3/2)} x^{L}   {}_2F_1(\frac{1}{2},L+1;L+\frac{3}{2};x^{2})\qquad\qquad \quad \,\mbox{for}\,\, x<1,
\end{eqnarray}
where $_2F_1$ is a generalised hypergeometric function. We can write $_2F_1$  in terms of a series expansion in the form
\begin{eqnarray}
_2F_1(a,b;c;z)=\sum_{n=0}^{\infty}\frac{(a)_n (b)_n}{(c)_n}\frac{z^n}{n!},
\end{eqnarray}
where $(p)_n=\Gamma(p+n)/\Gamma(p)$.
The conditions for convergence, namely that this series converges for $c$ a non-negative integer with $|z|<1$, are satisfied in this case. Using this decomposition we find 
\begin{align}
&\mathcal{T}^{l_1 l_2\rm{const}}_{l_3 l_4}(L)=\nonumber\\
&\frac{\tilde{\Delta}_{\Phi}^3 N}{3^4 } h_{l_1 l_2 L}h_{l_3 l_4 L}\frac{1}{{\pi}}\frac{1}{(2l_1+1)(2l_2+1)(2l_3+1)(2l_4+1)}\sum_{n=0}^{\infty}\frac{\Gamma(1/2+n) \Gamma(L+1+n)}{\Gamma(L+3/2+n)n!}\nonumber\\
&\times \Bigg[ \frac{1}{2n+3+l_3+l_4+L}\left( \frac{1}{\sum l_i+4}-\frac{1}{A_1}\right)+\frac{1}{2n+1+l_1+l_2+L}\left( \frac{1}{\sum l_i}+\frac{1}{A_1}\right) \nonumber\\
& + \frac{1}{2n+3+l_1+l_2+L}\left( \frac{1}{\sum l_i+4}-\frac{1}{A_2}\right)+\frac{1}{2n+1+l_3+l_4+L}\left( \frac{1}{\sum l_i}+\frac{1}{A_2}\right)\Bigg].
\end{align}
Notice that this sum is still finite if the denominators $A_1=l_1+l_2-l_3-l_4+2$ or $A_2=l_3+l_4-l_1-l_2+2$ are zero since in those cases the respective numerators vanish. Alternatively, we can integrate over the hypergeometric function directly and write the solution in the following closed form
\begin{align}
&\mathcal{T}^{l_1 l_2\rm{const}}_{l_3 l_4}(L)=\frac{\tilde{\Delta}_{\Phi}^3 N}{3^4} h_{l_1 l_2 L}h_{l_3 l_4 L}\frac{1}{2{\pi}}\frac{1}{(2l_1+1)(2l_2+1)(2l_3+1)(2l_4+1)} B(L+1,1/2)\nonumber \\
&\times\Bigg[\left( \frac{1}{\sum l_i+4}-\frac{1}{A_1}\right)B\left(\frac{C_1+2}{2},1\right) {}_3 F_2\left(\{\frac{C_1+2}{2},\frac{1}{2},L+1 \};\{ L+\frac{3}{2},\frac{C_1+4}{2}  \};1\right) \nonumber \\
&+\left( \frac{1}{\sum l_i}+\frac{1}{A_1}\right)B\left(\frac{C_2}{2},1\right) {}_3 F_2\left(\{\frac{C_2}{2},\frac{1}{2},L+1 \};\{ L+\frac{3}{2},\frac{C_2+2}{2}  \};1\right) \nonumber\\
&+\left( \frac{1}{\sum l_i+4}-\frac{1}{A_2}\right) B\left(\frac{C_2+2}{2},1\right) {}_3 F_2\left(\{\frac{C_2+2}{2},\frac{1}{2},L+1 \};\{ L+\frac{3}{2},\frac{C_2+4}{2}  \};1\right) \nonumber\\
&+\left( \frac{1}{\sum l_i}+\frac{1}{A_2}\right)B\left(\frac{C_1}{2},1\right) {}_3 F_2\left(\{\frac{C_1}{2},\frac{1}{2},L+1 \};\{ L+\frac{3}{2},\frac{C_1+2}{2}  \};1\right) \Bigg],
\end{align}
where $B(x,y)$ denotes the beta function and $C_1=1+l_3+l_4+L , C_2=1+l_1+l_2+L$.
\par
The equilateral shape has also received a lot of attention in the literature. As has been described in~\cite{aChen}, for the purposes of data analyses, there are two representative forms for the equilateral trispectra. These are given by the following shapes for the reduced trispectra
\begin{eqnarray}
S_{\mathcal{T}, c_1}^{\rm{equil}}(\mathbf{k}_1,\mathbf{k}_2,\mathbf{k}_3,\mathbf{k}_4;\mathbf{K})&\propto& K \frac{k_1 k_2 k_3 k_4}{(k_1 +k_2 +k_3 +k_4)^5},\label{c1}\\
S_{\mathcal{T}, s_1}^{\rm{equil}}(\mathbf{k}_1,\mathbf{k}_2,\mathbf{k}_3,\mathbf{k}_4;\mathbf{K})&\propto& \frac{k_1k_2k_3k_4 K^2}{(k_3+k_4 +K)^3}\bigg( \frac{1}{2(k_1+k_2+K)^3} +\frac{6(k_3+k_4 +K)^2}{(k_1+k_2+k_3+k_4)^5}\nonumber\\
&&+\frac{3(k_3+k_4 +K)}{(k_1+k_2+k_3+k_4)^4}+\frac{1}{(k_1+k_2+k_3+k_4)^3}\bigg)\label{s1},
\end{eqnarray}
where we use the notation $c_1$ and $s_1$ to correspond to~\cite{aChen}. These shapes are similar in most regions apart from the doubly squeezed limit ($k_3=k_4\rightarrow 0$). It has been observed that the first ansatz is factorisable by introducing the integral $1/M^n=(1/\Gamma(n))\int_0^{\infty} t^{n-1}e^{-M t} dt$ where $M=\sum k_i$. As we observe from Figures~\ref{fig:equilc1} and~\ref{fig:equils1}, it is clear that the shapes for the two representative forms are highly correlated. Therefore, for the purposes of analysis of the equilateral model, it may only be necessary to consider the $c_1$ model.

\begin{figure}[htp]
\centering 
\includegraphics[width=120mm]{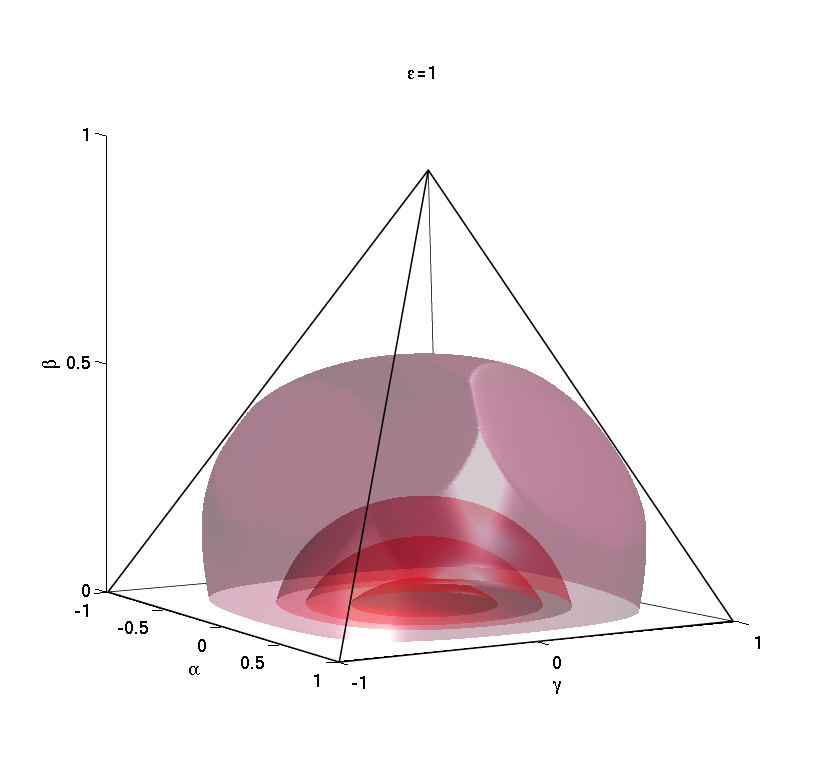}
\includegraphics[width=80mm]{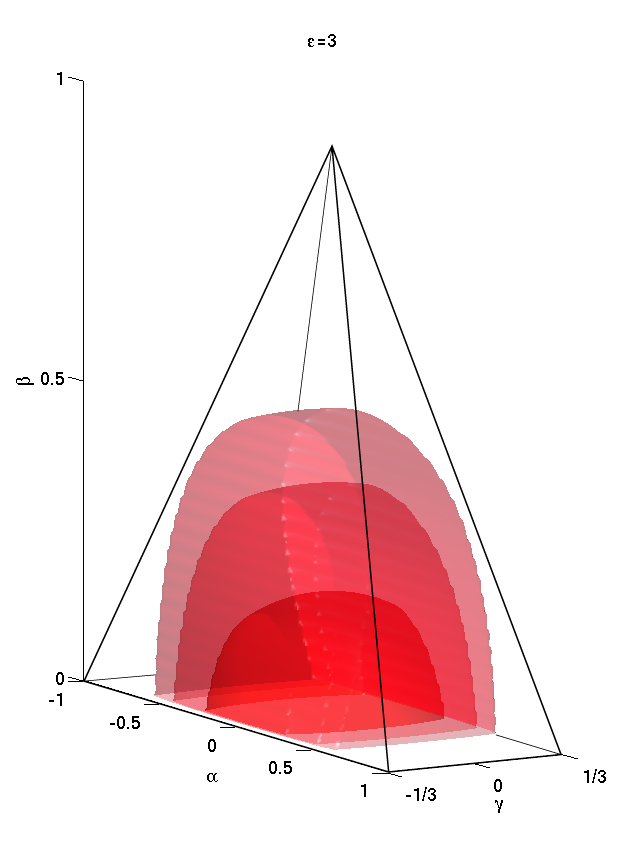}
\includegraphics[width=70mm]{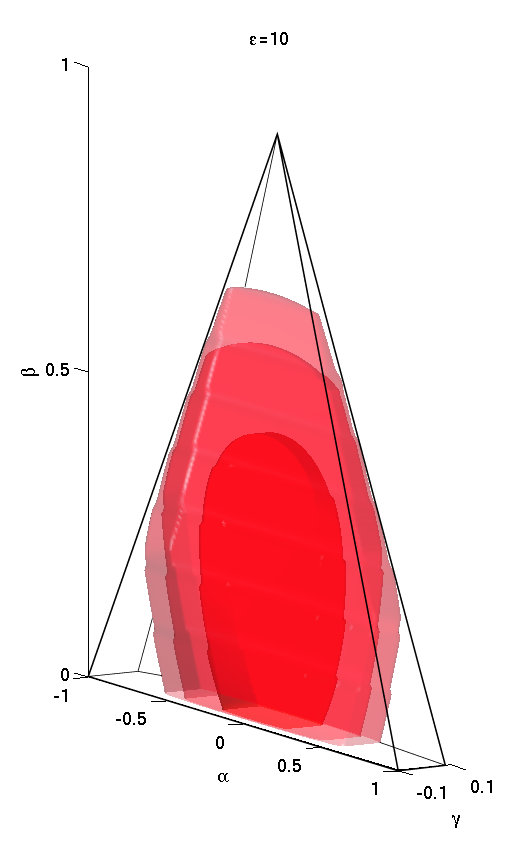}
\caption{Equilateral $c_1$ model~\eqref{c1}. The signal peaks at $\epsilon=1$ towards $\gamma=0,\alpha=0,\beta=0$, i.e. at $k_1=k_2=k_3=k_4=K/2$. For $\epsilon>1$ the signal similarly peaks for $k_1=k_2$, $k_3=k_4$ but since the triangles $(k_1,k_2,K)$ and $(k_3,k_4,K)$ are now unequal the peak position is less sharp and shifts to smaller values of $K$.}
\label{fig:equilc1}
\end{figure}

\begin{figure}[htp]
\centering 
\includegraphics[width=120mm]{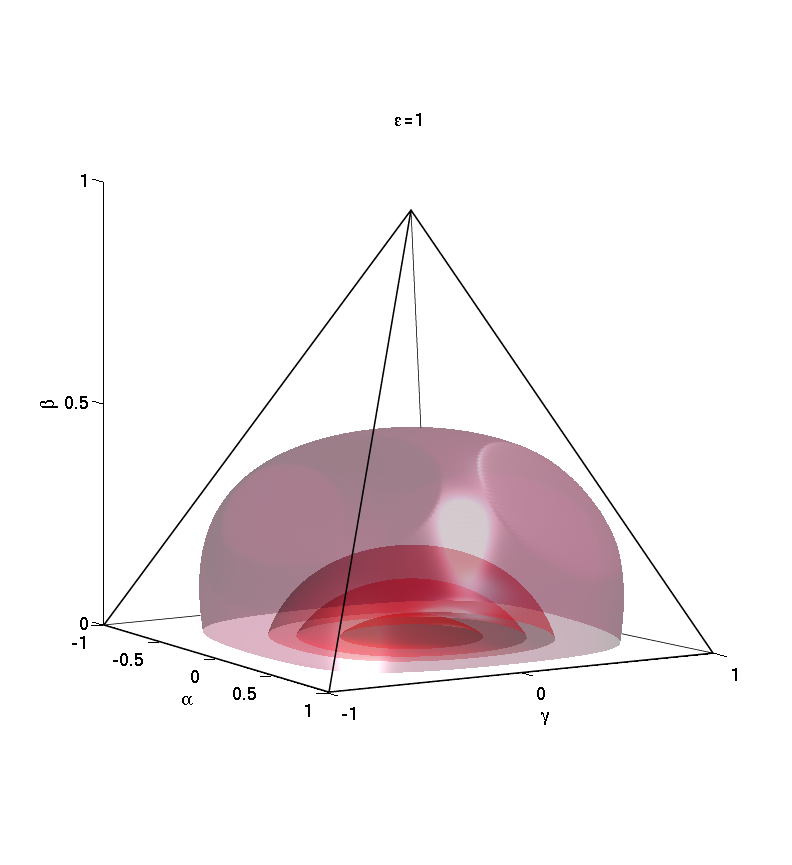}
\includegraphics[width=80mm]{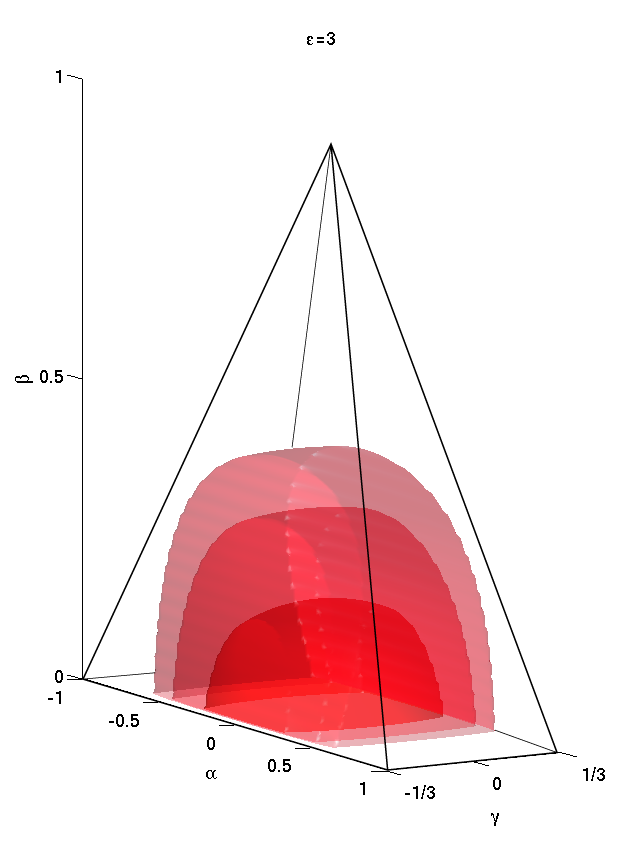}
\includegraphics[width=70mm]{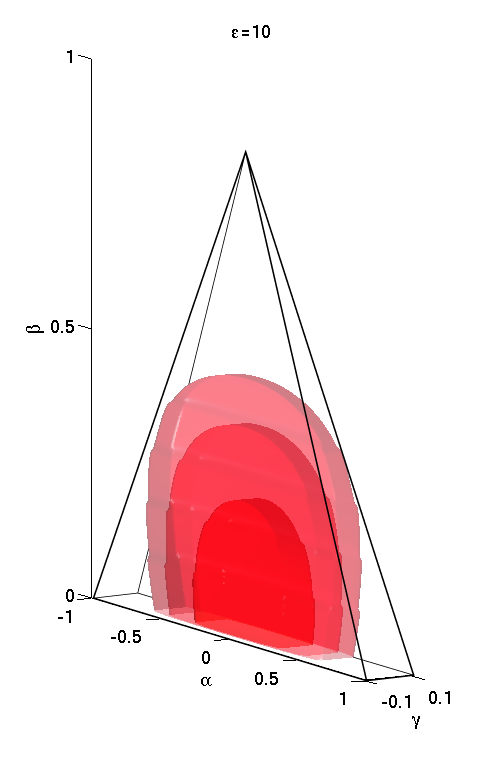}
\caption{Equilateral $s_1$ model~\eqref{s1}. As for the equilateral $c_1$ model the signal peaks for equal $k_i$ at $\epsilon$, while for $\epsilon>1$ the peak position becomes less sharp. As is clearly observable from the figures the equilateral $c_1$ and $s_1$ models are highly correlated.}
\label{fig:equils1}
\end{figure}

\section{Mode decomposition}\label{sec:V}
Our goal is to represent an arbitrary non-separable reduced primordial trispectrum (zero mode) $\mathcal{T}_{\Phi,0}(k_1,k_2,k_3,k_4;K)$ or reduced CMB trispectrum $\mathcal{T}^{l_1 l_2}_{l_3 l_4}(L)$ on their respective wavenumber or multipole domains using a rapidly convergent mode expansion. We need to achieve this in a separable manner, in order to make tractable the five dimensional integrals ($\sim dk_1 dk_2 dk_3 dk_4 dK$) required for trispectrum estimation by breaking them down into products of one-dimensional integrals. In particular, this means that we wish to expand an arbitrary non-separable primordial (reduced) shape function in the form
\begin{eqnarray}\label{decomp}
S_{\mathcal{T}}(k_1,k_2,k_3,k_4,K)
&=&\sum_{p, r, s,u,v} \alpha_{p r s u v} q_p(k_1)q_r(k_2)q_s(k_3)q_u(k_4) r_v(K), 
\end{eqnarray}
where the $q_p, r_v$ are appropriate basis mode functions which are convergent and complete, that is, they span the space of all functions on the wavenumber domain. The differing notation, $q,r$ is due to the different ranges of the variables - $k_i\in [0,k_{\rm{max}}]$ but $K\in [0,2 k_{\rm{max}}]$. In the case of more general sources of non-Gaussianity this is easily extended to include the other Legendre modes of equation~\eqref{expansion} by writing
\begin{eqnarray}
S(k_1,k_2,k_3,k_4,K,\theta_4)=\sum_n S_{\mathcal{T} n}(k_1,k_2,k_3,k_4,K)P_n(\cos\theta_4),
\end{eqnarray}
where $S$ is the shape function applied to the full Legendre expansion~\eqref{expansion}. The shape function of the $n$th Legendre mode, $S_{\mathcal{T}n}$, may be decomposed as in equation~\eqref{decomp}.

We will present one method for finding the basis functions $q,r$ below. We will achieve this objective in stages. First, we create examples of one dimensional mode functions which are orthogonal and well-behaved over the full wavenumber (or multipole) domain. We then construct five dimensional products of these wavefunctions $q_p(k_1)q_r(k_2)q_s(k_3)q_u(k_4) r_v(K) \rightarrow \mathcal{Q}_m$. This creates a complete basis for all possible reduced trispectra on the given domain. By orthonormalising these product basis functions $\mathcal{Q}_m\rightarrow \mathcal{R}_m$ we obtain a rapid and convenient method for calculating the expansion coefficients $ \alpha_{p r s u v}$
(or $ \alpha_m$). Here we use bounded symmetric polynomials as a concrete implementation of this methodology. Of course as outlined in the case of the bispectrum in~\cite{Ferg3} there are alternatives to using the polynomials $ \mathcal{Q}_m,\mathcal{R}_m$ but the shortcoming of these alternatives is that either (i) they can lead to overshooting at the domain boundaries or (ii) the choice may compromise separability. However, it is possible that an alternative to the polynomial expansion may be desirable to improve the rate of convergence. These should be able to conveniently represent functions in a separable form, and should be derived explicitly for the domain.

\subsection{Domain and weight functions}
In Fourier space, the primordial reduced trispectrum  zero mode $\mathcal{T}_{\Phi,0}(k_1,k_2,k_3,k_4;K)$ is defined when the wavevectors $\mathbf{k_1,k_2,K}$ and $\mathbf{k_3,k_4,K}$ close to form triangles subject to $\mathbf{k_1}+\mathbf{k_2}+\mathbf{K}=0=\mathbf{k_3}+\mathbf{k_4}-\mathbf{K}$. Each such triangle is uniquely defined by the lengths of the sides $k_1,k_2,K$ and $k_3,k_4,K$. In terms of these wavenumbers, the triangle conditions restricts the allowed combinations into a region defined by
\begin{align}\label{Domain1}
&k_1\leq K+k_2\,\,{\rm{for}}\,\, k_1\geq k_2,K,\quad {\rm{or}}\quad k_2\leq K+k_1\,\,{\rm{for}}\,\, k_2\geq k_1,K,\nonumber\\
& {\rm{or}}\quad K\leq k_1+k_2\,\,{\rm{for}}\,\, K\geq k_1,k_2,
\end{align}
and
\begin{align}\label{Domain2}
& k_3\leq K+k_4\,\,{\rm{for}}\,\, k_3\geq k_4,K,\quad {\rm{or}}\quad k_4\leq K+k_3\,\,{\rm{for}}\,\, k_4\geq k_3,K,\nonumber\\
& {\rm{or}}\quad K\leq k_3+k_4\,\,{\rm{for}}\,\, K\geq k_3,k_4.
\end{align}
Since the wavenumber $K$ is common to both triangles the region is a product of the tetrahedral domains defined by the conditions~\eqref{Domain1} and~\eqref{Domain2}. Considering each region individually we note that they each describe a regular tetrahedron for $k_1+k_2+K<2k_{\rm{max}}$ (or $k_3+k_4+K<2k_{\rm{max}}$). However, motivated by issues of separability and observation, it is more natural to extend the domain out to values given by a maximum wavenumber $k_{\rm{max}}$. In particular, we have $k_i<k_{\rm{max}}$ and $K<2k_{\rm{max}}$. In each case the allowed region is a hexahedron formed by the intersection of a tetrahedron and a rectangular parallelepiped. For brevity we will denote this configuration as a $tetrapiped$. This region is an extension of the tetrapyd referred to in~\cite{Ferg3} due to the extended range of $K$ and is shown in Figure~\ref{fig:tetrapiped}.
\begin{figure}[htp]
\centering 
\includegraphics[width=82mm]{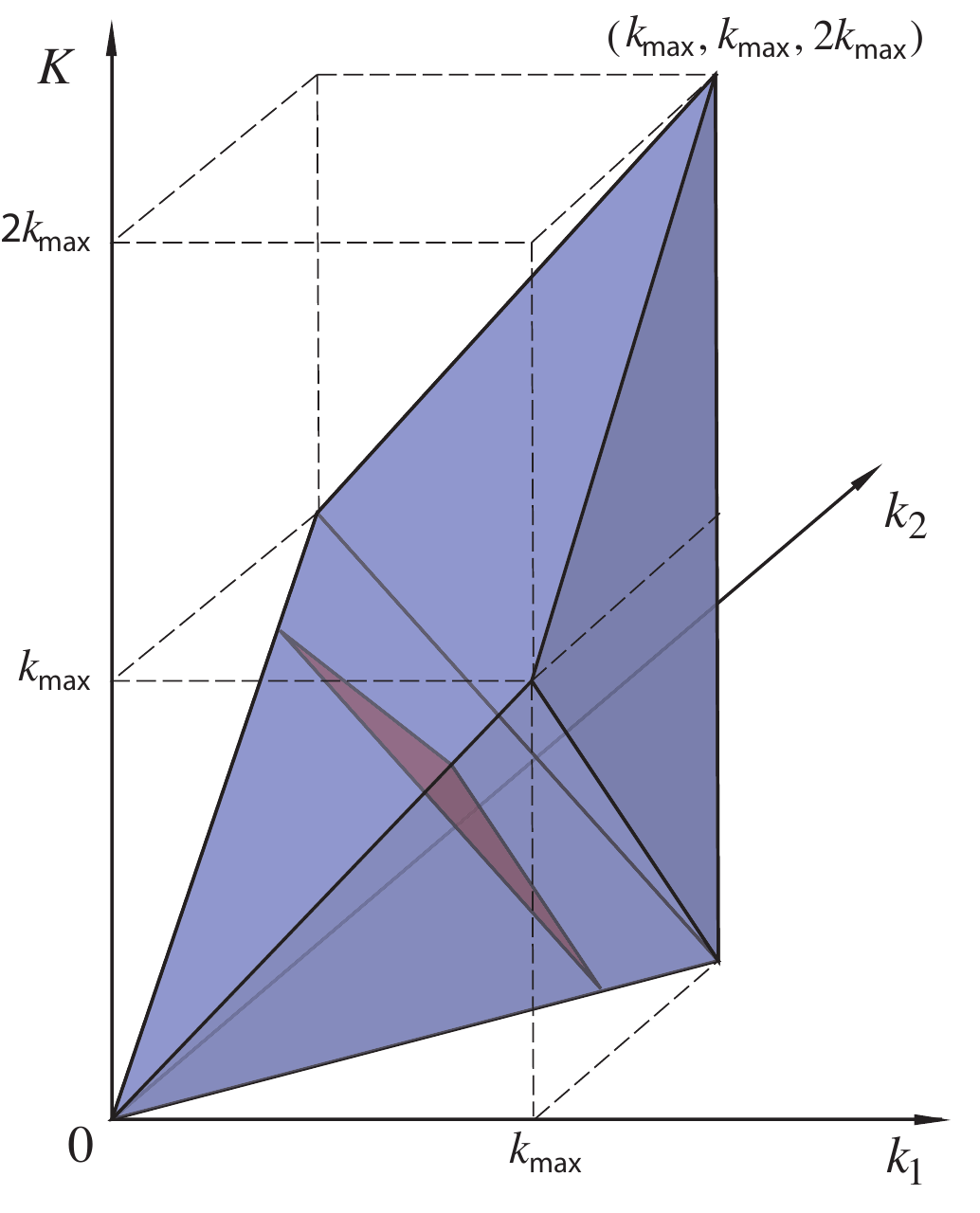}
\caption{`Tetrapiped' domain for allowed wavenumbers of the primordial reduced trispectrum $\mathcal{T}(k_1,k_2,k_3,k_4;K,\theta_4)$ imposed by the triangle created by $(k_1,k_2,K)$. There is a corresponding tetrapiped domain imposed by the triangle created by $(k_3,k_4,K)$. The region is an extension of the tetrapyd domain described in~\cite{Ferg3} due to the extended range of $K$. The same domain is valid for allowed multipole values $l_i,L$ in the case of the reduced CMB trispectrum $\mathcal{T}^{l_1 l_2}_{l_3 l_4}(L)$. The shaded area denotes the region described in Figure~\ref{fig:shape}.}
\label{fig:tetrapiped}
\end{figure}
\par
In order to integrate functions $f(k_1,k_2,k_3,k_4,K)$ over the tetrapiped domains, which we denote $\mathcal{V}_{\mathcal{T}}$, we note the presence of $K$ in both regions. We find that the integration is given explicitly by
\begin{align}
\mathcal{I}[f]\equiv& \int_{\mathcal{V}_{\mathcal{T}}}f(k_1,k_2,k_3,k_4,K) \omega(k_1,k_2,k_3,k_4,K)d \mathcal{V}_{\mathcal{T}}\nonumber\\
=&k_{\rm{max}}^5\Big\{   \int_0^{1/2}dt\left(\int_0^t ds \int_{t-s}^{t+s}dx+\int_t^{1-t}  ds \int_{s-t}^{t+s}dx+\int_{1-t}^1 ds \int_{s-t}^1 dx\right)\nonumber\\
&\times \left(\int_0^t dy \int_{t-y}^{t+y}dz+\int_t^{1-t}  dy \int_{y-t}^{t+y}dz+\int_{1-t}^1 dy \int_{y-t}^1 dz\right) FW \nonumber\\
&+ \int_{1/2}^1 dt  F W\left(\int_0^{1-t} ds \int_{t-s}^{t+s}dx+\int^t_{1-t}  ds \int_{t-s}^{t+s}dx+\int_{t}^1 ds \int_{s-t}^1 dx\right) \nonumber\\ 
&\times \left(\int_0^{1-t} dy \int_{t-y}^{t+y}dz+\int^t_{1-t}  dy \int_{t-y}^{t+y}dz+\int_{t}^1 dy \int_{y-t}^1 dz\right)\nonumber\\
&+\int_1^2 dt \int_{t-1}^1 ds\int_{t-s}^1 dx \int_{t-1}^1 dy \int_{t-y}^1    dz F  W  \Big\},
\end{align}
where  $\omega(k_1,k_2,k_3,k_4,K)$ is an appropriate weight function and we have made the transformation $t=K/k_{\rm{max}}, s=k_1/k_{\rm{max}},x=k_2/k_{\rm{max}}, y=k_3/k_{\rm{max}}, z=k_4/k_{\rm{max}}$ with $F(s,x,y,z,t)=f(k_{\rm{max}}\times(s,x,y,z,t))$ and $W(s,x,y,z,t)=\omega(k_{\rm{max}}\times(s,x,y,z,t))$. For integrals over the product of two functions $f$ and $g$ we can define the inner product $\langle f,g \rangle=\mathcal{I}[f g]$. This inner product essentially defines a Hilbert space of possible shape functions in the domain. The total volume of the domain is given by $\mathcal{I}[1]=k_{\rm{max}}^5/3$. Initially we will restrict attention to the case of weight $\omega=1$. However, it is important to incorporate a weight function for a variety of reasons which we will discuss later.
\par
Analysis of the CMB `extra'-reduced trispectrum, $t^{l_1 l_2}_{l_3 l_4}(L)$, is more straightforward than in the primordial case. This is because the CMB trispectrum, being defined on a two sphere, is an explicitly five dimensional quantity and therefore is defined completely in terms of the multipoles. We note here that the quantity $p^{l_1 l_2}_{l_3 l_4}(L)=t^{l_1 l_2}_{l_3 l_4}(L)+t^{l_2 l_1}_{l_3 l_4}(L)+t^{l_1 l_2}_{l_4 l_3}(L)+t^{l_2 l_1}_{l_4 l_3}(L)$ is probably a more elegant expression for this analysis since it is more symmetric, whilst being defined on the same domain and being subject to the same weighting over the domain. Nonetheless, we proceed in this chapter with analysis of $t^{l_1 l_2}_{l_3 l_4}(L)$ leaving exploration of this minor issue to an upcoming paper~\cite{Regan2}. As for the primordial case, we extend the tetrahedral domains to include multipoles out to $l_i, L/2<l_{\rm{max}}$. The respective tetrapiped domains for the extra-reduced trispectrum becomes the discrete $\{l_1,l_2,l_3,l_4,L\}$ satisfying
\begin{eqnarray}
&&l_1,l_2,l_3,l_4,L/2 <l_{\rm{max}},\,\, l_i,L\in \mathbb{N},\nonumber\\
&&l_1\leq l_2+L \,\, {\rm{for}}\,\, l_1\geq l_1,L, +\,\,\rm{cyclic \,perms}\nonumber\\
&&l_3\leq l_4+L \,\, {\rm{for}}\,\, l_3\geq l_4,L, +\,\,\rm{cyclic \,perms}\nonumber\\
&&l_1+l_2+L=2 n_1,\quad l_3+l_4+L=2 n_2,\quad n_1,n_2\in \mathbb{N}.
\end{eqnarray}
In multipole space, we will be primarily dealing with a summation over all possible  $\{l_1,l_2,l_3,l_4,L\}$  combinations in the correlator $\mathcal{C(T,T')}$. The appropriate weight function in the sum from~\eqref{ExtraTrispRed} is then
\begin{eqnarray}
\omega(l_1,l_2,l_3,l_4,L)=h_{l_1 l_2 L}^2 h_{l_3 l_4 L}^2.
\end{eqnarray}
A straightforward continuum version of this can be deduced by comparison of this `weight' formula to the bispectrum multipole weight function in~\cite{Ferg3}. Similarly to that analysis, we should eliminate a scaling in this weight such that the overall weight becomes very nearly constant. We can do this by using a separable weight function as
\begin{eqnarray}
\omega_s(l_1 l_2 l_3 l_4 L)=\frac{\omega(l_1,l_2,l_3,l_4,L)}{(2 l_1+1)^{1/3}(2 l_2+1)^{1/3}(2 l_3+1)^{1/3}(2 l_4+1)^{1/3}(2 L+1)^{2/3}}.
\end{eqnarray}
We note that there is also a freedom to absorb an arbitrary separable function $v_l$ into the weight functions. If we define a new weight $\overline{\omega}$ in the estimator as
\begin{eqnarray}
\overline{\omega}_{l_1 l_2 l_3 l_4 L}=\omega_{l_1 l_2 l_3 l_4 L}/(v_{l_1}v_{l_2}v_{l_3}v_{l_4}v_L)^2,
\end{eqnarray}
then we must rescale the estimator functions by the factor $v_{l_1}v_{l_2}v_{l_3}v_{l_4}v_L$. The important point is to use both the weight $\overline{\omega}$ and the estimator rescaling throughout the analysis, including the generation of appropriate orthonormal mode functions.

\subsection{Orthogonal polynomials on the domain}
We now construct some concrete realisations of mode functions which are orthogonal on the domain $\mathcal{V}_{\mathcal{T}}$ and which have the form required for a separable expansion. First, we will generate one-dimensional orthogonal polynomials $q_p(s), r_v(t)$ for unit weight $\omega=1$. Considering functions $q_p(s)$ depending only on the $s$-coordinate\footnote{We can consider $s$ as corresponding to any of the $k_i$.}, we integrate over the $t,x,y,z$-directions to yield the weight functions $\overline{\omega}(s)$ for $s\in [0,1]$ (for simplicity we take $k_{\rm{max}}=1$):
\begin{eqnarray}
\overline{\omega}(s)=s-s^3+\frac{5}{12}s^4,\quad {\rm{with}}\quad \mathcal{I}[f]=\int_0^1 f(s) \overline{\omega}(s) ds.
\end{eqnarray}
Therefore, the moments for each power of $s$ become
\begin{eqnarray}
\overline{\omega}_n\equiv\mathcal{I}[s^n]&=&\frac{1}{n+2}+\frac{5}{12(n+5)}-\frac{1}{n+4}\nonumber\\
&=&\frac{5n^2+54 n+160}{12(n+2)(n+4)(n+5)}.
\end{eqnarray}
For functions $r_v(t)$ (where $t$ corresponds to the diagonal $K$), we integrate over the $s,x,y,z$- directions to yield the weight functions $\hat{\omega}(t)$ for $t\in[0,2]$:
\begin{eqnarray}
\hat{\omega}(t)&=&\left(\frac{t}{2}(4-3t)\right)^2\,\,{\rm{for}}\,\, t\in [0,1]\nonumber\\
&=&\frac{(t-2)^4}{4}\qquad\quad {\rm{for}}\,\, t\in [1,2], \quad {\rm{with}}\quad \mathcal{I}[f]=\int_0^2 f(t) \hat{\omega}(t) dt.
\end{eqnarray}
With this choice of weight the moments of each power of $t$ become
\begin{align}
\hat{\omega}_n\equiv\mathcal{I}[t^n]=\frac{n^2+15n+68}{4(n+3)(n+4)(n+5)}+\frac{768\times 2^n-744-474 n-131n^2-18n^3-n^4}{4(n+1)(n+2)(n+3)(n+4)(n+5)}.
\end{align}
From these moments we can create orthogonal polynomials using the generating functions,
\begin{eqnarray}
q_n(s)=\frac{1}{\mathcal{N}_1} \left| \begin{array}{ccccc}
1/3 & 73/360 & 1/7&\dots&\overline{\omega}_n \\
73/360 & 1/7&367/3360&\dots&\overline{\omega}_{n+1}\\
\dots&\dots&\dots&\dots&\dots\\
\overline{\omega}_{n-1}&\overline{\omega}_{n}&\overline{\omega}_{n+1}&\dots&\overline{\omega}_{2n-1}\\
1&s&s^2&\dots&s^n \end{array} \right|
\end{eqnarray}
and
\begin{eqnarray}
r_n(t)=\frac{1}{\mathcal{N}_2} \left| \begin{array}{ccccc}
1/3 & 7/30 & 4/21&\dots&\hat{\omega}_n \\
7/30 & 4/21&73/420&\dots&\hat{\omega}_{n+1}\\
\dots&\dots&\dots&\dots&\dots\\
\hat{\omega}_{n-1}&\hat{\omega}_{n}&\hat{\omega}_{n+1}&\dots&\hat{\omega}_{2n-1}\\
1&t&t^2&\dots&t^n \end{array} \right|,
\end{eqnarray}
where we choose the normalisation factors $\mathcal{N}_1,\mathcal{N}_2$ such that $\mathcal{I}[q_n^2]=1$ and $\mathcal{I}[r_n^2]=1$ for all $n\in \mathbb{N}$, that is such that the $q_n(s)$ (or $r_n(t)$) are orthonormal
\begin{eqnarray}
\langle q_n,q_p\rangle\equiv\mathcal{I}[q_n q_p]&=&\int_{\mathcal{V}_{\mathcal{T}}}q_n(s)q_p(s) d\mathcal{V}_{\mathcal{T}}=\delta_{np},\\
\langle r_v,r_u\rangle\equiv\mathcal{I}[r_v r_u]&=&\int_{\mathcal{V}_{\mathcal{T}}}r_v(t)r_v(t) d\mathcal{V}_{\mathcal{T}}=\delta_{vu}.
\end{eqnarray}
The first few orthonormal polynomials on the domain $\mathcal{V}_{\mathcal{T}}$ are explicitly
\begin{eqnarray}
q_0(s)&=&\sqrt{3}\nonumber\\
q_1(s)&=&7.16103(-0.608333+s)\nonumber\\
q_2(s)&=&7.76759-33.2061 s+29.0098 s^2,\nonumber\\
q_3(s)&=&-11.7911+93.1318s-194.111s^2+116.964 s^3 ,\dots
\end{eqnarray}
and
\begin{eqnarray}
r_0(t)&=&\sqrt{3}\nonumber\\
r_1(t)&=&6.06977(-0.7+t)\nonumber\\
r_2(t)&=&7.65066-24.1493t+16.1942t^2\nonumber\\
r_3(t)&=&-12.2182+63.4315 t-91.6438 t^2+38.7091 t^3,\dots
\end{eqnarray}
\par
We note that the $q_n$ and $r_v$ are only orthogonal in one dimension (e.g. $\langle q_n (s)r_v(t)\rangle\neq \delta_{n v}$ and $\langle q_n (s)q_m(x)\rangle\neq \delta_{n m}$ in general). However, as product functions of $t,s,x,y$ and $z$ they form an independent and well-behaved basis which we will use to construct orthonormal five-dimensional eigenfunctions. In practice the $q_n$ and $r_v$ remain the primary calculation tools, notably in performing separable integrations. In Figure~\ref{fig:orthoModes} we plot the first few $q_n$ and $r_v$.
\begin{figure}[htp]
\centering 
\includegraphics[width=152mm]{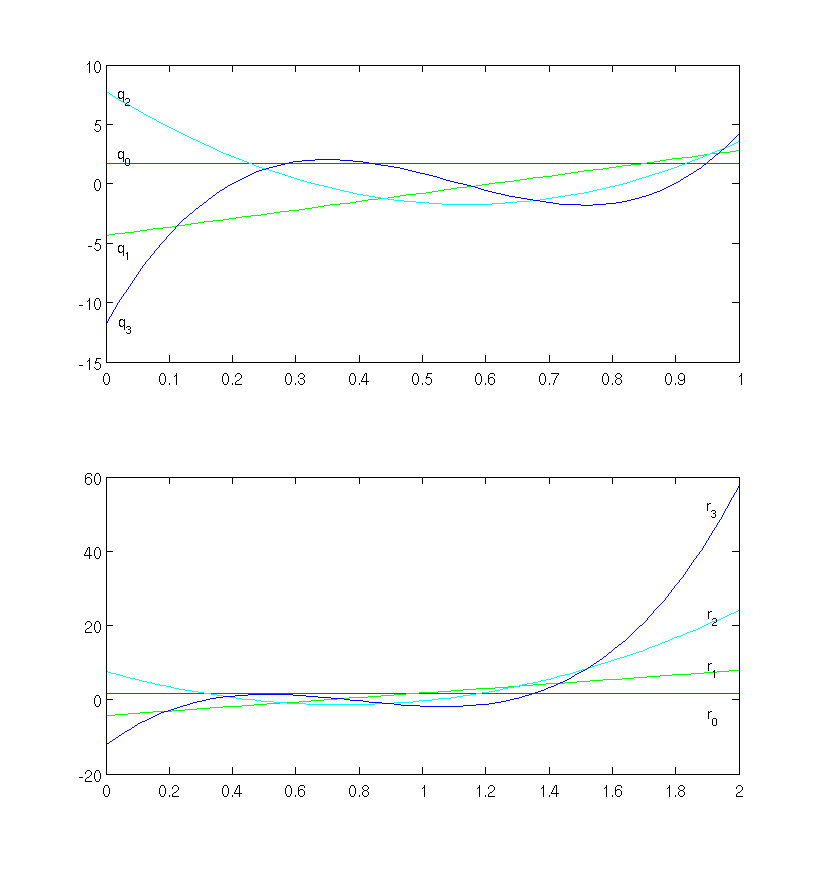}
\caption{The orthonormal one-dimensional eigenmodes $q_n, r_v$ plotted on their respective domains for $n,v=0,1,2,3$. The $q_n, r_v$ plotted are calculated for unit weight $\omega=1$ on the domain $\mathcal{V}_{\mathcal{T}}$. The shape of these eigenmodes alters for different choices of the weighting. }
\label{fig:orthoModes}
\end{figure}
\par
Now we turn to the polynomials $\overline{q}(s)$ and $\overline{r}(t)$ which are orthonormal on the multipole domain. Using the weight function $\omega=1$ we find the same polynomials as above. For the scaled weight function $\omega_s$ the polynomials will, of course, differ. While using either polynomial sets would suffice as independent basis functions on the multipole domain, the use of correctly weighted functions lead to improvements in the immediate orthogonality of the derived five dimensional polynomial sets. For definiteness we take $(t,s,x,y,z)\times l_{\rm{max}}=(L,l_1,l_2,l_3,l_4)$. The generating function is obtained as above but now using the moments $\overline{\omega}_n\equiv \mathcal{I}[s^n]=\int \omega(t,s,x,y,z)s^n d\mathcal{V}_{\mathcal{T}} $ and $\hat{\omega}_n\equiv \mathcal{I}[t^n]=\int \omega(t,s,x,y,z)t^n d\mathcal{V}_{\mathcal{T}}$.

\subsection{Five-dimensional basis functions}
We can represent arbitrary (reduced) trispectra Legendre modes on the domain $\mathcal{V}_{\mathcal{T}}$ using a suitable set of independent basis functions formed from products $q_p(s)q_r(x)q_s(y)q_u(z)r_v(t)$. (Here again we take $t=K/k_{\rm{max}},s=k_1/k_{\rm{max}},x=k_2/k_{\rm{max}},y=k_3/k_{\rm{max}},z=k_4/k_{\rm{max}}$ or $t=L/l_{\rm{max}}$, etc.) We denote the $5D$ basis function as
\begin{eqnarray}
\mathcal{Q}_m(t,s,x,y,z)= q_p(s)q_r(x)q_s(y)q_u(z) r_v(t). 
\end{eqnarray}
We can order these products linearly with a single index $m$ in a similar manner to that described in~\cite{Ferg3} for the bispectrum.
\par
While the $Q_m$ by construction are an independent set of five-dimensional functions on the domain $\mathcal{V}_{\mathcal{T}}$, they are not in general orthogonal. To construct an orthonormal set $\mathcal{R}_m$ from the $\mathcal{Q}_m$ we perform an iterative Gram-Schmidt orthogonalisation process such that
\begin{eqnarray}
\langle \mathcal{R}_n \mathcal{R}_m\rangle=\delta_{n m}.
\end{eqnarray}
In particular, we form the Gram matrix $\Gamma=(\langle  \mathcal{Q}_n \mathcal{Q}_m \rangle )$ which needs to be factorised as $\Gamma=\Lambda^T \Lambda$ where $\Lambda=(\langle  \mathcal{Q}_n \mathcal{R}_m \rangle )$ is triangular. This process is described in more detail in~\cite{Ferg3}.

\subsection{Mode decomposition of the trispectrum}
Having formed the orthonormal basis $\{\mathcal{R}_m\}$ we consider an arbitrary primordial reduced trispectrum (zero mode) $\mathcal{T}_{,0}(k_1,k_2,k_3,k_4;K)$ described by the shape function $S_{\mathcal{T}}$ and decompose it as follows
\begin{eqnarray}
S_{\mathcal{T}}(k_1,k_2,k_3,k_4,K)=\sum_{m} \alpha_{m}^{\mathcal{R}}\mathcal{R}_m (t,s,x,y,z),
\end{eqnarray}
where the expansion coefficients $\alpha_{m}^{\mathcal{R}}$ are given by
\begin{eqnarray}
\alpha_{m}^{\mathcal{R}}=\langle \mathcal{R}_m,S_{\mathcal{T}}\rangle=\int_{\mathcal{V}_{\mathcal{T}}}\mathcal{R}_m S_{\mathcal{T}}\omega d\mathcal{V}_{\mathcal{T}}
\end{eqnarray}
and $k_{\rm{max}}(t,s,x,y,z)=(K,k_1,k_2,k_3,k_4)$ on the domain $\mathcal{V}_{\mathcal{T}}$ defined in~\eqref{Domain1} and \eqref{Domain2}. In practice, we must always work with partial sums up to a given $N=n_{\rm{max}}$ with
\begin{eqnarray}
S_{\mathcal{T}}^N=\sum_{m=0}^N  \alpha_{m}^{\mathcal{R}}\mathcal{R}_m (t,s,x,y,z), \quad S_{\mathcal{T}}=\lim_{N\rightarrow \infty}S_{\mathcal{T}}^N.
\end{eqnarray}
Given the complete orthonormal basis $\mathcal{R}_m$, Parseval's theorem for the integrated product of two functions implies
\begin{eqnarray}
\langle S_{\mathcal{T} },S'_{\mathcal{T} }\rangle=\int_{\mathcal{V}_{\mathcal{T}}} S_{\mathcal{T} }S'_{\mathcal{T} }\omega d\mathcal{V}_{\mathcal{T}}=\lim_{N\rightarrow \infty}\sum_{m=0}^N \alpha_{m}^{\mathcal{R}}\alpha_{m}^{\mathcal{R}'},
\end{eqnarray}
which for the square of a mode $S_{\mathcal{T} }$ yields the sum of the squares of the expansion coefficients, $\mathcal{I}[S_{\mathcal{T} }^2]=\sum_m \alpha_{m}^{\mathcal{R}\,\,2}$.
\par
In order to accomplish the goal of a general separable expansion, we must transform backwards from the orthogonal sum $\mathcal{R}_m$ into an expansion over the separable product functions $\mathcal{Q}_m$ through
\begin{eqnarray}
S_{\mathcal{T}}^N=\sum_{m=0}^N  \alpha_{m}^{\mathcal{Q}}\mathcal{Q}_m (t,s,x,y,z).
\end{eqnarray}
The $ \alpha_{m}^{\mathcal{Q}}$ can be obtained from the  $ \alpha_{m}^{\mathcal{R}}$ via
\begin{eqnarray}
 \alpha_{m}^{\mathcal{Q}}=\sum_{p=0}^N (\lambda^T)_{m p} \alpha_{p}^{\mathcal{R}},
\end{eqnarray}
where the transformation matrix $\lambda_{np}$ was defined above. Using the inverse relation $ \alpha_{m}^{\mathcal{R}}=\sum_{p=0}^N (\lambda^T)_{m p}^{-1} \alpha_{p}^{\mathcal{Q}}$ we find that the matrices $\Gamma$ and $\Lambda$ are related by
\begin{eqnarray}
\left(\gamma^{-1}\right)_{np}=\sum_{r}^N \left(\lambda^T\right)_{n r} \lambda_{r p}.
\end{eqnarray}
This implies that 
\begin{eqnarray}
\langle S_{\mathcal{T}}^N,S_{\mathcal{T}}^{N}\rangle =\sum_{m=0}^N \alpha_{m}^{\mathcal{R}\,\,2}=\sum_{m=0}^N \sum_{p=0}^N \alpha_{m}^{\mathcal{Q}}\gamma_{m p}\alpha_{p}^{\mathcal{Q}}.
\end{eqnarray}
As we have already noted the separable $\mathcal{Q}_m$ expansion is most useful for practical calculations. However, its coefficients must be constructed from the orthonormal $\mathcal{R}_m$.
\par
We can expand the CMB extra-reduced trispectrum $t^{l_1 l_2}_{l_3 l_4}(L)$ at late times using the same polynomials. However, as noted previously the CMB trispectrum is an explicitly five dimensional quantity and, as such we do not require the extra step of finding the zero mode of the Legendre series expansion. In particular, the appropriate expansion is of the form $t^{l_1 l_2}_{l_3 l_4}(L)=\sum_m \overline{\alpha}_m^{\mathcal{R}}\mathcal{R}_m(t,s,x,y,z)(=\sum_m \overline{\alpha}_m^{\mathcal{Q}}\mathcal{Q}_m(t,s,x,y,z))$.

\section{Measures of $T_{NL}$}\label{sec:VI}
\subsection{Primordial estimator}
We have obtained related mode expansions for a general primordial shape function, one with the orthonormal basis $\mathcal{R}_m$ and the other with the separable basis functions $\mathcal{Q}_m$. Substitution of the (reduced) separable form into the expression for the `extra'-reduced trispectrum~\eqref{ExtraTrispRed} offers an efficient route to its direct calculation through
\begin{eqnarray}
t^{l_1 l_2}_{l_3 l_4}(L)&=&N \tilde{\Delta}_{\Phi}^3 \left(\frac{2}{\pi}\right)^5  \int r_1^2 dr_1 r_2^2 dr_2 dk_1 dk_2 dk_3 dk_4 dK K\\
&&\times  \sum_{m}\alpha_{m}^{\mathcal{Q}} \mathcal{Q}_m (k_1,k_2,k_3,k_4,K) j_L(K r_1) j_L(K r_2)[j_{l_1}(k_1 r_1)\Delta_{l_1}(k_1)]\nonumber\\
&&\times[j_{l_2}(k_2 r_1)\Delta_{l_2}(k_2)][j_{l_3}(k_3 r_2)\Delta_{l_3}(k_3)][j_{l_4}(k_4 r_2)\Delta_{l_4}(k_4)] \nonumber\\
&=&N \tilde{\Delta}_{\Phi}^3 \sum_{m}     \alpha_{m}^{\mathcal{Q}} \int r_1^2 dr_1 r_2^2 dr_2\mathcal{Q}^{l_1 l_2 l_3 l_4 L}_{m} (r_1,r_2),\nonumber
\end{eqnarray}
where
\begin{eqnarray}
\mathcal{Q}^{l_1 l_2 l_3 l_4 L}_{m} (r_1,r_2)=     q_p^{l_1}(r_1)q_r^{l_2}(r_1)q_s^{l_3}(r_2)q_{u}^{l_4}(r_2)r_v^{L}(r_1,r_2),
\end{eqnarray}
with
\begin{eqnarray}
q_p^{l}(r)&=&\frac{2}{\pi}\int dk q_p(k) \Delta_{l}(k)j_{l}(k r),\nonumber\\
r_v^{L}(r_1,r_2)&=&\frac{2}{\pi}\int dK K r_v(K) j_L(K r_1) j_L(K r_2).
\end{eqnarray}
Next, we note from~\eqref{TotalRedTrisp} that
\begin{align}
\mathcal{T}_{l_1 m_1 l_2 m_2 l_3 m_3 l_4 m_4}=\sum_{L M}(-1)^M  \left( \begin{array}{ccc}
l_1 & l_2 & L \\
m_1 &m_2 & -M \end{array} \right)    \left( \begin{array}{ccc}
l_3 & l_4 & L \\
m_3 & m_4 & M \end{array} \right)\mathcal{T}^{l_1 l_2}_{l_3 l_4}(L).
\end{align}
Therefore, using these formulae in the estimator~\eqref{Estimator} (we omit the normalisation factor $N_T$ here and return to it later in the section) we find
\begin{align}\label{EstimatorPrim1}
\mathcal{E}&=12 \sum_{l_i m_i} \mathcal{T}_{l_1 m_1 l_2 m_2 l_3 m_3 l_4 m_4}\frac{\left(a_{l_1 m_1}a_{l_2 m_2}a_{l_3 m_3}a_{l_4 m_4}\right)_c}{C_{l_1}C_{l_2}C_{l_3}C_{l_4}}\nonumber\\
\implies \mathcal{E}&=\sum_{l_i m_i}\sum_{LM}12 N \tilde{\Delta}_{\Phi}^3\Bigg[\left(\int d\hat{n}_1 Y_{l_1 m_1}(\hat{n}_1)Y_{l_2 m_2}(\hat{n}_1)Y_{L M}^*(\hat{n}_1)\right)\nonumber\\
&\times\left(\int d\hat{n}_2 Y_{l_3 m_3}(\hat{n}_2)Y_{l_4 m_4}(\hat{n}_2)Y_{L M}(\hat{n}_2)\right) \sum_{m}  \alpha_{ m}^{\mathcal{Q}} \int r_1^2 dr_1 r_2^2 dr_2\mathcal{Q}^{l_1 l_2 l_3 l_4 L}_{m} (r_1,r_2)\Big]\nonumber\\
&\times \frac{\left(a_{l_1 m_1} a_{l_2 m_2} a_{l_3 m_3} a_{l_4 m_4} \right)_c}{C_{l_1}C_{l_2}C_{l_3}C_{l_4}}.
\end{align}
Now using the notation $\mathcal{E}=\mathcal{E}_{\rm{tot}}-\mathcal{E}_{\rm{uc}}$ where $\rm{tot}$ refers to $a_{l_1 m_1} a_{l_2 m_2} a_{l_3 m_3} a_{l_4 m_4}$ and $uc$ refers to $\left(a_{l_1 m_1} a_{l_2 m_2} a_{l_3 m_3} a_{l_4 m_4} \right)_{\rm{uc}}$ in place of $\left(a_{l_1 m_1} a_{l_2 m_2} a_{l_3 m_3} a_{l_4 m_4} \right)_c$, we find
\begin{align}
\mathcal{E}_{\rm{uc}}
&=12 N\tilde{\Delta}_{\Phi}^3\sum_{ m}\alpha_{m}^{\mathcal{Q}}\int d\hat{n}_1 d\hat{n}_2\int dr_1 dr_2 r_1^2 r_2^2 N_v(\hat{n}_1,\hat{n}_2,r_1,r_2) \nonumber\\
&\times\Bigg(M^{\rm{uc}}_{p r}(\hat{n}_1,\hat{n}_1,r_1,r_1)M^{\rm{uc}}_{s u }(\hat{n}_2,\hat{n}_2,r_2,r_2)+M^{\rm{uc}}_{p s}(\hat{n}_1,\hat{n}_2,r_1,r_2)M^{\rm{uc}}_{r u }(\hat{n}_1,\hat{n}_2,r_1,r_2)\nonumber\\
&+M^{\rm{uc}}_{p u }(\hat{n}_1,\hat{n}_2,r_1,r_2)M^{\rm{uc}}_{r s}(\hat{n}_1,\hat{n}_2,r_1,r_2)\Bigg)
\end{align}
and
\begin{align}
\mathcal{E}_{\rm{tot}}=12 N\tilde{\Delta}_{\Phi}^3\sum_{ m}\alpha_{m}^{\mathcal{Q}}&\int d\hat{n}_1 d\hat{n}_2\int dr_1 dr_2  r_1^2 r_2^2 N_v(\hat{n}_1,\hat{n}_2,r_1,r_2)\nonumber\\
&\times M_{p}(\hat{n}_1,r_1)M_{r}(\hat{n}_1,r_1)M_{s}(\hat{n}_2,r_2)M_{u }(\hat{n}_2,r_2),
\end{align}
where
\begin{eqnarray}
M^{\rm{uc}}_{p s}(\hat{n}_1,\hat{n}_2,r_1,r_2)&=&\sum_{l_1 m_1}
\frac{Y_{l_1 m_1}(\hat{n}_1)Y_{l_1 m_1}^*(\hat{n}_2)q_p^{l_1}(r_1) q_s^{l_1}(r_2)}{C_{l_1}} ,\nonumber\\
M_{p}(\hat{n}_1,r_1)&=&\sum_{l_1 m_1}\frac{Y_{l_1 m_1}(\hat{n}_1) a_{l_1 m_1}q_p^{l_1}(r_1)}{C_{l_1}},\nonumber\\
N_v(\hat{n}_1,\hat{n}_2,r_1,r_2)&=&\sum_{L M}Y_{L M}^* (\hat{n}_1) Y_{LM}(\hat{n}_2) r_v^L (r_1,r_2).
\end{eqnarray}
We can summarise these results (substituting back in $N_T$) as
\begin{eqnarray}\label{EstimatorPrim2}
\mathcal{E}&=&\frac{12 N\tilde{\Delta}_{\Phi}^3}{N_T}\sum_{m}\alpha_{m}^{\mathcal{Q}}\int d\hat{n}_1 d\hat{n}_2\int dr_1 dr_2 r_1^2 r_2^2 \mathcal{M}_{m}^{\mathcal{Q}}(\hat{n}_1,\hat{n}_2,r_1,r_2)\nonumber\\
&=&\frac{N\tilde{\Delta}_{\Phi}^3}{N_T}\sum_{m}\alpha_{m}^{\mathcal{Q}}\beta_{m}^{\mathcal{Q}},
\end{eqnarray}
with
\begin{eqnarray}
\beta_{m}^{\mathcal{Q}}=12\int d\hat{n}_1 d\hat{n}_2\int dr_1 dr_2 r_1^2 r_2^2 \mathcal{M}_{m}^{\mathcal{Q}}(\hat{n}_1,\hat{n}_2,r_1,r_2)
\end{eqnarray}
and the form of $\mathcal{M}_{m}^{\mathcal{Q}}$ inferred from the above equations. The estimator has been reduced entirely to tractable integrals and sums which can be performed relatively quickly.
\par
We can estimate the computational time needed to evaluate this estimator. The multipole summation needed for each basis is $\mathcal{O}(l_{\rm{max}})$ (since the sum over the $m$'s can be precomputed). The integral $\sim \int d^2 \hat{n}$ is an $\mathcal{O}(l_{\rm{max}}^2)$ calculation, while the line of sight integral $\sim \int d r$ is conservatively estimated as an $\mathcal{O}(100)$ operation. Therefore, in total the estimated number of operations is $\mathcal{O}(10 000)\times \mathcal{O}(l_{\rm{max}}^5)$.
\par
In the case that the primordial trispectrum is independent of the diagonal $K$, the estimated number of operations reduces to $\mathcal{O}(100)\times \mathcal{O}(l_{\rm{max}}^3)$ as outlined in Section~\ref{sec:SpecialClass}.

\subsection{CMB estimator}
In the case of a precomputed CMB trispectrum or a late-time source of non-Gaussianity in the CMB, such as gravitational lensing or active models such as cosmic strings, we wish to find a late-time CMB estimator. For the late-time analysis we wish to expand the estimator functions using the separable $\overline{\mathcal{Q}}_m(l_1,l_2,l_3,l_4,L)$ mode functions created out of the $\overline{q}_p(l)$ and $\overline{r}_v(L)$ polynomials. (Note that we denote the multipole modes with a bar to distinguish from the primordial equivalents, and also that we have no need for a subscript for the zeroth Legendre mode since the CMB trispectrum is an explicitly five dimensional quantity as described earlier.) In order to effectively expand in mode functions modulated by the $C_l$ we choose to decompose the estimator functions directly as
\begin{eqnarray}\label{DecompExtra}
\frac{v_{l_1}v_{l_2}v_{l_3}v_{l_4}v_{L}}{\sqrt{C_{l_1}C_{l_2}C_{l_3}C_{l_4}}}t^{l_1 l_2}_{l_3 l_4}(L)=\sum_m \overline{\alpha}^{\mathcal{Q}}_m  \overline{\mathcal{Q}}_m, 
\end{eqnarray}
where the separable $v_l$ incorporates the freedom to make the weight function $\omega$ even more scale invariant. The estimator expansion with $C_l$ in~\eqref{EstimatorPrim1} is appropriate for primordial models, but it is expected that flatter choices will be more suitable for late-time anisotropy, such as from cosmic strings.
\par
Substituting this mode expansion into the estimator~\eqref{Estimator} (where again we omit the normalisation factor $N_T$ and return to it later in the section) we find,
\begin{eqnarray}
\mathcal{E}&=&12\sum_{l_i m_i}\sum_{L M}\sum_n \overline{\alpha}^{\mathcal{Q}}_n \overline{q}_p(l_1)\overline{q}_r(l_2)\overline{q}_s(l_3)\overline{q}_u(l_4)\overline{r}_v(L)\int d^2 \hat{n}_1 Y_{l_1 m_1}(\hat{n}_1)Y_{l_2 m_2}(\hat{n}_1)Y^*_{L M}(\hat{n}_1)\nonumber\\
&&\times\int d^2 \hat{n}_2 Y_{l_3 m_3}(\hat{n}_2)Y_{l_4 m_4}(\hat{n}_2)Y_{L M}(\hat{n}_2) \frac{\left(a_{l_1 m_1}a_{l_2 m_2}a_{l_3 m_3}a_{l_4 m_4}\right)_c}{v_{l_1}v_{l_2}v_{l_3}v_{l_4}v_{L}\sqrt{C_{l_1}C_{l_2}C_{l_3}C_{l_4}}}.
\end{eqnarray}
After some algebra we find
\begin{align}
\mathcal{E}^{\rm{tot}}&=12 N\tilde{\Delta}_{\Phi}^3\sum_n \overline{\alpha}_n^{\mathcal{Q}} \int d^2 \hat{n}_1 d^2 \hat{n}_2\overline{M}_p(\hat{n}_1) \overline{M}_r(\hat{n}_1) \overline{M}_s(\hat{n}_2) \overline{M}_u(\hat{n}_2)\overline{N}_v(\hat{n}_1,\hat{n}_2),\\
\mathcal{E}^{\rm{uc}}&=12 N\tilde{\Delta}_{\Phi}^3\sum_n \overline{\alpha}_n^{\mathcal{Q}}\int d^2 \hat{n}_1 d^2 \hat{n}_2 \overline{N}_v(\hat{n}_1,\hat{n}_2)\nonumber\\
&\times\left(\overline{M}^{\rm{uc}}_{p r}(\hat{n}_1,\hat{n}_1) \overline{M}^{\rm{uc}}_{s u}(\hat{n}_2,\hat{n}_2) +\overline{M}^{\rm{uc}}_{p s}(\hat{n}_1,\hat{n}_2) \overline{M}^{\rm{uc}}_{r u}(\hat{n}_1,\hat{n}_2)+\overline{M}^{\rm{uc}}_{p u}(\hat{n}_1,\hat{n}_2) \overline{M}^{\rm{uc}}_{r s}(\hat{n}_1,\hat{n}_2) \right) ,\nonumber
\end{align}
where
\begin{eqnarray}
\overline{M}_p(\hat{n}_1)&=&\sum_{l_1 m_1}\frac{a_{l_1 m_1}Y_{l_1 m_1}(\hat{n}_1) \overline{q}_p(l_1)}{v_{l_1}\sqrt{C_{l_1}}},\nonumber\\
\overline{M}^{\rm{uc}}_{p s}(\hat{n}_1,\hat{n}_2) &=&\sum_{l_1 m_1}\frac{Y_{l_1 m_1}(\hat{n}_1)Y_{l_1 m_1}^*(\hat{n}_2)   \overline{q}_p(l_1) \overline{q}_s(l_1)  }{v_{l_1}^2} ,\nonumber\\
 \overline{N}_v(\hat{n}_1,\hat{n}_2)&=& \sum_{L M}\frac{Y_{L M}^* (\hat{n}_1) Y_{LM}(\hat{n}_2) \overline{r}_v(L)}{v_L}.
\end{eqnarray}
Again we can summarise these results (substituting back in $N_T$) as
\begin{eqnarray}
\mathcal{E}=\frac{N\tilde{\Delta}_{\Phi}^3}{N_T}\sum_n  \overline{\alpha}_n^{\mathcal{Q}}  \overline{\beta}_n^{\mathcal{Q}}, 
\end{eqnarray}
with
\begin{eqnarray}
 \overline{\beta}_n^{\mathcal{Q}} =12\int d^2 \hat{n}_1 d^2 \hat{n}_2 \overline{\mathcal{M}}_n^{\mathcal{Q}}(\hat{n}_1,\hat{n}_2),
\end{eqnarray}
and the form of $\overline{\mathcal{M}}_n^{\mathcal{Q}}(\hat{n}_1,\hat{n}_2)$ can be deduced from the equations for $\mathcal{E}^{\rm{tot}}$ and $\mathcal{E}^{\rm{uc}}$.
\par
Since there are no line of sight integrals ($\sim \int d r$) for this estimator the number of operations required in this case is $\mathcal{O}(l_{\rm{max}}^5)$ suggesting that the late-time estimator is much more computationally efficient than the primordial version.
\par
Similarly to the primordial case, there is a reduction in complexity to $\mathcal{O}(l_{\rm{max}}^3)$ in the case that the late-time extra-reduced trispectrum is independent of the diagonal $L$. This is explained further in Section~\ref{sec:SpecialClass}.

\subsection{Special case of trispectrum independent of diagonal}\label{sec:SpecialClass}
Suppose that the primordial reduced trispectrum is independent of the diagonal $K$. In particular, we write $\mathcal{T}_{\Phi,0}(k_1,k_2,k_3,k_4;K)=\mathcal{T}_{\Phi,0}(k_1,k_2,k_3,k_4)$. In that case the `extra'-reduced trispectrum (see \eqref{TrispRed2} and \eqref{ExtraTrispRed}) becomes
\begin{align}
t^{l_1 l_2}_{l_3 l_4}(L)=&\left(\frac{2}{\pi}\right)^5 \int (k_1 k_2 k_3 k_4 K)^2 dk_1 dk_2 dk_3 dk_4 dK r_1^2 dr_1 r_2^2 dr_2 j_L(K r_1) j_L(K r_2)\nonumber\\
&\times[j_{l_1}(k_1 r_1)\Delta_{l_1}(k_1)][j_{l_2}(k_2 r_1)\Delta_{l_2}(k_2)][j_{l_3}(k_3 r_2)\Delta_{l_3}(k_3)][j_{l_4}(k_4 r_2)\Delta_{l_4}(k_4)] \nonumber\\
&\times\mathcal{T}_{\Phi,0}(k_1,k_2, k_3,k_4).
\end{align}
Next, using equation~\eqref{IdentitySpecial} we find
\begin{eqnarray}
\int dK K^2 j_L(K r_1) j_L(K r_2)=\frac{\pi}{2 r_2^2}\delta(r_2-r_1).
\end{eqnarray}
This implies that
\begin{align}
t^{l_1 l_2}_{l_3 l_4}(L)=&\left(\frac{2}{\pi}\right)^4 \int (k_1 k_2 k_3 k_4)^2 dk_1 dk_2 dk_3 dk_4 r_1^2 dr_1 [j_{l_1}(k_1 r_1)\Delta_{l_1}(k_1)][j_{l_2}(k_2 r_1)\Delta_{l_2}(k_2)]\nonumber\\
&\times[j_{l_3}(k_3 r_1)\Delta_{l_3}(k_3)][j_{l_4}(k_4 r_1)\Delta_{l_4}(k_4)] \mathcal{T}_{\Phi,0}(k_1,k_2, k_3,k_4),
\end{align}
i.e. we only have one line of sight integral. This expression also shows that, if the primordial trispectrum is independent of the diagonal $K$, then $t^{l_1 l_2}_{l_3 l_4}(L)$ is independent of $L$. We can exploit this property in our estimators. From equations~\eqref{RedTrisp},~\eqref{ExtraTrispRed} and equation~\eqref{Gaunt} for the Gaunt integral (which we denote here in the form $\mathcal{G}^{l_1 l_2 l_3}_{m_1 m_2 m_3}$) we have
\begin{eqnarray}
\mathcal{T}_{l_1 m_1 l_2 m_2 l_3 m_3 l_4 m_4}=\sum_{L M}(-1)^M \mathcal{G}^{l_1 l_2 L}_{m_1 m_2 -M} \mathcal{G}^{l_3 l_4 L}_{m_3 m_4 M} t^{l_1 l_2}_{l_3 l_4}(L).
\end{eqnarray}
If the extra-reduced trispectrum is independent of $L$, we can use equation~\eqref{eq:UsefulIdentity} to write
\begin{eqnarray}
\mathcal{T}_{l_1 m_1 l_2 m_2 l_3 m_3 l_4 m_4}=\int d\Omega_{\hat{n}} Y_{l_1 m_1}(\hat{\mathbf{n}})Y_{l_2 m_2}(\hat{\mathbf{n}})Y_{l_3 m_3}(\hat{\mathbf{n}})Y_{l_4 m_4}(\hat{\mathbf{n}}) t^{l_1 l_2}_{l_3 l_4}.
\end{eqnarray}
where we drop the label $L$ from the extra-reduced trispectrum. The extra reduced trispectrum now has the following mode expansion
\begin{eqnarray}
t^{l_1 l_2}_{l_3 l_4}=N\tilde{\Delta}_{\Phi}^3\sum_{m}\alpha_{m}^{\mathcal{Q}}\int dr r^2 \mathcal{Q}_m^{l_1 l_2 l_3 l_4}(r),
\end{eqnarray}
where now we have
\begin{eqnarray}
\mathcal{Q}^{l_1 l_2 l_3 l_4}_{m} (r)=     q_p^{l_1}(r)q_r^{l_2}(r)q_s^{l_3}(r)q_{u}^{l_4}(r),
\end{eqnarray}
with
\begin{eqnarray}
q_p^{l}(r)&=&\frac{2}{\pi}\int dk q_p(k) \Delta_{l}(k)j_{l}(k r).
\end{eqnarray}
The mode decomposition is similar to that described in Section \ref{sec:V} with $r_v = \rm{constant}$. In this case we use the following primordial decomposition
\begin{eqnarray}
(k_1 k_2 k_3 k_4)^2\mathcal{T}_{\Phi,0}(k_1,k_2,k_3,k_4)=\sum_{m}\alpha_{m}^{\mathcal{Q}}q_p(k_1)q_r(k_2)q_s(k_3)q_t(k_4).
\end{eqnarray}
Using this in the expression for the general estimator~\eqref{eq:Estimator2}, which can be re-expressed as
\begin{eqnarray}
\mathcal{E}=\frac{12}{N_T}\sum_{l_i m_i}\mathcal{T}_{l_1 m_1 l_2 m_2 l_3 m_3 l_4 m_4} \left(a_{l_1 m_1}	a_{l_2 m_2}	a_{l_3 m_3}	a_{l_4 m_4}		\right)^{\rm{obs}}_c,
\end{eqnarray}
with
\begin{align}
 \left(a_{l_1 m_1}	a_{l_2 m_2}	a_{l_3 m_3}	a_{l_4 m_4}		\right)^{\rm{obs}}_c=& a_{l_1 m_1}^{\rm{obs}}	a_{l_2 m_2}^{\rm{obs}}	a_{l_3 m_3}^{\rm{obs}}	a_{l_4 m_4}^{\rm{obs}}\nonumber\\
 &- \Big((-1)^{m_1}C_{l_1}\delta_{l_1 l_2}\delta_{m_1 -m_2}a^{\rm{obs}}_{l_3 m_3}a^{\rm{obs}}_{l_4 m_4}+\rm{5\,perms}\Big)\nonumber\\
&+\Big((-1)^{m_1+m_3}\delta_{l_1 l_2}\delta_{m_1 -m_2} \delta_{l_3 l_4}\delta_{m_3 -m_4}C_{l_1}C_{l_3}	+\rm{2\,perms}\Big),
\end{align}
we find
\begin{align}
\mathcal{E}=\frac{12N \tilde{\Delta}_{\Phi}^3}{N_T}\sum_{m}\alpha_{m}^{\mathcal{Q}} \int d \hat{n}\int dr &r^2\Big[  M_p(\hat{n},r) M_r(\hat{n},r) M_s(\hat{n},r) M_t(\hat{n},r)\nonumber\\
&-\Big(M^{\rm{uc}}_{pr}(\hat{n},r) M_s(\hat{n},r)M_t(\hat{n},r)+\rm{5\,perms}\Big)   \nonumber\\
&+ \Big(M^{\rm{uc}}_{pr}(\hat{n},r) M_{s t}^{\rm{uc}}(\hat{n},r)+\rm{2\,perms}\Big)   \Big],
\end{align}
where
\begin{eqnarray}
M_p(\hat{n},r)&=&\sum_{l m}\frac{a_{l m}Y_{l m}(\hat{n})}{C_l} q_p^{l}(r),\nonumber\\
M_{p r}^{\rm{uc}}(\hat{n},r)&=&\sum_{l m}\frac{Y^*_{l m}(\hat{n})Y_{l m}(\hat{n})}{C_l} q_p^{l}(r)q_r^{l}(r).
\end{eqnarray}
\par
We again can estimate the computational time needed to find this estimator. Since we now have only one line of sight integral and one integral over the sky $\sim \int d\hat{n}$ we use the prescription outlined in Section \ref{sec:VI} to estimate the complexity conservatively as $\mathcal{O}(100)\times \mathcal{O}(l_{\rm{max}}^3)$.
\par
The implementation in the case of the late-time estimator for which the extra-reduced trispectrum is independent of $L$ can be found similarly. Since this estimator no longer requires a line of sight integral the complexity of the calculation can be estimated as  $\mathcal{O}(l_{\rm{max}}^3)$.

\subsection{$T_{NL}$ Estimator}
As with shortcomings of normalising the quantity $f_{NL}$ of the bispectrum that was addressed in~\cite{Ferg3}, the current method~\cite{aChen}  of normalising the level of non-Gaussianity due to the trispectrum, $t_{NL}$ poses problems. In particular, the level of non-Gaussianity is found by normalising the shape function against a central point. More specifically we can identify this method as setting $S_{\mathcal{T}}(k,k,k,k,k)=1$ and identifying the normalisation $N$ of equation~\eqref{eq:Shape} as $(50/27)t_{NL}$. In the case of the local model this gives
\begin{eqnarray}
t_{NL}^{\rm{loc}A}=2.16 f_{NL}^2=1.5 \tau_{NL}\,,\qquad t_{NL}^{\rm{loc}B}=1.08 g_{NL},
\end{eqnarray}
where we note again that the relationship between $\tau_{NL}$ and $f_{NL}$ is only strictly true for single field inflation.
This approach assumes scale invariance and therefore will produce inconsistent results between models peaking or dipping at this central point. Also, this definition is not well-defined for models which are not scale-invariant, such as feature models, and it is simply not applicable to non-Gaussian signals created at late times, such as those induced by cosmic strings or secondary anisotropies. An alternative measure of the non-Gaussianity is given by comparison of the primordial trispectrum to the local primordial trispectrum but this approach is not well-defined and is essentially an order of magnitude estimation~\cite{Valen}.
\par
Therefore, we propose a universally defined trispectrum non-Gaussianity parameter $T_{NL}$ which (i) is a measure of the total observational signal expected for the trispectrum of the model in question and (ii) is normalised for direct comparison with the canonical local model (with $g_{NL}=0$). We define $\tilde{T}_{NL}$ from an adapted version of the estimator~\eqref{Estimator} with
\begin{eqnarray}
\tilde{T}_{NL}=\frac{1}{N_T \overline{N}_{T\rm{loc}A}}\sum_{l_i m_i}\frac{\langle a_{l_1m_1}a_{l_2m_2}a_{l_3m_3}a_{l_4m_4}\rangle_c \left(a_{l_1 m_1}^{\rm{obs}}a_{l_2 m_2}^{\rm{obs}}a_{l_3 m_3}^{\rm{obs}}a_{l_4 m_4}^{\rm{obs}}\right)_c}{C_{l_1}C_{l_2}C_{l_3}C_{l_4}},
\end{eqnarray}
where $N_T$ is the appropriate normalisation factor for the given model,
\begin{eqnarray}
N_T^2=\sum_{l_i, L}\frac{\left(T^{l_1 l_2}_{l_3 l_4}(L)\right)^2}{(2L+1)C_{l_1}C_{l_2}C_{l_3}C_{l_4}},
\end{eqnarray}
and $ \overline{N}_{T\rm{loc}A}$ is the normalisation for the local model with $\tau_{NL}=1, \,g_{NL}=0$,
\begin{eqnarray}
 \overline{N}_{T\rm{loc}A}^2&=&\sum_{l_i, L}\frac{\left(T^{l_1 l_2}_{l_3 l_4}(L)^{\rm{loc (\tau_{NL}=1, g_{NL}=0)} }\right)^{2}   }{(2L+1)C_{l_1}C_{l_2}C_{l_3}C_{l_4}}.
 \end{eqnarray}
The $\tilde{T}_{NL}$ estimator will recover $\tau_{NL}$ for the local model with $g_{NL}=0$, while it gives $( \overline{N}_{T\rm{loc}B}/ \overline{N}_{T\rm{loc}A})g_{NL}$ for the local model with $\tau_{NL}=0$ where $ \overline{N}_{T\rm{loc}B}$ is the normalisation for the local model with $\tau_{NL}=0, \,g_{NL}=1$. This coefficient is dependent on $l_{\rm{max}}$ but is a number of order unity.
\par
Results for primordial models should not depend strongly on the multipole cut-off $l_{\rm{max}}$. However, diffusion due to Silk damping in the transfer functions ensures that the primordial signal is exponentially suppressed for $l\gtrsim 2000$. Therefore, an appropriate choice for a canonical cut-off is $l_{\rm{max}}=2000$. Late-time anisotropies, such as cosmic strings, do not generically fall off exponentially for $l\gtrsim 2000$ but, nonetheless, in the domain $l\lesssim 2000$ we can make a meaningful comparison to the local $\tau_{NL}=1,g_{NL}=0$ model. Alternative measures must be proposed beyond this domain. As indicated in Section~\ref{sec:idealestimator} the normalisation factor $N_T^2$ is a computationally intensive calculation. Instead we use the approximation $N_T\approx \overline{N}_{T\rm{loc}A} (N_{\mathcal{T}}/\overline{N}_{\mathcal{T}\rm{loc}A})$ where the subscript $\mathcal{T}$ instead of $T$ refers to using the reduced trispectrum instead of the full trispectrum in the above calculations. With these approximations we need only accurately calculate the full normalisation factor in the case of the model with $\tau_{NL}=1, g_{NL}=0$. Regardless of the accuracy we propose that given the vastly increased speed of the calculation we adopt this latter convention and define $T_{NL}$, i.e.
\begin{eqnarray}\label{TNL}
T_{NL}=\frac{ \overline{N}_{\mathcal{T}\rm{loc}A}}{N_\mathcal{T} \overline{N}_{T\rm{loc}A}^2}\sum_{l_i m_i}\frac{\langle a_{l_1m_1}a_{l_2m_2}a_{l_3m_3}a_{l_4m_4}\rangle_c \left(a_{l_1 m_1}^{\rm{obs}}a_{l_2 m_2}^{\rm{obs}}a_{l_3 m_3}^{\rm{obs}}a_{l_4 m_4}^{\rm{obs}}\right)_c}{C_{l_1}C_{l_2}C_{l_3}C_{l_4}}.
\end{eqnarray}
The relation between $T_{NL}$ and $\tilde{T}_{NL}$ as well as the accuracy of the above approximation for $N_T$ - which is only a conjecture at present- will be explored further in an upcoming paper. However we note here that the $T_{NL}$ estimator also recovers $\tau_{NL}$ in the case of the local model with $g_{NL}=0$.
\par
If the CMB trispectrum is not known precisely for the primordial model under study, then we can make an estimate for the normalisation factor $N_{\mathcal{T}}$ in~\eqref{TNL}  using the shape function for the reduced trispectrum $S_{\mathcal{T}}(k_1,k_2,k_3,k_4,K)$. One can obtain a fairly accurate approximation to the relative normalisations in~\eqref{TNL} from 
\begin{eqnarray}
\hat{N}^2=F(S_{\mathcal{T}},S_{\mathcal{T}})=\int d\mathcal{V}_k S_{\mathcal{T}}^2(k_1,k_2,k_3,k_4,K) \omega(k_1,k_2,k_3,k_4,K),
\end{eqnarray}
where the appropriate weight function was found in~\eqref{Weightk} and the domain $\mathcal{V}_k$ is given
 by \eqref{Domain1}, \eqref{Domain2}.  Using $N_{\mathcal{T}}/\overline{N}_{\mathcal{T}\rm{loc}A}\approx \hat{N}/\hat{N}_{\rm{loc}A}$ to approximate $N_{\mathcal{T}}$ we can make a fairly accurate estimate of the level of non-Gaussianity or can renormalise $\tau_{NL}$ constraints for different models into compatible constraints in a similar manner to the analysis  of $f_{NL}$ constraints in the case of the bispectrum in~\cite{Ferg3}.

\section{Recovering the trispectrum}\label{sec:VII}
\subsection{Recovering the primordial trispectrum}
The form of the estimator in~\eqref{EstimatorPrim1} suggests that further information may be extracted from the observed trispectrum beyond the $\tau_{NL}$ for one specific theoretical model. This is because, through the coefficients $\beta_{m}^{\mathcal{Q}}$, we have obtained some sort of mode decomposition of the trispectrum of the observational map. Consider the expectation value of $\beta_{m}^{\mathcal{Q}}$ obtained from an ensemble of maps generated for a particular theoretical model with shape function $S_{\mathcal{T}}=\sum_{m}\alpha_{m}^{\mathcal{Q}}\mathcal{Q}_m$. Since the shape function is in terms of the zeroth mode of the Legendre expansion of the primordial trispectrum, we can only hope to recover information about this mode via recovery of the shape function\footnote{This is to be somewhat expected since, although the primordial (reduced) trispectrum may be a six dimensional quantity, the CMB (reduced) trispectrum is explicitly five dimensional.}. Using the expression
\begin{align}
\langle \beta_{m}^{\mathcal{Q}}\rangle=&12\sum_{l_i m_i}\sum_{LM}\Bigg[\left(\int d\hat{n}_1 Y_{l_1 m_1}(\hat{n}_1)Y_{l_2 m_2}(\hat{n}_1)Y_{L M}^*(\hat{n}_1)\right)\nonumber\\ &\times\left(\int d\hat{n}_2 Y_{l_3 m_3}(\hat{n}_2)Y_{l_4 m_4}(\hat{n}_2)Y_{L M}(\hat{n}_2)\right)\nonumber\\
&\times \int r_1^2 dr_1 r_2^2 dr_2\mathcal{Q}^{l_1 l_2 l_3 l_4 L}_{m} (r_1,r_2)\Big] \frac{\langle a_{l_1 m_1} a_{l_2 m_2} a_{l_3 m_3} a_{l_4 m_4} \rangle_c}{C_{l_1}C_{l_2}C_{l_3}C_{l_4}},
\end{align}
as well as the identity for the Wigner $6j$ symbol in Section~\ref{sec:idealestimator} we find, after some algebra,
\begin{align}
\langle \beta_{m}^{\mathcal{Q}}\rangle&\equiv\sum_{m'}\Gamma_{m m'}\alpha^{\mathcal{Q}}_{m'}=\nonumber\\
12 \sum_{l_i,L}&\int dr_1 dr_2 r_1^2 r_2^2 \mathcal{Q}_{m}^{l_1 l_2 l_3 l_4 L}(r_1, r_2)\sum_{m'}\alpha^{\mathcal{Q}}_{m'}\Bigg[h_{l_1 l_2 L}^2 h_{l_3 l_4 L}^2\int dr_1 dr_2 r_1^2 r_2^2 \mathcal{P}_{m'}^{l_1 l_2 l_3 l_4 L}(r_1, r_2)\nonumber\\
+\sum_{L'}& h_{l_1 l_2 L}h_{l_3 l_4 L}h_{l_1 l_3 L'}h_{l_2 l_4 L'}(-1)^{l_2+l_3} \left\{ \begin{array}{ccc}
l_1 & l_2 & L \\
l_4 &l_3 & L' \end{array} \right\}\int dr_1 dr_2 r_1^2 r_2^2 \mathcal{P}_{m'}^{l_1 l_3 l_2 l_4 L'}(r_1, r_2)\nonumber\\
+\sum_{L'}& h_{l_1 l_2 L}h_{l_3 l_4 L}h_{l_1 l_4 L'}h_{l_2 l_3 L'}(-1)^{L+L'} \left\{ \begin{array}{ccc}
l_1 & l_2 & L \\
l_3 &l_4 & L' \end{array} \right\}\int dr_1 dr_2 r_1^2 r_2^2 \mathcal{P}_{m'}^{l_1 l_4 l_2 l_3 L'}(r_1, r_2)\Bigg],
\end{align}
where $\mathcal{P}_{m'}^{l_1 l_2 l_3 l_4 L}=\mathcal{Q}_{m'}^{l_1 l_2 l_3 l_4 L}+\mathcal{Q}_{m'}^{l_2 l_1 l_3 l_4 L}+\mathcal{Q}_{m'}^{l_1 l_2 l_4 l_3 L}+\mathcal{Q}_{m'}^{l_2 l_1 l_4 l_3 L}$. The quantity $\Gamma_{m m'}$ represents a matrix with positions labelled by $m,m'$ and can be inferred readily by the above equation. Inverting the relationship we can recover the $\alpha_{m}^{\mathcal{Q}}$ via
\begin{eqnarray}
\alpha_{m}^{\mathcal{Q}}=\sum_{m'}(\Gamma^{-1})_{m m'}\langle \beta_{m'}^{\mathcal{Q}}\rangle.
\end{eqnarray}
Therefore, if the decomposition coefficients are found with adequate significance, we can reconstruct the shape function through the expansion
\begin{eqnarray}
S_{\mathcal{T}}=\sum_{m}\sum_{ m'}(\Gamma^{-1})_{m m'} \beta_{m'}^{\mathcal{Q}}\mathcal{Q}_m.
\end{eqnarray}
This reconstruction will be sufficient to uniquely define the planar case (independent of $\theta_4$) (as well as the 
general CMB case in the next section).  However, as already discussed the CMB shape function only gives five dimenional information whereas the primordial trispectrum may be six dimensional (non-planar). Therefore, recovery of the full primordial trispectrum is compromised by this degeneracy. However, as we have discussed in Section \ref{sec:II}, this degeneracy may be broken by using other probes of non-Gaussianity such as galaxy surveys and $21$ cm observations. We note also that the calculation of the matrix $\Gamma$ is computationally intensive due to the presence of the Wigner $6j$ symbols. Nonetheless we include the discussion here for completeness. 

\subsection{Recovering the CMB trispectrum}
The recovery of the CMB trispectrum from a given observational map is more straightforward
(as for the bispectrum).  
In a similar fashion to the calculation in the case of the primordial trispectrum, we obtain the result that
\begin{align}
\langle \overline{\beta}_{n}^{\mathcal{Q}}\rangle=&12\sum_{l_i m_i}\sum_{LM}\Bigg[\left(\int d\hat{n}_1 Y_{l_1 m_1}(\hat{n}_1)Y_{l_2 m_2}(\hat{n}_1)Y_{L M}^*(\hat{n}_1)\right)\nonumber\\
&\times \left(\int d\hat{n}_2 Y_{l_3 m_3}(\hat{n}_2)Y_{l_4 m_4}(\hat{n}_2)Y_{L M}(\hat{n}_2)\right) \int r_1^2 dr_1 r_2^2 dr_2\overline{\mathcal{Q}}^{l_1 l_2 l_3 l_4 L}_{n} (r_1,r_2)\Big] \nonumber\\
&\times\frac{\langle a_{l_1 m_1} a_{l_2 m_2} a_{l_3 m_3} a_{l_4 m_4} \rangle_c}{v_{l_1}v_{l_2}v_{l_3}v_{l_4}v_{L}\sqrt{C_{l_1}C_{l_2}C_{l_3}C_{l_4}}}\nonumber\\
\implies \langle \overline{\beta}_{n}^{\mathcal{Q}}\rangle =&12 \sum_{l_i,L}\overline{Q}^{l_1 l_2 l_3 l_4 L}_n\sum_{p} \overline{\alpha}^{\mathcal{Q}}_p\Bigg[ h_{l_1 l_2 L}^2 h_{l_3 l_4 L}^2 \mathcal{P}_{p}^{l_1 l_2 l_3 l_4 L}\nonumber\\
&+\sum_{L'} h_{l_1 l_2 L}h_{l_3 l_4 L}h_{l_1 l_3 L'}h_{l_2 l_4 L'}(-1)^{l_2+l_3} \left\{ \begin{array}{ccc}
l_1 & l_2 & L \\
l_4 &l_3 & L' \end{array} \right\} \mathcal{P}_{p}^{l_1 l_3 l_2 l_4 L'}\nonumber\\
&+\sum_{L'} h_{l_1 l_2 L}h_{l_3 l_4 L}h_{l_1 l_4 L'}h_{l_2 l_3 L'}(-1)^{L+L'} \left\{ \begin{array}{ccc}
l_1 & l_2 & L \\
l_3 &l_4 & L' \end{array} \right\} \mathcal{P}_{p}^{l_1 l_4 l_2 l_3 L'}    \Bigg]\nonumber\\
=&\sum_{p}\overline{\Gamma}_{n p}\alpha^{\mathcal{Q}}_{p},
\end{align}
where $\overline{Q}^{l_1 l_2 l_3 l_4 L}=\overline{q}_p(l_1)\overline{q}_r(l_2)\overline{q}_s(l_3)\overline{q}_u(l_4)\overline{r}_v(L)$ and $\mathcal{P}_{p}^{l_1 l_2 l_3 l_4 L}=\mathcal{Q}_{p}^{l_1 l_2 l_3 l_4 L}+\mathcal{Q}_{p}^{l_2 l_1 l_3 l_4 L}+\mathcal{Q}_{p}^{l_1 l_2 l_4 l_3 L}+\mathcal{Q}_{p}^{l_2 l_1 l_4 l_3 L}$. Inverting the matrix $\overline{\Gamma}_{n p}$ we find 
\begin{eqnarray}
\alpha^{\mathcal{Q}}_{n}=\sum_p \overline{\Gamma}^{-1}_{n p} \langle \overline{\beta}_{p}^{\mathcal{Q}}\rangle.
\end{eqnarray}
Therefore, if we can measure the coefficients $\overline{\beta}_{p}^{\mathcal{Q}}$ with significance from a particular experiment, we can reconstruct the trispectrum map using~\eqref{DecompExtra},
\begin{eqnarray}
t^{l_1 l_2}_{l_3 l_4}(L)=\frac{\sqrt{C_{l_1}C_{l_2}C_{l_3}C_{l_4}}}{v_{l_1}v_{l_2}v_{l_3}v_{l_4}v_{L}}\sum_{n p}\overline{\Gamma}^{-1}_{n p}  \overline{\beta}_{p}^{\mathcal{Q}}\overline{\mathcal{Q}}_n.
\end{eqnarray}
The calculation of the matrix $\overline{\Gamma}$ remains computationally intensive but it is 
tractable.    We can, in principle, extract the full CMB trispectrum which, together with the extracted
CMB bispectrum~\cite{Ferg3},  will prove to be a key test of the Gaussianity of the Universe. 

\section{Map Making}\label{sec:VIII}
In this section we derive an algorithm for creating a non-Gaussian map with given trispectrum,
developing methods presented for the bispectrum in ref.~\cite{Smith} and generalised in ref.~\cite{Ferg3}.
 This algorithm is valid in the limit of weak non-Gaussianity. 
\par
We define the function
\begin{eqnarray}
T_2[a^G]=\frac{1}{24}\sum_{l_i m_i}T_{l_1 m_1 l_2 m_2 l_3 m_3 l_4 m_4}a^G_{l_1 m_1}a^G_{l_2 m_2}a^G_{l_3 m_3}a^G_{l_4 m_4},
\end{eqnarray}
where $a^G_{lm}$ is the Gaussian part of the CMB multipoles, generated using the angular power spectrum $C_l$, while $T_{l_1 m_1 l_2 m_2 l_3 m_3 l_4 m_4}$ is the given trispectrum of the theoretical model for which simulations are required.
\par
Setting
\begin{align}
a_{l m}'&=a_{l m}^G +\frac{1}{6}\sum_{l_i m_i}b_{l l_2 l_3}\mathcal{G}^{l l_2 l_3}_{m m_2 m_3}\frac{a_{l_2 m_2}^{* G}}{C_{l_2}} \frac{a_{l_3 m_3}^{* G}}{C_{l_3}}
+\frac{1}{4}\frac{\partial}{\partial a_{l m}^*}T_2[C^{-1} a^G]\nonumber\\
&=a_{l m}^G +\frac{1}{6}\sum_{l_i m_i}b_{l l_2 l_3}\mathcal{G}^{l l_2 l_3}_{m m_2 m_3}\frac{a_{l_2 m_2}^{* G}}{C_{l_2}} \frac{a_{l_3 m_3}^{* G}}{C_{l_3}}
+\frac{1}{24}\sum_{l_i m_i}T_{l m l_2 m_2 l_3 m_3 l_4 m_4}\frac{a_{l_2 m_2}^{* G}}{C_{l_2}}\frac{a_{l_3 m_3}^{* G}}{C_{l_3}}\frac{a_{l_4 m_4}^{* G}}{C_{l_4}},
\end{align}
we recover the bispectrum from $\langle a_{l_1 m_1}' a_{l_2 m_2}' a_{l_3 m_3}'\rangle$ (as described in \cite{Smith}). Next, we calculate the four-point correlator of the $a_{lm}'$s in the absence of a bispectrum and find
\begin{align}
\langle a_{l_1 m_1}' a_{l_2 m_2}' a_{l_3 m_3}'a_{l_4 m_4}'\rangle=&\langle a_{l_1 m_1}^G a_{l_2 m_2}^G a_{l_3 m_3}^G a_{l_4 m_4}^G \rangle\nonumber\\
&+\langle\frac{1}{24}\sum_{l_j m_j}T_{l_1 m_1 l_b m_b l_c m_c l_d m_d}\frac{a_{l_b m_b}^{* G}}{C_{l_b}}\frac{a_{l_c m_c}^{* G}}{C_{l_c}}\frac{a_{l_d m_d}^{* G}}{C_{l_d}}a_{l_2 m_2}^G a_{l_3 m_3}^G a_{l_4 m_4}^G\rangle\nonumber\\
&+\rm{permutations},
\end{align}
where $j\in (b,c,d)$. The first term clearly gives the desired unconnected component of the four-point correlator. We note that the contribution from the bispectrum term (which we denote as $a_{lm}^{NG,B}$) does not vanish due to contributions of the form $\langle a^G a^G a^{NG,B} a^{NG,B}\rangle$. Therefore, in order to recover the correct trispectrum, it is necessary to subtract this contribution. This issue will be discussed further later in the section. 

Evaluating the correlators in the second term on the right hand side and adding up the different permutations we find
\begin{eqnarray}
\langle a_{l_1 m_1}' a_{l_2 m_2}' a_{l_3 m_3}'a_{l_4 m_4}'\rangle=
\langle a_{l_1 m_1}^G a_{l_2 m_2}^G a_{l_3 m_3}^G a_{l_4 m_4}^G \rangle+ T_{l_1 m_1 l_2 m_2 l_3 m_3 l_4 m_4}.
\end{eqnarray}
This verifies the validity of the use of $a_{lm}'$ to make maps including the Gaussian and trispectrum contributions to the model under study.
\par
We observe, using \eqref{TotalRedTrisp}, that we may write $T_2[a^G]$ in the form
\begin{eqnarray}
T_2[a^G]=\frac{1}{2}\sum_{l_i m_i}\mathcal{T}_{l_1 m_1 l_2 m_2 l_3 m_3 l_4 m_4}a^G_{l_1 m_1} a^G_{l_2 m_2}a^G_{l_3 m_3}a^G_{l_4 m_4}.
\end{eqnarray}
Using this formula we may also rewrite the trispectrum contribution to $a_{l m}'$, which we denote $a_{l m}^{NG,T}$ as
\begin{align}\label{almT}
a_{l m}^{NG,T}=&\frac{1}{24}\sum_{l_i m_i}T_{l m l_2 m_2 l_3 m_3 l_4 m_4}\frac{a_{l_2 m_2}^{* G}}{C_{l_2}}\frac{a_{l_3 m_3}^{* G}}{C_{l_3}}\frac{a_{l_4 m_4}^{* G}}{C_{l_4}}\\
=&\frac{1}{8}\sum_{l_i m_i}\left(\mathcal{T}_{l m l_2 m_2 l_3 m_3 l_4 m_4}+\mathcal{T}_{l_2 m_2 l m l_3 m_3 l_4 m_4}+\mathcal{T}_{l_2 m_2 l_3 m_3 l m l_4 m_4}+\mathcal{T}_{l_2 m_2 l_3 m_3 l_4 m_4 l m}\right)\nonumber\\
&\times\frac{a_{l_2 m_2}^{* G}}{C_{l_2}}\frac{a_{l_3 m_3}^{* G}}{C_{l_3}}\frac{a_{l_4 m_4}^{* G}}{C_{l_4}}\nonumber.
\end{align}
Using the formulae for the extra-reduced trispectrum \eqref{ExtraTrispRed} and the Gaunt integral \eqref{Gaunt}, we note that
\begin{align}
\mathcal{T}_{l_1 m_1 l_2 m_2 l_3 m_3 l_4 m_4}=&\sum_{L M}\left(\int d\Omega_{\hat{n}_1}Y_{l_1 m_1}(\hat{\mathbf{n}}_1)Y_{l_2 m_2}(\hat{\mathbf{n}}_1)Y_{L M}^*(\hat{\mathbf{n}}_1)\right)\nonumber\\
&\times\left(\int d\Omega_{\hat{n}_2}Y_{l_3 m_3}(\hat{\mathbf{n}}_2)Y_{l_4 m_4}(\hat{\mathbf{n}}_2)Y_{L M}(\hat{\mathbf{n}}_2)\right)t^{l_1 l_2}_{l_3 l_4}(L).
\end{align}
As an aside, we note that if $t^{l_1 l_2}_{l_3 l_4}(L)$ is independent of the diagonal $L$ then we can use equation~\eqref{eq:UsefulIdentity} to write
\begin{eqnarray}\label{SpecialId}
\mathcal{T}_{l_1 m_1 l_2 m_2 l_3 m_3 l_4 m_4}&=&\int d\Omega_{\hat{n}}Y_{l_1 m_1}(\hat{\mathbf{n}})Y_{l_2 m_2}(\hat{\mathbf{n}})Y_{l_3 m_3}(\hat{\mathbf{n}})Y_{l_4 m_4}(\hat{\mathbf{n}})t^{l_1 l_2}_{l_3 l_4},
\end{eqnarray}
where we drop the label $L$ from the extra-reduced trispectrum. This special class of trispectra is explored further in Section~\ref{sec:SpecialClass}.
\par
Next, we return to the issue of a spurious trispectrum contribution due to terms of the form $\langle a^G a^G a^{NG,B} a^{NG,B}\rangle$. Setting the trispectrum term $a_{lm}^{NG,T}$ to zero, the four-point (connected) correlator may be shown to give
\begin{align}\label{trispexpression1}
\langle a_{l_1 m_1}'a_{l_2 m_2}'a_{l_3 m_3}'a_{l_4 m_4}'\rangle_c=\frac{4}{9}\sum_{L M}(-1)^M\Big[\frac{b_{l_1 l_2 L}b_{l_3 l_4 L}}{C_L}&\mathcal{G}^{l_1 l_2 L}_{m_1 m_2 -M} \mathcal{G}^{l_3 l_4 L}_{m_3 m_4 M} \nonumber\\
&+(2\leftrightarrow 3)+(2\leftrightarrow 4)\Big].
\end{align}
Now consider the expression
\begin{align}
\hat{a}_{lm}^{NG,T}=\frac{1}{18}\sum_{l_i m_i}\sum_{L M}(-1)^M\frac{b_{l l_2 L}b_{l_3 l_4 L}}{C_{l_2}C_{l_3}C_{l_4}C_L}\mathcal{G}^{l l_2 L}_{m m_2 -M} \mathcal{G}^{l_3 l_4 L}_{m_3 m_4 M} a_{l_2 m_2}^{G*}a_{l_3 m_3}^{G*}a_{l_4 m_4}^{G*}.
\end{align}
It may easily be shown that evaluation of the quantity $\langle \hat{a}_{l_1 m_1}\hat{a}_{l_2 m_2} \hat{a}_{l_3 m_3} \hat{a}_{l_4 m_4}\rangle_c$, where $ \hat{a}_{l m}=a_{lm}^G+\hat{a}_{lm}^{NG,T}$, yields exactly the right hand side of equation \eqref{trispexpression1}. Therefore, in the case of a non-zero bispectrum, the replacement of trispectrum contribution by the expression
\begin{align}
a^{NG,T}_{lm}\rightarrow a^{NG,T}_{lm}-\hat{a}^{NG,T}_{lm}
\end{align}
yields the correct trispectrum\footnote{Of course, this replacement does not alter the three point correlator.}.
\par
Denoting the bispectrum contribution to $a_{lm}'$ as $a_{l m}^{NG, B}$, we have verified the following prescription for forming maps including the bispectrum and trispectrum contributions
\begin{eqnarray}
a_{l m}'=a_{lm}^G+f_{NL }\tilde{a}_{lm}^{NG, B}+\tau_{NL}\tilde{a}_{lm}^{NG, T},
\end{eqnarray}
where we have written $a_{lm}^{NG, B}=f_{NL }\tilde{a}_{lm}^{NG, B}$ and $a_{lm}^{NG, T}=\tau_{NL }\tilde{a}_{lm}^{NG, T}$ to make the size of the respective non-Gaussian components more explicit. In the remainder of the paper we shall only consider map-making in the absence of a bispectrum contribution.
\par
Since the computation of the reduced trispectrum is more efficient using the late time expression (due to the absence of the line of sight integrals) we write out the formula for $a_{l m}^{NG,T}$ using the late-time mode decomposition. It is straightforward to find the equivalent formula using the primordial expression. Later in the section we present a particular application using the primordial local model of this formalism.
\par
The late-time mode decomposition of the extra-reduced trispectrum $t^{l_1 l_2}_{l_3 l_4}(L)$ - as detailed at the end of Section \ref{sec:V} - may be written as
\begin{eqnarray}\label{Special}
t^{l_1 l_2}_{l_3 l_4}(L)=\sum_n \overline{\alpha}_n^{\mathcal{Q}}\overline{q}_p(l_1)\overline{q}_r(l_2)\overline{q}_s(l_3)\overline{q}_u(l_4)\overline{r}_v(L).
\end{eqnarray}
Using these expressions we have
\begin{align}
\sum_{l_i m_i}&\mathcal{T}_{l m l_2 m_2 l_3 m_3 l_4 m_4}\frac{a_{l_2 m_2}^{* G}}{C_{l_2}}\frac{a_{l_3 m_3}^{* G}}{C_{l_3}}\frac{a_{l_4 m_4}^{* G}}{C_{l_4}}\nonumber\\
&=\sum_n \overline{\alpha}_n^{\mathcal{Q}}\int d\Omega_{\hat{\mathbf{n}}_1}d\Omega_{\hat{\mathbf{n}}_2}Y_{l m}(\hat{\mathbf{n}}_1)\overline{q}_p(l)\overline{\mathcal{M}}_r(\hat{\mathbf{n}}_1)\overline{\mathcal{M}}_s(\hat{\mathbf{n}}_2)\overline{\mathcal{M}}_u(\hat{\mathbf{n}}_2)\overline{\mathcal{N}}_v(\hat{\mathbf{n}}_1,\hat{\mathbf{n}}_2),
\end{align}
where
\begin{eqnarray}
\overline{\mathcal{M}}_r(\hat{\mathbf{n}}_1)&=&\sum_{l_2 m_2}\frac{Y_{l_2 m_2}(\hat{\mathbf{n}}_1)a_{l_2 m_2}^{* G}}{C_{l_2}}\overline{q}_r(l_2),\nonumber\\
\overline{\mathcal{N}}_v(\hat{\mathbf{n}}_1,\hat{\mathbf{n}}_2)&=&\overline{\mathcal{N}}_v(\hat{\mathbf{n}}_2,\hat{\mathbf{n}}_1)=\sum_{L M}Y_{L M}^*(\hat{\mathbf{n}}_1)Y_{L M}(\hat{\mathbf{n}}_2)\overline{r}_v(L).
\end{eqnarray}
Evaluating, in a similar way, the other terms in equation~\eqref{almT} we find
\begin{align}
a_{l m}^{NG,T}=&\frac{1}{8}\sum_n \overline{\alpha}_n^{\mathcal{Q}}\int d\Omega_{\hat{\mathbf{n}}_1}d\Omega_{\hat{\mathbf{n}}_2}Y_{l m}(\hat{\mathbf{n}}_1)\Big[\left(\overline{q}_p(l)\overline{\mathcal{M}}_r(\hat{\mathbf{n}}_1)+\overline{q}_r(l)\overline{\mathcal{M}}_p(\hat{\mathbf{n}}_1) \right)\overline{\mathcal{M}}_s(\hat{\mathbf{n}}_2) \overline{\mathcal{M}}_u(\hat{\mathbf{n}}_2) \nonumber\\
&+\left(\overline{q}_s(l)\overline{\mathcal{M}}_u(\hat{\mathbf{n}}_1)+\overline{q}_u(l)\overline{\mathcal{M}}_s(\hat{\mathbf{n}}_1) \right)\overline{\mathcal{M}}_p(\hat{\mathbf{n}}_2) \overline{\mathcal{M}}_r(\hat{\mathbf{n}}_2) 
\Big]\overline{\mathcal{N}}_v(\hat{\mathbf{n}}_1,\hat{\mathbf{n}}_2).
\end{align}
As emphasised in~\cite{Ferg3} the condition that the map has the power spectrum $C_l$ specified in the imput will only be satisfied if the power spectrum of the non-Gaussian components $C_l^{NG}$ is small. Therefore, one has to ascertain that spuriously large $C_l^{NG}$ contributions do not affect the overall power spectrum significantly. We will discuss the implementation of the algorithm presented here in an upcoming paper \cite{Regan2}.

\subsection{Application to the Local Model}
The reduced trispectrum for the local model, as shown in Section \ref{sec:IV}, is made up of two terms which we denoted $\rm{locA}$ and $\rm{locB}$. As shown in \cite{Okamoto} - and can be deduced from Section \ref{sec:IV} - the extra reduced trispectra may be expressed as
\begin{eqnarray}
{t^{l_1 l_2}_{l_3 l_4}(L)}^{\rm{locA}}&=&\frac{25}{9}\tau_{NL}\int d r_1 d r_2 r_1^2 r_2^2 F_L (r_1,r_2) \alpha_{l_1}(r_1)\beta_{l_2}(r_1)\alpha_{l_3}(r_2)\beta_{l_4}(r_2),\\
{t^{l_1 l_2}_{l_3 l_4}(L)}^{\rm{locB}}&=&g_{NL}\int d r r^2 \beta_{l_2}(r) \beta_{l_4}(r)\left(\mu_{l_1}(r)\beta_{l_3}(r)+\beta_{l_1}(r)\mu_{l_3}(r)\right),
\end{eqnarray}
where
\begin{eqnarray}
F_L (r_1,r_2)&=&\frac{2}{\pi}\int K^2 dK P_{\Phi}(K)j_L(K r_1)j_L(K r_2),\nonumber\\
\alpha_{l}(r)&=&\mu_{l}(r)=\frac{2}{\pi}\int k^2 dk \Delta_{l}(k) j_l(k r),\nonumber\\
\beta_{l}(r)&=&\frac{2}{\pi}\int k^2 dk P_{\Phi}(k) \Delta_{l}(k) j_l(k r).
\end{eqnarray}
Using these formulae, and exploiting that the $\rm{locB}$ trispectrum is independent of the diagonal $L$ with equation~\eqref{SpecialId}, we find
\begin{align}
\sum_{l_i m_i}\mathcal{T}_{l m l_2 m_2 l_3 m_3 l_4 m_4}^{\rm{locA}}\frac{a_{l_2 m_2}^{* G}}{C_{l_2}}\frac{a_{l_3 m_3}^{* G}}{C_{l_3}}\frac{a_{l_4 m_4}^{* G}}{C_{l_4}}=&\frac{25}{9}\tau_{NL}\int d r_1 dr_2 r_1^2 r_2^2 \alpha_l(r_1)\int d\Omega_{\hat{\mathbf{n}}_1}d\Omega_{\hat{\mathbf{n}}_2}Y_{l m}(\hat{\mathbf{n}}_1)\nonumber\\
\times \mathcal{M}_F (\hat{\mathbf{n}}_1,\hat{\mathbf{n}}_2,r_1,r_2)&\mathcal{M}_{\beta}(\hat{\mathbf{n}}_1,r_1)\mathcal{M}_{\alpha}(\hat{\mathbf{n}}_2,r_2)\mathcal{M}_{\beta}(\hat{\mathbf{n}}_2,r_2),\\
\sum_{l_i m_i}\mathcal{T}_{l m l_2 m_2 l_3 m_3 l_4 m_4}^{\rm{locB}}\frac{a_{l_2 m_2}^{* G}}{C_{l_2}}\frac{a_{l_3 m_3}^{* G}}{C_{l_3}}\frac{a_{l_4 m_4}^{* G}}{C_{l_4}}=&g_{NL}\int d r r^2\int d \Omega_{\hat{\mathbf{n}}}Y_{l m}(\hat{\mathbf{n}})\nonumber\\
\times\Bigg[ \mu_l (r)\mathcal{M}_{\beta}(\hat{\mathbf{n}},r)+ &\beta_l (r)\mathcal{M}_{\mu}(\hat{\mathbf{n}},r)\Bigg]\mathcal{M}_{\beta}(\hat{\mathbf{n}},r)\mathcal{M}_{\beta}(\hat{\mathbf{n}},r),
\end{align}
where
\begin{eqnarray}
\mathcal{M}_{\alpha}(\hat{\mathbf{n}},r)&=&\mathcal{M}_{\mu}(\hat{\mathbf{n}},r)=\sum_{l m}\alpha_l(r)\frac{Y_{l m}(\hat{\mathbf{n}}) a_{l m}^{* G}}{C_{l}},\nonumber\\
\mathcal{M}_{\beta}(\hat{\mathbf{n}},r)&=&\sum_{l m}\beta_l(r)\frac{Y_{l m}(\hat{\mathbf{n}}) a_{l m}^{* G}}{C_{l}},\nonumber\\
\mathcal{M}_F(\hat{\mathbf{n}}_1,\hat{\mathbf{n}}_2,r_1,r_2)&=& \sum_{L M}Y_{L M}^*(\hat{\mathbf{n}}_1)Y_{L M}(\hat{\mathbf{n}}_2)F_L(r_1,r_2).
\end{eqnarray}
Similarly, we evaluate the other terms in~\eqref{almT} to get
\begin{align}
(a_{l m}^{NG,T})_{\rm{locA}}=&\frac{25}{36}\tau_{NL}\int d r_1 dr_2 r_1^2 r_2^2 \mathcal{M}_{\alpha}(\hat{\mathbf{n}}_2,r_2)\mathcal{M}_{\beta}(\hat{\mathbf{n}}_2,r_2)\mathcal{M}_{F}(\hat{\mathbf{n}}_1,\hat{\mathbf{n}}_2,r_1,r_2)\nonumber\\
\times \Big[\alpha_l(r_1)\int & d\Omega_{\hat{\mathbf{n}}_1}d\Omega_{\hat{\mathbf{n}}_2}Y_{l m}(\hat{\mathbf{n}}_1)\mathcal{M}_{\beta}(\hat{\mathbf{n}}_1,r_1)+\beta_l(r_1)\int d\Omega_{\hat{\mathbf{n}}_1}d\Omega_{\hat{\mathbf{n}}_2}Y_{l m}(\hat{\mathbf{n}}_1)\mathcal{M}_{\alpha}(\hat{\mathbf{n}}_1,r_1)\Big],\\
(a_{l m}^{NG,T})_{\rm{locB}}=&\frac{1}{4}g_{NL}\int d r r^2 \Bigg[\mu_l(r)\int d\Omega_{\hat{\mathbf{n}}}Y_{l m}(\hat{\mathbf{n}})\mathcal{M}_{\beta}(\hat{\mathbf{n}},r)\mathcal{M}_{\beta}(\hat{\mathbf{n}},r)\mathcal{M}_{\beta}(\hat{\mathbf{n}},r)\nonumber
\\&+         \beta_l(r)\int d\Omega_{\hat{\mathbf{n}}}   Y_{l m}(\hat{\mathbf{n}})\times\Bigg(3 \mathcal{M}_{\beta}(\hat{\mathbf{n}},r)\mathcal{M}_{\mu}(\hat{\mathbf{n}},r)\mathcal{M}_{\beta}(\hat{\mathbf{n}},r)\Bigg)\Bigg].
\end{align}
In the case of the bispectrum, direct implementation of the explicitly separable local shape results in spuriously large $C_l^{NG}$ contributions. However, it was found that using the eigenmode expansion in ref.~\cite{Ferg3} was much more robust circumventing such effects because of the bounded nature
of the polynomial eigenmodes. This improvement is expected to occur for the trispectrum. An alternative method is to regularise the expressions given here by eliminating pathological terms, while leaving the final trispectrum of the map unchanged.  For arbitrary separable trispectra (unlike the eigenmode
expansion),  convergence must be achieved by hand on a case-by-case basis.

\section{Conclusions}\label{sec:IX}
We have described in this chapter two comprehensive pipelines for the analysis of general primordial or CMB trispectra. The methods are based on mode expansions, exploiting a complete orthonormal eigenmode basis to efficiently decompose arbitrary trispectra into a separable polynomial expansion. These separable mode expansions allow for a reduction of the computational overhead to tractable levels, regardless of whether the reduced trispectrum is being computed at Planck resolution, or we are directly finding an estimator for the size of the trispectrum from a real data set. A shape decomposition has been described allowing for a visualisation of a scale invariant reduced trispectrum on particular slices.
\par
We have presented a correlator for comparing trispectra. We have also defined a correlator for comparing the shape functions that is expected to closely approximate the former. However, the main purpose of this chapter was to present a detailed theoretical framework for finding an estimator for the size of the trispectrum using separable eigenmode expansions. Using this efficient method for finding an estimator for the trispectrum we have defined a universal measure $T_{NL}$ which will allow for consistent comparison between theoretical models. This measure can be calculated for both primordial models and late-time models, e.g. due to active models such as cosmic strings. The completeness of the orthogonal eigenmodes should allow for a reconstruction of the full CMB trispectrum from the data, assuming the presence of a sufficiently significant non-Gaussian signal. We have also detailed an algorithm for producing non-Gaussian simulations with a given power spectrum, bispectrum and trispectrum. The implementation of these methods will be discussed in Chapter~\ref{chapter:evaluation}. Clearly, the full implementation of the primordial and late-time pipelines represents a significant challenge. However the generality and robustness of the methodology described here indicates that there is an intriguing possibility of exploring and constraining a wide class of non-Gaussian models using the trispectrum.

\newpage

\thispagestyle{empty}
\mbox{}
\newpage
\chapter{Implementation of CMB Trispectrum Analysis}
\label{chapter:evaluation}
\textbf{Summary}\\
\textit{In this chapter we present the application of a separable mode expansion estimator with WMAP
data to estimate the trispectrum for a special class of non-Gaussian models. This special class
are explicitly those models for which the trispectrum is independent of the diagonal. This includes
the cubic term of local model inflation, the equilateral model and the so-called `constant' model.
We review the estimator methodology which may be applied to any non-separable primordial and CMB diagonal-free trispectrum. We also demonstrate how to reconstruct the estimator using (single field) local map simulations. Constraints from the single field local model $g_{NL}=(1.62\pm 6.98)\times 10^5$ are broadly consistent with results from large scale structure. The constraints on the equilateral model, $t_{NL}^{{\rm{equil}}}=(-3.11\pm 7.5)\times 10^6 $, are a notable new result, as are the constraints on the constant model $t_{NL}^{\rm{const}}=(-1.33\pm 3.62)\times 10^6 $. The results are consistent with Gaussianity. We also present the $2\sigma$ constraint on cosmic strings, $G\mu\lesssim 1.1\times 10^{-6} $, and establish that the trispectrum is expected to provide, perhaps, the best probe for cosmic strings in the near future. We discuss the importance of constraining the more general class of trispectra, as well as the possibility of obtaining constraints with higher precision using the Planck satellite.
}
\vspace{50pt}

\section{Introduction}
Measurements of the cosmic microwave background (CMB) and large scale structure, such as those provided by the WMAP satellite or the Sloan Digital Sky Survey (SDSS) agree well with the predictions of standard single field slow-roll inflation. In particular the power spectrum verifies the prediction of a nearly scale invariant spectrum of adiabatic perturbations with a nearly Gaussian distribution. However, there remains the possibility that large non-Gaussianities may be produced by models consistent with these measurements. In order to measure this non-Gaussianity we require higher order correlators, beyond the two-point function or power spectrum.
In \cite{FLS10} a wide range of models were constrained using their three-point correlators or bispectra. That paper exploited the use of a separable expansion in order to investigate a much wider class of models than had previously been considered. Their results were consistent with a Gaussian distribution. However, it is possible that the bispectrum remains small while the trispectrum (four-point correlator) is large (see, for example, some 
inflationary models  \cite{aChen} and cosmic strings \cite{Regan:2009hv}). Therefore, measurement of the trispectrum is an important test of the standard paradigm. 
\par
In this chapter we consider the class of trispectra which are independent of the diagonal. Such models include the cubic term of the local model, the equilateral model and the so-called `constant' model. Therefore, bounds on the local model will constrain single field local models to a very strong degree. Given the high degree of correlation between DBI models and equilateral models the constraints presented here represent an especially interesting novel result. We calculate the correlation between DBI inflation, k-inflation and the equilateral model used in this chapter. We also analyse the trispectrum due to cosmic strings which can be reduced to a closely correlated diagonal-free case.
\par
The approach adopted here is the implementation of the methodology outlined in Chapter \ref{chapter:methodology} to the aforementioned special class of trispectra. Exploiting the use of a separable expansion ensures the 
computation is tractable, reducing the complexity from $\mathcal{O}(l_{\rm{max}}^7)$ to $\mathcal{O}(l_{\rm{max}}^4)$, and also ensures the stability of the algorithms used in the analysis without the need to correct for pathological terms commonly present in other approaches. 
\par
This work, we believe, is an important first step in the comprehensive classification of all possible trispectra. In addition to recent results on the CMB bispectrum \cite{FLS10} this work will provide a thorough examination of possible deviations from Gaussianity in the very early universe.  The constraints obtained here on the cubic term of the local model, the equilateral model, the `constant' model and cosmic strings are as a result of comparison to WMAP5 data out to $l=500$ together with a pseudo-optimal analysis of noise and masking contributions. The estimators and map making algorithms employed were outlined in detail in Chapter \ref{chapter:methodology}, but are expressed in this chapter with the simplifying assumption that the trispectrum considered is independent of the diagonal term. We validate our results by using known analytic results in the large angle limit (Chapter \ref{chapter:methodology}) where the signal-to-noise can be calculated explicitly.  This is important because we are able to show that previous trispectrum forecasts using this Sachs-Wolfe approximation were over-optimistic (see, for example, ref~\cite{DS2}).
Despite this, the results presented here are significant, not least because the constraint on the equilateral model is an entirely novel result.
\par
  It should also be observed that future galaxy and other surveys of large scale structure may be analysed using similar techniques to those outlined in this chapter (see \cite{FRS10}).  Some alternative approaches to extract information from the trispectrum have been explored in the literature. Desjacques and Seljak \cite{DS2} have used large scale structure to constrain local non-Gaussianity. In \cite{vielva2} the local CMB trispectrum was analysed using the N-point probability distribution of CMB anisotropies, assuming a local non-linear perturbative model, $\Phi=\Phi_L+f_{NL}(\Phi_L^2+\langle \Phi_L^2\rangle)+g_{NL}\Phi_L^3+\mathcal{O}(\Phi_L^4)$, to characterise the large scale anisotropies. Smidt et al \cite{Cooray2} have analysed the local trispectrum using a pseudo-$C_l$ estimator. However, as we will point out in this chapter, these latter two approaches appear to contain some inconsistencies in their analysis. Other alternatives include the wavelet approach \cite{0301220} and the possibility of using a set of orthonormal estimators which was explored in \cite{0111250}.
\par
In section \ref{sec:diagfree} we review results relating primordial and CMB trispectra and their optimal estimation. The 
approach adopted here is to assume the primordial (and hence CMB) trispectrum is diagonal-free. The more general case was explored in detail in Chapter \ref{chapter:methodology}. We describe how to measure the correlation between two primordial trispectra. A naive approximation to the signal to noise is also calculated in order to derive an estimate for the one-sigma error bars expected. However, as we shall see, the effect of Silk damping renders this approximation inaccurate. The eigenmode decomposition of the trispectrum constitutes the basis of our method and is reviewed in section \ref{sec:decomp}. The application of the eigenmode expansion to reconstruct the full trispectrum from observations is shown in section \ref{sec:reconTrisp}. The accuracy of this approach is verified in section \ref{sec:WMAPTRISP}. In section \ref{sec:constraints} we obtain constraints on the cubic local model term, the equilateral model, the constant model, as well as on cosmic strings. In order to verify the use of the equilateral model adopted we apply our correlation measure to compare with alternatives. We also discuss forecasts for cosmic strings for Planck data. Finally we discuss our results and present our conclusions in section \ref{sec:concluImplement}.

\section{CMB Trispectrum Estimation (Diagonal-Free Class)}\label{sec:diagfree}
\subsection{Primordial and CMB trispectrum}
 The temperature anisotropies
may be represented using the $a_{lm}$ coefficients of a spherical harmonic decomposition of the cosmic microwave sky,
\begin{eqnarray}\label{DeltaT}
\frac{\Delta T}{T}(\hat{\mathbf{n}})=\sum_{l,m} a_{lm}Y_{lm}(\hat{\mathbf{n}}).
\end{eqnarray}
The primordial potential $\Phi$ induces the multipoles $a_{lm}$ via a convolution with the transfer functions $\Delta_l(k)$ through the relation
\begin{eqnarray}
a_{lm}=4\pi (-i)^l \int \frac{d^3k}{(2\pi)^3} \Delta_l(k) \Phi(\mathbf{k}) Y_{lm}(\hat{\mathbf{k}}).
\end{eqnarray}
The connected part of the four-point correlator of the $a_{lm}$ gives us the trispectrum. In particular, 
\begin{align}\label{Tconnls}
T_{l_1 m_1 l_2 m_2 l_3 m_3 l_4 m_4}=&\langle a_{l_1 m_1}	 a_{l_2 m_2} a_{l_3 m_3} a_{l_4 m_4}	\rangle_c \nonumber\\
=& (4\pi)^4 (-i)^{\sum_i l_i}\int  \frac{d^3 k_1 d^3 k_2 d^3 k_3 d^3 k_4}{(2\pi)^{12}}\Delta_{l_1}(k_1)\Delta_{l_2}(k_2)\Delta_{l_3}(k_3)\Delta_{l_4}(k_4)\times \nonumber\\
&\langle \Phi(\mathbf{k_1})\Phi(\mathbf{k_2})\Phi(\mathbf{k_3})\Phi(\mathbf{k_4}) \rangle_c Y_{l_1 m_1}(\hat{\mathbf{k_1}}) Y_{l_2 m_2}(\hat{\mathbf{k_2}}) Y_{l_3 m_3}(\hat{\mathbf{k_3}}) Y_{l_4 m_4}(\hat{\mathbf{k_4}}),
\end{align}
where $k_i=|\mathbf{k}_i|$ and the subscript $c$ is used to denote the connected component. In this chapter we consider the class of trispectra which are depend only on the wavenumbers $k_1,k_2,k_3,k_4$ and so are classified as `diagonal-free'. Using this assumption we may write the primordial trispectrum as
\begin{eqnarray}\label{Tconn}
\langle \Phi(\mathbf{k_1})\Phi(\mathbf{k_2})\Phi(\mathbf{k_3})\Phi(\mathbf{k_4}) \rangle_c=(2\pi)^3 \delta (\mathbf{k_1+k_2+k_3+k_4})T_{\Phi}(k_1,k_2,k_3,k_4).
\end{eqnarray}
Substituting this into \eqref{Tconnls} it may easily be shown (see Chapter \ref{chapter:methodology}) that
\begin{eqnarray}\label{TconnlsNew}
T_{l_1 m_1 l_2 m_2 l_3 m_3 l_4 m_4}&=& \int d\Omega_{\hat{\bn}}Y_{l_1 m_1}(\hat{\bn})Y_{l_2 m_2}(\hat{\bn})Y_{l_3 m_3}(\hat{\bn})Y_{l_4 m_4}(\hat{\bn}) t^{l_1 l_2}_{l_3 l_4},
\end{eqnarray}
where the `extra'-reduced trispectrum $t^{l_1 l_2}_{l_3 l_4}$ is given by\footnote{This is identical to the definition of the extra-reduced trispectrum given in Chapter \ref{chapter:methodology} modulo a factor of $12$.}
\begin{align}\label{extraRed}
t^{l_1 l_2}_{l_3 l_4}=&\left(\frac{2}{\pi}\right)^4 \int x^2 dx \int (k_1 k_2 k_3 k_4)^{2}T_{\Phi}(k_1,k_2,k_3,k_4)\Delta_{l_1}(k_1)\Delta_{l_2}(k_2)\Delta_{l_3}(k_3)\Delta_{l_4}(k_4)\nonumber\\
&\times j_{l_1}(k_1 x) j_{l_2}(k_2 x) j_{l_3}(k_3 x) j_{l_4}(k_4 x).
\end{align}
As with the bispectrum analysis it is simpler to analyse the trispectrum - especially for scale-invariant models - in terms of a shape function, i.e. a scale-invariant version of the trispectrum. In this chapter we use the following shape function
\begin{align}\label{Shape}
S(k_1,k_2,k_3,k_4)=\frac{(k_1 k_2 k_3 k_4)^{9/4}}{\tilde{\Delta}_{\Phi}^3 N}T_{\Phi}(k_1,k_2,k_3,k_4)
\end{align}
where $\tilde{\Delta}_{\Phi}$ is defined in terms of the power spectrum as $P_{\Phi}(k)=\tilde{\Delta}_{\Phi}/k^3$ (i.e. $\tilde{\Delta}_{\Phi}=2\pi^2\Delta_{\Phi}$) and $N$ is a normalisation factor which will be specified later.
\par
In terms of this shape function~\eqref{extraRed} becomes
\begin{align}\label{extraRed2}
t^{l_1 l_2}_{l_3 l_4}=&\left(\frac{2}{\pi}\right)^4 \int x^2 dx \int (k_1 k_2 k_3 k_4)^{-1/4}S(k_1,k_2,k_3,k_4)\Delta_{l_1}(k_1)\Delta_{l_2}(k_2)\Delta_{l_3}(k_3)\Delta_{l_4}(k_4)\nonumber\\
&\times j_{l_1}(k_1 x) j_{l_2}(k_2 x) j_{l_3}(k_3 x) j_{l_4}(k_4 x).
\end{align}
Naively, this appears to be an extremely numerically challenging calculation involving a line of sight integral over a 4D integral involving highly oscillatory functions. However, the integral breaks down into a product of one-dimensional integrals if the shape function can be represented in the form $S(k_1,k_2,k_3,k_4)=W(k_1)X(k_2)Y(k_3)Z(k_4)$. 

\subsection{Correlation Measure} \label{sec:correl}
In order to compare primordial trispectra a correlation measure was described in~Chapter \ref{chapter:methodology}. Here we review the argument, with minor variations such as a different choice of weight function. 
\par
A general trispectrum $T_{\Phi}(\mathbf{k}_1,\mathbf{k}_2,\mathbf{k}_3,\mathbf{k}_4;\mathbf{K})$ as defined in equation~\eqref{eq:Tconn} may be decomposed into the following three contributions
signifying the three different diagonals that may be formed by the quadrilateral with sides $\bk_1,\bk_2,\bk_3,\bk_4$,
\begin{align}
T_{\Phi}(\mathbf{k}_1,\mathbf{k}_2,\mathbf{k}_3,\mathbf{k}_4;\mathbf{K})=&P_{\Phi}(\mathbf{k}_1,\mathbf{k}_2,\mathbf{k}_3,\mathbf{k}_4;\mathbf{K})\nonumber\\
&+\int d^3 K' [\delta (\mathbf{k_3-k_2-K+K'})P_{\Phi}(\mathbf{k}_1,\mathbf{k}_3,\mathbf{k}_2,\mathbf{k}_4;\mathbf{K'})\nonumber \\
&\qquad\qquad+\delta (\mathbf{k_4-k_2-K+K'})P_{\Phi}(\mathbf{k}_1,\mathbf{k}_4,\mathbf{k}_3,\mathbf{k}_2;\mathbf{K'}) ].
\end{align}
As with most of the models described in the literature we shall assume that $P_{\Phi}$ depends only on the magnitude of its arguments, i.e. $P_{\Phi}(\mathbf{k}_1,\mathbf{k}_2,\mathbf{k}_3,\mathbf{k}_4;\mathbf{K})=P_{\Phi}(k_1,k_2,k_3,k_4,K)$. The shape function describing this term is denoted, as in equation~\eqref{Shape}, by
\begin{align}\label{ShapeP}
S_P(k_1,k_2,k_3,k_4,K)=\frac{(k_1 k_2 k_3 k_4)^{9/4}}{\tilde{\Delta}_{\Phi}^3 N}P_{\Phi}(k_1,k_2,k_3,k_4,K).
\end{align}
The correlation between two models is then given by
\begin{align}
F(S_{P},S_{P'})=\int d\mathcal{V}_k  S_{P}(k_1,k_2,k_3,k_4,K) S_{P'}(k_1,k_2,k_3,k_4,K) \omega(k_1,k_2,k_3,k_4,K)
\end{align}
where $\omega$ is an appropriate weight function, and $d\mathcal{V}_k$ corresponds to the area inside the region $k_i,K/2\in[0,k_{\rm{max}}]$ allowed by the triangle conditions. With this choice of weight the primordial shape correlator then takes the form
\begin{align}\label{eq:correl}
\mathcal{C}(S_{P},S_{P'})=\frac{F(S_{P},S_{P'})}{\sqrt{F(S_{P},S_{P})F(S_{P'},S_{P'})}}.
\end{align}
The weight is usually chosen such that $F$ has the same scaling as the estimator $T^2/((2L+1)C^4)$ in $l$- space. In this chapter we make the following choice of weight function
\begin{align}
\omega(k_1,k_2,k_3,k_4,K)=\frac{1}{(k_1+k_2+K)^{3/2}(k_3+k_4+K)^{3/2}}.
\end{align}
It should be noted, however, that the correlation measure will be independent of the choice of weight chosen.
The calculation may be further simplified, in the case of scale invariant models, by the following parametrisation of the wavenumbers
\begin{align}
&K=k(1-\beta),\qquad k_1=\frac{k}{2}(1+\alpha+\beta),\qquad k_2=\frac{k}{2}(1-\alpha+\beta),\qquad \nonumber\\
&k_3=\frac{\epsilon k}{2}\left(2+\gamma-\frac{1-\beta}{\epsilon} \right),\qquad k_4=\frac{\epsilon k}{2}\left(2-\gamma-\frac{1-\beta}{\epsilon} \right),
\end{align}
where $2k=(k_1+k_2+K)$, and the variables $\alpha,\beta,\gamma,\epsilon$ have the following ranges
\begin{align}
1\leq\epsilon<\infty,\quad 0\leq \beta \leq 1,\quad -(1-\beta)\leq \alpha \leq 1-\beta,\quad -\frac{1-\beta}{\epsilon} \leq\gamma\leq \frac{1-\beta}{\epsilon}.
\end{align}
In terms of these variables the shape function of a scale invariant model may be written as $S_P(k_1,k_2,k_3,k_4,K)=f(k)S_P(\alpha,\beta,\gamma,\epsilon)$, and the volume function becomes $d\mathcal{V}_k=\epsilon k^4 dk d\alpha d\beta d\gamma d\epsilon $.
For such models we may find the correlation by setting $k=\rm{constant}$ and integrating with respect to the variables $\{\alpha,\beta,\gamma,\epsilon\}$ only.

\subsection{CMB trispectrum estimator}
The trispectrum signal is too weak to measure individual multipoles directly. Thus, in order to compare theory with observations it is necessary to use an estimator which sums over the range of multipoles probed. Estimators may be considered here as performing a least squares fit of the data to the theoretical trispectrum $\langle \aone^{\rm{th}} \atwo^{\rm{th}} \athree^{\rm{th}} \afour^{\rm{th}} \rangle_c$. In Chapter \ref{chapter:methodology} the general optimum trispectrum estimator was derived giving
\begin{align}\label{Estimator2}
\mathcal{E}=&\frac{f_{\rm{sky}}}{\tilde{N}^2}\sum_{l_i m_i} \langle a_{l_1m_1}a_{l_2m_2}a_{l_3m_3}a_{l_4m_4}\rangle_c  \Bigg[(C^{-1} a^{\rm{obs}})_{l_1 m_1}(C^{-1} a^{\rm{obs}})_{l_2 m_2} (C^{-1} a^{\rm{obs}})_{l_3 m_3} (C^{-1} a^{\rm{obs}})_{l_4 m_4}\nonumber\\
&-6(C^{-1})_{l_1 m_1,l_2 m_2} (C^{-1} a^{\rm{obs}})_{l_3 m_3} (C^{-1} a^{\rm{obs}})_{l_4 m_4}+3 (C^{-1})_{l_1 m_1,l_2 m_2}(C^{-1})_{l_3 m_3,l_4 m_4} \Bigg],
\end{align}
where
\begin{align}\label{normalis}
\tilde{N}^2= \sum_{l_i m_i} &\langle \aone^{\rm{th}} \atwo^{\rm{th}} \athree^{\rm{th}} \afour^{\rm{th}} \rangle_c (C^{-1})_{l_1 m_1,l_1' m_1'} (C^{-1})_{l_2 m_2,l_2' m_2'}\nonumber\\
&\times (C^{-1})_{l_3 m_3,l_3' m_3'} (C^{-1})_{l_4 m_4,l_4' m_4'} \langle a_{l_1' m_1'}a_{l_2' m_2'}a_{l_3' m_3'}a_{l_4' m_4'}\rangle_c,
\end{align}
$f_{\rm{sky}}$ is the fraction of the sky outside the mask, and where the covariance matrix $C_{l_1 m_1,l_2 m_2}\equiv\langle \aone \atwo\rangle$ is non-diagonal due to mode-mode coupling introduced by the mask and anisotropic noise. In this chapter we follow \cite{WMAP5,YadavWandelt2009} by assuming a nearly diagonal covariance matrix ($C_{l_1 m_1,l_2 m_2}\approx (-1)^{m_1} C_{l_1}\d_{l_1 l_2}\d_{m_1 -m_2}$) and account for the noise and instrument beam effects by setting 
\begin{align}\label{NoiseRelations}
C_l\rightarrow\tilde{C}_{l}=b_l^2 C_l+N_l, \qquad \mathrm{and}\qquad {t}^{l_1 l_2}_{l_3 l_4}\rightarrow\tilde{t}^{l_1 l_2}_{l_3 l_4}=b_{l_1}b_{l_2}b_{l_3 }b_{l_4}{t}^{l_1 l_2}_{l_3 l_4}.
\end{align}
With this identification the estimator becomes
\begin{eqnarray}\label{estim}
\mathcal{E}&=&\frac{f_{\rm{sky}}}{N_T^2}\sum_{l_i m_i}  \frac{\langle a_{l_1m_1}a_{l_2m_2}a_{l_3m_3}a_{l_4m_4}\rangle_c }{\tilde{C}_{l_1} \tilde{C}_{l_2} \tilde{C}_{l_3} \tilde{C}_{l_4}}\Big[a^{\rm{obs}}_{l_1 m_1}a^{\rm{obs}}_{l_2 m_2}a^{\rm{obs}}_{l_3 m_3}a^{\rm{obs}}_{l_4 m_4}-6C_{l_1 m_1, l_2 m_2}^{\rm{sim}}a^{\rm{obs}}_{l_3 m_3}a^{\rm{obs}}_{l_4 m_4}\nonumber\\
&&+3C_{l_1 m_1, l_2 m_2}^{\rm{sim}}C_{l_3 m_3, l_4 m_4}^{\rm{sim}}\Big],
\end{eqnarray}
where $N_T^2$ is derived in section \ref{sec:normalisationfactor} and is given by
\begin{align}\label{Normal}
N_T^2=&\frac{1}{2 (4\pi)^2}\sum_{l_i}(2l_1+1)(2l_2+1)(2l_3+1)(2l_4+1)\frac{\tilde{t}^{l_1 l_2}_{l_3 l_4}\tilde{t}^{l_1 l_2}_{l_3 l_4}}{\tilde{C}_{l_1}\tilde{C}_{l_2}\tilde{C}_{l_3}\tilde{C}_{l_4}}\nonumber\\
&\times\int_{-1}^1 P_{l_1}(\mu) P_{l_2}(\mu) P_{l_3}(\mu) P_{l_4}(\mu)d\mu.
\end{align}
We note that the signal to noise squared is given by $(S/N)^2=f_{sky}N_T^2/24$. The second and third terms in \eqref{estim} ensures subtraction of spurious inhomogeneous noise and masking by using the covariance matrix $C_{l_1 m_1, l_2 m_2}^{\rm{sim}}$ from an ensemble average of Gaussian maps in which these effects are incorporated. To simplify notation we shall assume inclusion of noise, beam and mask throughout and henceforth drop any special notation.
\par
It should be noted that if the theoretical trispectrum $t^{l_1 l_2}_{l_3 l_4}$ has the property of primordial separability then the summations in \eqref{estim} and \eqref{Normal} become much more tractable, taking only $\mathcal{O}(l_{\rm{max}}^4)$ operations. 

\subsection{Normalisation Factor}\label{sec:normalisationfactor}
As was shown in Chapter \ref{chapter:methodology} the normalisation factor \eqref{normalis} for an ideal experiment becomes
\begin{align}\label{norms}
\tilde{N}^2\rightarrow N_T^2=\sum_{l_i, L}\frac{h_{l_1 l_2 L}^2 h_{l_3 l_4 L}^2 t^{l_1 l_2}_{l_3 l_4} t^{l_1 l_2}_{l_3 l_4}}{(2L+1)C_{l_1}C_{l_2}C_{l_3}C_{l_4}},
\end{align}
where the geometric factor $h_{l_1 l_2 L}$ is given in terms of the Wigner 3J symbol by
\begin{align*}
h_{l_1 l_2 L}=\sqrt{\frac{(2l_1+1)(2l_2+1)(2L+1)}{4\pi}}\left( \begin{array}{ccc}
l_1 & l_2 & L \\
0 & 0 & 0 \end{array} \right).
\end{align*}
In this case, where the extra-reduced trispectrum is independent of the diagonal $L$, we may simplify the expression further. First we note the identity given in Chapter \ref{chapter:methodology}
\begin{align}\label{UsefulIdentity}
\int d\Omega_{\hat{n}} Y_{l_1 m_1}(\hat{\mathbf{n}})Y_{l_2 m_2}(\hat{\mathbf{n}})Y_{l_3 m_3}(\hat{\mathbf{n}})Y_{l_4 m_4}(\hat{\mathbf{n}})=&\nonumber\\
\sum_{L' M'}(-1)^{M'}h_{l_1 l_2 L'}h_{l_3 l_4 L'}&\left( \begin{array}{ccc}
l_1 & l_2 & L' \\
m_1 & m_2 & -M' \end{array}\right)\left( \begin{array}{ccc}
l_3 & l_4 & L' \\
m_3 & m_4 & M' \end{array} \right),
\end{align}
and the relation between the spherical harmonics and Legendre polynomials $P_l=\\ \sqrt{4\pi/(2l+1)}Y_{l 0}$. It is a simple calculation to check that these identities imply
\begin{align}
\sum_L \frac{h_{l_1 l_2 L}^2 h_{l_1 l_2 L}^2 }{2L+1}=&\frac{(2l_1+1)(2l_2+1)(2l_3+1)(2l_4+1)}{2(4\pi)^2}\nonumber\\
&\times\int_{-1}^1 P_{l_1}(\cos\theta)P_{l_2}(\cos\theta)P_{l_3}(\cos\theta)P_{l_4}(\cos\theta)d\cos\theta.
\end{align}
Substituting this into \eqref{norms} (and setting $x = \cos\theta$) gives
\begin{align}
N_T^2=&\frac{1}{2 (4\pi)^2}\sum_{l_i}(2l_1+1)(2l_2+1)(2l_3+1)(2l_4+1) \frac{{t}^{l_1 l_2}_{l_3 l_4}{t}^{l_1 l_2}_{l_3 l_4}}{{C}_{l_1}{C}_{l_2}{C}_{l_3}{C}_{l_4}}\nonumber\\
&\times\int_{-1}^1 P_{l_1}(\mu) P_{l_2}(\mu) P_{l_3}(\mu) P_{l_4}(\mu)d\mu.
\end{align}

The generalisation to include inhomogeneous noise and beam effects follows the prescription given in \eqref{NoiseRelations}.

\subsection{$G_{NL}$ normalisation}
It should be noted that is desirable to compare measures of the trispectra of different models. The general measure, $t_{NL}$, is defined in the equilateral limit for which the wavenumbers $k_i$ and the diagonals of the quadrilateral (formed by the wavevectors $\mathbf{k}_i$) have equal values, i.e. $k_1=k_2=k_3=k_4=|\mathbf{k}_1+\mathbf{k}_2|=|\mathbf{k}_1+\mathbf{k}_3|=k$. We define
\begin{align}\label{tnldef}
t_{NL}=\frac{9}{200}\frac{T_{\Phi}(k,k,k,k;k,k)}{P_{\Phi}(k)^3}.
\end{align}
With this definition we find that for the local model $t_{NL}^{\rm{local}}=1.5 \tau_{NL}+1.08 g_{NL}$ (see \cite{aChen}). 
\par
As with the shortcomings of using the parameter $f_{NL}$ to compare different bispectra consistently (see \cite{FLS09}), the method of comparing trispectra using their value at a central point is problematic.
Choosing a different choice of normalisation factor to $N_T^2$ allows us to define an integrated measure with which a consistent comparison between models may be made. In Chapter \ref{chapter:methodology} the following choice of normalisation factor was adopted $N^2\equiv N_{T}N_{T\rm{locB}}$, where $\rm{locB}$ refers to the $g_{NL}$ expression of the local model trispectrum (with $g_{NL}=1$ and $\tau_{NL}=0$). The $g_{NL}$ term is chosen since it is independent of the diagonal term and therefore may be determined more quickly than the normalisation for the $\tau_{NL}$ term. Hence we adopt the following definition for a general measure of the trispectrum
\begin{align} \label{eq:GNLdef}
G_{NL}=\frac{N_T}{N_{T{\rm{loc}}\,\,g_{NL}=1,\tau_{NL}=0}}\mathcal{E}.
\end{align}
As has been emphasised in the case of the bispectrum in \cite{FLS10}, this choice of measure is more democratic than other measures in the literature which generally involve the comparison of models at a particular choice of wavenumbers in the shape function. Such a choice can result in a huge disparity between the quoted constraints. A further advantage of the integrated measure is that it allows for non-scale invariant models and trispectra induced by late-time processes such as gravitational lensing and cosmic strings. An approximation scheme for evaluating $N_T$ using primordial shape correlators has been outlined in Chapter \ref{chapter:methodology}.

\subsection{Naive Signal to Noise Approximation} \label{sec:sachswolfe}
In order to estimate the error bounds for an ideal experiment which may be expected for the trispectrum, we adapt an argument used in the case of the bispectrum by Babich and Zaldarriaga \cite{0408455}. In that paper it was shown that the physical scales in the radiative transfer functions are largely irrelevant in the calculation of non-Gaussian bispectrum constraints in the multipole range of WMAP ($l\lesssim 500$) and Planck ($l\lesssim 2000$). 
\par
Consider first  the flat-sky power spectrum and trispectrum (assumed to depend only on its wavenumbers) given by
\begin{align}
\langle a(\lb_1)a(\lb_2)\rangle&=(2\pi)^2 \d^{(2)}(\lb_1+\lb_2)C(l_1)\nonumber\\
\langle a(\lb_1)a(\lb_2)a(\lb_3)a(\lb_4)\rangle_c&=(2\pi)^2 \d^{(2)}(\lb_1+\lb_2+\lb_3+\lb_4)T(l_1,l_2,l_3,l_4).
\end{align}
The signal to noise, $S/N$, may be found by similar arguments to \cite{0408455} to be given by
\begin{align}
(S/N)^2=\frac{\d^{(2)}(0)}{(2\pi)^4}\int d^2\lb_1d^2\lb_2 d^2\lb_3d^2\lb_4 \d^{(2)}(\lb_1+\lb_2+\lb_3+\lb_4)\frac{T^2(l_1,l_2,l_3,l_4)}{24 C(l_1)C(l_2)C(l_3)C(l_4)},
\end{align}
where $\d^{(2)}(0)$ may also be written as $f_{\textrm{sky}}/\pi$.
With the assumption of a flat sky with no radiative transfer, the Fourier coefficients of the temperature anisotropies can be expressed as
\begin{align}
a(\lb)=(2\pi)^2\int \frac{d^3\bk}{(2\pi)^3}\Phi(\bk) e^{i k^z r_D}\d^{(2)}(\lb-\bk^{\parallel} r_D),
\end{align}
where $r_D$ is the distance to the last scattering surface, $\bk^{\parallel}$ denotes the component of $\bk$ parallel to the last scattering surface and $k^z$ the component orthogonal. The flat sky power spectrum is calculated to give $l^2 C(l)=\tilde{\Delta}_{\Phi}/(9\pi)$. The flat sky trispectrum for the $g_{NL}$ local model (for which $T(k_1,k_2,k_3,k_4)=6g_{NL}[P_{\Phi}(k_1)P_{\Phi}(k_2)P_{\Phi}(k_3)+\rm{3\,perms}]$) is similarly found to give
\begin{align}
T(l_1,l_2,l_3,l_4)=\frac{6 g_{NL}}{3^4\pi^3}\tilde{\Delta}_{\Phi}^3\left( \frac{1}{(l_1 l_2 l_3)^2}+\rm{3\,perms}\right).
\end{align}
Substituting these expressions into the formula for the signal to noise and integrating between $l_{\rm{min}}$ and $l_{\rm{max}}$ we find
\begin{align}\label{eq:sachswolfe}
\frac{S}{N}=\frac{3}{\pi^2}g_{NL}\tilde{\Delta}_{\Phi} l_{\textrm{max}} \ln\left(\frac{l_{\textrm{max}} }{l_{\textrm{min}} }\right)\sqrt{f_{\rm{sky}}}.
\end{align}
Substituting the values $\tilde{\Delta}_{\Phi}\approx 1.71\times 10^{-8}$, $f_{sky}=0.7$, $l_{\textrm{min}}=2$ and the value of $l_{\textrm{max}}$ relevant for WMAP and Planck (i.e. $l_{\textrm{max}}\approx 750,2000$, respectively)  we find the expected error bound for an ideal experiment by setting $\Delta g_{NL}=1/(S/N)_{g_{NL}=1}$. For WMAP we find $\Delta g_{NL}^{WMAP}\approx 8.3\times 10^4$, while for Planck we get $\Delta g_{NL}^{Planck}\approx 1.6\times 10^4$. As we shall see, however, this analysis leads to an overestimate of the signal for the trispectrum and, hence, an underestimate of the error bounds. We include the argument here for later discussion.
\par
Since the measure $G_{NL}$ is normalised to give $g_{NL}$ when compared to the $g_{NL}$ local model, these values are also expected to give the error bars for $G_{NL}$ for any diagonal-free model.

\section{Separable Mode Expansion}\label{sec:decomp}
In Chapter \ref{chapter:methodology} a separable mode expansion was derived for a general (reduced) trispectrum. Since in this chapter we consider trispectra which are independent of the diagonal, the complexity in finding a general separable expansion reduces from $\mathcal{O}(l_{\rm{max}}^5)$ to $\mathcal{O}(l_{\rm{max}}^4)$. In order to elucidate this point we briefly outline the steps involved in computing the expansion for the particular class of models considered in this work.
\par
The four dimensional domain of allowed wavenumbers bounded by $k_i\in[0,k_{\rm{max}}]$ is defined by the following quadrilateral conditions
\begin{align}
k_1\leq k_2+k_3+k_4 \quad \rm{for}\quad k_1\geq k_2,k_3,k_4,\quad +\quad \rm{cyclic\,perms}.
\end{align}
We may integrate arbitrary functions $f(k_1,k_2,k_3,k_4)$ over this domain, which we denote $d\mathcal{V}_T$, i.e. $\mathcal{I}[f]\equiv\int_{\mathcal{V}_T}f(k_1,k_2,k_3,k_4)\omega(k_1,k_2,k_3,k_4)d\mathcal{V}_T$, where $\omega$ is a given weight function which we choose to be unity in this chapter. This integral defines a measure $\langle f,g\rangle=\mathcal{I}[f g]$.
The construction of bounded one-dimensional polynomials on this domain follows a similar argument to that described in Chapter \ref{chapter:methodology}. The first few polynomials are found to be
\begin{align}
q_0(k)&=1.09545,\nonumber\\
q_1(k)&=-2.10925+3.97971\konkmax,\nonumber\\
q_2(k)&=2.90417-15.78084 \konkmax+15.30751\konkmaxb^2,\nonumber\\
q_3(k)&=-3.53306+38.26522\konkmax-92.03652\konkmaxb^2+60.18822\konkmaxb^3 ,\dots
\end{align}
These basis functions may be used to create a separable mode expansion of the shape function \eqref{Shape} in the form
\begin{align}\label{eq:shapedecomp}
S(k_1,k_2,k_3,k_4)=\sum_n \alpha_n^{Q}Q_n(k_1,k_2,k_3,k_4),
\end{align}
where
$Q_n(k_1,k_2,k_3,k_4)=q_{(p}(k_1) q_r(k_2) q_s(k_3) q_{t)} (k_4)$
and $n=\{p,r,s,t\}$, with $(p r s t)$ representing the $24$ permutations of the symbols reflecting the underlying symmetry of the trispectra considered here. The task therefore is the determination of the $\alpha_n^Q$s.
\par
The $Q_n$s though independent and separable are not, in general, orthonormal, i.e. $\langle Q_n Q_p\rangle=\gamma_{np}\neq \delta_{np}$. Often it is more useful to work in terms of orthonormal basis functions and we may perform a Gram-Schmidt process to obtain a set of such functions $R_n$ with $\langle R_n R_p\rangle= \delta_{np}$. The relation between the $R_n$s and $Q_n$s is given by
\begin{align}
R_n=\sum_{p=0}^n \lambda_{n p}Q_p, \quad\textrm{where}\quad (\lambda_{np})^{-1}=\langle R_n  Q_p\rangle.
\end{align}
The matrices $\Gamma$ and $\Lambda$ may be related by
\begin{align}
\gamma_{np}^{-1}=\sum_r \lambda_{r n}\lambda_{r p}\quad \textrm{or}\quad \Gamma^{-1}=\Lambda^T\Lambda.
\end{align}
From this expression it is clear that a more numerically stable method for finding the matrix $\Lambda$ than calculating the Gram-Schmidt orthogonalisation is to perform a Cholesky factorisation of the matrix $\Gamma$.
The shape function may be expanded in terms of the orthonormal basis as
\begin{align}
S(k_1,k_2,k_3,k_4)=\sum_n \alpha_n^{R}R_n(k_1,k_2,k_3,k_4),
\end{align}
where $\alpha_n^R=\langle R_n S\rangle$. These expressions allow us to find $\alpha_n^Q$. Explicitly,
\begin{align}\label{eq:relQR}
\alpha_n^Q=\sum_p \lambda_{p n}\alpha_p^{R}.
\end{align}
It should be noted that the sums are bounded by the number of modes considered.
\par
A similar decomposition may be applied to the late-time trispectrum $t^{l_1l_2}_{l_3 l_4}$ with separable product and orthonormal modes, denoted 
with a bar as $\overline{Q}_n(l_1,l_2,l_3,l_4)\equiv$ \\$\overline{q}_{\{p}(l_1) \overline{q}_r(l_2) \overline{q}_s(l_2) \overline{q}_{t\}}(l_4)$ and $\overline{R}_n(l_1,l_2,l_3,l_4)$ respectively (with the $\overline {q}_p(l)$ defined on the allowed multiple domain and a chosen ordering $n\leftrightarrow \{prst\}$).  We expand the trispectrum  signal-to-noise in the estimator as
\begin{align}\label{eq:latetime}
\frac{(2l_1+1)^{1/4}(2l_2+1)^{1/4}(2l_3+1)^{1/4}(2l_4+1)^{1/4}}{\sqrt{C_{l_1}C_{l_2}C_{l_3}C_{l_4}}}t^{l_1l_2}_{l_3 l_4}=\sum_n \overline{\alpha}_n^Q \overline{Q}_n(l_1,l_2,l_3, l_4).
\end{align}
The coefficients, $\overline{\alpha}_n$, are denoted with a bar to distinguish them as late-time. 
\par
In particular, given a (diagonal-free) primordial trispectrum we may decompose the shape as in \eqref{eq:shapedecomp}. This gives the late-time trispectrum 
\begin{align}\label{earlytime}
\frac{(2l_1+1)^{1/4}(2l_2+1)^{1/4}(2l_3+1)^{1/4}(2l_4+1)^{1/4}}{\sqrt{C_{l_1}C_{l_2}C_{l_3}C_{l_4}}}t^{l_1l_2}_{l_3 l_4}=\tilde{\Delta}_{\Phi}^3 N\sum_n \alpha_n^Q\int dx x^2 \tilde{Q}_n^{l_1 l_2 l_3 l_4}(x), 
\end{align}
where $\tilde{Q}_n^{l_1 l_2 l_3 l_4}(x)=\tilde{q}_r^{l_1}(x) \tilde{q}_s^{l_2}(x) \tilde{q}_t^{l_3}(x) \tilde{q}_u^{l_4}(x)$, with $\tilde{q}^{l}_r(x)=q^l_r(x)(2 l+1)^{1/4}/\sqrt{C_l}$. 
\par
The late-time trispectrum may itself be decomposed in the form \eqref{eq:latetime}. The late-time decomposition is then given by
\begin{align}\label{eq:latetimedecomp}
\frac{(2l_1+1)^{1/4}(2l_2+1)^{1/4}(2l_3+1)^{1/4}(2l_4+1)^{1/4}}{\sqrt{C_{l_1}C_{l_2}C_{l_3}C_{l_4}}}t^{l_1l_2}_{l_3 l_4}&=\nonumber\\
\tilde{\Delta}_{\Phi}^3 N\sum_{n p v} \alpha_n^{Q}\int dx x^2 \tilde{\gamma}_{n p}(x) &\left(\overline{\gamma}^{}\right)^{-1}_{p v}\overline{Q}_v(l_1,l_2,l_3,l_4),
\end{align}
where $\tilde{\gamma}_{n p}(x)= \langle\tilde{Q}_n^{l_1 l_2 l_3 l_4}(x) \overline{Q}_p(l_1,l_2,l_3,l_4)\rangle_{\text{late}}$, $\overline{\gamma}_{n m}^{}=\langle \overline{Q}_n \overline{Q}_m\rangle_{\text{late}}$ and where we define
the late-time correlator such that $\langle f g \rangle_{\text{late}}$ is given by
\begin{align}
\langle f g \rangle_{\text{late}}=\sum_{l_i}& f(l_1,l_2,l_3,l_4)g(l_1,l_2,l_3,l_4)\frac{\sqrt{(2l_1+1)(2l_2+1)(2l_3+1)(2l_4+1)}}{2(4\pi)^2}\nonumber\\
&\times\int_{-1}^1 d\mu P_{l_1}(\mu)P_{l_2}(\mu)P_{l_3}(\mu)P_{l_4}(\mu).
\end{align}
In particular, with $\tilde{Q}_n^{l_1 l_2 l_3 l_4}(x)=\tilde{q}^{l_1}_{n_1}(x)\tilde{q}^{l_2}_{n_2}(x)\tilde{q}^{l_3}_{n_3}(x)\tilde{q}^{l_4}_{n_4}(x)$ and $\overline{Q}_p(l_1,l_2,l_3,l_4)=\\
\overline{q}_{p_1}(l_1)\overline{q}_{p_2}(l_2)\overline{q}_{p_3}(l_3)\overline{q}_{p_4}(l_4)$, we have
\begin{align}
\tilde{\gamma}_{n p}(x)=\int_{-1}^1 \frac{d\mu}{2(4\pi)^2}\Pi_{i=1}^4\left(\sum_{l_i}\sqrt{2l_i+1}\tilde{q}_{n_i}^{l_i}(x)\overline{q}_{p_i}P_{l_i}(\mu)\right).
\end{align}
 Thus, we identify $\overline{\alpha}_v^Q$ as
\begin{align}
\overline{\alpha}_v^Q=\sum_{n p}  \alpha_n^{Q}\int dx x^2 \tilde{\gamma}_{n p}(x) \left(\overline{\gamma}^{}\right)^{-1}_{p v}.
\end{align}
As usual the correlation measure may be used to estimate the accuracy of using the expansion $\sum_n \overline{\alpha}^Q_n \overline{Q}_n$ in place of \eqref{earlytime}.
The advantage of using the late-time decomposition is that the line-of-sight integration is captured in the expansion
coefficients $\overline{\alpha}_v^Q$. Therefore, such an expansion is desirable for efficiency in calculations of
the Fisher matrix and estimator, with the complexity passed to the calculation of these coefficients. For example,
equation \eqref{Normal} may be written in terms of the early-time expansion as
\begin{align}\label{normEarly}
N_T^2&=\tilde{\Delta}_{\Phi}^6 N^2 \sum_{n m}{\alpha}^Q_n {\alpha}^Q_m \int_{-1}^1 \frac{d\mu}{2(4\pi)^2} \int dx dy x^2 y^2 \Pi_{i=1}^4\left( \frac{2 l_i+1}{C_{l_i}}q_{n_i}^{l_i}(x) q_{m_i}^{l_i}(y)P_{l_i}(\mu)\right)\nonumber\\
&=\tilde{\Delta}_{\Phi}^6 N^2 \sum_{n m}{\alpha}^Q_n {\alpha}^Q_m \gamma_{nm}^{},
\end{align}
where in the second line we define the matrix $\gamma_{nm}^{}$ to simplify notation.
In terms of the late-time decomposition, the normalisation may be calculated efficiently as
\begin{align}\label{normLate}
N_T^2=\tilde{\Delta}_{\Phi}^6 N^2 \sum_{n m}\overline{\alpha}^Q_n \overline{\alpha}^Q_m \overline{\gamma}^{}_{n m}.
\end{align}

\section{Reconstructing the CMB Trispectrum}\label{sec:reconTrisp}
Substituting the mode expansion \eqref{eq:shapedecomp} into the estimator \eqref{estim}, while exploiting 
equation~\eqref{TconnlsNew} which is valid for diagonal-free trispectra, we find
\begin{align}
\mathcal{E}=\frac{\tilde{\Delta}_{\Phi}^3 N}{N_T^2}\sum_n \alpha^Q_n \int dx x^2\Bigg(&M_{\{n_1}(\hat{\bf{n}},x)M_{n_2}(\hat{\bf{n}},x)M_{n_3}(\hat{\bf{n}},x)M_{n_4\}}(\hat{\bf{n}},x)\nonumber\\
&-6\langle M_{\{n_1}(\hat{\bf{n}},x)M_{n_2}(\hat{\bf{n}},x)\rangle M_{n_3}(\hat{\bf{n}},x)M_{n_4\}}(\hat{\bf{n}},x)\nonumber\\
&+3\langle M_{\{n_1}(\hat{\bf{n}},x)M_{n_2}(\hat{\bf{n}},x)\rangle \langle M_{n_3}(\hat{\bf{n}},x)M_{n_4\}}(\hat{\bf{n}},x)\rangle \Bigg).
\end{align}
Here the $M_{p}(\hat{\bf{n}},x)$ are versions of the CMB map filtered with the polynomial $q^l_p(x)$ with the weight function $1/C_l$, i.e.
\begin{align}
M_{p}(\hat{\bf{n}},x)=\sum_{l m}\frac{q_p^l(x)a_{l m}Y_{l m}(\hat{\bf{n}})}{C_l}.
\end{align}
Defining the integrated expression in brackets as $\beta_n^Q$ we may write the above expression succinctly as
$\mathcal{E}=(\tilde{\Delta}_{\Phi}^3 N/N_T^2)\sum_n \alpha^Q_n \beta_n^Q$. Alternatively, we may express the estimator in terms of the late-time mode expansion \eqref{eq:latetimedecomp} as
\begin{align}
\mathcal{E}=\frac{\tilde{\Delta}_{\Phi}^3 N}{N_T^2}\sum_n \overline{\alpha}^Q_n &\Bigg(M_{\{n_1}(\hat{\bf{n}})M_{n_2}(\hat{\bf{n}})M_{n_3}(\hat{\bf{n}})M_{n_4\}}(\hat{\bf{n}})-6\langle M_{\{n_1}(\hat{\bf{n}})M_{n_2}(\hat{\bf{n}})\rangle M_{n_3}(\hat{\bf{n}})M_{n_4\}}(\hat{\bf{n}})\nonumber\\
&+3\langle M_{\{n_1}(\hat{\bf{n}})M_{n_2}(\hat{\bf{n}})\rangle \langle M_{n_3}(\hat{\bf{n}})M_{n_4\}}(\hat{\bf{n}})\rangle \Bigg),
\end{align}
where
\begin{align}
M_{p}(\hat{\bf{n}})=\sum_{l m}\frac{\overline{q}_p(l)a_{l m}Y_{l m}(\hat{\bf{n}})}{\sqrt{C_l}(2l+1)^{1/4}}.
\end{align}
Expressing the term in brackets as $\overline{\beta}^Q_n$ we arrive at the simple formula for the estimator
\begin{align}\label{lateestim}
\mathcal{E}=\frac{\tilde{\Delta}_{\Phi}^3 N}{N_T^2}\sum_n \overline{\alpha}^Q_n \overline{\beta}^Q_n.
\end{align}
Explicitly, there are quartic, quadratic and constant (unconnected) parts  $  \overline{\beta}^Q_n =  \overline{\beta}^{(4)}_n - 6 \overline{\beta}^{(2)}_n +  3 \overline{\beta}^{(0)}_n$ defined by 
\begin{align}
&\overline{\beta}^{(4)}_n \equiv \int d^2{\hat {\bf n}}\; M_{\{p}\,M_{r}\,M_{s}\,M_{t\}}\,,\quad \overline{\beta}^{(2)}_n \equiv \int d^2{\hat {\bf n}}\; \langle M_{\{p}\,M_{r}\rangle 
\,M_{s}\,M_{t\}} ~~\hbox{and}~~\nonumber\\
&\overline{\beta}^{(0)}_n \equiv \int d^2{\hat {\bf n}}\; \langle M_{\{p}\,M_{r}\rangle \langle M_{s}\,M_{t\}}\rangle\,,
\end{align}
where the angled brackets represent map products averaged over many Monte Carlo simulations incorporating realistic anisotropic noise, beam and masking effects. Remarkably, once the mode coefficients and transformations have been calculated,  trispectrum estimation using (\ref{lateestim}) collapses down to only ${\cal O}(\lmax^2)$ operations. 
\par
Since $\langle \mathcal{E}\rangle=1$ we deduce, by comparison of equations \eqref{lateestim} and \eqref{normLate}, that
\begin{align}
\langle\overline{\beta}^Q_n\rangle=\tilde{\Delta}_{\Phi}^3 N \sum_{m}\overline{\alpha}^Q_m \overline{\gamma}^{}_{n m}.
\end{align}
Therefore, assuming that we can extract the $\overline{\beta}^Q_n$ coefficients with sufficient significance from a particular experiment, we may directly reconstruct the CMB trispectrum using the expansion \eqref{eq:latetime}. Similarly by using the early-time expression for the estimator and equation \eqref{normEarly} we find
\begin{align}
\langle {\beta}^Q_n\rangle=\tilde{\Delta}_{\Phi}^3 N \sum_{m}{\alpha}^Q_m \gamma^{}_{n m},
\end{align}
thus allowing for a reconstruction of the primordial trispectrum (again given sufficient significance of the signal). In the orthonormal frame $\{ R_n\}$ we may express the estimator \eqref{lateestim} in the form
\begin{align}\label{lateestimrot}
\mathcal{E}=\frac{\tilde{\Delta}_{\Phi}^3 N}{N_T^2}\sum_n \overline{\alpha}^R_n \overline{\beta}^R_n,
\end{align}
where we define $\beta_n^R=\sum_{m}\overline{\lambda}_{n m}\beta_{m}^Q$ (c.f. equation \eqref{eq:relQR}). With respect to this frame we have
\begin{align}
\langle \overline{\beta}^R_n\rangle=\tilde{\Delta}_{\Phi}^3 N \overline{\alpha}^R_n.
\end{align}
 Alternatively, we may reconstruct the primordial trispectrum with the observation
\begin{align}
\langle\overline{\beta}^Q_n\rangle=\tilde{\Delta}_{\Phi}^3 N \sum_{m}\alpha_m^Q \int dx x^2 \tilde{\gamma}_{m n}(x).
\end{align}

\section{The WMAP Trispectrum}\label{sec:WMAPTRISP}
In this section we apply the mode decomposition techniques described in this chapter to the analysis of WMAP$5$ data. We aim to estimate $G_{NL}$ (equivalently $t_{NL}$) from different primordial shapes as well as from cosmic strings. We also aim to provide a reconstruction of the trispectrum from the data, using the methods described in the previous sections to recover the modes $\overline{\beta}_n^R$. The analysis presented here is intended as a first step towards the implementation of this formalism to analysing the CMB trispectrum, rather than its completion. We use multipoles up to $l_{\rm{max}}=500$ and work with WMAP$5$ data. The analysis here uses the pseudo-optimal weighting rather than the full inverse covariance weighting in the estimator \eqref{estim}.
\par
In order to validate the algorithm used to extract the mode coefficients from the WMAP$5$ data we perform $400$ Gaussian map simulations with WMAP-realistic noise. The results are plotted in Figure~\ref{fig:gaussbetas}. The mean of the recovered mode coefficients are shown, with error bars showing twice the standard error of the mean ($=2\sigma/\sqrt{400}$). It is clearly evident that the coefficients are consistent (at $95\%$ level of confidence) with zero as expected for Gaussian maps. It should be noted that, as with the bispectrum \cite{FLS10}, the mode decomposition may be used to characterise anisotropic contributions, such as inhomogeneous noise.
\par
Next, we perform our WMAP$5$ analysis. After coadding the V and W band data (using the same weights as in the WMAP$5$ analysis), our first step was to extract the $\overline{\beta}_n^Q$ mode coefficients from the data, as detailed in Section \ref{sec:reconTrisp}. This coefficients are used to create the rotated mode coefficients $\overline{\beta}_n^R$ which, being defined on the orthonormal basis $\{ R_n\}$, are a more useful quantity with which to compare to the theoretical data. We chose to compute the first $50$ modes of the modal expansion because this proved sufficient to describe the theoretical CMB trispectra on the observational domain $l_{\rm{max}}=500$. We will present the constraints on these models in the next section. 
\par
In the next section we will establish that deviations from Gaussianity are not apparent from the WMAP$5$ analysis we present in this chapter. It also appears that for the primordial models investigated in this chapter, such as the local $g_{NL}$ model, that the sensitivity to primordial non-Gaussianity will not be improved greatly using Planck data. This suggests that such models may be better constrained by using other probes of non-Gaussianity such as large scale structure. However, we note that there exists other primordial models, in particular the $\tau_{NL}$ local model, that are predicted to have a more significant signal to noise dependence on the multipole range \cite{Kogo}. Constraints on such models are, therefore, expected to improve considerably using Planck data. In an upcoming paper we will investigate the application of the formalism outlined here to such models (which have a diagonal dependence) \cite{FRS11}.
\begin{figure}[htp]
\centering 
\includegraphics[width=162mm]{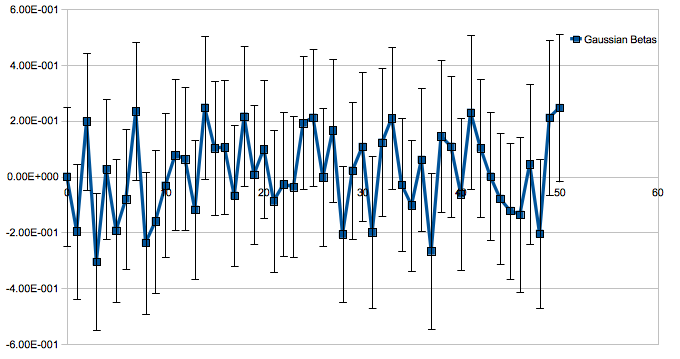}
\caption{Mean of the recovered (Gaussian) mode coefficients $\overline{\beta}^R_n$ with error bars showing the $95\%$ confidence interval (i.e. $2\sigma/\sqrt{400}$), as estimated from $400$ Gaussian map simulations in the same WMAP-realistic context. The results, consistent with zero, are shown to validate the estimation algorithm presented in Section~\ref{sec:reconTrisp}.}
\label{fig:gaussbetas}
\end{figure}

\section{Constraints on Nearly Scale-Invariant Models}\label{sec:constraints}
Due to the numerical complexity in calculating the trispectrum using standard approaches, the only constraints that exist in the literature are for the local $g_{NL}$ model trispectrum. Desjacques and Seljak \cite{DS2} have used the bias parameter of dark matter halos to constrain $g_{NL}$. They found the 2$\sigma$ result $-3.5\times 10^5 <g_{NL}<8.2\times 10^5 $, although, as the authors caution, a better theoretical modelling of non-Gaussian halo bias is needed for an accurate analysis of non-Gaussianity using large scale structure. Vielva and Sanz in \cite{vielva2} also presented constraints on this parameter. Their analysis adopted a local model in the Sachs-Wolfe regime which is likely to be inaccurate for $l\gtrsim 100$. The only other constraint, that we are aware of, has been presented by Smidt et al. \cite{Cooray2}. Their work uses a pseudo-$C_l$ estimator and presents a joint analysis on the parameters $g_{NL}$ and $\tau_{NL}$. Unfortunately their study assumes that there is just one solid angle present in the trispectrum estimator, which is not valid for the $\tau_{NL}$ model they have analysed (see Chapter \ref{chapter:methodology} for a general analysis of the trispectrum). Essentially their trispectrum estimators, $\mathcal{K}^{(4)}$ and $\mathcal{K}^{(3,1)}$, make use of the (generally invalid) replacement of the expression $\sum_{L M} (-1)^M\mathcal{G}^{l_1 l_2 L}_{m_1 m_2 -M}\mathcal{G}^{l_3 l_4 L}_{m_3 m_4 M}t^{l_1 l_2}_{l_3 l_4}(L)$ - which as the product of two Gaunt integrals is a function of two solid angles - by the single solid angle expression $\int d\Omega_{\hat{\mathbf{n}}}Y_{l_1 m_1}({\hat{\mathbf{n}}})Y_{l_2 m_2}({\hat{\mathbf{n}}})Y_{l_3 m_3}({\hat{\mathbf{n}}})Y_{l_4 m_4}({\hat{\mathbf{n}}})t^{l_1 l_2}_{l_3 l_4}$. However, as we have shown in Chapter \ref{chapter:methodology} this assumption does hold for the $g_{NL}$ model (which is independent of $L$). Their $2\sigma$ constraint on the $g_{NL}$ model is $-7.4\times 10^5<g_{NL}<8.2 \times 10^5$ at $l_{\rm{max}}=600$. Since these bounds were obtained using a joint analysis with the $\tau_{NL}$ model, there may be some inaccuracies due to the aforementioned problem. Another issue with their approach is that it does not directly subtract the effect of anisotropic noise and other systematic effects using the quadratic terms in the optimal trispectrum estimator (Chapter \ref{chapter:methodology}); we know from the present work that these are important in obtaining an accurate and optimized result. 
\par 
In this section, we apply the general mode expansion estimator \eqref{lateestim} to obtain the first constraints on the constant and equilateral models using the trispectrum. We also present a fully consistent treatment of the $g_{NL}$ model, identifying why the naive treatment of the signal to noise using a Sachs-Wolfe estimator leads to inaccuracies. We also present the first bounds on cosmic strings using the trispectrum. As well as calculating the optimal Fisher bounds, we evaluate the constraints on these models by comparing to WMAP$5$ year data. We note that, for the primordial models considered,  
the models are first decomposed at primordial times in the form \eqref{eq:shapedecomp} and subsequently this expansion is written in the form of a late-time expansion using \eqref{eq:latetimedecomp}. The correlation between the early and late-time expansions has been verified to exceed $99\%$ for the primordial models considered here.
\par
In this section we shall assume the primordial shape is normalised to give unity at the equal $k$ limit. To achieve this we set the normalisation constant $N=24$. In our analysis, inhomogeneous noise obtained by coadding WMAP V and W channels was included, along with the use of a KQ$75$ mask.

\subsection{Constant Model}
The constant model with shape given by
\begin{align}
S_T^{\rm{const}}(k_1,k_2,k_3,k_4)=\frac{(k_1 k_2 k_3 k_4)^{9/4}}{24 \tilde{\Delta}_{\Phi}^3}T_{\rm{const}}(k_1,k_2,k_3,k_4)=1
\end{align}
represents the simplest possible primordial shape. This shape results in a CMB trispectrum $t^{l_1 l_2}_{l_3 l_4}$ with features entirely due to the transfer functions. In a similar manner to which the constant primordial bispectrum was shown to correspond to quasi-single field inflation \cite{Chen2}, we expect this shape to have an explicit physical motivation. The primordial decomposition is accurate to $100\%$, while the late-time decomposition recovers the full shape to $\gtrsim 99\%$ using $50$ modes. In Figure~\ref{fig:constmodes} we plot a comparison of the mode coefficients $\overline{\alpha}^R_n$ from the constant model CMB trispectrum with the WMAP coefficients $\overline{\beta}_n^R$, showing little correlation. Of particular interest is the negative zeroth WMAP mode, which, as we shall see, is at odds with the various models considered in this chapter.

The optimal (Fisher) bound is given by 
\begin{align}
\Delta {t_{NL\, \mathrm{opt}}^{\mathrm{const}}}=\frac{1.08}{(S/N)}=1.08\frac{\sqrt{24 }}{\sqrt{f_{sky}}N_T^{\mathrm{const}}}=2.67\times 10^6.
\end{align}
By comparison of the constant model with WMAP5 data we obtain the constraint
\begin{align}
t_{NL}^{\rm{const}}=(-1.33\pm 3.62)\times 10^6,
\end{align}
where the $t_{NL}$ parameter was defined in \eqref{tnldef}.
In terms of $G_{NL}$, defined in \eqref{eq:GNLdef}, this reads
\begin{align}
G_{NL}=(-2.64\pm 7.20 )\times 10^5.
\end{align}
The variance is determined from $400$ Gaussian simulations in the same WMAP-realistic context. We conclude that there is no current evidence for a significant constant primordial non-Gaussian signal.
\begin{figure}[htp]
\centering 
\includegraphics[width=152mm]{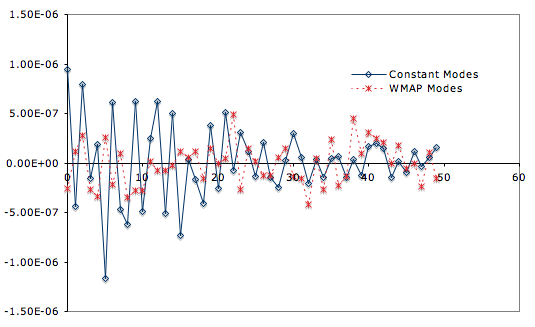}
\caption{Comparison between constant model expansion coefficients $\overline{\alpha}_n^R$ and recovered WMAP$5$ modes, $\beta_n^R$. The features present for the constant model are mainly due to the transfer functions.}
\label{fig:constmodes}
\end{figure}

\subsection{Equilateral Model}
Equilateral type models are produced through the amplification of nonlinear effects around the time the modes exit the horizon. A non-standard kinetic term allows for such a possibility.
In \cite{aChen} it was shown that the leading order trispectrum for single field inflation models is described by a combination of three scalar-exchange trispectra, $T_{s_1,s_2,s_3}$, and three contact-interaction trispectra, $T_{c_1,c_2,c_3}$. We omit the detailed formulae for these models referring the reader to \cite{aChen}. These trispectra generally have a diagonal dependence. Therefore, we consider the decomposition in terms of these diagonals and present a measure of the correlation as described in Section \ref{sec:correl}. The equilateral model is defined in this chapter as $T_{c_1}$, with shape given by
\begin{align}\label{eq:shapeequil}
S_T^{c_1}(k_1,k_2,k_3,k_4)=\frac{(k_1 k_2 k_3 k_4)^{9/4}}{24 \tilde{\Delta}_{\Phi}^3}T_{c_1}(k_1,k_2,k_3,k_4)=\frac{(k_1 k_2 k_3 k_4)^{5/4}}{\left((k_1+k_2+k_3+k_4)/4\right)^{5}}.
\end{align}
The trispectrum for $k$-inflation \cite{PDM99,WY08,ELL08} is given by 
\begin{align}
P_k\propto P_{s_3}.
\end{align}
The trispectrum for DBI inflation is given by evaluating the combination
\begin{align}
P_{DBI}=-P_{c_1}+\frac{1}{4}P_{s_1}+\frac{1}{2}P_{s_2}+P_{s_3}.
\end{align}
In order to estimate the accuracy of using the $c_1$ model to approximate DBI inflation we evaluate the correlators $\mathcal{C}(P_{c_1},P_{DBI})$ and $\mathcal{C}(P_{c_1},P_{c_3})$ using a Monte-Carlo integration method. We find
\begin{align}
\mathcal{C}(P_{c_1},P_{DBI})\approx 83\%,\qquad \mathrm{and}\qquad \mathcal{C}(P_{c_1},P_{k})\approx 82\%.
\end{align}
Thus all single field inflation models of interest are well approximated by considering the $c_1$ model.
For completeness we include the correlation of the $c_1$ model and the other contact-interaction and scalar-exchange models in Table~\ref{tablecorrel}.
\begin{table}
\vskip-5pt
\centering
\begin{tabular}{|c||c|}
 \hline

(Model A, Model B) &       $\mathcal{C}$(Model A, Model B)       \\

    \hline
    \hline

$(c_1,s_1)$ &$0.95$  \\
 \hline
$(c_1,s_2)$ &$0.88$  \\
 \hline
$(c_1,s_3)$ equivalently $(c_1,k)$ &$0.82$  \\
 \hline
$(c_1,c_2)$ &$0.70$  \\
 \hline
$(c_1,c_3)$ &$0.56$  \\
 \hline
$(c_1,DBI)$ &$0.83$  \\
\hline
  \end{tabular}
  \caption{Correlation of trispectra of single field inflation models with the equilateral ($c_1$) model. The correlation measure is defined in Section~\ref{sec:correl}.}
\label{tablecorrel}
\end{table}
The primordial eigenmode expansion \eqref{eq:shapedecomp} with $50$ modes correlates with the shape \eqref{eq:shapeequil} to $\sim 99.7\%$. We then project this primordial shape forward to the CMB 
as in equation \eqref{earlytime}. Then the late-time eigenmode expansion is found using the prescription described in Section~\ref{sec:decomp}. In Figure~\ref{fig:equilmodes} we plot a comparison between the equilateral model and recovered mode coefficients for the WMAP5 data. We also plot a comparison between the equilateral modes and the modes of the constant model, establishing that their late-time behaviour is very similar. The Fisher matrix analysis reveals the optimal bound
\begin{align}
\Delta {t_{NL\,\mathrm{opt}}^{\mathrm{equil}}}=\frac{1.08}{(S/N)}=1.08\frac{\sqrt{24 }}{\sqrt{f_{sky}}N_T^{\mathrm{equil}}}=5.52\times 10^6.
\end{align}
By comparison of the equilateral model with WMAP$5$ data we obtain the constraint
\begin{align}
t_{NL}^{\rm{equil}}=(-3.11\pm 7.5)\times 10^6 \quad\Longleftrightarrow \quad G_{NL}=(-3.02\pm 7.27)\times 10^5.
\end{align}
Again we conclude that there is little evidence in favour of the equilateral model given current CMB observations.
\begin{figure}[htp]
\centering 
\includegraphics[width=162mm]{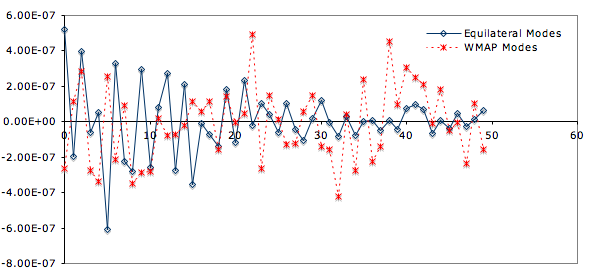}
\includegraphics[width=162mm]{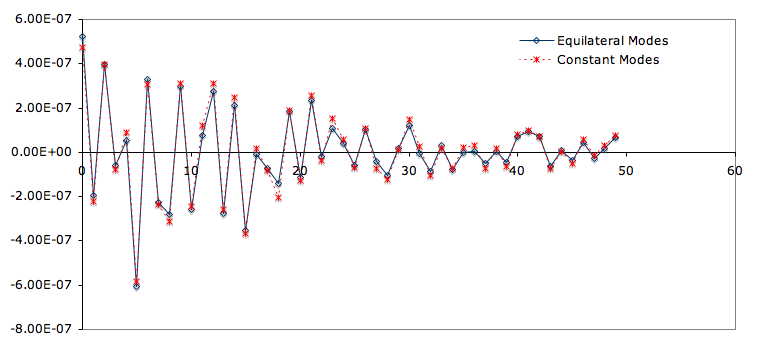}
\caption{Equilateral model expansion coefficients $\overline{\alpha}_n^R$ compared with WMAP$5$ modes, $\beta_n^R$ (top panel) and compared with constant model expansion coefficients (bottom panel). In the bottom panel the coefficients are normalised for a more direct comparison.}
\label{fig:equilmodes}
\end{figure}

\subsection{Local Model (cubic/$g_{NL}$ term)}
The local model refers to a range of models where the non-Gaussianity is produced by local interactions. Examples include single-field slow-roll inflation, which produces only tiny levels of non-Gaussianity. Nonetheless large local non-Gaussianity may be produced in multifield inflation models \cite{RigopoulosShellardvanTent2006A,SeeryLidsey2005,VernizziWands2006} and curvaton models \cite{LindeMukhanov2006,LythUngarelliWands2003,BartoloMatarreseRiotto2004}, amongst others. For a comprehensive review see \cite{Chen2010}.
The $g_{NL}$ term of the local model relates to the cubic term in the (local) Taylor expansion of the potential,
\begin{align}
\Phi=\Phi_L+f_{NL}(\Phi_L^2-\langle \Phi_L^2\rangle)+g_{NL}\Phi_L^3+\mathcal{O}(\Phi_L^4).
\end{align}
This term induces the following trispectrum shape (see \cite{Okamoto}):
\begin{align}\label{eq:shapelocalgnl}
S_T^{g_{NL}}(k_1,k_2,k_3,k_4)=\frac{(k_1 k_2 k_3 k_4)^{9/4}}{24 \tilde{\Delta}_{\Phi}^3}T_{g_{NL}}(k_1,k_2,k_3,k_4)=\frac{k_1^3+k_2^3 +k_3 ^3 +k_4^3}{4(k_1 k_2 k_3 k_4)^{3/4}}.
\end{align}
A further verification of the accuracy of the modal decomposition to the correlation measure may be obtained by comparison of the CMB trispectrum calculated using a Sachs-Wolfe approximation with the full CMB trispectrum. The Sachs-Wolfe approximation is obtained by calculating equation~\eqref{extraRed2} with the transfer function replaced by a spherical Bessel function,
\begin{align}
\Delta_l(k)=\frac{1}{3}j_l((\tau_0-\tau_{\mathrm{dec}})k).
\end{align}
Using, as in Chapter \ref{chapter:methodology}, the results
\begin{align}
\int d k k^2 j_l(k) j_l(x k)&=\frac{\pi}{2x^2}\delta(x-1),\\
 \int d k k^{-1} j_l(k) j_l(x k)&=\frac{1}{2 l(l+1)},
\end{align}
we obtain the Sachs-Wolfe approximation
\begin{align}\label{eq:SachsWolfegnl}
t^{l_1 l_2}_{l_3 l_4}=\frac{2\tilde{\Delta}_{\Phi}^3 }{27 \pi^3}\left(\frac{1}{l_2(l_2+1)l_3(l_3+1)l_4(l_4+1)}	+3\,\mathrm{perms}\right).
\end{align}
In Figure~\ref{fig:equalLtrisp} we plot the ratio of the full result with the Sachs-Wolfe approximation. In the Sachs-Wolfe regime, $l\lesssim 60$, the modal decomposition shows extraordinary agreement with the analytic approximation, thus further validating the approach. Beyond this limit, we observe the imprint of the first acoustic peak around $l\approx 200$. Naively, we would expect to observe a second peak near $l\approx 400$, as observed for the equal $l$ bispectrum (see \cite{FLS10}). It appears that the extra $1/k$ in the calculation of the trispectrum \eqref{extraRed2} leads to an additional damping effect for the trispectrum. This leads to large discrepancies between approximations for the signal-to-noise calculated using a Sachs-Wolfe approximation as in Section~\ref{sec:sachswolfe} and the real signal-to-noise. To emphasise this point, we plot in Figure~\ref{fig:signoise} the signal-to-noise as calculated using the full implementation of the modal decomposition versus the signal-to-noise as calculated naively using the Sachs-Wolfe approximation. As expected, there is good agreement deep in the Sachs-Wolfe regime, $l\ll100$. However, for higher multipole values the discrepancies increase. Due to the increased damping effect the signal to noise increases approximately as $\log(l)$, rather than the $l \log(l)$ behaviour suggested by the Sachs-Wolfe result \eqref{eq:sachswolfe}. The correlation measure shows that the eigenmode expansion \eqref{eq:shapedecomp} with $50$ modes correlates with the shape \eqref{eq:shapelocalgnl} to $\sim 90\%$.
\begin{figure}[htp]
\centering 
\includegraphics[width=162mm]{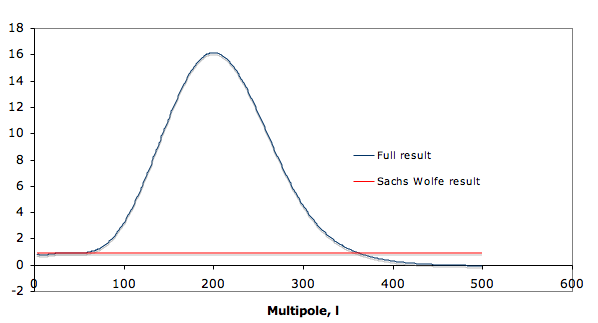}
\caption{Plot of the full modal decomposition trispectrum divided by the approximate analytic Sachs-Wolfe result \eqref{eq:SachsWolfegnl} for the local $g_{NL}$ model in the equal $l$ limit, i.e. $l_1=l_2=l_3=l_4=l$. The comparison to the normalised Sachs-Wolfe result (unity) for $l\lesssim 60$ shows the accuracy of the formalism adopted in this chapter. }
\label{fig:equalLtrisp}
\end{figure}

\begin{figure}[htp]
\centering 
\includegraphics[width=172mm]{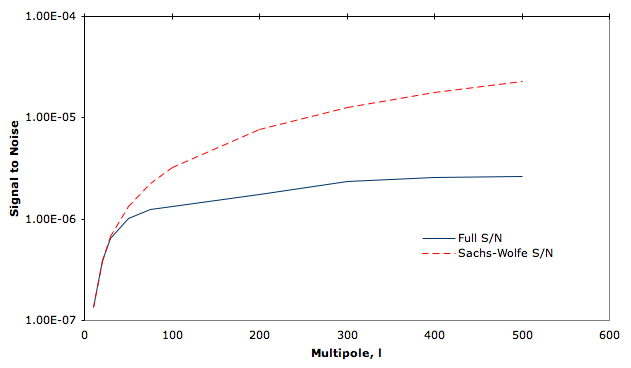}
\caption{Plot of the signal to noise for the $g_{NL}$ local model, as calculated using the modal expansion with $50$ modes and using the Sachs-Wolfe approximation.}
\label{fig:signoise}
\end{figure}
\begin{figure}[htp]
\centering 
\includegraphics[width=162mm]{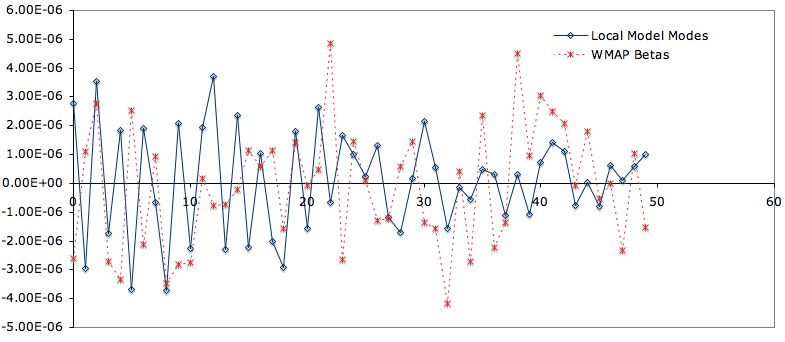}
\caption{Comparison between local $g_{NL}$ model expansion coefficients, $\overline{\alpha}_n^R$, and recovered modes for the WMAP$5$ data. The apparently slow convergence of the local model may be observed. There appears to only be a weak correlation between the local $g_{NL}$ model and WMAP$5$ data. }
\label{fig:modeslocalwmap}
\end{figure}
The optimal (Fisher) bound is given by 
\begin{align}
\Delta {g_{NL\,\mathrm{opt}}}(\equiv \Delta t_{NL\,\mathrm{opt}}^{g_{NL}}/1.08)=\frac{1}{(S/N)}=\frac{\sqrt{24 }}{\sqrt{f_{sky}}N_T^{g_{NL}}}=5.35\times 10^5.
\end{align}
These bounds are at $1\sigma$ significance. This establishes that results quoted in \cite{vielva2,Cooray2} are somewhat super-optimal. The reasons for this discrepancy have been discussed at the beginning of the section.
\par
In order to compare the local model trispectrum with the recovered WMAP trispectrum we plot the modes $\overline{\alpha}^R_n$ and $\overline{\beta}^R_n$ in Figure~\ref{fig:modeslocalwmap}. There appears to only be a weak correlation between the two sets of modes. Calculating the mode estimator~\eqref{lateestim}, we obtain the constraint,
\begin{align}
g_{NL}(\equiv  t_{NL}^{g_{NL}}/1.08)=G_{NL}=(1.62\pm 6.98)\times 10^5.
\end{align}
The result establishes that the correlation between the local $g_{NL}$ model and the WMAP trispectrum is quite weak. From this bound, obtained from $400$ Gaussian realisations with WMAP$5$ year specifications, we observe that our implementation of the estimator is accurate to within $\sim 30\%$. In order to achieve a result closer to the optimal bound, an estimator incorporating the full inverse covariance matrix, as given by equation~\eqref{estim}, must be calculated.

\subsection{Cosmic Strings}
A significant advantage of the formalism developed here (and in Chapter \ref{chapter:methodology}) is that it may be readily applied to late-time models as well as primordial models. Here, we apply the methodology to the trispectrum of cosmic strings. Cosmic strings are line-like discontinuities which may be formed during a phase transition in the very early universe \cite{KibbleMech}\footnote{ Cosmic strings may also be formed at the end of brane inflation \cite{TyeSarangi} but have somewhat different network properties and are generally referred to as cosmic superstrings.}. Cosmic strings are characterised by their tension $G\mu$. Our intention, therefore, is to use the trispectrum induced by cosmic strings to constrain $G\mu$.
\par
Firstly, we note that a diagonal dependent CMB trispectrum may be decomposed in a similar form to the primordial decomposition in Section~\ref{sec:correl} (see Chapter \ref{chapter:methodology}) with
\begin{align}
\langle a_{l_1 m_1}	 a_{l_2 m_2} a_{l_3 m_3} a_{l_4 m_4}	\rangle_c=\sum_{L M}(-1)^M \mathcal{G}^{l_1 l_2 L}_{m_1 m_2 -M}\mathcal{G}^{l_3 l_4 L}_{m_3 m_4 M} p^{l_1 l_2}_{l_3 l_4}(L)+(l_2\leftrightarrow l_3)+(l_2\leftrightarrow l_4),
\end{align}
where the Gaunt integral is given by
\begin{align}
\mathcal{G}^{l_1 l_2 l_3}_{m_1 m_2 m_3}=\int d\hat{\mathbf{n}}Y_{l_1 m_1}(\hat{\mathbf{n}})Y_{l_2 m_2}(\hat{\mathbf{n}})Y_{l_3 m_3}(\hat{\mathbf{n}}).
\end{align}
For primordial models the combination $P_{\Phi}(k_1,k_2,k_3,k_4,K)$ corresponds directly to $p^{l_1 l_2}_{l_3 l_4}(L)$. In the diagonal-free case we have
\begin{align}\label{eq:trispcombi}
t^{l_1 l_2}_{l_3 l_4}=p^{l_1 l_2}_{l_3 l_4}+p^{l_1 l_3}_{l_2 l_4}+p^{l_1 l_4}_{l_2 l_3}.
\end{align}
\par
In Chapter~\ref{chapter:stateoftheart} (see also \cite{hind10}) the trispectrum induced by cosmic strings was derived. The analysis assumed that the temperature discontinuity produced by cosmic strings is given entirely by the Gott-Kaiser-Stebbins effect \cite{Gott,KAISERSTEBBINS}. Contributions due to decoupling were neglected in the analysis. On angular scales relevant for WMAP, i.e. $l\lesssim 500$, the cosmic string CMB trispectrum was derived. The quantity derived was inclusive of the unconnected term. Accounting for the unconnected term requires altering the late-time formulae in Sections \ref{sec:decomp} and \ref{sec:reconTrisp}. However, we have verified that regarding this quantity as independent of the unconnected term leads to little quantitative difference (of order less than a few percent). Therefore, we neglect this subtlety for the purposes of clarity in this section and write
\begin{align}\label{eq:pcombi}
(l_1 l_2 l_3 l_4)^{3/2}p^{l_1 l_2}_{l_3 l_4}(L)=&(8\pi G\mu)^4\frac{2\overline{v}^4 \pi }{s^2 }\frac{ l_2^2}{(l_1 l_2 l_3 l_4)^{1/2}}   \frac{1}{(0.63+L_1\tilde{\xi})} \frac{\left( l_1\tilde{\xi}\right)^2}{(0.63+l_1\tilde{\xi})} \nonumber\\
&\times\ln\left(\frac{1+\eta_0/\eta_{lss}}{2}\right)\left( \frac{2}{1+500/l_m}\right)^{2.3},
\end{align}
where we note that $L_1=\sqrt{l_1^2 l_2^2 -l_{12}^2 }/l_1$, $2 l_{12}=L^2-l_1^2-l_2^2$, $l_m=\min(500,l_i)$, $\tilde{\xi}=1/l_m$ and $\eta_0/\eta_{lss}\approx 50$. The values of the parameters $\overline{v}^2$ and $s^2$ are found by simulations \cite{MartShell,BBS,AllShel} to be given by the numerical values $\overline{v}^2=0.365$ and $s^2=0.42$. Of course, since \eqref{eq:pcombi} possesses a diagonal dependence we must make an approximation in order to avail of the formalism outlined in this chapter. In particular, we make the underestimation by noting $L_1\tilde{\xi}\leq 1$ and replace
\begin{align}
\frac{1}{0.63+L_1\tilde{\xi}}\longrightarrow \frac{1}{1.63}.
\end{align} 
The maximum error in the estimation of $G\mu$ by the use of this approximation is of order $25\%$, although the actual error incurred is likely to be much less. Calculating the correlation using equation \eqref{eq:correl} we find that this approximation is accurate to $\sim 90\%$.
\par
The approximation gives the quantity $p^{l_1 l_2}_{l_3 l_4}$ which is independent of the diagonal $L$. The CMB trispectrum is then given by equation~\eqref{eq:trispcombi}. We then decompose the cosmic string trispectrum in the form \eqref{eq:latetime}. Using $50$ modes the expansion is found to be accurate to $\sim 99\%$. The mode expansion coefficients are plotted in Figure~\ref{fig:cosmicstrings}. The modes are clearly distinguishable from the primordial models previously discussed due to the lack of acoustic peaks.
\begin{figure}[htp]
\centering 
\includegraphics[width=122mm]{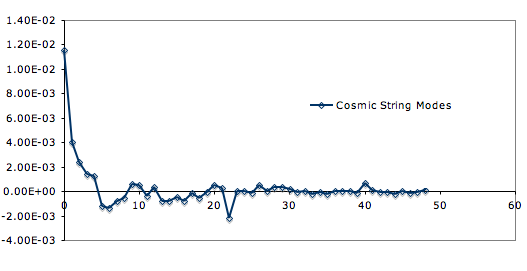}
\caption{Plot of the modal expansion coefficients, $\overline{\alpha}_n^R$, of cosmic strings. Notice the relative lack of structure, which is due to the absence of acoustic peaks. }
\label{fig:cosmicstrings}
\end{figure}
The signal to noise $S/N=N_T\sqrt{f_{sky}/24}$ gives the following optimal bound on the cosmic string tension achievable using the CMB trispectrum at WMAP$5$ resolution,
\begin{align}
G\mu \lesssim 8.8\times 10^{-7}.
\end{align}
This should be compared to current constraints on Abelian-Higgs strings $7\times 10^{-7}$ \cite{Bevis}, and on Nambu-Goto strings $2.5\times 10^{-7}$ \cite{Battye}. We have also obtained a forecast for the optimal tension that may be probed using Planck data $G\mu_{\rm{opt}}^{\rm{Planck}}=1.8\times 10^{-7}$. In Figure~\ref{fig:signoisecstring}  we plot the signal to noise of cosmic strings as a function of multipole. Unlike the local $g_{NL}$ model the signal to noise for cosmic strings is unaffected by Silk damping. Thus, we expect the trispectrum to provide a competitive probe, as a test for cosmic strings, to the power spectrum given the increased resolution of the Planck satellite and may provide the best probe for cosmic strings in future surveys.
\begin{figure}[htp]
\centering 
\includegraphics[width=122mm]{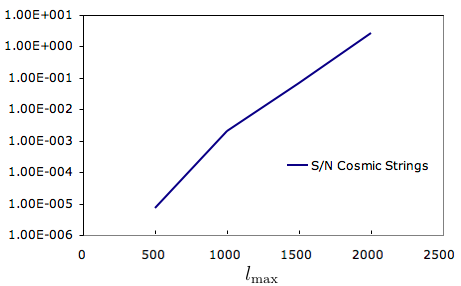}
\caption{Plot of the signal to noise against maximum multipole for the trispectrum of cosmic strings. The signal to noise is expressed in units $(G\mu/2\times 10^{-7})^4$. }
\label{fig:signoisecstring}
\end{figure}
Upon comparison with WMAP$5$ data, by calculating the mode estimator~\eqref{lateestim}, we obtain the $1\sigma$ bound
\begin{align}
\left(\frac{G\mu}{2\times 10^{-7}}\right)^4= -852\pm 870.
\end{align}
Since the tension must be positive, we deduce that cosmic strings appear to be disfavoured by current CMB data. The $2\sigma$ bound on cosmic strings is given by $1.1\times 10^{-6}$.

\section{Discussion and Conclusion}\label{sec:concluImplement}
We have implemented a separable mode expansion to investigate a class of models that are independent of the diagonal, i.e. only depend on the wavenumbers $k_1,k_2,k_3,k_4$. We have obtained constraints on the cubic term for the local model, the constant model and a notable new constraint on the equilateral model. The results for these models are summarised in Table~\ref{tableconstraints}. We found no evidence for significant deviations from Gaussianity in these models (at 95\%) confidence. 
The constraints on the parameter $g_{NL}^{local}$ represent a constraint on the self-interaction term of the local model and, in addition, allows for a qualitative classification of models of the local type \cite{1009.1979}. The constraints presented here on the equilateral model are entirely new results. The importance of finding bounds on such models is, not least, as a consequence of the high correlation between the equilateral model and DBI inflation. We have also obtained the $95\%$ bound on cosmic strings $G\mu\lesssim 1.1\times 10^{-6} $. Using forecasts for the signal to noise at Planck resolution we establish that the trispectrum of cosmic strings is expected to give comparable constraints to the power spectrum in the near future. The advantage of using such a probe to search for cosmic strings is that, unlike, for example, gravitational waves, this test is largely background independent.
\par
Aside from obtaining constraints on these models, the approach adopted in this chapter allows us to directly reconstruct the CMB trispectrum given observations of the mode coefficients $\beta_n^Q$. The general mode expansion allows for a characterisation of noise and foregrounds which much must be subtracted to obtain an estimation measure.
This work marks a significant first step in a general analysis of trispectrum models. The advantage of the modal approach is the absence of pathologies and the opportunity to investigate models which were previously deemed intractable.
\par
While the constraints published in this chapter are consistent with Gaussianity, the extension of the analysis to general trispectra represents an important next step \cite{FRS11}. Such an analysis will, for instance, allow a measurement of the local trispectrum parameter $\tau_{NL}$. This importance of this quantity is that it allows for a test of local inflation which requires $\tau_{NL}\geq(6f_{NL}/5)^2$.
An implementation of the late-time modal estimator outlined in Chapter \ref{chapter:methodology} to include trispectra dependent on the diagonal term will allow identification of any trispectrum whether generated at primordial times like inflation or late-times like gravitational lensing or second-order gravitational effects. This, in conjunction with recent results classifying CMB bispectrum constraints, offers the hope of a comprehensive test for non-Gaussianity. 
\begin{table}
\vskip-5pt
\centering
\begin{tabular}{|c||c|c|}
 \hline

Model &      $t_{NL}$    &$G_{NL}$  \\

    \hline
    \hline

Constant Model &$(-1.33\pm 3.62)\times 10^6$&$(-2.64\pm 7.20 )\times 10^5$  \\
 \hline
Equilateral Model &$(-3.11\pm 7.5)\times 10^6$&$(-3.02\pm 7.27)\times 10^5$  \\
 \hline
Local $g_{NL}$  Model&$(1.75\pm 7.54)\times 10^5$&$(1.62\pm 6.98)\times 10^5$  \\
\hline
  \end{tabular}
  \caption{$1\sigma$ constraints from WMAP$5$ data for the constant, equilateral and local $g_{NL}$ models. It is evident that the integrated measure $G_{NL}$ allows for fairer comparison between models than the primordial measure $t_{NL}$.}
\label{tableconstraints}
\end{table}

\newpage

\thispagestyle{empty}
\mbox{}
\newpage
\chapter{Large Scale Structure}
\label{chapter:implementation}
\textbf{Summary}\\
\textit{We present an efficient separable approach to the estimation and
reconstruction of the bispectrum and the trispectrum from observational
(or simulated) large scale structure data.  This is developed from
general CMB (poly-)spectra methods which exploit the
fact that the bispectrum and trispectrum in the literature
can be represented by
a separable mode expansion which converges rapidly (with $\nmax=
{\cal{O}}(30)$ terms).  With an effective grid
resolution $\lmax$ (number of particles/grid points $N=\lmax^3$),
we present a bispectrum estimator which requires only ${\cal O}(\nmax \times \lmax^3)$ operations, along with a corresponding method
for direct bispectrum reconstruction. This method is extended to the trispectrum revealing an estimator which requires only  ${\cal O}(\nmax^{4/3} \times \lmax^3)$ operations. The complexity in calculating the trispectrum in this method is now involved in the original decomposition and orthogonalisation process which need only be performed once for each model. However, for non-diagonal trispectra these processes present little extra difficulty and may be performed in ${\cal O}(\lmax^4)$ operations. A discussion of how the methodology may be applied to the quadspectrum is also given. An efficient algorithm for the generation of arbitrary non-Gaussian initial conditions for use in N-body codes using this separable approach is described. This prescription allows for the production of non-Gaussian initial conditions for arbitrary bispectra and trispectra. A brief outline of the key issues involved in parameter estimation, particularly in the non-linear regime, is also given. }
\vspace{50pt}

\section{Introduction}

In previous work (\cite{FLS09}, Chapter \ref{chapter:methodology}, Chapter \ref{chapter:evaluation}) we developed and implemented a methodology for the efficient and  
general analysis of non-Gaussianity in the cosmic microwave sky.   Our purpose here is to apply  
these separable mode methods to  large-scale structure, making tractable a fast general 
analysis of all bispectra and trispectra, rather than the few special cases studied to date.   
Calculation of the three-point correlator or bispectrum $\langle \delta_{{\bf k}_1}\delta_{{\bf k}_2}\delta_{{\bf k}_3}\rangle$
using 3D large-scale structure data naively appears to require a computationally
intensive $\lmax^6$ operations, or $\lmax^9$ for the trispectrum, where $\lmax$ is the 
effective observational or simulated grid resolution (i.e.\ the volume sidelength $L$ over the averaged 
galaxy or grid spacing $\Delta x$, giving a particle number $N\approx\lmax^3$). However, if - as in the
CMB - predicted non-Gaussianity can be described by rapidly convergent and separable mode expansions, then
there is a dramatic reduction  to only ${\cal O}(\nmax\times\lmax^3)$ operations for estimating any bispectrum, 
where $\nmax$ is the (small) number of modes required for an accurate representation ($\nmax \approx 30$ for
WMAP analysis \cite{FLS10}).   The relative impact on trispectrum estimation is even more dramatic, reducing 
again to $\sim {\cal O}(\nmax^{4/3} \times \lmax^3)$ operations.   
Direct reconstruction of the bispectrum today then allows for the decomposition into its 
constituent  and independent shapes, including contributions directly from the primordial bispectrum, 
from next-to-leading order terms in nonlinear gravitational collapse, from the convolved primordial 
trispectrum, etc.   These methods equally can be applied to 
generating simulation initial conditions with arbitrary given bispectrum and trispectrum, 
again using a simple separable mode algorithm requiring only ${\cal O}(\nmax\times\lmax^3)$ or ${\cal O}(\nmax^{4/3}\times\lmax^3)$ operations respectively.

Our purpose here is not to review the many important contributions made to the study of higher-order correlators 
in large-scale 
structure, for which there are some comprehensive recent reviews available (\cite{LigSef2010,DS}).   However, we
note that the 
field is well-motivated because non-Gaussianity is recognised as a critical test of the simplest standard inflationary 
scenario.   Moreover, there are a growing number of alternative inflationary scenarios where deviations from non-Gaussianity can be large (see \cite{Chen2010} for a review). The most stringent constraints on primordial non-Gaussianity so far have come from CMB bispectrum measurements (e.g.\ \cite{WMAP7, FLS10}, see \cite{LigSef2010}) with relatively weak constraints coming from the large-scale structure galaxy bispectrum \cite{0312286} due to complications in dealing with non-linear evolution. While it appears to be possible also to derive competitive constraints using the abundance 
of rare objects or scale-dependent bias (e.g.\ \cite{08053580}), these complementary approaches generally assume 
a local-type non-Gaussianity (see the review \cite{Verde2010}).   With improving galaxy and other surveys 
covering a growing fraction of the sky, it is reasonable to expect measurements of higher order correlators from this three-dimensional data to provide the best and most comprehensive information about non-Gaussianity.   
These large-scale structure (poly-)spectra should allow us to discriminate between different non-Gaussian shapes, 
notably between primordial and late-time sources, ultimately complementing 
CMB measurements and exceeding them in precision.

In this chapter we present a method for quickly calculating the bispectrum from a given density perturbation in section \ref{sec:bispectrum}.  Next we show how to extend this analysis to the trispectrum in section \ref{sec:trispectrum}. As any estimator would require non-Gaussian simulations for testing and error analysis, we present an approach in section \ref{sec:intcon} for including a general bispectrum and trispectrum in the initial conditions for $N$-body simulations. We then go on to show in section \ref{sec:fnlestimation} how a general estimator for constraining primordial non-Gaussianity can be constructed, when the bispectrum can be approximated using a simple ansatz, and in the completely general case. Finally we present our concluding remarks.

\section{Large scale structure bispectrum calculation}\label{sec:bispectrum}

\subsection{General bispectrum estimator}

Higher-order correlators of the galaxy or matter density distribution can be expected to exhibit a low signal-to-noise
for individual combinations of wavenumbers (as for multipoles in the CMB).  A useful strategy for the comparison 
between observations and theoretical models (or simulated numerical models) is the use of an estimator which 
tests for consistency by summing over all multipoles using an optimal signal-to-noise weighting.  
The general estimator for the galaxy or density bispectrum, when searching for a given theoretical 
three-point correlator $\<\d_{\bk_1}\d_{\bk_2}\d_{\bk_3}\>$, is
\begin{align}\label{eq:bestimatorgen}
\curl{E} =& \int \frac{d^3k_1}{(2\pi)^3} \frac{d^3k_2}{(2\pi)^3} \frac{d^3k_3}{(2\pi)^3} 
\<\d_{\bk_1}\d_{\bk_2}\d_{\bk_3}\>
\Big[C^{-1}(\d^{obs}_{\bk_1})C^{-1}(\d^{obs}_{\bk_2})C^{-1}(\d^{obs}_{\bk_3}) \nonumber\\
&- 3 C^{-1}(\d^{obs}_{\bk_1}\d^{obs}_{\bk_2})C^{-1}(\d^{obs}_{\bk_3})\Big]
\end{align}
where $\delta^{obs}_{{\bf k}}$ represents a noisy measurement of the galaxy or density perturbation with 
signal plus noise covariance $C$ given by 
\begin{align}
C^{-1}(\d^{obs}_{\bk}) = \int \frac{d^3k'}{(2\pi)^3} \<\d_{\bk}\d_{\bk'}\>^{-1}\d^{obs}_{\bk'}.
\end{align}
We will discuss the normalisation necessary for parameter estimation in section~\ref{sec:fnlestimation}. Here, we have added a linear term to the cubic estimator in order to account for inhomogeneous effects from 
incomplete survey coverage (e.g.\ due to dust extinction), sampling bias, shot noise, 
and other known systematics, which together can substantially increase the experimental variance.  

If we assume that the density field is statistically isotropic, as it is in most well-motivated theoretical models, then
the bispectrum $ B(k_1,k_2,k_3)$ is defined by  
\begin{align}
\<\d_{\bk_1}\d_{\bk_2}\d_{\bk_3}\> &= (2\pi)^3 \d_D(\bk_1+\bk_2+\bk_3) B(k_1,k_2,k_3)\,,
\end{align}
where $ \d_D ({\bf k})$ is the three-dimensional Dirac $\delta$-function enforcing a triangle condition on the wavevectors ${\bf k}_i$, for which it is sufficient to use only the wavenumbers $k_i = |\bk_i|$.   
For simplicity, let us suppose we are only in a mildly nonlinear regime 
with good observational coverage over a modest redshift range, so that we can make 
the approximation that the covariance matrix is nearly diagonal 
$C^{-1}(\d^{obs}_{\bk}) \approx {\d^{obs}_{\bk}}/{P(k)}$.
With these replacements, the estimator (\ref{eq:bestimatorgen}) becomes
\begin{align}\label{eq:bestimator}
\curl{E} = \int \frac{d^3k_1}{(2\pi)^3} \frac{d^3k_2}{(2\pi)^3} \frac{d^3k_3}{(2\pi)^3} &\frac{(2\pi)^3 \d_D(\bk_1+\bk_2+\bk_3)B(k_1,k_2,k_3)}{P(k_1)P(k_2)P(k_3)}\nonumber\\
&\times\left[\d^{obs}_{\bk_1}\d^{obs}_{\bk_2}\d^{obs}_{\bk_3}  - 3\langle
\d^{sim}_{\bk_1}\d^{sim}_{\bk_2}\rangle \d^{obs}_{\bk_3}\right]\,,
\end{align}
where $\d^{sim}_\bk$ represents simulated data with the known inhomogeneous systematic effects included, while 
we also assume that shot noise is incorporated in the power spectrum $P+N \rightarrow \tilde P$, along with 
incomplete sample coverage (though we will drop the tilde).  We note that, although this 
galaxy estimator with a linear term (\ref{eq:bestimator}) has not been given in this form explicitly before, 
the bispectrum scaling and signal-to-noise ratios here and in what follows are consistent with the pioneering discussions in  refs.~\cite{9704075,0312286} (see also the analogous CMB bispectrum 
estimator discussed in ref.~\cite{Crem} and elsewhere).
In any case, this large-scale structure bispectrum estimator (\ref{eq:bestimator}) does not appear to be 
particularly useful because its brute force evaluation would require at least $\lmax^6$ operations for a single measurement (after 
imposing the triangle condition).  The problem is compounded by the many simulated realizations of the 
observational set-up which are required to obtain an accurate linear term in (\ref{eq:bestimator}).  In fact, if the
 theoretical bispectrum $ B(k_1,k_2,k_3)$ is computed numerically, then this is even more computationally intensive,  
 since it requires  many $N$-body simulations and bispectrum evaluations to achieve statistical precision.       

Nevertheless, let us now suppose that we have a large set of simulated non-Gaussian realisations $\d^{obs}_k$ generated 
with the same theoretical bispectrum $B(k_1,k_2,k_3)$  (and the same power spectrum $P(k)$).   If we take the expectation 
value of the estimator (\ref{eq:bestimator}) by summing over these realisations, then we find the average to be  
\eqa\label{eq:bestav}
\<\curl{E}\> & =&  \int \frac{d^3k_1}{(2\pi)^3} \frac{d^3k_2}{(2\pi)^3} \frac{d^3k_3}{(2\pi)^3} (2\pi)^6 \d_D^2(\bk_1+\bk_2+\bk_3)  \frac{B^2(k_1,k_2,k_3)}{P(k_1)P(k_2)P(k_3)} \cr
&= & \frac{V}{\pi}\int_{\curl{V}_B}  dk_1 dk_2 dk_3 \,\frac{ k_1 k_2 k_3 \,B^2(k_1,k_2,k_3)}{P(k_1)P(k_2)P(k_3)}\,,
\qea
where ${\curl{V}_B}$ is the tetrahedral region allowed by the triangle condition.   The averaged estimator (\ref{eq:bestav}) is
an important expression, so it is instructive for subsequent calculations
to outline the explicit steps that take us between these two lines.   First, the second 
Dirac $\delta$-function contributes only a volume factor $\delta({\bf 0}) = V/(2\pi)^3$.   Secondly, we complete the angular 
integration by expanding the integral form of the remaining $\delta$-function in spherical Bessel functions and harmonics,
\begin{align}\label{eq:deltaint}
\d_D(\bk) &= \frac{1}{(2 \pi)^3} \int d^3x e^{i \bk \cdot \bx},\\\label{eq:exptobessel}
 e^{i \bk \cdot \bx} &=4\pi \sum_{l m} i^l j_l(kx) Y_{lm}(\uk)Y^*_{lm}(\ux)\,.
\end{align}
Thirdly, each $\uk_i$ integration involves just a single spherical harmonic and contributes a factor 
$2\sqrt{\pi}\,\d_{l0}\,\d_{m0}$, so we end up with only a constant term from the Gaunt integral $G^{000}_{000} = 1/2\sqrt{\pi}$
(i.e.\ the integration over the three remaining $Y_{lm}({\bf x})$). 
Finally, the last integral over
the three Bessel functions $j_0(k_1)j_0(k_2)j_0(k_3)$ yields $\pi/4 k_1k_2k_3$ and simultaneously imposes a triangle condition on $k_1,\,k_2,\,k_3$ which we denote by the restricted domain of integration ${\curl{V}_B}$.  

The estimator average (\ref{eq:bestav}) leads naturally to a weighted cross-correlator or inner product between two different  
bispectra $B_i(k_1,k_2,k_3)$ and $B_j(k_1,k_2,k_3)$, that is,
\begin{align}\label{eq:innerprod}
\mathcal{C}(B_i,B_j)=\frac{\langle B_i, \, B_j\rangle}{\sqrt{\langle B_i, \, B_i\rangle \langle B_j, \, B_j\rangle}}\,,
\end{align}
where   
 \begin{align}
\langle B_i, \, B_j\rangle ~\equiv~
\frac{V}{\pi}\int_{\curl{V}_B}  dk_1 dk_2 dk_3 \,\frac{ k_1 k_2 k_3 \,B_i(k_1,k_2,k_3)\,B_j(k_1,k_2,k_3)}{P(k_1)P(k_2)P(k_3)}\,.
 \end{align}
The estimator (\ref{eq:bestimator}) is thus proportional to the Fisher matrix of the bispectrum, $F_{ij} = \mathcal{C}(B_i,B_j) / 6\pi$ (see ref.~\cite{0312286}). 

The fiducial model for non-Gaussianity is the $\fnl=1$ local model.  For the CMB, where the final CMB bispectrum $\Blll$ is linearly related to the primordial bispectrum $B_0(\kall)$, it is straightforward to define a normalisation which yields a universal $\Fnl$, representing the total integrated bispectrum for a particular theoretical model relative to that from the $\fnl=1$ local model (see ref.~\cite{FLS10}).  However, with bispectrum contributions from gravitational collapse and nonlinear bias arising even with Gaussian initial conditions, a universal normalisation is a more subtle issue which we will defer to section~\ref{sec:fnlestimation}.

Finally, we point out that the bispectrum estimator (\ref{eq:bestimatorgen}) can be applied in any three-dimensional 
physical context where we wish to test for a particular non-Gaussian model.   It can be applied at primordial times, with 
potential fluctuations (i.e. replacing $\d_\bk \rightarrow \Phi_\bk$), in the late-time linear regime on large scales 
where the density perturbation is simply related by a transfer function $\delta_{\bf k} = T(k,z) \,\Phi_k$ (as in the CMB), 
in the mildly non-linear regime where next-to-leading order corrections are known, or
deep in the nonlinear regime on small scales 
where we must rely  on $N$-body and hydrodynamic simulations.   However, for a useful implementation, we
must rewrite (\ref{eq:bestimatorgen}) in a separable form.

\subsection{Separable mode expansions and bispectrum reconstruction}

The averaged estimator (\ref{eq:bestav}) gives a natural measure for defining separable mode functions 
\begin{align}\label{eq:Qexp}
Q_n(\kall) = {\textstyle \frac{1}{6}}[q_r(k_1)\,q_s(k_2)\, q_t(k_3) + 5 \hbox {perms}] \equiv q_{\{r}(k_1)\,q_s(k_2)\, q_{t\}}(k_3)\,,
\end{align}
 which we can use to decompose an arbitrary bispectrum (here, for convenience, 
the label $n$ denotes a linear ordering of the 3D products $n\leftrightarrow\{rst\}$).   
We choose to expand the bispectrum  $B(\kall)$ in its noise-weighted form (see ref.~\cite{FLS09}),
\begin{align}\label{eq:bseparable}
\frac{B(k_1,k_2,k_3)\, v(k_1)v(k_2)v(k_3)}{\sqrt{P(k_1) P(k_2) P(k_3)}} = \sum \aQn \Qn(\kall)\,,
\end{align}
where we have used the freedom to introduce a separable modification to the weight function $w(\kall) = 
k_1k_2k_3/v^2(k_1)v^2(k_2)v^2(k_3)$ in (\ref{eq:bestav}).  Series convergence usually 
can be improved with scale-invariance, suggesting 
the choice $v(k) = \sqrt{k}$.    The exact form of the one-dimensional basis functions $q_r(k)$ is not important, 
except that they should be bounded and well-behaved on the bispectrum domain ${\cal V}_B$.  Some $q_r(k)$ examples
which are orthogonal on ${\cal V}_B$ were given explicitly in ref.~\cite{FLS09}, 
analogues of Legendre polynomials $P_n(k)$.    

The product functions $\Qn$ are independent  but not necessarily orthogonal, so it is convenient 
from these to generate an orthonormal set of mode functions $\Rn$, such that, $\langle \Rn,\,\Rm\rangle =\delta_{nm} $ (achieved using Gram-Schmidt orthogonalisation 
with the inner product (\ref{eq:innerprod})).   We distinguish the expansion coefficients $\aQn$ and $\aRn$ 
by the superscripts for the separable `$Q$' and orthonormal `$R$' modes respectively; these are related to each 
other by a rotation involving the matrices $\langle \Qm,\,\Qn\rangle$ and $\langle \Qm,\,\Rn\rangle$(see ref.~\cite{FLS09}).    The orthonormal modes $\Rn$ are convenient for finding the expansion coefficients of an arbitrary bispectrum $B(\kall)$ from the inner product (\ref{eq:innerprod}) through $\aRn = \langle B,\,\Rn\rangle$ which are then rotated to the
more explicitly separable form $\aQn$.  Of course, there is some computational effort ${\cal O}(\nmax\times\lmax^3)$
to achieve this orthogonalisation and decomposition, but it is a modest initial computation which creates 
a framework for the subsequent data and error analysis. 
 
Now consider the effect of substituting the expansion (\ref{eq:bseparable}) into the bispectrum estimator (\ref{eq:bestimator}). 
It collapses to the simple summation
\begin{align}\label{eq:bestsum}
\curl{E} =\sum_n \aQn\, \bQn\,,
\end{align}
where the observed $\bQn$ coefficients are defined by 
\begin{align}
\bQn = \int d^3x \,M_r(\bx)\, M_s(\bx)\, M_t(\bx)\,,
\end{align}
with $M_r(\bx)$ the observed density perturbation convolved in Fourier space with the mode functions $q_r(k)$, that is, 
\begin{align}\label{eq:bfiltered}
M_r(\bx) = \int d^3k \frac{\d^{obs}_{\bk}q_r(k)\,e^{i \bk \cdot \bx}}{\sqrt{k P(k)}}\,.
\end{align}
Including the linear term in (\ref{eq:bestimator}) to account for systematic inhomogeneous effects we have 
\begin{align}
\bQn = \int d^3x  \(\,M_r(\bx)\, M_s(\bx)\, M_t(\bx) -  [\langle M_r(\bx)\, M_s(\bx)\rangle M_t(\bx)+ \hbox{2 perms}]\)\,.
\end{align}
Furthermore, rotating to the orthonormal frame with $\Rn$, it is straightforward to demonstrate that 
the averaged observed coefficient will be $\aRn= \langle \bRn\rangle$, given a set of realizations with 
the bispectrum $B(\kall)$ in (\ref{eq:bseparable}).  Thus we can directly reconstruct the bispectrum from a single realization (with 
sufficient single-to-noise) using
\begin{align}\label{eq:breconstruct}
B(\kall) = \frac{\sqrt{P(k_1) P(k_2) P(k_3)}}{\sqrt{k_1k_2k_3}}\,\sum_n \bRn \, \Rn(\kall)\,.
\end{align}
This reconstruction yields the full bispectrum shape in a model independent manner.   One can also consider a model independent measure of the total integrated non-Gaussian signal, using Parseval's theorem in the orthonormal frame (see ref.~\cite{FLS10} for a discussion of the quantity $\bar \Fnl^2 = \sum_n\bRn{}^2$). However, the  bispectrum estimator (\ref{eq:bestsum}) provides an immediate means to determine the significance of an observation of a particular type of non-Gaussianity with specific coefficients $\aQn$, e.g.\ by comparison with the $\bRn$ extracted from Gaussian simulations. We note that an initial implementation of the bispectrum reconstruction method (\ref{eq:breconstruct}) indicates its efficacy in recovering local non-Gaussianity.

We emphasise that the  bispectrum reconstruction (\ref{eq:breconstruct}) provides an extremely efficient method for calculating 
the bispectrum from any given density field $\d_\bk$ with optimum noise weighting.    Moreover, these separable mode 
expansion methods have been thoroughly tested in a CMB context \cite{FLS10}.   In essence, the $\lmax ^6$ 
operations required with the original estimator (or for a direct bispectrum calculation such as that described in 
ref.~\cite{9704075}) have been reduced to a series of $\lmax^3$ integrations given by (\ref{eq:bfiltered}).   
Of course, the number of mode coefficients 
depends on the rate of convergence of the expansion (\ref{eq:bseparable}) which is usually remarkably rapid.
For the CMB, a comprehensive survey of most theoretical bispectra in the literature required only 30 eigenmodes
for an accurate description at WMAP resolution \cite{FLS10}.  Even for a separable bispectrum in the linear regime (i.e. a terminating sum), we shall explain the advantages of using the well-behaved mode expansion
(\ref{eq:bseparable}).    The form of the next-to-leading order corrections for 
large-scale structure show no obvious pathologies which would alter  this 
convergence significantly in the mildly nonlinear regime (see later), and substantial efficiencies will remain even in highly nonlinear contexts.    This reconstruction approach (\ref{eq:breconstruct}) is ideally suited for 
$N$-body simulations where the bispectrum can be predicted at high precision by efficiently extracting it from 
multiple realizations 
using both Gaussian and non-Gaussian initial conditions (see later).    In an observational context, sparse sampling 
or poor survey strategies could reduce the effectiveness of the estimator (\ref{eq:bestimator}) in Fourier space, 
so care must be taken in large scale structure survey design to ensure good 
coverage so that higher order correlator measurements exploit these efficiencies.  

\section{Extension to the trispectrum and beyond}\label{sec:trispectrum}

\subsection{General trispectrum estimator}

In Chapter \ref{chapter:methodology} we discussed general CMB estimators for the trispectrum, where 
the decomposition of a trispectrum (non-diagonal or single diagonal) is sufficient to study the majority of 
cases described in the literature. While this projection depends explicitly on five parameters (or four in the 
non-diagonal case), in order to study other probes of non-Gaussianity, particularly for nonlinear large-scale 
structure, it may be necessary to consider the general trispectrum depending on the full six parameters. 
This is further motivated by  the study of the galaxy 
bispectrum, which may contain an enhanced contribution due to the trispectrum (see, e.g., ref.~\cite{09040497}). 
Clearly, then, we should also include a non-zero trispectrum to obtain non-Gaussian initial 
conditions suitable for a general bispectrum analysis using $N$-body codes.

The form of the general trispectrum estimator, for the connected part of a given four-point correlator $\<\d_{\bk_1}\d_{\bk_2}\d_{\bk_3}\d_{\bk_4}\>_c$, is directly analogous
to that presented already in Chapter \ref{chapter:methodology} for the CMB:
\begin{align}\label{eq:testimator}
\curl{E} =& \int \frac{d^3\bk_1}{(2\pi)^3} \frac{d^3\bk_2}{(2\pi)^3} \frac{d^3\bk_3}{(2\pi)^3}\frac{d^3\bk_4}{(2\pi)^3} \frac{\<\d_{\bk_1}\d_{\bk_2}\d_{\bk_3}\d_{\bk_4}\>_c}{P(k_1)P(k_2)P(k_3)P(k_4)}\nonumber\\
&\times\(  \d^{\rm{obs}}_{\bk_1}\d^{\rm{obs}}_{\bk_2}\d^{\rm{obs}}_{\bk_3}\d^{\rm{obs}}_{\bk_4}-6\< \d^{\rm{sim}}_{\bk_1}\d^{\rm{sim}}_{\bk_2}\>\d^{\rm{obs}}_{\bk_3}\d^{\rm{obs}}_{\bk_4} +3\< \d^{\rm{sim}}_{\bk_1}\d^{\rm{sim}}_{\bk_2}\>\<\d^{\rm{sim}}_{\bk_3}\d^{\rm{sim}}_{\bk_4}\> \),
\end{align}
where the notation $\<\dots \>_c$ denotes the connected component of the correlator. Note that this formula includes the quadratic term necessary to generalise to the case of incomplete sample coverage and inhomogeneous noise in a similar fashion to the CMB trispectrum estimator (see the discussion after (\ref{eq:bestimator})).  We omit the covariance-weighted version of the expression which is obvious from a comparison with (\ref{eq:bestimatorgen}).  
Imposing the $\d$-function appears to leave an intractable $\lmax^9$ operations for a full trispectrum estimator evaluation,
but, as with the bispectrum, this can be reduced dramatically using a separable approach. 

Assuming statistical isotropy, we can choose to parametrise the trispectrum using the lengths of four of its sides 
and two of its diagonals. In particular, we can exhibit these dependencies explicitly by representing the $\d$-function
imposing the quadrilateral condition, as a product of triangle conditions using the diagonals:
\begin{align}
 \<\d_{\bk_1}\d_{\bk_2}\d_{\bk_3}\d_{\bk_4}\>_c =& (2\pi)^3 \d_D(\bk_1+\bk_2+\bk_3+\bk_4) T(\bk_1,\bk_2,\bk_3,\bk_4)\nonumber\\
=&(2\pi)^3\int d^3\mathbf{K}_1 d^3\mathbf{K}_2 \d_D(\bk_1+\bk_2-\mathbf{K}_1)\d_D(\bk_3+\bk_4+\mathbf{K}_1)\nonumber\\
&\times \d_D(\bk_1+\bk_4-\mathbf{K}_2)T(k_1,k_2,k_3,k_4,K_1,K_2),
\end{align}
The decomposition of the trispectrum $T(k_1,k_2,k_3,k_4,K_1,K_2)$ is similar to that described in Chapter \ref{chapter:methodology}, but in 
which the trispectrum is assumed to depend on the first five parameters only. In the interest of completeness we evaluate a suitable weight function necessary for
evaluation of the more general decomposition from the expectation value of the estimator (\ref{eq:testimator}).
Similarly to the case of the bispectrum (\ref{eq:bestav}), the expectation value for the estimator is found to take the 
following simple form:
\eqa\label{eq:testav}
\< \curl{E}\> &=& \frac{V}{(2\pi)^3}\int \frac{d^3\bk_1}{(2\pi)^3} \frac{d^3\bk_2}{(2\pi)^3} \frac{d^3\bk_3}{(2\pi)^3}\frac{d^3\bk_4}{(2\pi)^3} \frac{(2\pi)^6 \d_D(\bk_1+\bk_2+\bk_3+\bk_4) T^2(\bk_1,\bk_2,\bk_3,\bk_4)}{P(k_1)P(k_2)P(k_3)P(k_4)}\nonumber\\
& &\\
&=& \frac{V}{(2\pi)^3}\frac{1}{2\pi^4}\int_{\curl{V}_T} d k_1 dk_2 dk_3 dk_4 dK_1 dK_2 \frac{k_1k_2 k_3 k_4 K_1 K_2}{\sqrt{g_1}}   \frac{T^2(k_1,k_2,k_3,k_4,K_1,K_2)}{P(k_1)P(k_2)P(k_3)P(k_4)},\nonumber\\
& &\label{eq:TrispEstim}
\qea
where  the function $g_1$ is given by the expression  
\begin{align}\label{eq:gdef}
g_1=K_1^2 K_2^2 (\sum_i k_i^2 -K_1^2-K_2^2)-K_1^2 \kappa_{23}\kappa_{14}+K_2^2 \kappa_{1 2} \kappa_{3 4}-(k_1^2k_3^2-k_2^2k_4^2)(\kappa_{12}+\kappa_{34}),
 \end{align}
and we denote $\kappa_{ij}=k_i^2-k_j^2$.  Here, we note that $\curl{V}_T$ is the region allowed by the quadrilateral condition which is 
described in some detail in  Chapter \ref{chapter:methodology}, noting the different ranges for the wavenumbers $k_i<k_{\rm{max}}$ and
diagonals $K_i<2 k_{\rm{max}}$.  By considering two different trispectra $T^2 \rightarrow T_iT_j$ in the 
estimator average (\ref{eq:testav}), we can use this expression
to define a noise-weighted cross-correlator and inner product (or Fisher matrix, see the discussion after (\ref{eq:bestav})).
\par
This derivation of \eqref{eq:TrispEstim} is instructive for the calculation of many of the results presented in this chapter and so we elucidate the calculation here. In a similar manner to the case of the bispectrum, the expectation value for the estimator is found to give
\begin{align}
\< \curl{E}\> = \frac{V}{(2\pi)^3}\int \frac{d^3\bk_1}{(2\pi)^3} \frac{d^3\bk_2}{(2\pi)^3} \frac{d^3\bk_3}{(2\pi)^3}\frac{d^3\bk_4}{(2\pi)^3} \frac{(2\pi)^6 \d_D(\bk_1+\bk_2+\bk_3+\bk_4) T^2(\bk_1,\bk_2,\bk_3,\bk_4)}{P(k_1)P(k_2)P(k_3)P(k_4)}.
\end{align}
Using the parametrisation in terms of $(k_1,k_2,k_3,k_4,K_1,K_2)$ and expanding the Dirac delta functions using \eqref{eq:deltaint} and \eqref{eq:exptobessel} we find
\begin{align}
\< \curl{E}\> =& \frac{V}{(2\pi)^3}\int \frac{(k_1 k_2 k_3 k_4 K_1 K_2)^2 d k_1 dk_2 dk_3 dk_4 dK_1 dK_2}{(2\pi)^{15}}  \frac{ T^2(k_1,k_2,k_3,k_4,K_1,K_2)}{P(k_1)P(k_2)P(k_3)P(k_4)}\nonumber\\
&\times (4\pi)^9\sum_{l_1}(2l_1 +1)\left(\int dx_1 x_1^2 j_{l_1}(k_1 x_1) j_0(k_2 x_1)j_{l_1}(K_1 x_1)\right)\nonumber\\
&\times\left(\int dx_2 x_2^2 j_{0}(k_3 x_2) j_{l_1}(k_4 x_2)j_{l_1}(K_1 x_2)\right)
 \left(\int dx_3 x_3^2 j_{l_1}(k_1 x_3) j_{l_1}(k_4 x_3)j_{0}(K_2 x_3)\right),
\end{align}
where the expression on the second and third lines arises from the integration over the angular variables. Next, we use the following identity from \cite{watson,9309023}
\begin{align}\label{sphbessel}
\int_0^{\infty}r^2 dr j_l(k r) j_l(k' r)j_0(\rho r)=\Theta(k,k',\rho) \frac{\pi}{4 k k' \rho}P_l\left(\frac{k^2+{k'}^2-\rho^2 }{2k k'}\right),
\end{align}
where $\Theta$ imposes the triangle condition on wavenumbers $(k,k',\rho)$ which is automatically satisfied for the trispectrum estimator at all points of the quadrilateral due to the Dirac delta functions, and $P_l$ represents the $l$th Legendre polynomial. Finally we may further simplify using the following result from \cite{Vinti1951},
\begin{align}
\sum_{l=0}^{\infty}(2l+1)P_l(x)P_l(y)P_l(z)&=\frac{2}{\pi\sqrt{g}},\qquad g=1+2xyz-x^2-y^2-z^2>0\nonumber\\
 &= 0,\qquad \mbox{otherwise}.
\end{align}
For the case of the trispectrum estimator we have
\begin{align}
x=\frac{k_1^2+K_1^2-k_2^2}{2 k_1 K_1},\qquad y=\frac{k_4^2+K_1^2-k_3^2}{2 k_4 K_1},\qquad z=\frac{k_1^2+k_4^2-K_2^2}{2 k_1 k_4},
\end{align}
and the condition $g>0$ is again satisfied for all points within the quadrilateral.

Using these expressions the expectation value of the estimator takes the following simple form
\begin{align}
\< \curl{E}\> =& \frac{V}{(2\pi)^3}\frac{1}{2\pi^4}\int_{\curl{V}_T} d k_1 dk_2 dk_3 dk_4 dK_1 dK_2 \frac{k_2 k_3 K_2}{2\sqrt{g}}   \frac{T^2(k_1,k_2,k_3,k_4,K_1,K_2)}{P(k_1)P(k_2)P(k_3)P(k_4)}.
\end{align}
In writing this expression we set $\d_D({\bf{0}})=V/(2\pi)^3$.
Therefore a suitable weight for the mode decomposition, which is a simple generalisation of the discussion in Chapter \ref{chapter:methodology} to include an extra diagonal is given by $w(k_1,k_2,k_3,k_4,K_1,K_2)=$\\ $k_2 k_3 K_2/(\sqrt{g}P(k_1)P(k_2)P(k_3)P(k_4))$. We note that the factor
$k_2 k_3 K_2/(2\sqrt{g})$ may be written as
\begin{align}\label{g1def}
 &\frac{k_2 k_3 K_2}{2\sqrt{g}}=\frac{ k_1 k_2 k_3 k_4 K_1 K_2}{\sqrt{g_1}} \equiv\nonumber\\
 &\frac{ k_1 k_2  k_3 k_4 K_1 K_2}{\sqrt{K_1^2 K_2^2 (\sum_i k_i^2 -K_1^2-K_2^2)-K_1^2 \kappa_{23}\kappa_{14}+K_2^2 \kappa_{1 2} \kappa_{3 4}-(k_1^2k_3^2-k_2^2k_4^2)(\kappa_{12}+\kappa_{34})}},
 \end{align}
 where we denote $\kappa_{ij}=k_i^2-k_j^2$ and we denote the denominator $\sqrt{g_1}$ for brevity.

\subsection{Separable mode expansions and the trispectrum estimator}

Using the weight \eqref{eq:testav}, a simple extension of the argument outlined in Chapter \ref{chapter:methodology} to include two diagonals instead of one reveals a similar eigenmode to the case of the bispectrum. In particular, we could expand the trispectrum as $\w T(k_1,k_2,k_3,k_4,K_1,K_2)=\sum_n \alpha_n Q_n(k_1,k_2,k_3,k_4,K_1,K_2)$, where $Q_n=q_{\{r}(k_1)q_{s}(k_2)q_t(k_3) q_{u\}}(k_4) r_v(K_1) r_w(K_2)$, $n$ represents $\{r s t u v w\}$\footnote{The diagonals and the wavenumbers are described by different eigenmodes due to their differing range, i.e. $k_i<k_{\rm{max}}$ while $K_i<2 k_{\rm{max}}$. } and $\w$, here and subsequently, is shorthand for an appropriate separable weighting. As we will see in the estimator below, however,  it is simpler to achieve a separable form by parametrising our bispectrum using  angles rather than diagonals. To achieve this, we may make a coordinate transformation from $(K_1,K_2)\rightarrow (\mu=\hat{\bk}_1.\hat{\bk}_2,\nu=\hat{\bk}_1.\hat{\bk}_4)$, where we use $K_1=\sqrt{k_1^2+k_2^2+2k_1 k_2 \mu}$ and $K_2=\sqrt{k_1^2+k_4^2+2k_1 k_4\nu}$. The Jacobian of this transformation is $k_1^2 k_2 k_4/(K_1 K_2)$. Thus (\ref{eq:testav}) becomes
\begin{align}\label{eq:tinnerprod}
\< \curl{E}\> =& \frac{V}{(2\pi)^3}\frac{1}{2\pi^4}\int_{\curl{V}_T} d k_1 dk_2 dk_3 dk_4 d\mu d\nu \frac{k_1^3 k_2^2 k_3 k_4^2}{\sqrt{g_1}}   \frac{T^2(k_1,k_2,k_3,k_4,\mu,\nu)}{P(k_1)P(k_2)P(k_3)P(k_4)},
\end{align}
where $g_1$ is given by equation~\eqref{eq:gdef} but now must be expressed in terms of $\mu,\nu$. We may use this weight to form an eigenmode expansion of the trispectrum where we use Legendre polynomials to describe the angular part. Explicitly we may expand the trispectrum in noise-weighted form as
\begin{align}\label{eq:tdecomp}
\frac{v(k_1)v(k_2)v(k_3)v(k_4)}{\sqrt{P(k_1)P(k_2)P(k_3)P(k_4)}}T(k_1,k_2,k_3,k_4,\mu,\nu)=\sum_{n l_1 l_2} \alpha_{n l_1 l_2} Q_n(k_1,k_2,k_3,k_4)P_{l_1}(\mu)P_{l_2}(\nu),
\end{align}
where $n=\{r,s,t,u\}$ and $Q_n(k_1,k_2,k_3,k_4)= q_{\{r} (k_1) q_s(k_2)  q_t (k_3) q_{u\}}(k_4)$ in an analogous manner to equation \eqref{eq:Qexp}. Scale invariance suggests the choice $v(k)=k^{3/4}$. In order to make this expression separable in terms of the vectors $\bk_i$ we note the following expansion of the Legendre polynomials
\begin{align}\label{Legendre}
 P_l(\hat{\bk}_1.\hat{\bk}_2)=\frac{4\pi}{2 l+1}\sum_{m=-l}^{l} Y_{l m}(\hat{\bk}_1)Y_{l m}^*(\hat{\bk}_2).
\end{align}
Using equations \eqref{eq:deltaint} and \eqref{eq:exptobessel} we can now write the estimator as expressed in \eqref{eq:testimator} in the form 
\begin{align}\label{eq:testsum}
\curl{E} = \sum_{n l_1 l_2} \baQ_{n l_1 l_2}\bbQ_{n l_1 l_2}\,,
\end{align} 
where the extracted trispectrum coefficients are given by 
\begin{align}
\bbQ_{n l_1 l_2}=& \frac{(4\pi)^2}{(2l_1+1)(2l_2+1)}\sum_{m_1 m_2}  \int d^3\bx  \Bigg[ M_{r l_1 l_2}^{m_1 m_2}(\bx)M_{s l_1}^{m_1*}(\bx)M_{t}(\bx)M_{u l_2}^{m_2 *}(\bx)\nonumber\\
&-\left(M_{r l_1 l_2}^{m_1 m_2}(\bx)M_{s l_1}^{m_1*}(\bx)\langle M_{t}(\bx)M_{u l_2}^{m_2 *}(\bx)\rangle+\rm{5\,perms}\right)\nonumber\\
&+\left(\langle M_{r l_1 l_2}^{m_1 m_2}(\bx)M_{s l_1}^{m_1*}(\bx)\rangle \langle M_{t}(\bx)M_{u l_2}^{m_2 *}(\bx)\rangle+\rm{2\,perms}\right)\Bigg],
\end{align}
where the permutations are with respect to the indices $\{r,s,t,u\}$.
In the above we define the filtered density perturbations $M^{...}_{...}$ by 
\begin{align}\label{eq:mapstrisp}
M_{r l_1 l_2}^{m_1 m_2}(\bx)&=\int \frac{d^3 \bk}{(2\pi)^3} e^{i\bk.\bx} \frac{q_r(k)\d_{\bk}^{\rm{obs}}}{\sqrt{P(k)}k^{3/4}} Y_{l_1 m_1}(\hat{\bk})Y_{l_2 m_2}(\hat{\bk}),\qquad\nonumber\\
M_{s l_1}^{m_1 *}(\bx)&=\int \frac{d^3 \bk}{(2\pi)^3} e^{i\bk.\bx} \frac{q_s(k)\d_{\bk}^{\rm{obs}}}{\sqrt{P(k)}k^{3/4}}Y_{l_1 m_1}^*(\hat{\bk}),\qquad \nonumber \\
M_t (\bx)&= \int \frac{d^3 \bk}{(2\pi)^3} e^{i\bk.\bx} \frac{q_t(k)\d_{\bk}^{\rm{obs}}}{\sqrt{P(k)}k^{3/4}},
\end{align}
with $*$ denoting a filtered map using $Y_{l m}^*$.

The algorithm (\ref{eq:testsum}) provides a highly efficient method for  estimating any trispectrum from a given 
 density field.  It requires only ${\cal O}(\nmax^{4/3}\times 
\lmax^3)$ operations, which makes feasible the intractable naive brute force calculation 
requiring ${\cal O}(\lmax^9)$ operations.  
In making this rough numerical estimate, we assume that the number of modes in each 
of the six dimensions is equal (and small), while noting that we have to perform a double summation for the two 
angle parameters $\mu,\,\nu$ over the indices $l_1,\,m_1,\,l_2,\,m_2$.   

As for the bispectrum, it is possible from the separable $\barQ_{nl_1l_2}$ modes to create a set of orthonormal 
 $\barR_{nl_1l_2}$ modes using the inner product (\ref{eq:tinnerprod}). 
Like the original decomposition of a theoretical trispectrum (\ref{eq:tdecomp}), 
orthogonalisation is a computationally intensive task
requiring up to ${\cal O}(\lmax^6)$ operations.  However, it  need only be performed once at the outset to set up 
the calculation framework, with the 
resulting rotation matrices being available for all the repetitive  subsequent analysis ($\sim \lmax^3$
operations).   We can realistically envisage, then, reconstructing the complete trispectrum directly from the observational data using the rotated $\bbQ_{n l_1 l_2}$ coefficients (as in  (\ref{eq:breconstruct}).   It is interesting to note 
that almost all theoretical trispectra presented to date in the literature are `planar', that is, either depending on only 
one diagonal or none.  We treat the latter special case below, but we leave the simplifications
arising from the single
diagonal case for discussion elsewhere \cite{Regan2011}.

\subsection{Non-diagonal trispectrum and quadspectrum estimation}

In the case that the trispectrum is independent of the diagonals $K_1,K_2$ (or angles $\mu$, $\nu$) we get a simpler 
expression for the averaged estimator (\ref{eq:testimator}):
\begin{align}
\langle \mathcal{E}\rangle=\frac{V}{ (2\pi)^6}\int_{\curl{V}_T} d k_1 dk_2 dk_3 dk_4 k_1 k_2 k_3 k_4 &\Big(\sum_i k_i-|\tilde{k}_{34}|-|\tilde{k}_{24}|-|\tilde{k}_{23}|\Big)\nonumber\\
&\times\frac{T^2(k_1,k_2,k_3,k_4)}{P(k_1)P(k_2)P(k_3)P(k_4)},
\end{align}
where $\tilde{k}_{34}=k_1+k_2-k_3-k_4$, etc. We may use the weighting this suggests to decompose the trispectrum into the form $\w T=\sum_n \alpha_n Q_n$ where $Q_n=q_{\{r} q_s q_t q_{u\}}$. The estimator is simpler to calculate since there are no cross terms between integrals. We find the extracted observational coefficients simplify to  
\begin{align}\label{eq:tndest}
 \beta_n=&\int d^3\bx \Bigg[ M_r (\bx) M_s (\bx) M_t(\bx) M_u(\bx)-\left( M_r (\bx) M_s (\bx) \langle M_t(\bx) M_u(\bx)\rangle +\rm{5\,perms}\right)\nonumber\\
 & +\left(\langle M_r (\bx) M_s (\bx)\rangle \langle M_t(\bx) M_u(\bx)\rangle +\rm{2\,perms}\right) \Bigg],
\end{align}
where $M_t$ was defined in \eqref{eq:mapstrisp}.   Here, we see that the trispectrum estimation scales once 
again as only ${\cal O}(\nmax\times\lmax^3)$ operations.  The extraction of expansion coefficients $\baQ$ 
from a given non-separable theoretical trispectrum appears to require up to $\lmax^4$ operations, but it is a one-off
calculation amenable to many shortcuts.   A practical implementation reveals that non-diagonal 
trispectra given in the literature require only $\nmax \approx {\cal O}(10)$ modes for accurate representation. 
As an example, even the pathological local model with diverging squeezed states requires only $\nmax =20$ for the 
expansion (\ref{eq:tdecomp}) to achieve a 95\% correlation with the primordial shape.   It is clear that there is 
no inherent impediment to direct  estimation and evaluation of trispectra from survey data of adequate quality.  

This separable methodology can be applied to correlators beyond the trispectrum, such as the quadspectrum
$\tilde {\cal Q}(\bk_1,\bk_2,\bk_3,\bk_4,\bk_5)$ defined from
\begin{align}
\langle \d_{\bk_1} \d_{\bk_2} \d_{\bk_3} \d_{\bk_4} \d_{\bk_5}\rangle=(2\pi)^3\d(\bk_1+\bk_2+\bk_3+\bk_4+\bk_5)\tilde{\cal Q}(\bk_1,\bk_2,\bk_3,\bk_4,\bk_5)\,.
\end{align}
For simplicity, however, we restrict attention here to quadspectra that are non-diagonal, depending only on the wavenumbers $k_1,\dots,k_5$, that is,  $\tilde{\cal Q}(\bk_1,\bk_2,\bk_3,\bk_4,\bk_5)=$ \newline
$\tilde{\cal Q}(k_1,k_2,k_3,k_4,k_5)$. 
The expectation value of the quadspectrum estimator is then given by 
\begin{align}\label{eq:qinnerprod}
\langle\mathcal{E}\rangle&=\frac{V}{(2\pi)^3}\int \left(\Pi_{i=1}^5 \frac{d^3 \bk_i}{(2\pi)^3}\right)\frac{(2\pi)^6 \d(\bk_1+\bk_2+\bk_3+\bk_4+\bk_5)\tilde{Q}^2(k_1,k_2,k_3,k_4,k_5)}{P(k_1)P(k_2)P(k_3)P(k_4)P(k_5)}\nonumber\\
&=\frac{V}{(2\pi^3)^3}\int dk_1dk_2 dk_3 dk_4 dk_5 (k_1 k_2 k_3 k_4 k_5)^2\nonumber\\
&\times\left( \int dx x^2 j_0 (k_1 x)  j_0 (k_2 x) j_0 (k_3 x) j_0 (k_4 x) j_0 (k_5 x)\right)
\frac{\tilde{Q}^2(k_1,k_2,k_3,k_4,k_5)}{P(k_1)P(k_2)P(k_3)P(k_4)P(k_5)}\,,
\end{align}
where the integral over the five spherical Bessel functions serves also to define the allowed quadspectrum domain 
${\cal V}_Q$.   
The expression (\ref{eq:qinnerprod}) may be used to derive a weight to decompose the quadspectrum in the form $\Big[\Pi_{i=1}^5 v(k_i)/\sqrt{P(k_i)}\Big]\tilde{\cal Q}(k_1,k_2,k_3,k_4,k_5)=\sum_n \alpha_n Q_n(k_1,k_2,k_3,k_4,k_5)$ where $n\leftrightarrow \{r,s,t,u,v\}$ and  $Q_n(k_1,k_2,k_3,k_4,k_5)=\\q_{\{r}(k_1)q_s(k_2)q_t(k_3)q_u(k_4)q_{v\}}(k_5)$, and where imposing scale invariance sets $v(k)=k^{9/10}$.   The resulting separable estimator is directly analogous to that for the non-diagonal trispectrum (\ref{eq:tndest}), but for brevity we will only discuss initial
conditions with a non-trivial quadspectrum.

\section{Efficient generation of arbitrary non-Gaussian initial conditions}\label{sec:intcon}

The generation of non-Gaussian initial conditions for $N$-body simulations with a given primordial bispectrum has been 
achieved to date only for bispectra which have a simple separable form (see, e.g., \cite{0701131,07072516,07104560,
10065793}).  For $N$-body codes to efficiently produce non-Gaussian initial conditions for an arbitrary non-separable bispectrum, will require a well-behaved separable mode decomposition, as achieved for CMB map simulations 
 in ref.~\cite{FLS09}.   However, we can do even better by simulating initial data given both an arbitrary 
 bispectrum and trispectrum, as shown for the CMB in Chapter \ref{chapter:methodology}.  As we have discussed already, 
 this is of particular interest for measurements of the large-scale structure bispectrum, because of nonlinear contributions
expected from the trispectrum.   We describe the non-Gaussian primordial potential perturbation as 
\begin{align}\label{eq:intcondexp}
\O = \O^G +\frac{1}{2} \Fnl \O^B +\frac{1}{6} G_{NL} \O^T,
\end{align}
where $\O^G$ is a Gaussian random field with the required power spectrum $P(k)$. It should be noted that this definition introduces two trispectrum terms of the form $\langle \O^T \O^G \O^G \O^G \rangle$ and $\langle \O^B \O^B \O^G \O^G \rangle$ (similar to the local trispectrum terms with coefficients $g_{NL}$ and $\tau_{NL}$ respectively). Therefore, it may be desirable to cancel this extra contribution. This issue will be addressed at the end of the section. Following ref.~\cite{FLS09} for 
the primordial bispectrum $B(\kall)$ with separable expansion
\begin{align}\label{symmexpan}
\frac{B(k,k^{'},k^{''}) }{P(k^{'}) P(k^{''})+P(k^{}) P(k^{'})+P(k^{}) P(k^{''})}=\sum_{n}\alpha_n^Q Q_n(k,k',k''),
\end{align}
the  bispectrum contribution to the primordial perturbation $\O$ becomes simply
\begin{align}\label{eq:icbi}
\O^B(\bk) &= \int \frac{d^3\bk^{'}}{(2\pi)^3}\frac{d^3\bk^{''}}{(2\pi)^3} \frac{(2\pi)^3 \d(\bk+\bk^{'}+\bk^{''}) B(k,k^{'},k^{''}) \O^G(\bk^{'}) \O^G(\bk^{''})}{P(k^{'}) P(k^{''})+P(k^{}) P(k^{'})+P(k^{}) P(k^{''})}\,,\\
=&\sum_{n}\alpha_{n} q_{\{r}(k)\int d^3 \bx e^{i\bk.\bx}M_s(\bx)M_{t\}}(\bx),
\end{align}
where the filtered density perturbations $M_s(\bx)$ are now defined by
\begin{align}\label{eq:b2filtered}
M_s(\bx) = \int \frac{d^3\bk}{(2\pi)^3} \Phi^{G}({\bk}) q_s(k)\,e^{i \bk \cdot \bx}\,.
\end{align}

 (We note that the bispectrum algorithm in ref.~\cite{FLS09} used here is a generalization of the 
CMB bispectrum algorithm presented in ref.~\cite{0612571}\footnote{The definition of $\O^B$ in terms of the expression given in equation \eqref{symmexpan} - as opposed to the asymmetric quantity $B(k,k^{'},k^{''}) /(P(k^{'}) P(k^{''}))$ - was advocated in \cite{VerdeWag}. It should be noted that, with this prescription, the definition agrees identically with the expansion $\O=\O^G+\Fnl \O^G*\O^G$ in the case of the local model.}.)  Of course, we normalise $B(\kall)$ such that it has $\Fnl =1$.
Like the estimator, this requires only ${\cal O}(\nmax\times\lmax^3)$ operations for every realization of new initial conditions, as opposed to a brute force approach which requires $\lmax^6$.  Note also that once the $\nmax$ filtered density perturbations $\int d^3 \bx e^{i\bk.\bx}M_s(\bx)M_{t\}}(\bx)$ have been obtained for a given $\O^B$, they can be applied to an arbitrary number of different shaped bispectra represented by $\aQn$s.

We can similarly find a relatively simple and highly efficient expression to compute initial conditions for the 
trispectrum $\Phi^T$. 
Following Chapter \ref{chapter:methodology},
the primordial trispectrum $T(\kall,\,k_4,\,\mu,\,\nu)$ is represented and expanded using wavenumber $q_r(k)$ and angle $P_u(\mu)$ modes in a similar fashion to equation (\ref{eq:tdecomp}), 
\begin{align}\label{eq:tdecomp2}
\frac{ T(k_1,k_2,k_3,k_4,\mu,\nu)}{{P(k_1)P(k_2)P(k_3)P(k_4)}+3\,{\mathrm{perms}}}=\sum_{n l_1 l_2} \alpha_{n l_1 l_2} Q_n(k_1,k_2,k_3,k_4)P_{l_1}(\mu)P_{l_2}(\nu).
\end{align}

The trispectrum contribution to $\O$ then becomes
\begin{align}\label{eq:ictri}
\Phi^T(\bk)=& \int \frac{d^3\bk^{'}d^3\bk^{''}d^3\bk^{'''}}{(2\pi)^6} \frac{ \d(\bk+\bk^{'}+\bk^{''}+\bk^{'''}) T(\bk,\bk^{'},\bk^{''},\bk^{'''}) \O^G(\bk^{'}) \O^G(\bk^{''})\O^G(\bk^{'''})}{P(k^{'}) P(k^{''}) P(k^{'''})+3\,{\mathrm{perms}}}
\\
=&\sum_{n l_1 l_2}\baQ_{n l_1 l_2}\frac{(4\pi)^2}{(2l_1+1)(2l_2+1)}\sum_{m_1 m_2}Y_{l_1 m_1}(\hat{\bk})Y_{l_2 m_2}(\hat{\bk}) q_r(k)\nonumber\\
&\times\int d^3 \bx e^{i\bk.\bx}M_{s l_1}^{m_1*}(\bx)M_t(\bx)M_{u l_2}^{m_2*}(\bx),
\end{align}
where the filtered density perturbations $M_{s l_1}^{m_1*}$ and $M_t$ are now given by
\begin{align}\label{eq:mapstrisp}
M_{s l_1}^{m_1 *}(\bx)&=\int \frac{d^3 \bk}{(2\pi)^3} e^{i\bk.\bx} {q_s(k)\O^G({\bk})}Y_{l_1 m_1}^*(\hat{\bk}),\qquad \nonumber\\
M_t (\bx)&= \int \frac{d^3 \bk}{(2\pi)^3} e^{i\bk.\bx} q_t(k)\O^G({\bk}).
\end{align}

For the particular case that the trispectrum is independent of the angles $\mu,\,\nu$ (or diagonals $K_1,\,K_2$) the decomposition is somewhat simpler:
\begin{align}
 \Phi^T(\bk) =\sum_n \baQ_n q_r(k)\int d^3\bx e^{i\bk.\bx} M_s (\bx) M_t(\bx) M_u(\bx)\,.
\end{align}
This applies to many cases in the literature, including constant, local and equilateral models.   
This simplification 
will also apply to initial conditions with non-diagonal quadspectra.
The expression for quadspectrum perturbation $\Phi^{\tilde{Q}}$ is very similar to the expressions above with 
\begin{align}
\Phi^{\tilde{Q}}=\sum_n \tilde\alpha^{\scriptstyle Q}_n \,q_r(k) \int d^3 \bx e^{i\bk.\bx}M_s(\bx)M_t(\bx)M_u(\bx)M_v(\bx).
\end{align}
It is clear that it is possible, given separable expansions of an arbitrary bispectrum and trispectrum, to efficiently 
generate multitudes of realizations, with each requiring only ${\cal O}(\nmax\times\lmax^3)$ operations.

We have shown in Chapter \ref{chapter:methodology}
that the bispectrum (\ref{eq:icbi}) and trispectrum (\ref{eq:ictri}) contributions are not independent. It may be necessary to subtract out an unwanted `bispectrum' contribution to the trispectrum. The bispectrum contribution induces a trispectrum given by
\begin{align}
\langle\O(\bk_1)\O(\bk_2)\O(\bk_3)\O(\bk_4)\rangle_c & \nonumber\\
=(2\pi)^3 F_{NL}^2 \int d^3\bK& \Big[\tilde{T}(k_1,k_2,k_3,k_4,K)\delta_D(\bk_1+\bk_2-\bK)\delta_D (\bk_3+\bk_4+\bK)\nonumber\\
&+\tilde{T}(k_1,k_3,k_2,k_4,K)\delta_D(\bk_1+\bk_3-\bK)\delta_D (\bk_2+\bk_4+\bK)\nonumber\\
&+\tilde{T}(k_1,k_4,k_2,k_3,K)\delta_D(\bk_1+\bk_4-\bK)\delta_D (\bk_2+\bk_3+\bK)\Big]
\end{align}
where 
\begin{align}
\tilde{T}(k_1,k_2,k_3,k_4,K)=&\frac{B(k_1,k_2,K)}{P(k_1)P(k_2)+2\,{\mathrm{perms}}} \frac{B(k_3,k_4,K)}{P(k_3)P(k_4)+2\,{\mathrm{perms}}}P(K)\nonumber\\
&\times\left(P(k_1)P(k_3)+P(k_1)P(k_4)+P(k_2)P(k_3)+P(k_2)P(k_4)\right).
\end{align}
Cancellation of this spurious `trispectrum' may be achieved by altering the algorithm given by equation \eqref{eq:intcondexp} to the form
\begin{align}\label{eq:intcondexp2}
\O = \O^G +\frac{1}{2} \Fnl \O^B +\frac{1}{6} G_{NL} \O^T-\frac{1}{2} \Fnl^2 \tilde{\O}^T,
\end{align}
where
\begin{align}
\tilde{\O}^T(\bk)=&\int \frac{d^3 \bk_2}{(2\pi)^3} \frac{d^3 \bk_3}{(2\pi)^3} \frac{d^3 \bk_4}{(2\pi)^3}d^3 \bK(2\pi)^3 \delta_D(\bk+\bk_2-\bK) \delta_D(\bk_3+\bk_4+\bK)\nonumber\\
&\times\frac{\tilde{T}(k,k_2,k_3,k_4,K)}{P(k)P(k_2)P(k_3)+3\,{\mathrm{perms}}}\O^{G}(\bk_2)\O^{G}(\bk_3)\O^{G}(\bk_4).
\end{align}
With this prescription it is found that 
\begin{align}
\langle \O(\bk_1)\O(\bk_2)\O(\bk_3) \rangle&=(2\pi)^3 \delta_D(\bk_1+\bk_2+\bk_3)B(k_1,k_2,k_3)\, ,\nonumber\\
\langle \O(\bk_1)\O(\bk_2)\O(\bk_3) \O(\bk_4)\rangle_c&=(2\pi)^3 \delta_D(\bk_1+\bk_2+\bk_3+\bk_4)T(\bk_1,\bk_2,\bk_3,\bk_4),
\end{align}
as desired. We shall leave a detailed analysis of this issue to a future work.

Very recently, ref.~\cite{10065793} proposed an alternative approach to creating non-Gaussian 
initial conditions from bispectra by integrating directly the expression
\begin{align}
\O^B(\bk) &= \int \frac{d^3\bk^{'}}{(2\pi)^3}\frac{B(k,k^{'},k^{''}) \O^G(\bk^{'}) \O^G(\bk+\bk^{'})}{P(k^{'}) P(|\bk+\bk^{'}|)}\,.
\end{align}
For explicitly separable bispectra,  using convolutions they were able to exploit the same efficiencies 
described above to reduce the problem from ${\cal O}(\lmax^6)$ to ${\cal O}(\lmax^3)$ operations.  The key
differences are that the method does not apply to non-separable bispectra and that, even for separable bispectra, 
the functions by which they are factorized must be well-behaved.
As observed in the CMB, for example,  the usual separable form of  
the equilateral model must be treated carefully to prevent non-Gaussian contributions distorting
the power spectrum (see \cite{FLS09}).   Similar effects appear to occur in ref.~\cite{10065793}
in one particular model (orthogonal).   The robust general prescription described here using well-behaved bounded 
mode functions ensures bispectrum (as well as trispectrum) scaling in the initial conditions which is designed 
to avoid such pathologies, even in the separable case.

\section{Non-Gaussian parameter estimation}\label{sec:fnlestimation}

Fast separable methods for estimating arbitrary bispectra or trispectra
in large scale structure observations or simulated data greatly improve 
the prospect of using higher order correlators as an important cosmological
 diagnostic.  This is particularly pertinent for testing the Gaussian hypothesis of 
 the inflationary scenario.   The complication is that even Gaussian initial 
fluctuations receive non-Gaussian contributions through late-time gravitational collapse (see reviews \cite{LigSef2010,10035020} and the references therein). 
Here, we briefly sketch some key issues facing parameter estimation in this context. 

There has been much recent progress describing next-to-leading order contributions to non-Gaussianity
from gravity.   A simple example of this is the matter density power spectrum which contains
several contributions, including those from an enhanced primordial bispectrum $\Fnl B_0(\kall)$ \cite{08084085}:
\begin{align}\label{eq:PNG}
P^B(k)&=\frac{\Fnl}{(2\pi)^3}\int d^3 \by B_0(\bk,\by,\bk-\by)F_2(\by,\bk-\by)\nonumber\\
&=\frac{\Fnl}{(2\pi)^3}\int d^3 \by d^3 \bk_2 \d(\bk_2-\bk+\by) B_0(k,y,k_2)F_2(\by,\bk_2),
\end{align}
where the gravitational kernel for this convolution is given by 
\begin{align}\label{eq:F2exp}
F_2(\by,\bk_2)&=\frac{17}{21}+P_1(\mu)\left(\frac{y}{k_2}+\frac{k_2}{y}\right)+\frac{4}{21}P_2(\mu)\,,
\end{align}
where $\mu=\hat{\by}.\hat{\bk}_2$.
Taking the separable expansion (\ref{eq:bseparable}) for $B_0(\kall)$ and substituting into eqn (\ref{eq:PNG}), we
find the simple integral over the mode functions $q_r(k)$: 
\begin{align}\label{eq:PB}
P^B(k)=&\Fnl\sum_n \frac{\alpha_n}{2\pi^2} \frac{q_r(k)\sqrt{P(k)}}{k^{3/2}}\int_{\mathcal{V}_B} dy dk_2 \, \sqrt{yP(y)} \, q_s(y) \,\sqrt{k_2P(k_2)} \,q_t(k_2)\nonumber\\
&\times\Big[ \frac{5}{7}+\frac{2}{7}\left(\frac{k_2^2+y^2-k^2}{2 k_2 y}\right)^2-\left(\frac{y}{k_2}+\frac{k_2}{y}\right)\left(\frac{k_2^2+y^2-k^2}{2 k_2 y}\right)\Big],
\end{align}
where $\mathcal{V}_B$ represents the domain for which the triangle condition holds for the wavenumbers $(k_2,y,k)$. 
Note that this integral breaks down into products of one dimensional integrals over $y$ and $k_2$ which can be 
evaluated easily.   Here, the calculation steps leading to (\ref{eq:PB}) are very similar to those used to obtain (\ref{eq:bestav}).

In the mildly nonlinear regime, the matter density bispectrum similarly 
contains nonlinear contributions  from gravitational collapse, from the 
primordial bispectrum $\Fnl B_0$,  and from the primordial trispectrum $\t_{\rm NL}T_0$ \cite{09040497,09050717}:
\begin{align}\label{eq:BT}
B(k_1,k_2,k_3)&=[2 F_2(\bk_1,\bk_2) P_0(k_1) P_0(k_2) + \hbox{2 perms}]+   \Fnl B_0(\kall) ]\\
&~~~+\frac{\t_{\rm NL}}{(2\pi)^3}\int d^3 \by  T_0(\bk_1,\bk_2,\by,\bk_3-\by)F_2(\by,\bk_3-\by)+\hbox{2 perms}\,,\nonumber\\
&\equiv B^G(\kall) + \Fnl B_0(\kall) +\t_{NL}B^T(\kall).\nonumber
\end{align}
Next, we substitute the separable expansion for the trispectrum (\ref{eq:tdecomp}) into (\ref{eq:BT}) to 
find integral expressions for the resulting bispectrum.  
The contribution to the galaxy bispectrum due to the primordial trispectrum is given by
\begin{align}
B_g^T(k_1,k_2,k_3)&=\frac{1}{(2\pi)^3}\int d^3 \by  T(\bk_1,\bk_2,\by,\bk_3-\by)F_2(\by,\bk_3-\by)+\rm{2\,perms}\nonumber\\
&=\frac{1}{(2\pi)^3}\int d^3 \by d^3 \bk_4  T(\bk_1,\bk_2,\by,\bk_4)F_2(\by,\bk_4)\delta_D(\bk_4-\bk_3+\by)+\rm{2\,perms},
\end{align}
where $F_2$ is given by equation \eqref{eq:F2exp} and the permutations are cyclic in $(k_1,k_2,k_3)$. First we consider the special case that the trispectrum depends only on the wavenumbers $k_1,k_2,y,k_4$ such that we may write $T(k_1,k_2,y,k_4)=\sum_n\alpha_n q_r(k_1)q_s(k_2)q_t(y)q_u(k_4)$. The calculation is very similar to the power spectrum case and we find
\begin{align}\label{eq:bisplss}
&B_g^T(k_1,k_2,k_3)=\sum_n \frac{\alpha_n}{4\pi^2}\frac{\sqrt{P(k_1)P(k_2)}}{(k_1 k_2)^{3/4}}\frac{q_r(k_1)q_s(k_2)}{ k_3}\int_{\mathcal{V}} dy dk_4 (y \, k_4)^{1/4} \sqrt{P(y)P(k_4)}  \nonumber\\
&\times q_t(y)  q_u(k_4)\Big[ \frac{5}{7}+\frac{2}{7}\left(\frac{k_4^2+y^2-k_3^2}{2 k_4 y}\right)^2-\left(\frac{y}{k_4}+\frac{k_4}{y}\right)\left(\frac{k_4^2+y^2-k_3^2}{2 k_4 y}\right)\Big]+\rm{2\,perms},
\end{align}
where $\mathcal{V}$ represents to domain for which the wavenumbers $(y,k_4,k_3)$ satisfy the triangle condition. The integral, we note again, may be written as a sum of products of one dimensional integrals over $y$ and $k_4$.
\par
Next, we consider the more general case where the trispectrum depends also on two diagonals or equivalently the angles $\mu=\hat{\bk}_1.\hat{\bk}_2$ and $\nu=\hat{\bk}_1.\hat{\bk}_4$. In this case we may decompose the trispectrum as
\begin{align}
\frac{(k_1\, k_2\, y \,k_4)^{3/4}}{\sqrt{P(k_1)P(k_2)P(y)P(k_4)}}T(\bk_1,\bk_2,\by,\bk_4)=\sum_{n l_1 l_2}\alpha_{n l_1 l_2}q_r(k_1)q_s(k_2)q_t(y)q_u(k_4)  P_{l_1}(\mu)P_{l_2}(\nu),
\end{align}
where $n\equiv\{r,s,t,u\}$. The calculation follows much the same lines as the special case with simplification of the formulae in this case achieved using equation \eqref{sphbessel}, the following identity, as described in \cite{Seaton62,BurgWhelan87},
\begin{align}
\int dx x^2 j_{l}(kx)j_{l'}(k'x) j_n(\rho x)=\Theta(k,k',\rho)\frac{\pi}{2k k' \rho^{n+1} }\sum_L Q_{nL}(k,l,k',l') P_L\left(\frac{k^2 +{k'}^2-\rho^2}{2 k k'}\right) 
\end{align}
(where the $\Theta$ function imposes the triangle condition on the three wavenumbers, $P_L$ is a Legendre polynomial and the functions $Q_{nL}$ may be found in \cite{Seaton62,BurgWhelan87}) and the identity
\begin{align}
\sum_{m_1,m_2}\left( \begin{array}{ccc}
l_1 & l_2 & L \\
m_1 & m_2 & M \end{array} \right)\left( \begin{array}{ccc}
l_1 & l_2 & L' \\
m_1 & m_2 & M' \end{array} \right)=\frac{\delta_{L L'}\delta_{M M'}}{2L+1}.
\end{align}
With these considerations we find
\begin{align}
&B^T_g(k_1,k_2,k_3)=
\sum_{n l_1 l_2}\frac{\alpha_{n l_1 l_2}}{2\pi^2}\frac{\sqrt{P(k_1)P(k_2)}}{(k_1 k_2)^{3/4}}\frac{q_r(k_1)q_s(k_2)}{k_3}P_{l_1}(\hat{\bk}_1.\hat{\bk}_2)P_{l_2}(\hat{\bk}_1.\hat{\bk}_3)\nonumber\\
&\times \int_{\mathcal{V}} dy dk_4 (y \, k_4)^{1/4} \sqrt{P(y)P(k_4)} q_t(y)q_u(k_4)
\Bigg[\frac{17}{42}P_{l_2}\left(\frac{k_4^2+k_3^2-y^2}{2k_3 k_4}\right)\nonumber\\
&+\frac{4\pi}{3}\sum_{l_4}\frac{(-1)^{(l_4-l_2+1)/2}h_{l_2 l_4 1}^2}{(2l_2+1)}\frac{1}{y}\left(\frac{y}{k_4}+\frac{k_4}{y}\right)\sum_L Q_{1L}(k_4,l_4,k_4,l_2)P_L\left(\frac{k_4^2+k_3^2-y^2}{2k_3 k_4}\right)\nonumber\\
&\nonumber\\
&+\frac{16\pi}{105}\sum_{l_4}\frac{(-1)^{(l_4-l_2+2)/2}h_{l_2 l_4 2}^2}{(2l_2+1)y^2}\sum_{L'} Q_{2L'}(k_4,l_4,k_4,l_2)P_{L'}\left(\frac{k_4^2+k_3^2-y^2}{2k_3 k_4}\right)\Bigg]+\rm{2\,perms}.
\end{align}
\par
For non-diagonal trispectra, the result \eqref{eq:bisplss} is simple and 
very similar to the power spectrum modification (\ref{eq:PB}).   The result is three distinct contributions 
to the late-time bispectrum $\w B(\kall) = \sum_n\alpha_nQ_n$ with the bispectrum approximated as in 
separable form as 
\begin{align}
\w B(\kall)  = \sum_n (\alpha^G_n +\Fnl \alpha^B_n +\t_{NL} \alpha^T_n)\,\Rn(\kall)\,,
\end{align}
with the coefficients $\alpha^i_n$ representing distinct shapes in the orthonormal frame.   Here, the primordial $\alpha^B$ coefficients are normalised
such that in the initial conditions $\Fnl = 1$, and similarly for the primordial trispectrum $\t_{\rm NL}=1$. 

Setting aside the trispectrum contribution, if we can remove the Gaussian part from $\alpha_n,\,\beta_n$ 
then we have an optimal estimator for the non-Gaussianity parameter $\Fnl$ ,
\begin{align}\label{eq:estimatorfnl}
\curl{E} = \frac{1}{{N}^2}\sum \a^{B}_n \b^{B}_n\,,
\end{align}
where we have defined the predicted $\a^{B}_n$ and measured $\b^{B}_n$ by
\begin{align}
\a^{B}_n = \a_n - \bar{\a}^G_n\,, \qquad
\b^{B}_n = \b_n - \bar{\b}^G_n\,, \qquad
{N}^2 =  \sum {\a^{B}_n}^2\,.
\end{align}
Here $\bar{\a}^G_n$ refers to the decomposition coefficients for Gaussian initial conditions, calculated either from theory (as above in (\ref{eq:BT})) or obtained from $N$-body simulations (note $\bar{\a}^G_n=\bar{\b}^G_n$)  and the $\a_n$  are calculated from initial conditions with $\Fnl=1$. The variance of the estimator can then be calculated by applying it to a large set of Gaussian simulations. This is directly analogous to the CMB estimator used in \cite{FLS09} (where of course $\bar{\a}^G_n = 0$).  

However, in the nonlinear regime, and with significant bias affecting the galaxy distribution, it will not be possible to  approximate non-Gaussianity in this simple way. We need to approach parameter estimation for $F^{NL}$ (or $\t_{\rm NL}$) quite differently. The estimator (\ref{eq:estimatorfnl}) can be thought of as a least squares fit of the theory to the data. As the relative size of the individual $\a^{B}_n$ are constant, we can only change the amplitude, $\Fnl$. Thus, we must simply choose a $\Fnl$ which minimises
\begin{align}
\curl{E} = \sum \( \a^{B}_n \Fnl - \b^{B}_n \)^2
\end{align}
for a given form of $\a^{B}_n$. In the general case we expect the ratios of the individual coefficients to change as we change $\Fnl$. As a result we must consider the $\a_n$ to be an arbitrary function of $\Fnl$ and so we now wish to minimise
\begin{align}
\curl{E}(\Fnl) = \sum \( \a_n(\Fnl) - \b_n \)^2
\end{align}
with respect to $\Fnl$. We will assume that it will not be possible in general to determine $\a_n(\Fnl)$ analytically so that we could then try to solve $\p \curl{E} / \p \Fnl = 0$. This means that to minimise $\curl{E}$ requires extracting the $\a_n$ from sets of $N$-body simulations each with different non-Gaussian initial conditions which correspond to a particular $\Fnl$. We then reconstruct the dependence of $\curl{E}$ on $\Fnl$ and find the best-fit $\Fnl$ for the
given observations.
One also must be careful calculating the variance on such a measurement of $\Fnl$. In general this would entail applying the same approach to each density distribution in the set of simulations with the estimated $\Fnl$ and then determining the distribution of the recovered $\Fnl$.  Of course, Gaussian simulations may be substituted if $\Fnl$ is sufficiently small that the effect on the error bars is negligible.  

Finally, we note that in general the galaxy bispectrum will take contributions from both the bispectrum and trispectrum of the curvature perturbation \cite{09040497} (which is why we cannot in general connect $\Fnl$ with its CMB counterpart in a simple way).  The amplitudes  of $\Fnl$ and $\t_{\rm NL}$ can be determined by consistency conditions for certain models  or they can vary independently. In this case we must constrain the amplitude of both $\Fnl$ and $\t_{\rm NL}$ contributions marginalising over these two parameters. Such a computationally intensive analysis becomes much more feasible 
with an efficient bispectrum extraction method (\ref{eq:breconstruct}) and with 
non-Gaussian initial conditions which include the specification of the trispectrum (\ref{eq:intcondexp}).

\section{Conclusion}
While the CMB is an ideal observable for tests of primordial non-Gaussianity since the perturbations remain in the linear regime, the prospects for achieving comparable, and ultimately superior, constraints on non-Gaussianity in the near future using large-scale structure appears encouraging due to recent advancements in the analysis and development of N-body codes.

In this chapter we have described how methods developed for the analysis of non-Gaussianity in the CMB may be applied to surveys of large-scale structure. These methods are based on mode expansions, exploiting a complete orthonormal eigenmode basis to efficiently decompose arbitrary poly-spectra into a separable polynomial expansion.

Applying the methodology to the bispectrum reveals a vast improvement in computational speed for finding a general estimator and correlator, reducing complexity from $\mathcal{O}(l_{\rm{max}}^6)$ to $\mathcal{O}(n_{\rm{max}}\times l_{\rm{max}}^3)$. As we use a complete orthonormal basis we are also able to efficently calculate the bispectrum from simulations and, assuming sufficient signal to noise, observations. Of particular interest is the application to the generation of non-Gaussian initial conditions for N-body codes. The approach can be used to create initial conditions with arbitrary independent poly-spectra. With this method calculation of the bispectrum contribution requires a similar number of operations as decomposition. This improvement to the brute force approach opens up the opportunity of investigating a far wider range of models using large-scale structure than has hitherto been considered.

The extension of the approach to the trispectrum has also been described in some detail.  As with the bispectrum computational speed is vastly improved using the separable method. However, for trispectra that depend on the diagonals as well as the wavenumbers, the decomposition into separable modes is still a computationally intensive operation requiring up to $\mathcal{O}(l_{\rm{max}}^6)$ operations. Nonetheless, this decomposition need only be performed once for each model. In the particular case that the trispectra is independent of the diagonals the decomposition process may be performed efficiently in $\mathcal{O}(l_{\rm{max}}^4)$ operations. It should also be noted that the general trispectrum may be divided into contributions denoted as `reduced' trispectra. Since, for almost all theoretical trispectra presented to date in the literature, the reduced trispectra depends on five parameters (i.e. the four wavenumbers and one diagonal) a reduction in complexity for this wide range of models may also be achieved. This class of models will be discussed in a subsequent article \cite{Regan2011}.

As in the case of the bispectrum, this approach can also be used to recover trispectra from simulations and produce non-Gaussian initial conditions with arbitrary trispectra for N-body codes. Once the trispectrum has been decomposed into separable modes the calculation of the trispectrum contribution to non-Gaussian initial conditions is an extremely efficient operation which may be performed in $\mathcal{O}(n_{\rm{max}}^{4/3}\times l_{\rm{max}}^3)$ operations. In this chapter we have also briefly outlined how the method may be extended to higher order correlators such as the quadspectra, revealing a highly efficient algorithm in the case that the quadspectrum depends only on its wavenumbers. 

The estimation of non-Gaussian parameters using large-scale structure is complicated due to non-linear evolution. In this chapter we have outlined some of the issues involved. The application of the separable approximation to finding the contribution to the matter density power spectrum due to the bispectrum (as well as the matter density bispectrum contribution due to the trispectrum) has been derived. In addition a prescription for parameter estimation in the fully nonlinear regime has been described.

While observational problems connected to surveys, such as because of redshift distortion and photometric errors, have not been addressed here, the generality and robustness of the methodology described in this chapter suggests that a vast improvement on the scope of models investigated using large-scale structure is possible, offering a significant test of the initial conditions of the Universe. However, different large scale structure survey strategies affect the quality of the higher order correlators that can be extracted.  Given that these poly-spectra can be determined efficiently and their strong scientific motivation, this should become an issue of growing importance in survey design.

\newpage


\thispagestyle{empty}
\mbox{}
\newpage
\chapter{Discussion and Conclusions}
\markright{Conclusion}

The characterisation of the inflationary era of the universe represents one of the most active areas of research in current-day cosmology. The standard paradigm predicts a nearly Gaussian spectrum of primordial density perturbations. Testing this hypothesis involves the measurement of cosmic microwave background radiation or using probes of large scale structure. In order to distinguish competing theories via their deviation from Gaussianity, the skewness and kurtosis parameters must be calculated. Building on recent work exploiting the use of a separable eigenmode expansion to perform a stable, accurate and efficient analysis of the CMB bispectrum, we present in this thesis a general framework for the study of the CMB trispectrum. The methodology presented may be applied to general trispectra and ensures that models, previously deemed intractable, may be analysed and compared to data readily. An optimal estimator which accounts for inhomogeneous noise and incomplete sky coverage has also been developed. Using this estimator a general integrated measure of the trispectrum may be performed.
\par
Cosmic strings, once seen as a viable source for the seeding of large-scale structure, have experienced a resurgence in interest in recent years due, in part, to the discovery that such topological defects may be formed at the end of brane inflation. A large part of this thesis is devoted to the derivation of analytic estimates of the power spectrum, bispectrum and trispectrum produced by a network of cosmic strings. The analysis, which neglects decoupling and assumes the signal produced by strings is due to the temperature discontinuity induced by the conical nature of spacetime surrounding such objects, is valid on the full sky. Therefore, a prediction for the non-Gaussian signal produced by cosmic strings, which may be tested against current CMB probes such as WMAP and Planck, is made. A somewhat surprising result is that the trispectrum appears to offer, perhaps, the best opportunity for a detection of cosmic strings in the near future. A major advantage of the formalism developed for the study of the CMB trispectrum is that it may be applied to  late-time as well as primordial trispectra.
\par 
The methodology for analysing the CMB trispectrum has been applied in this thesis to a class of models which are diagonal-free, i.e. are explicity four dimensional quantities. Such models include the local $g_{NL}$ model, the equilateral model and a so-called `constant' model. An accurate approximation to the cosmic string trispectrum may also be made which is diagonal-free. Such models have been compared to WMAP $5$-year data to produce constraints on their respective measures. The WMAP trispectrum has also been extracted. The code used to carry out this analysis has been made fully parallel using an implementation of MPI. The constraints derived represent the main observational work carried out in this thesis. Constraints on the equilateral and constant models, as well as cosmic strings, from the trispectrum presented here represent the first such analysis performed on these models. In fact, the only existing constraints in the literature are for the local model, although the methods developed here are advantageous for this model also since they offer a more optimal and stable analysis than carried out elsewhere.
\par
While the CMB offers the best current observational test for models of the very early universe, recent advances in N-body codes and modelling of galaxy distributions indicate that, in the near future, large-scale structure may be expected to provide a competitive and, eventually, superior probe. The primary complication in carrying out analyses of large-scale structure surveys is solely due to the inherent non-linear evolution. However, this disadvantage is countered by the fact that, unlike the CMB, the data set is three-dimensional. Application of the separable eigenmode expansion, which has proven very successful in making CMB analysis more efficient, is expected to greatly improve the efficiency of creating arbitrary non-Gaussian initial conditions for use in N-body codes. The formalism to carry out such an analysis has been also developed within this thesis.
\par
There remains much scope for application of the work presented in this thesis. A full implementation of the formalism to analyse CMB trispectrum which have a diagonal dependence represents an immediate goal. Such models include the $\tau_{NL}$ local model. This model is particularly interesting since detection of a signal with $\tau_{NL}<(6 f_{NL}/5)^2$ would rule out all local multifield inflation models \cite{WANDS}. The implementation of the separable mode expansion to calculate the power-spectrum induced by gravitational lensing represents another particularly interesting application. In \cite{Hanson} it was shown that the full lensed trispectrum (to fourth order) is necessary to accurately characterise the lensing potential power spectrum. The development and implementation of the study presented on large-scale structure may allow for more a accurate analysis of non-Gaussianity using galaxy surveys. As such, this represents an exciting area for future research.

\newpage

\appendix
\addcontentsline{toc}{chapter}{Bibliography}
\bibliographystyle{hunsrt}
\bibliography{database}
\newpage


\end{document}